\documentclass{article}

\usepackage{arxiv}
\usepackage{amsmath,amssymb,amsfonts}
\usepackage[utf8]{inputenc}       
\usepackage[T1]{fontenc}          
\usepackage[hidelinks]{hyperref}  
\usepackage{url}                  
\usepackage{booktabs}             
\usepackage{amsfonts}             
\usepackage{nicefrac}             
\usepackage{microtype}            
\usepackage{graphicx}
\usepackage{natbib}
\usepackage{caption}

\title{DeFi's Concentrated Liquidity From Scratch}

\author{ 
    Mark B. Richardson\thanks{To whom correspondence should be addressed.} \\
    Bancor Protocol \\
    Zug, Switzerland \\
    \texttt{mark@bancor.network} \\
    \And
    Stefan Loesch \\
    Topaze Blue \\
    London, United Kingdom \\
    \texttt{stefan@topaze.blue}
}

\date{
    Edited: August 25, 2024\\
    Original: July 4, 2024
}

\renewcommand{\headeright}{Review}
\renewcommand{\undertitle}{Review}

\hypersetup{
pdftitle={DeFi's Concentrated Liquidity From Scratch},
pdfsubject={q-fin.MF},
pdfauthor={Mark B. Richardson, Stefan Loesch},
pdfkeywords={Bonding Curve, Concentrated Liquidity, Automatic Market Maker, Decentralized Exchange, Blockchain, Smart Contracts, Bancor, Uniswap, Carbon DeFi},
}

\begin{document}
\maketitle

\begin{abstract}
The scope of this article includes the three preeminent descriptions of concentrated liquidity from Bancor (2020 and 2022), and Uniswap (2021), as well as three additional descriptions informed by trigonometric analysis of the same. The purpose of this contribution is to organize the seminal and derivative forms of this cornerstone DeFi technology, and algebraically and geometrically elaborate these descriptions to achieve an authoritative and near-exhaustive overview of the underlying theory powering the current state-of-the-art in decentralized exchange infrastructure.           

This material was created for the Token Engineering Academy Study Season 2024,\footnote{tokenengineering.net} a cohort-based online program scheduled for April–July 2024. The Study Season offers access to a bachelor-level online learning program, and complementary live tracks with the most influential practitioners and researchers in the sector – all provided as free, public goods.
\end{abstract}


\clearpage
\tableofcontents
\clearpage

\section{Introduction}\label{sec1}

The objective with this piece is to formally derive amplified (aka "concentrated liquidity") bonding curves from first principles, including their hyperbolic trigonometric descriptions. There are a plethora of white papers and research articles on the subject that do a fine job of elucidating these concepts with rigor. However, as often happens in pursuits of an academic nature, the literature has become somewhat piecemeal, and the establishing theory is no longer in vogue. Those steeped in the industry for several years have the benefit of experiencing the evolution of the technology over time, whereas new talent is left to contend with the patchy, unmaintained, and disorganized record that was left for them. Worse still, the lack of an editorial process combined with the conflation of research and marketing efforts has produced a sometimes confused, and often flawed knowledge base. More to the point of this article, even the high-quality papers are written for an exclusive audience, as evidenced by the skilled density of their composition which often sacrifices expository prose for brevity. As it should.

The impetus here is to take the opposite approach – one where the reader’s erudition in DeFi is not assumed. Of course, these are technical concepts and so command a minimum working knowledge of algebra, calculus, and maybe a few other things. However, the presentation format is designed to resemble a study guide more than a technical disclosure. Therefore, those equipped with the necessary tools (e.g. undergraduate math, maybe high school) can quickly orient themselves and get familiar with the theory and save some wasted effort. 

The concentrated liquidity construction will begin from the description first published by Bancor in 2020, with an exclusive focus on the now familiar two-dimensional, equal weights variety, as this has now reached market saturation thanks largely to Uniswap v3. By itself, this is nothing to write home about. However, an additional step further is taken to arrive at a trigonometric description of these concepts and derive what we refer to as the three "natural" concentrated liquidity invariant equations. While these invariant equations have appeared previously in the Carbon DeFi Whitepaper\footnote{resources.carbondefi.xyz/pages/CarbonWhitepaper.pdf} and Invention Disclosure\footnote{resources.carbondefi.xyz/pages/CarbonPatent.pdf}, this may be the first time their trigonometric origins have been elaborated in a public forum. 

\section{The Basics: Elementary Analysis of the Reference Curve}\label{sec2}

\begin{figure}[ht]
    \centering
    \includegraphics[width=\textwidth]{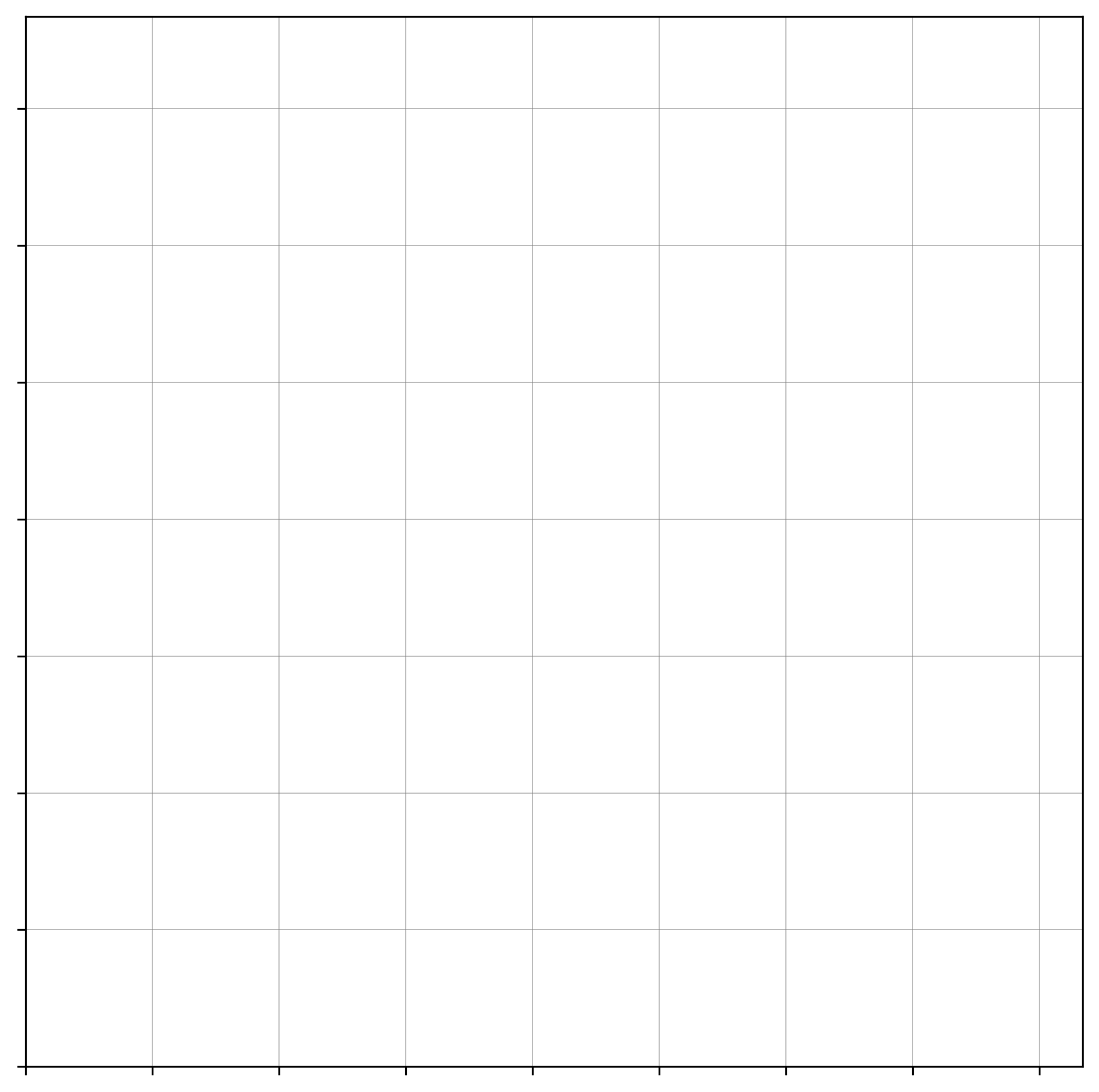}
    \captionsetup{
        justification=raggedright,
        singlelinecheck=false,
        font=small,
        labelfont=bf,
        labelsep=quad,
        format=plain
    }
    \caption{An empty Cartesian plane.}
    \label{fig1}
\end{figure}

This exercise begins truly from scratch, with an empty Cartesian plane (Figure \ref{fig1}).

\begin{figure}[ht]
    \centering
    \includegraphics[width=\textwidth]{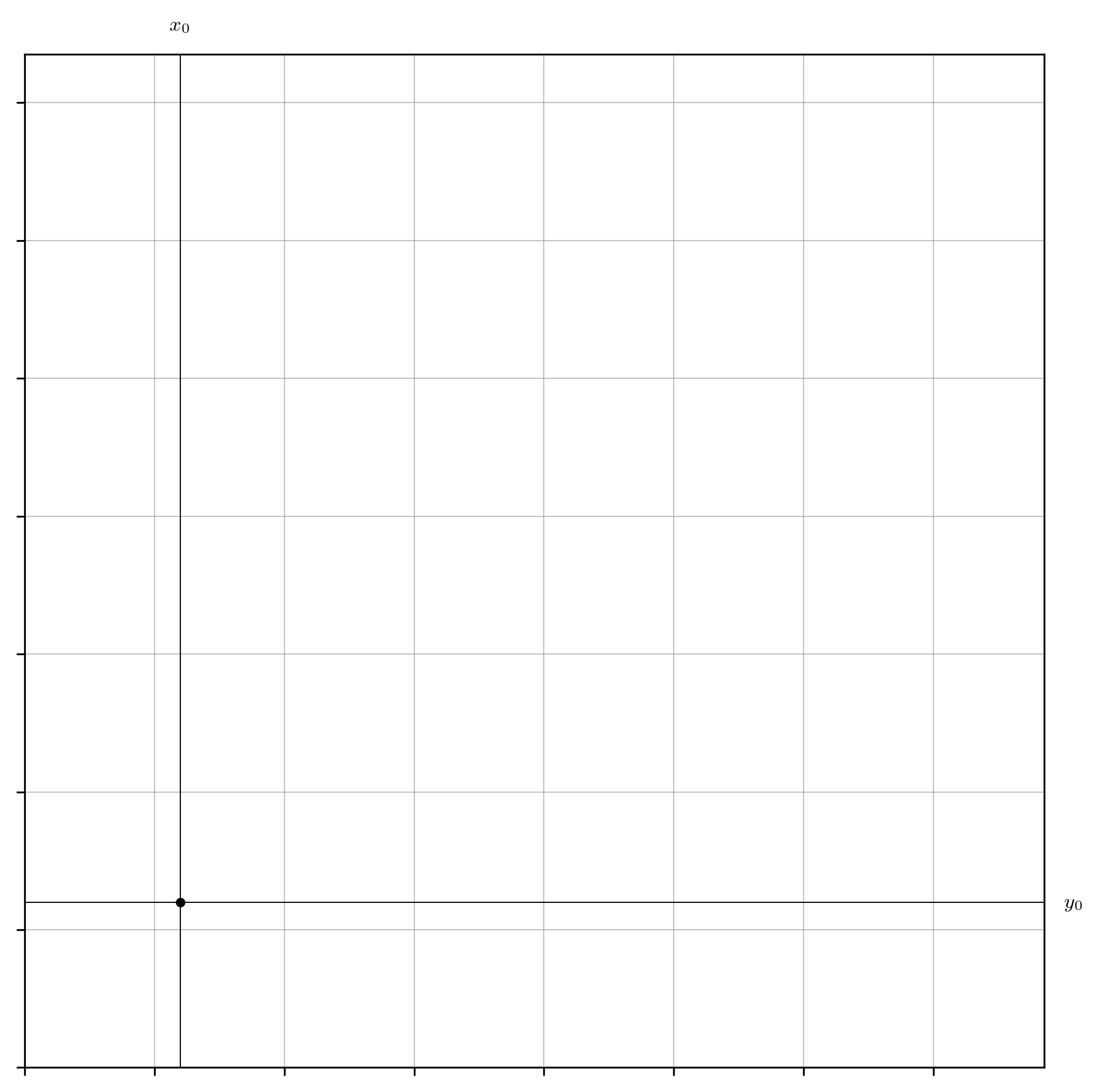}
    \captionsetup{
        justification=raggedright,
        singlelinecheck=false,
        font=small,
        labelfont=bf,
        labelsep=quad,
        format=plain
    }
    \caption{The point $x = x_{0},y = y_{0}$.}
    \label{fig2}
\end{figure}

Choose any point ($x = x_{0}, y = y_{0}$). This point represents your beginning token balances. The choice of these token balances is not incidental, and we will revisit this in a moment. For now, let $x_{0}$ and $y_{0}$ be any positive real number (Figure \ref{fig2}). 

Then, draw a rectangular hyperbola through this point. In its original 2017 form,\footnote{Hertzog, E.; Benartzi, G.; Benartzi, G.; Levi, Y. Methods for Exchanging and Evaluating Virtual Currency} the invariant was general, allowing for the $x$- and $y$-axis to be scaled independently (Equation \ref{eq1}).

\begin{equation} \label{eq1}
x^{\alpha} \cdot y^{\beta} = x_{0}^{\alpha} \cdot y_{0}^{\beta}
\end{equation}

Most of the industry has settled on the requirement for equal exponents, $\alpha = \beta$, which simplifies the invariant function a little (Equation \ref{eq2}).

\begin{flalign}
& \text{\renewcommand{\arraystretch}{0.66}
    \begin{tabular}{@{}c@{}}
    \scriptsize from \\
    \scriptsize (\ref{eq1})
  \end{tabular}} 
  & 
  x \cdot y = x_{0} \cdot y_{0}
  &  
  \label{eq2} 
  &
\end{flalign}

While the general and multi-dimensional form is still very much in operation, the remainder of this exercise will focus exclusively on the two-dimensional equal-weights variety. 

\begin{figure}[ht]
    \centering
    \includegraphics[width=\textwidth]{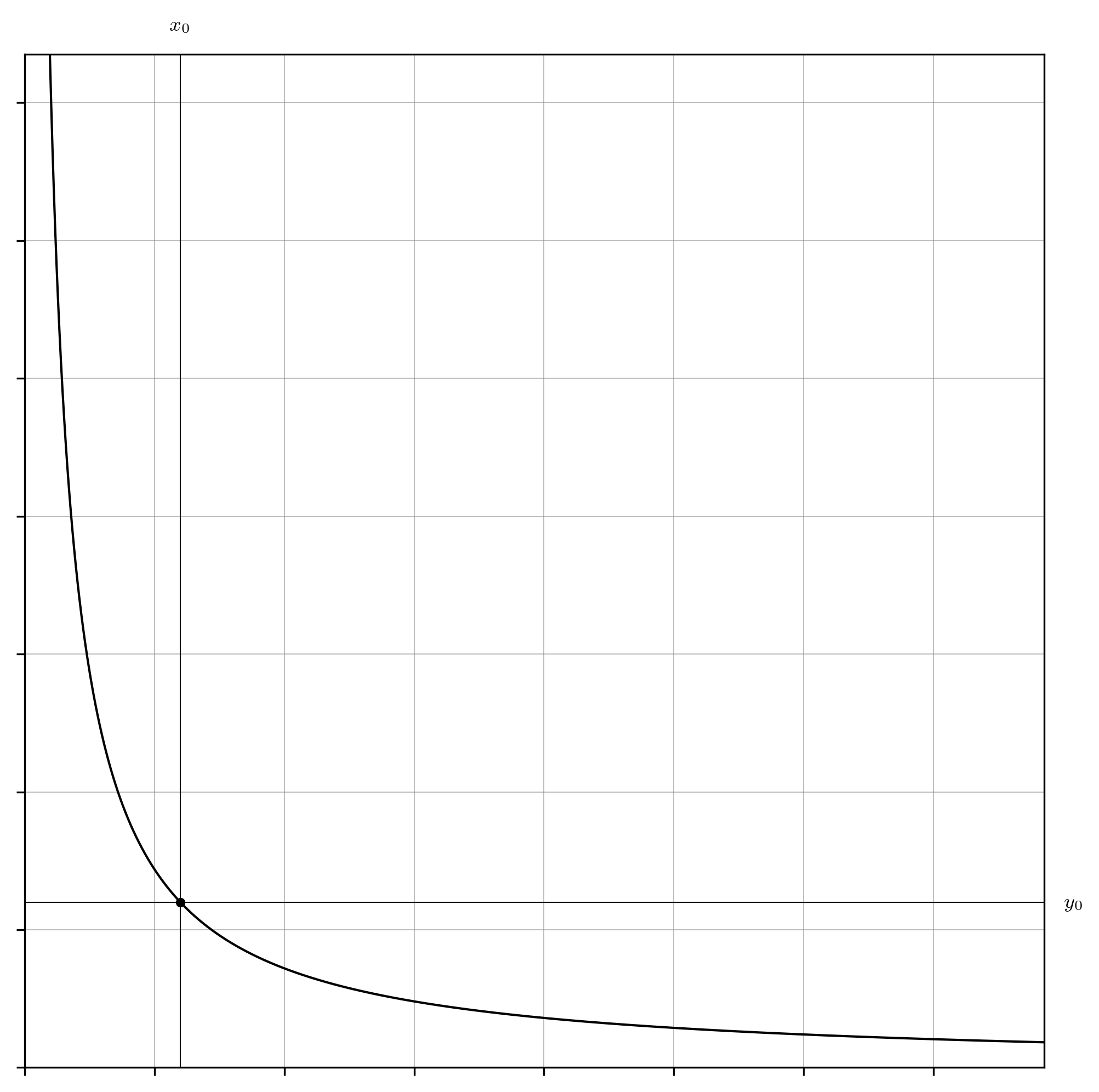}
    \captionsetup{
        justification=raggedright,
        singlelinecheck=false,
        font=small,
        labelfont=bf,
        labelsep=quad,
        format=plain
    }
    \caption{The rectangular hyperbola $ x \cdot y = x_{0} \cdot y_{0}$ (Equation \ref{eq2}), necessarily passing through the point $x = x_{0}, y = y_{0}$.}
    \label{fig3}
\end{figure}

To constrain the complexity of our discussion, we will ignore the [poorly named] “fee” parameter, which deserves a dedicated exposé. Therefore, this curve represents all the allowed token balances for your liquidity pool, and anyone can change the token balances so long as those balances satisfy the predicate $x \cdot y = x_{0} \cdot y_{0}$. This means that someone can add additional tokens to the liquidity pool in one dimension (e.g. $\mathrm{\Delta}x$) and remove tokens from the other dimension (e.g. $\mathrm{\Delta}y$), thus performing a trade with you. To satisfy the predicate, the token trade amounts $\mathrm{\Delta}x$ and $\mathrm{\Delta}y$ can be easily calculated as follows (Equations \ref{eq3}, \ref{eq4} and \ref{eq5}).

\begin{flalign}
& \text{\renewcommand{\arraystretch}{0.66}
    \begin{tabular}{@{}c@{}}
    \scriptsize from \\
    \scriptsize (\ref{eq2})
  \end{tabular}} 
  & 
  \left( x + \mathrm{\Delta}x \right) \cdot \left( y + \mathrm{\Delta}y \right) = x_{0} \cdot y_{0}
  &  
  \label{eq3} 
  &
\end{flalign}

\begin{flalign}
& \text{\renewcommand{\arraystretch}{0.66}
    \begin{tabular}{@{}c@{}}
    \scriptsize from \\
    \scriptsize (\ref{eq3})
  \end{tabular}} 
  & 
  \mathrm{\Delta}x = \displaystyle \frac{x_{0} \cdot y_{0}}{y + \mathrm{\Delta}y} - x
  &  
  \label{eq4} 
  &
\end{flalign}

\begin{flalign}
& \text{\renewcommand{\arraystretch}{0.66}
    \begin{tabular}{@{}c@{}}
    \scriptsize from \\
    \scriptsize (\ref{eq3})
  \end{tabular}} 
  & 
  \mathrm{\Delta}y = \displaystyle \frac{x_{0} \cdot y_{0}}{x + \mathrm{\Delta}x} - y
  &  
  \label{eq5} 
  &
\end{flalign}

The calculation can be made dependent on only one of either the $x$ or $y$ coordinates by substituting their identities in terms of each other (Equations \ref{eq6} and \ref{eq8}), and the $x_{0}$ and $y_{0}$ constants (Equations \ref{eq7} and \ref{eq9}).

\begin{flalign}
& \text{\renewcommand{\arraystretch}{0.66}
    \begin{tabular}{@{}c@{}}
    \scriptsize from \\
    \scriptsize (\ref{eq2})
  \end{tabular}} 
  & 
  x = \displaystyle \frac{x_{0} \cdot y_{0}}{y}
  &  
  \label{eq6} 
  &
\end{flalign}

\begin{flalign}
& \text{\renewcommand{\arraystretch}{0.66}
    \begin{tabular}{@{}c@{}}
    \scriptsize from \\
    \scriptsize (\ref{eq4})\\\scriptsize (\ref{eq6})
  \end{tabular}} 
  & 
  \mathrm{\Delta}x = \displaystyle \frac{x_{0} \cdot y_{0}}{y + \mathrm{\Delta}y} - \displaystyle \frac{x_{0} \cdot y_{0}}{y} = - \displaystyle \frac{\mathrm{\Delta}y \cdot x_{0} \cdot y_{0}}{y \cdot \left( y + \mathrm{\Delta}y \right)}
  &  
  \label{eq7} 
  &
\end{flalign}

\begin{flalign}
& \text{\renewcommand{\arraystretch}{0.66}
    \begin{tabular}{@{}c@{}}
    \scriptsize from \\
    \scriptsize (\ref{eq2})
  \end{tabular}} 
  & 
  y = \displaystyle \frac{x_{0} \cdot y_{0}}{x}
  &  
  \label{eq8} 
  &
\end{flalign}

\begin{flalign}
& \text{\renewcommand{\arraystretch}{0.66}
    \begin{tabular}{@{}c@{}}
    \scriptsize from \\
    \scriptsize (\ref{eq5})\\\scriptsize (\ref{eq8})
  \end{tabular}} 
  & 
  \mathrm{\Delta}y = \displaystyle \frac{x_{0} \cdot y_{0}}{x + \mathrm{\Delta}x} - \displaystyle \frac{x_{0} \cdot y_{0}}{x} = - \displaystyle \frac{\mathrm{\Delta}x \cdot x_{0} \cdot y_{0}}{x \cdot \left( x + \mathrm{\Delta}x \right)}
  &  
  \label{eq9} 
  &
\end{flalign}

\begin{figure}[ht]
    \centering
    \includegraphics[width=\textwidth]{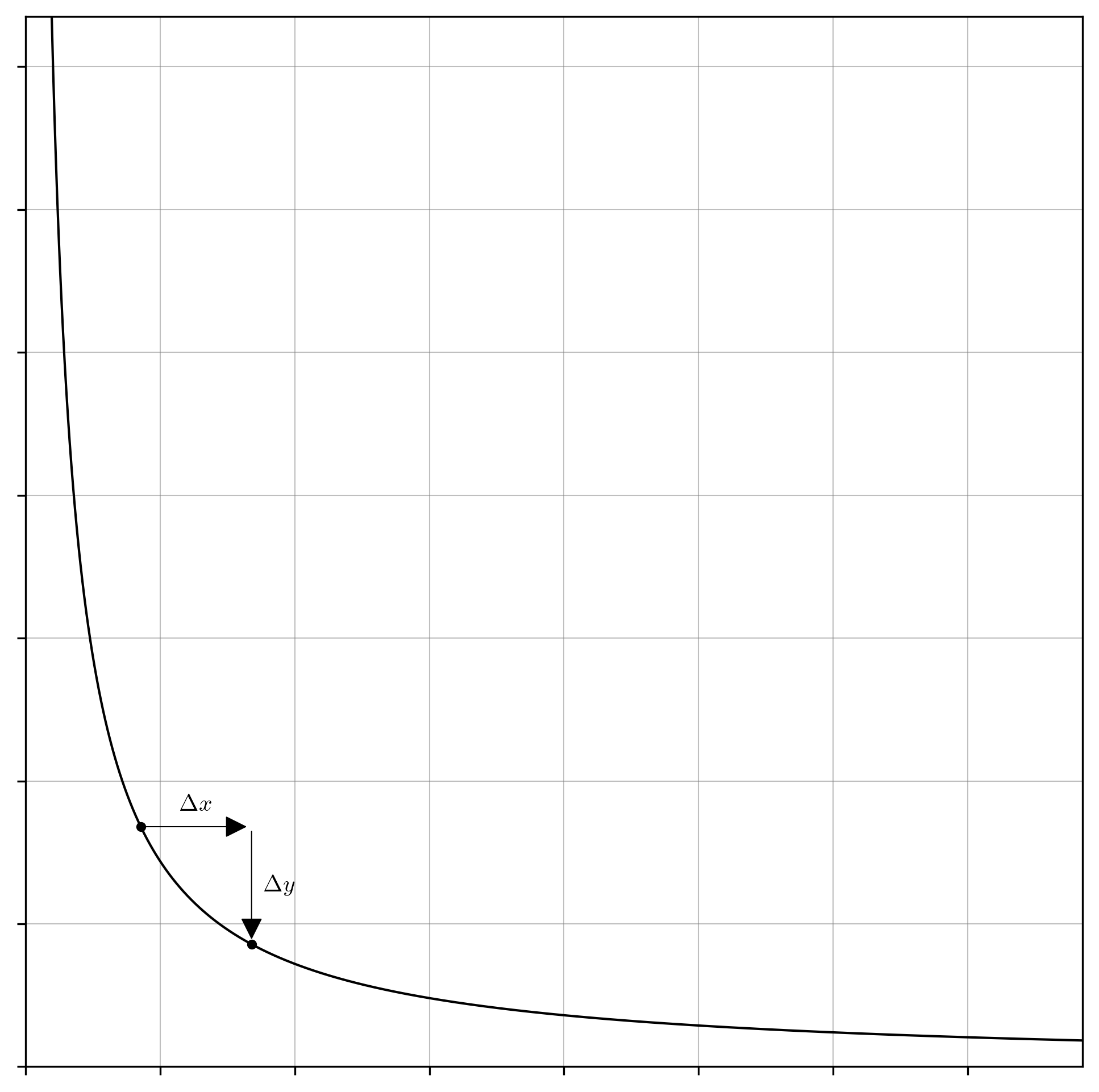}
    \captionsetup{
        justification=raggedright,
        singlelinecheck=false,
        font=small,
        labelfont=bf,
        labelsep=quad,
        format=plain
    }
    \caption{The traversal upon the rectangular hyperbola $x \cdot y = x_{0} \cdot y_{0}$ (Equation \ref{eq2}) representing a token swap against the liquidity pool, where $\mathrm{\Delta}x > 0$ and $\mathrm{\Delta}y < 0$.}
    \label{fig4}
\end{figure}

There are a few important notes here. First, it is important to remember that $\mathrm{\Delta}x$ and $\mathrm{\Delta}y$ have opposite signs with respect to each other; a positive change in one dimension is always coupled with a negative change in the opposite dimension. If $\mathrm{\Delta}x > 0$, then $\mathrm{\Delta}y < 0$ and vice-versa. Secondly, these are not the familiar swap equations you will see elsewhere, but rest assured they are redundant forms. The representations chosen here have explicit $x_{0}$ and $y_{0}$ constants to help to maintain consistency with the rest of the theory as it is presented below. To derive the more common forms, consider that the $x_{0} \cdot y_{0}$ term in Equations \ref{eq4}, and \ref{eq5} can be substituted for $x \cdot y$, as implied by the invariant function (Equation \ref{eq2}), which allows for further simplification (Equations \ref{eq10}, and \ref{eq11}). Note that this refactoring makes the sign inversion of $\mathrm{\Delta}x$ and $\mathrm{\Delta}y$ explicit, which is appropriate given the liquidity pool’s frame of reference and preserves consistency with some of the other manipulations we are going to explore. However, it is common to present these equations with a moving frame of reference, where both $\mathrm{\Delta}x > 0$ and $\mathrm{\Delta}y > 0$. While this is likely a reflection of their unsigned implementation in smart contracts, it also breaks a pleasing symmetry in the algebra and so is not the convention used here. Traversal upon the implicit curve representing a token swap is depicted in Figure \ref{fig4}.

\begin{flalign}
& \text{\renewcommand{\arraystretch}{0.66}
    \begin{tabular}{@{}c@{}}
    \scriptsize from \\
    \scriptsize (\ref{eq7})
  \end{tabular}} 
  & 
  \mathrm{\Delta}x = - \displaystyle \frac{\mathrm{\Delta}y \cdot x \cdot y}{y \cdot \left( y + \mathrm{\Delta}y \right)} = - \displaystyle \frac{\mathrm{\Delta}y \cdot x}{y + \mathrm{\Delta}y}
  &  
  \label{eq10} 
  &
\end{flalign}

\begin{flalign}
& \text{\renewcommand{\arraystretch}{0.66}
    \begin{tabular}{@{}c@{}}
    \scriptsize from \\
    \scriptsize (\ref{eq9})
  \end{tabular}} 
  & 
  \mathrm{\Delta}y = - \displaystyle \frac{\mathrm{\Delta}x \cdot x \cdot y}{x \cdot \left( x + \mathrm{\Delta}x \right)} = - \displaystyle \frac{\mathrm{\Delta}x \cdot y}{x + \mathrm{\Delta}x}
  &  
  \label{eq11} 
  &
\end{flalign}

From here, we can derive the marginal rate of exchange in two convenient ways. First, we can rearrange Equations \ref{eq10} and \ref{eq11} to get the \textit{effective rate of exchange} (Equations \ref{eq12} and \ref{eq14}), then take the limit as the denominator goes to zero (Equations \ref{eq13} and \ref{eq15}) to determine the \textit{instantaneous rate of exchange} (i.e. the marginal price).

\begin{flalign}
& \text{\renewcommand{\arraystretch}{0.66}
    \begin{tabular}{@{}c@{}}
    \scriptsize from \\
    \scriptsize (\ref{eq10})
  \end{tabular}} 
  & 
  \displaystyle \frac{\mathrm{\Delta}x}{\mathrm{\Delta}y} = - \displaystyle \frac{x}{y + \mathrm{\Delta}y}
  &  
  \label{eq12} 
  &
\end{flalign}

\begin{flalign}
& \text{\renewcommand{\arraystretch}{0.66}
    \begin{tabular}{@{}c@{}}
    \scriptsize from \\
    \scriptsize (\ref{eq12})
  \end{tabular}} 
  & 
  \displaystyle \frac{\partial x}{\partial y} = \lim_{\mathrm{\Delta}y \rightarrow 0}\displaystyle \frac{\mathrm{\Delta}x}{\mathrm{\Delta}y} = - \displaystyle \frac{x}{y}
  &  
  \label{eq13} 
  &
\end{flalign}

\begin{flalign}
& \text{\renewcommand{\arraystretch}{0.66}
    \begin{tabular}{@{}c@{}}
    \scriptsize from \\
    \scriptsize (\ref{eq11})
  \end{tabular}} 
  & 
  \displaystyle \frac{\mathrm{\Delta}y}{\mathrm{\Delta}x} = - \displaystyle \frac{y}{x + \mathrm{\Delta}x}
  &  
  \label{eq14} 
  &
\end{flalign}

\begin{flalign}
& \text{\renewcommand{\arraystretch}{0.66}
    \begin{tabular}{@{}c@{}}
    \scriptsize from \\
    \scriptsize (\ref{eq14})
  \end{tabular}} 
  & 
  \displaystyle \frac{\partial y}{\partial x} = \lim_{\mathrm{\Delta}x \rightarrow 0}\displaystyle \frac{\mathrm{\Delta}y}{\mathrm{\Delta}x} = - \displaystyle \frac{y}{x}
  &  
  \label{eq15} 
  &
\end{flalign}

Alternatively, rearrangement of Equation \ref{eq2} to make either $x$ or $y$ the subject (Equations \ref{eq6} and \ref{eq8}), followed by differentiation with respect to the other term while treating $x_{0}$ and $y_{0}$ as constants delivers an equivalent result (Equations \ref{eq16} and \ref{eq17}). Again, this form is less familiar, but also more analytically robust and consistent with the bulk of the following presentation.

\begin{flalign}
& \text{\renewcommand{\arraystretch}{0.66}
    \begin{tabular}{@{}c@{}}
    \scriptsize from \\
    \scriptsize (\ref{eq6})
  \end{tabular}} 
  & 
  \displaystyle \frac{\partial x}{\partial y} = - \displaystyle \frac{x_{0} \cdot y_{0}}{y^{2}};\ \ \ \displaystyle \frac{\partial y}{\partial x} = - \displaystyle \frac{y^{2}}{x_{0} \cdot y_{0}}
  &  
  \label{eq16} 
  &
\end{flalign}

\begin{flalign}
& \text{\renewcommand{\arraystretch}{0.66}
    \begin{tabular}{@{}c@{}}
    \scriptsize from \\
    \scriptsize (\ref{eq8})
  \end{tabular}} 
  & 
  \displaystyle \frac{\partial y}{\partial x} = - \displaystyle \frac{x_{0} \cdot y_{0}}{x^{2}};\ \ \ \displaystyle \frac{\partial x}{\partial y} = - \displaystyle \frac{x^{2}}{x_{0} \cdot y_{0}}
  &  
  \label{eq17} 
  &
\end{flalign}

It is important to recognize that the marginal rates of exchange, $\partial x / \partial y$ and $\partial y / \partial x$, are commensurate with price quotes. These are the onchain oracle prices for the token pair (simplified, but close enough) and represent the current price for a trade value of zero, whereas the effective price experienced during a trade is dependent on the trade amounts and the token balances inside the liquidity pool (Equations \ref{eq12} and \ref{eq14}). 

\begin{figure}[ht]
    \centering
    \includegraphics[width=\textwidth]{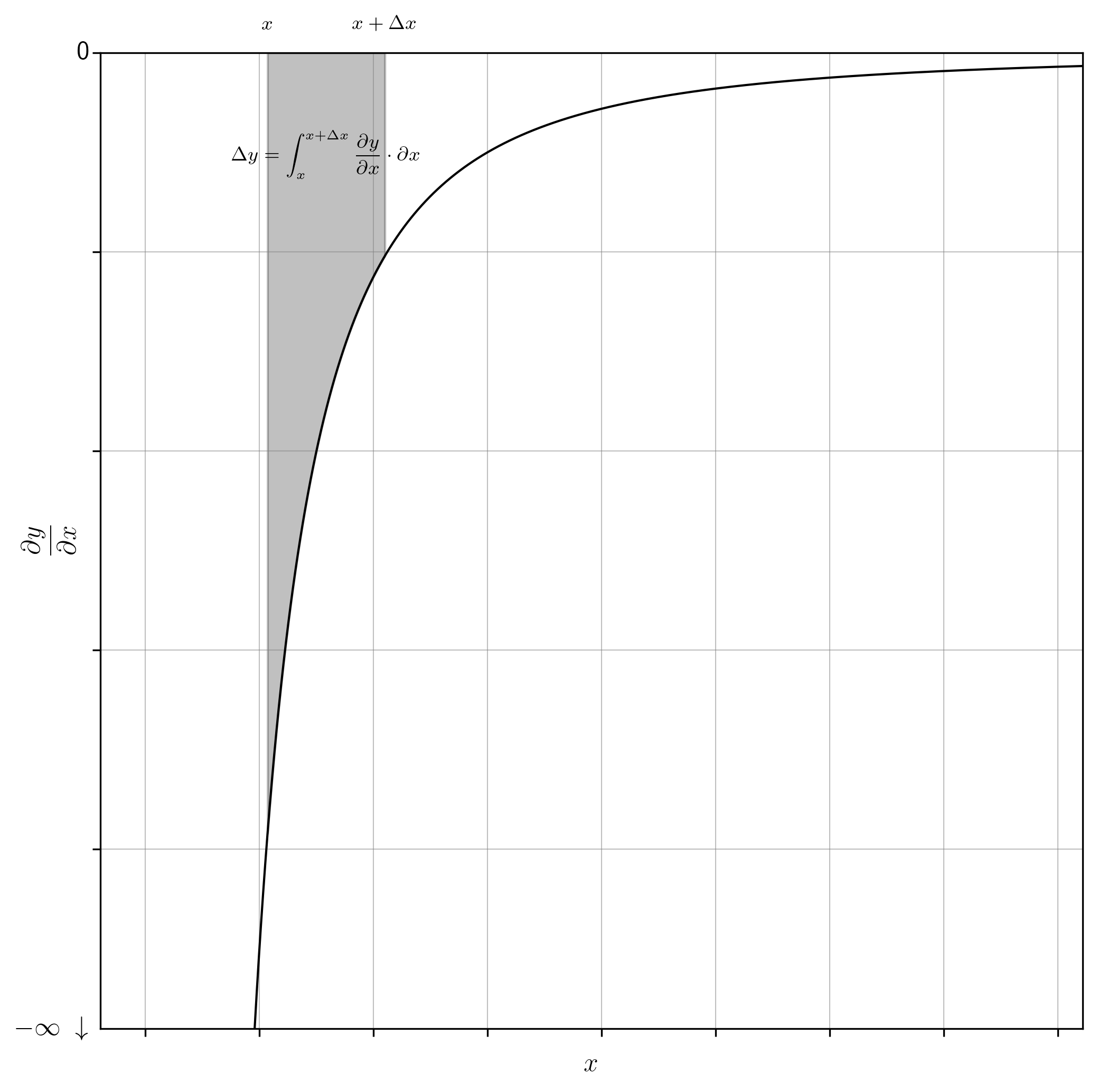}
    \captionsetup{
        justification=raggedright,
        singlelinecheck=false,
        font=small,
        labelfont=bf,
        labelsep=quad,
        format=plain
    }
    \caption{The integration above $\partial y / \partial x = - x_{0} \cdot y_{0} / x ^ {2}$ (Equation \ref{eq17}) over the interval $x \rightarrow x + \mathrm{\Delta}x$ representing a token swap against the liquidity pool, where $\mathrm{\Delta}x > 0$ and $\mathrm{\Delta}y < 0$.}
    \label{fig5}
\end{figure}

To complete this part of the exercise, consider that the familiar implicit curve of the invariant function, or bonding curve, is the integrated form of the price function. Therefore, the swap equations can also be derived from explicit integration of Equations \ref{eq16} and \ref{eq17} over the interval representing the number of tokens being swapped (Equations \ref{eq18} and \ref{eq19}, which yield results identical to Equations \ref{eq7} and \ref{eq9}). The integration above $\partial y / \partial x = - x_{0} \cdot y_{0} / x ^ {2}$ over the interval $x \rightarrow x + \mathrm{\Delta}x$ representing a token swap is depicted in Figure \ref{fig5}. 

\begin{flalign}
& \text{\renewcommand{\arraystretch}{0.66}
    \begin{tabular}{@{}c@{}}
    \scriptsize from \\
    \scriptsize (\ref{eq16})
  \end{tabular}} 
  & 
  \mathrm{\Delta}x = - \int_{y}^{y + \mathrm{\Delta}y}{\displaystyle \frac{x_{0} \cdot y_{0}}{y^{2}}} \cdot \partial y = \left\lbrack \displaystyle \frac{x_{0} \cdot y_{0}}{y} \right\rbrack_{y}^{y + \mathrm{\Delta}y} \Rightarrow \mathbf{Eqn.\ 7}
  &  
  \label{eq18} 
  &
\end{flalign}

\begin{flalign}
& \text{\renewcommand{\arraystretch}{0.66}
    \begin{tabular}{@{}c@{}}
    \scriptsize from \\
    \scriptsize (\ref{eq17})
  \end{tabular}} 
  & 
  \mathrm{\Delta}y = - \int_{x}^{x + \mathrm{\Delta}x}{\displaystyle \frac{x_{0} \cdot y_{0}}{x^{2}}} \cdot \partial x = \left\lbrack \displaystyle \frac{x_{0} \cdot y_{0}}{x} \right\rbrack_{x}^{x + \mathrm{\Delta}x} \Rightarrow \mathbf{Eqn.\ 9}
  &  
  \label{eq19} 
  &
\end{flalign}

Take special note that integration of the more familiar price functions over the same interval requires a more considered approach, as $x$ and $y$ vary with each other. Beginning from the previously obtained differential equations (Equations \ref{eq13} and \ref{eq15}), first separate the $x$ and $y$ variables across the equality (Equation \ref{eq332}) then evaluate their definite integrals over $x \rightarrow x + \mathrm{\Delta}x$ and $y \rightarrow y + \mathrm{\Delta}y$, independently (Equations \ref{eq333} and \ref{eq334}). Following that, the minuends and subtrahends resulting from integration of each variable are transformed into their corresponding quotients via the properties of logarithms (Equation \ref{eq335}). Then, rearrangement of the expressions thus obtained to isolate either $\mathrm{\Delta}x$ or $\mathrm{\Delta}y$ gives Equations \ref{eq336} and \ref{eq337}, which can be further simplified to recreate the identities previously obtained (Equations \ref{eq10} and \ref{eq11}).

\begin{flalign}
& \text{\renewcommand{\arraystretch}{0.66}
    \begin{tabular}{@{}c@{}}
    \scriptsize from \\
    \scriptsize (\ref{eq13}) \\
    \scriptsize (\ref{eq15})
  \end{tabular}} 
  & 
  \frac{1}{y} \cdot \partial y = - \frac{1}{x} \cdot \partial x
  &  
  \label{eq332} 
  &
\end{flalign}

\begin{flalign}
& \text{\renewcommand{\arraystretch}{0.66}
    \begin{tabular}{@{}c@{}}
    \scriptsize from \\
    \scriptsize (\ref{eq332})
  \end{tabular}} 
  & 
  \int_{y}^{y + \mathrm{\Delta}y} \frac{1}{y} \cdot \partial y = - \int_{x}^{x + \mathrm{\Delta}x} \frac{1}{x} \cdot \partial x = \left\lbrack \ln \left( y \right) \right\rbrack_{y}^{y + \mathrm{\Delta}y} = - \left\lbrack \ln \left( x \right) \right\rbrack_{x}^{x + \mathrm{\Delta}x}  
  &  
  \label{eq333} 
  &
\end{flalign}

\begin{flalign}
& \text{\renewcommand{\arraystretch}{0.66}
    \begin{tabular}{@{}c@{}}
    \scriptsize from \\
    \scriptsize (\ref{eq333})
  \end{tabular}} 
  & 
  \ln \left( y + \mathrm{\Delta}y \right) - \ln \left( y \right) = \ln \left( x \right) - \ln \left( x + \mathrm{\Delta}x \right)  
  &  
  \label{eq334} 
  &
\end{flalign}

\begin{flalign}
& \text{\renewcommand{\arraystretch}{0.66}
    \begin{tabular}{@{}c@{}}
    \scriptsize from \\
    \scriptsize (\ref{eq334})
  \end{tabular}} 
  & 
  \ln \left( \displaystyle \frac{y + \mathrm{\Delta}y }{ y } \right) = \ln \left( \displaystyle \frac{x}{x + \mathrm{\Delta}x} \right) \Rightarrow \displaystyle \frac{y + \mathrm{\Delta}y}{y} = \displaystyle \frac{x}{x + \mathrm{\Delta}x}  
  &  
  \label{eq335} 
  &
\end{flalign}

\begin{flalign}
& \text{\renewcommand{\arraystretch}{0.66}
    \begin{tabular}{@{}c@{}}
    \scriptsize from \\
    \scriptsize (\ref{eq335})
  \end{tabular}} 
  & 
  \mathrm{\Delta}x = - x \cdot \left( 1 - \displaystyle \frac{y}{y + \mathrm{\Delta}y} \right) \Rightarrow \mathbf{Eqn.\ 10}  
  &  
  \label{eq336} 
  &
\end{flalign}

\begin{flalign}
& \text{\renewcommand{\arraystretch}{0.66}
    \begin{tabular}{@{}c@{}}
    \scriptsize from \\
    \scriptsize (\ref{eq335})
  \end{tabular}} 
  & 
  \mathrm{\Delta}y = - y \cdot \left( 1 - \displaystyle \frac{x}{x + \mathrm{\Delta}x} \right) \Rightarrow \mathbf{Eqn.\ 11}  
  &  
  \label{eq337} 
  &
\end{flalign}

Therefore, the implicit curve of the invariant function and its first derivative are equally valid expressions of the underlying theory, and the preference for presenting only one or the other is an arbitrary choice. You should expect to see a combination of both throughout the DeFi whitepaper literature. 

Something important to note here is that the price equation is asymptotic at $\partial y / \partial x = 0$ and $\partial y / \partial x = -\infty$ for all real positive token balances, $\forall x, y \in \mathbb{R} ^ {+}$. Therefore, the $x \cdot y = x_{0} \cdot y_{0}$ invariant can quote any price, regardless of the relative values of the underlying tokens. This property is sacrificed in the concentrated liquidity extension.

\section{The Seminal Concentrated Liquidity Invariant}\label{sec3}

The original motivation for concentrated liquidity has been articulated in Bancor’s announcement blog \footnote{Hertzog, E.; Levi, Y.; Manos, B.; Shachaf, A.; Benartzi, G. Smart Contract of a Blockchain for Management of Cryptocurrencies}\textsuperscript{,}\footnote{blog.bancor.network/announcing-bancor-v2-2f56b515e9d8} on April 29, 2020, and later in Uniswap’s announcement blog\footnote{blog.uniswap.org/uniswap-v3} on March 22, 2021. 

Here is the short version. Two different liquidity pools will offer the same \textit{marginal exchange rate} if the ratio of their token balances is identical (Equations \ref{eq13}, \ref{eq15}, \ref{eq16} and \ref{eq17}); however, the liquidity pool with the larger absolute quantity of tokens can offer a better \textit{effective exchange rate} for the same trade amount than the smaller pool (Equations \ref{eq12} and \ref{eq14}). The improvement in the overall exchange rate on the larger pool for the same trade quantities is the same as observing that the larger pool has a reduced slippage compared to the smaller pool (these are equivalent statements). 

Similarly, for the same effective exchange rate, or for the same relative move in the market price of the two tokens inside the liquidity pool, the larger pool can process a larger trade volume than the smaller one. In other words, it takes a larger trade volume to move the marginal price of the larger pool compared to the smaller one. If these characteristics of the larger pool are desirable, then the smaller pool can \textit{pretend} to be equivalent in size, thereby achieving the same exchange rate profile and supporting larger trade volumes. The caveat is that emulation of the larger curve also restricts the price interval over which it operates.

\subsection{The Bancor v2 Virtual Curve}\label{subsec3.1}

\begin{figure}[ht]
    \centering
    \includegraphics[width=\textwidth]{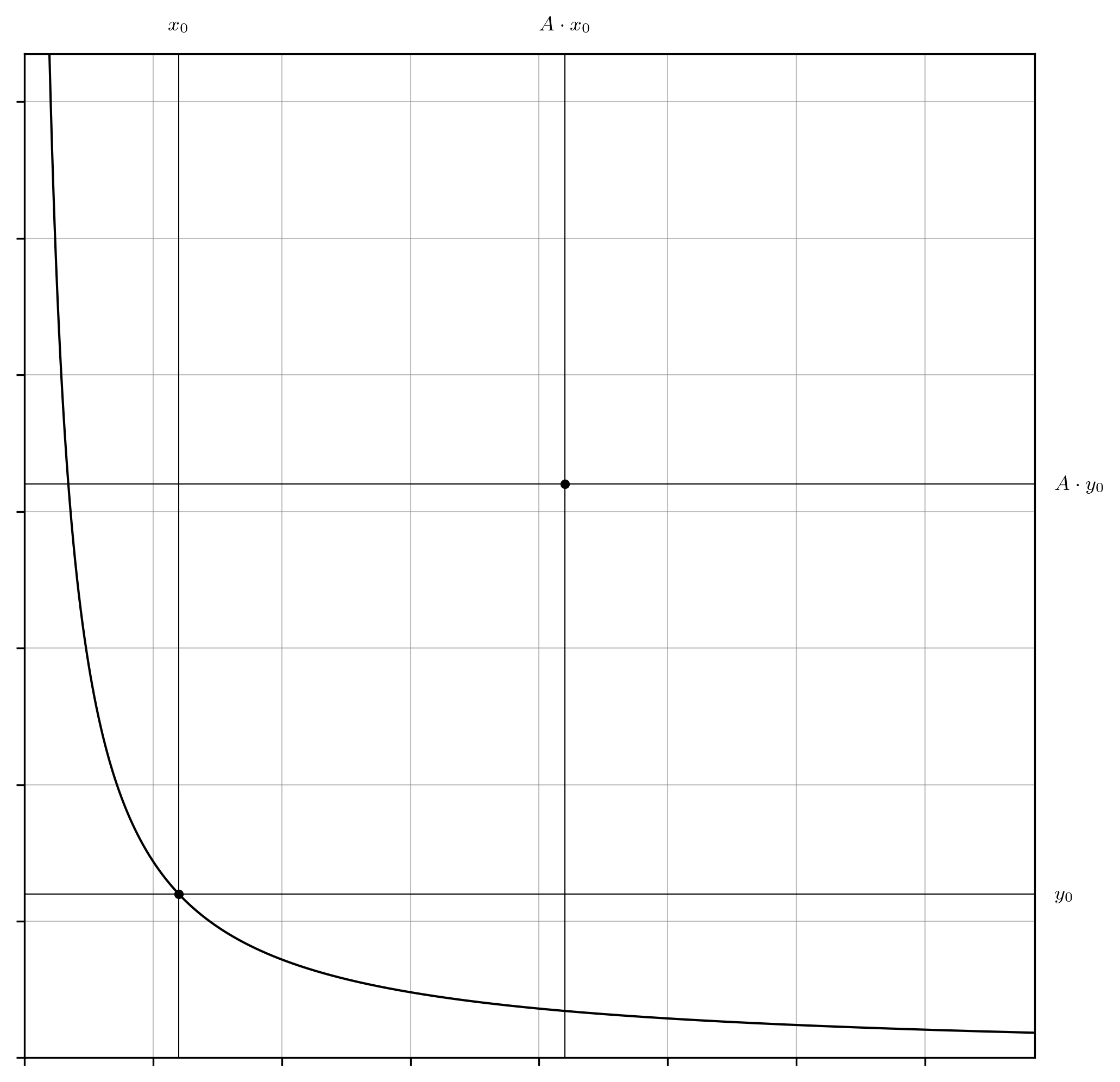}
    \captionsetup{
        justification=raggedright,
        singlelinecheck=false,
        font=small,
        labelfont=bf,
        labelsep=quad,
        format=plain
    }
    \caption{The point $\left( x_{\text{v}} = A \cdot x_{0}, y_{\text{v}} = A \cdot y_{0} \right)$, appended to the plot depicted in Figure \ref{fig3}.}
    \label{fig6}
\end{figure}

\begin{figure}[ht]
    \centering
    \includegraphics[width=\textwidth]{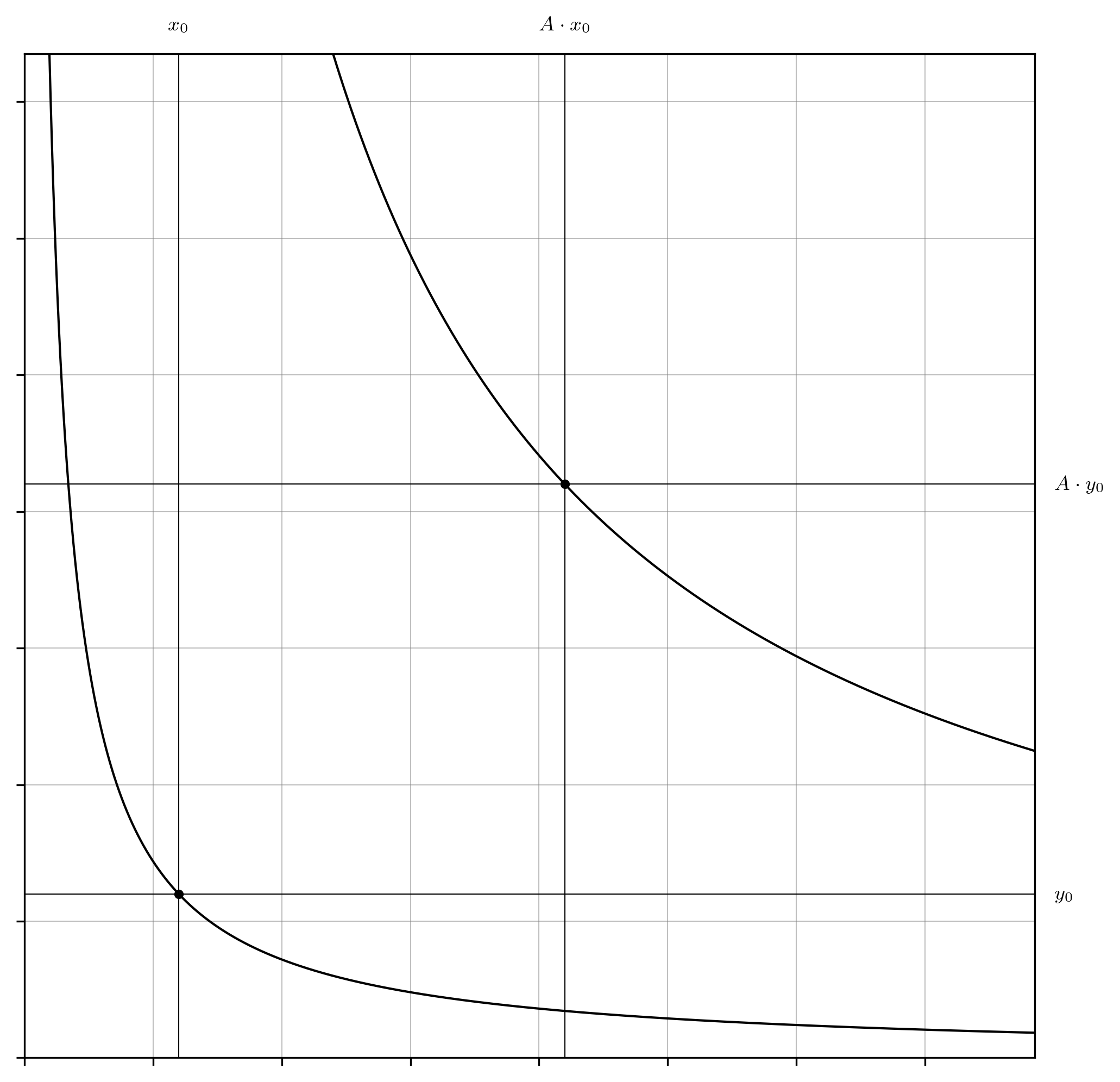}
    \captionsetup{
        justification=raggedright,
        singlelinecheck=false,
        font=small,
        labelfont=bf,
        labelsep=quad,
        format=plain
    }
    \caption{The rectangular hyperbola $x_{\text{v}} \cdot y_{\text{v}} = A^{2} \cdot x_{0} \cdot y_{0}$ (Equation \ref{eq20}), necessarily passing through the point $\left( x_{\text{v}} = A \cdot x_{0}, y_{\text{v}} = A \cdot y_{0} \right)$.}
    \label{fig7}
\end{figure}

To begin this part of the exercise, suppose that the starting coordinates chosen for Figure \ref{fig3} are multiplied by an amplification constant, $A$, to give new virtual coordinates, $x_{\text{v}}$ and $y_{\text{v}}$ (Figure \ref{fig6}). Then, draw through this new point the same rectangular hyperbola as we have before (Equation \ref{eq20}) (Figure \ref{fig7}). The only difference between Equations \ref{eq2} and \ref{eq20} is that the invariant part of the latter is increased by the square of the amplification constant. 

\begin{flalign}
& \text{\renewcommand{\arraystretch}{0.66}
    \begin{tabular}{@{}c@{}}
    \scriptsize from \\
    \scriptsize (\ref{eq2})
  \end{tabular}} 
  & 
  x_{\text{v}} \cdot y_{\text{v}} = \left( {A \cdot x_{0}} \right) \cdot \left( A \cdot y_{0} \right) = A^{2} \cdot x_{0} \cdot y_{0}
  &  
  \label{eq20} 
  &
\end{flalign}

The amplified $x_{\text{v}} \cdot y_{\text{v}} = A^{2} \cdot x_{0} \cdot y_{0}$ curve has the same properties as its $x \cdot y = x_{0} \cdot y_{0}$ counterpart. The expressions above that explicitly refer to the constants $x_{0}$ and $y_{0}$ (Equations \ref{eq3}, \ref{eq4}, \ref{eq5}, \ref{eq6}, \ref{eq7}, \ref{eq8}, \ref{eq9}, \ref{eq16}, and \ref{eq17}) can be adapted easily by introducing the amplification constant as appropriate (Equations \ref{eq21}, \ref{eq22}, \ref{eq23}, \ref{eq24}, \ref{eq25}, \ref{eq26}, \ref{eq27}, \ref{eq28} and \ref{eq29}). 

\begin{flalign}
& \text{\renewcommand{\arraystretch}{0.66}
    \begin{tabular}{@{}c@{}}
    \scriptsize from \\
    \scriptsize (\ref{eq8})
  \end{tabular}} 
  & 
  y_{\text{v}} = \displaystyle \frac{A^{2} \cdot x_{0} \cdot y_{0}}{x_{\text{v}}}
  &  
  \label{eq21} 
  &
\end{flalign}

\begin{flalign}
& \text{\renewcommand{\arraystretch}{0.66}
    \begin{tabular}{@{}c@{}}
    \scriptsize from \\
    \scriptsize (\ref{eq6})
  \end{tabular}} 
  & 
  x_{\text{v}} = \displaystyle \frac{A^{2} \cdot x_{0} \cdot y_{0}}{y_{\text{v}}}
  &  
  \label{eq22} 
  &
\end{flalign}

\begin{flalign}
& \text{\renewcommand{\arraystretch}{0.66}
    \begin{tabular}{@{}c@{}}
    \scriptsize from \\
    \scriptsize (\ref{eq3})
  \end{tabular}} 
  & 
  \left( x_{\text{v}} + \mathrm{\Delta}x \right) \cdot \left( y_{\text{v}} + \mathrm{\Delta}y \right) = A^{2} \cdot x_{0} \cdot y_{0}
  &  
  \label{eq23} 
  &
\end{flalign}

\begin{flalign}
& \text{\renewcommand{\arraystretch}{0.66}
    \begin{tabular}{@{}c@{}}
    \scriptsize from \\
    \scriptsize (\ref{eq4})
  \end{tabular}} 
  & 
  \mathrm{\Delta}x = \displaystyle \frac{A^{2} \cdot x_{0} \cdot y_{0}}{y_{\text{v}} + \mathrm{\Delta}y} - x_{\text{v}}
  &  
  \label{eq24} 
  &
\end{flalign}

\begin{flalign}
& \text{\renewcommand{\arraystretch}{0.66}
    \begin{tabular}{@{}c@{}}
    \scriptsize from \\
    \scriptsize (\ref{eq7})
  \end{tabular}} 
  & 
  \mathrm{\Delta}x = - \displaystyle \frac{\mathrm{\Delta}y \cdot A^{2} \cdot x_{0} \cdot y_{0}}{y_{\text{v}} \cdot \left( y_{\text{v}} + \mathrm{\Delta}y \right)}
  &  
  \label{eq25} 
  &
\end{flalign}

\begin{flalign}
& \text{\renewcommand{\arraystretch}{0.66}
    \begin{tabular}{@{}c@{}}
    \scriptsize from \\
    \scriptsize (\ref{eq5})
  \end{tabular}} 
  & 
  \mathrm{\Delta}y = \displaystyle \frac{A^{2} \cdot x_{0} \cdot y_{0}}{x_{\text{v}} + \mathrm{\Delta}x} - y_{\text{v}}
  &  
  \label{eq26} 
  &
\end{flalign}

\begin{flalign}
& \text{\renewcommand{\arraystretch}{0.66}
    \begin{tabular}{@{}c@{}}
    \scriptsize from \\
    \scriptsize (\ref{eq9})
  \end{tabular}} 
  & 
  \mathrm{\Delta}y = - \displaystyle \frac{\mathrm{\Delta}x \cdot A^{2} \cdot x_{0} \cdot y_{0}}{x_{\text{v}} \cdot \left( x_{\text{v}} + \mathrm{\Delta}x \right)}
  &  
  \label{eq27} 
  &
\end{flalign}

\begin{flalign}
& \text{\renewcommand{\arraystretch}{0.66}
    \begin{tabular}{@{}c@{}}
    \scriptsize from \\
    \scriptsize (\ref{eq16})
  \end{tabular}} 
  & 
  \displaystyle \frac{\partial x_{\text{v}}}{\partial y_{\text{v}}} = - \displaystyle \frac{A^{2} \cdot x_{0} \cdot y_{0}}{y_{\text{v}}^{2}};\ \ \ \displaystyle \frac{\partial y_{\text{v}}}{\partial x_{\text{v}}} = - \displaystyle \frac{y_{\text{v}}^{2}}{A^{2} \cdot x_{0} \cdot y_{0}}
  &  
  \label{eq28} 
  &
\end{flalign}

\begin{flalign}
& \text{\renewcommand{\arraystretch}{0.66}
    \begin{tabular}{@{}c@{}}
    \scriptsize from \\
    \scriptsize (\ref{eq17})
  \end{tabular}} 
  & 
  \displaystyle \frac{\partial y_{\text{v}}}{\partial x_{\text{v}}} = - \displaystyle \frac{A^{2} \cdot x_{0} \cdot y_{0}}{x_{\text{v}}^{2}};\ \ \ \displaystyle \frac{\partial x_{\text{v}}}{\partial y_{\text{v}}} = - \displaystyle \frac{x_{\text{v}}^{2}}{A^{2} \cdot x_{0} \cdot y_{0}}
  &  
  \label{eq29} 
  &
\end{flalign}

More significantly, the expressions that ignore the $x_{0}$ and $y_{0}$ terms (Equations \ref{eq10}, \ref{eq11}, \ref{eq12}, \ref{eq13}, \ref{eq14}, and \ref{eq15}) can be used verbatim; they are intrinsically self-referential with respect to the size of the curve, emulated or not, and so no correction is required. However, both the explicitly amplified and uncorrected expressions make no attempt to compensate for the fact that the emulated curve is only pretending to have the liquidity it represents. In the common vernacular, these emulated token balances are referred to as \textit{virtual token balances}, and the amplified curve as the \textit{virtual curve}.

Since the virtual curve has only the original tokens that were used to create it, $x_{0}$ and $y_{0}$, then it is possible to completely deplete its real token balances, regardless of its virtual token balances. Therefore, there are two boundaries to discover: the virtual token balances when the real $x$ token balance is depleted, and when the real $y$ token balance is depleted. 

\begin{figure}[ht]
    \centering
    \includegraphics[width=\textwidth]{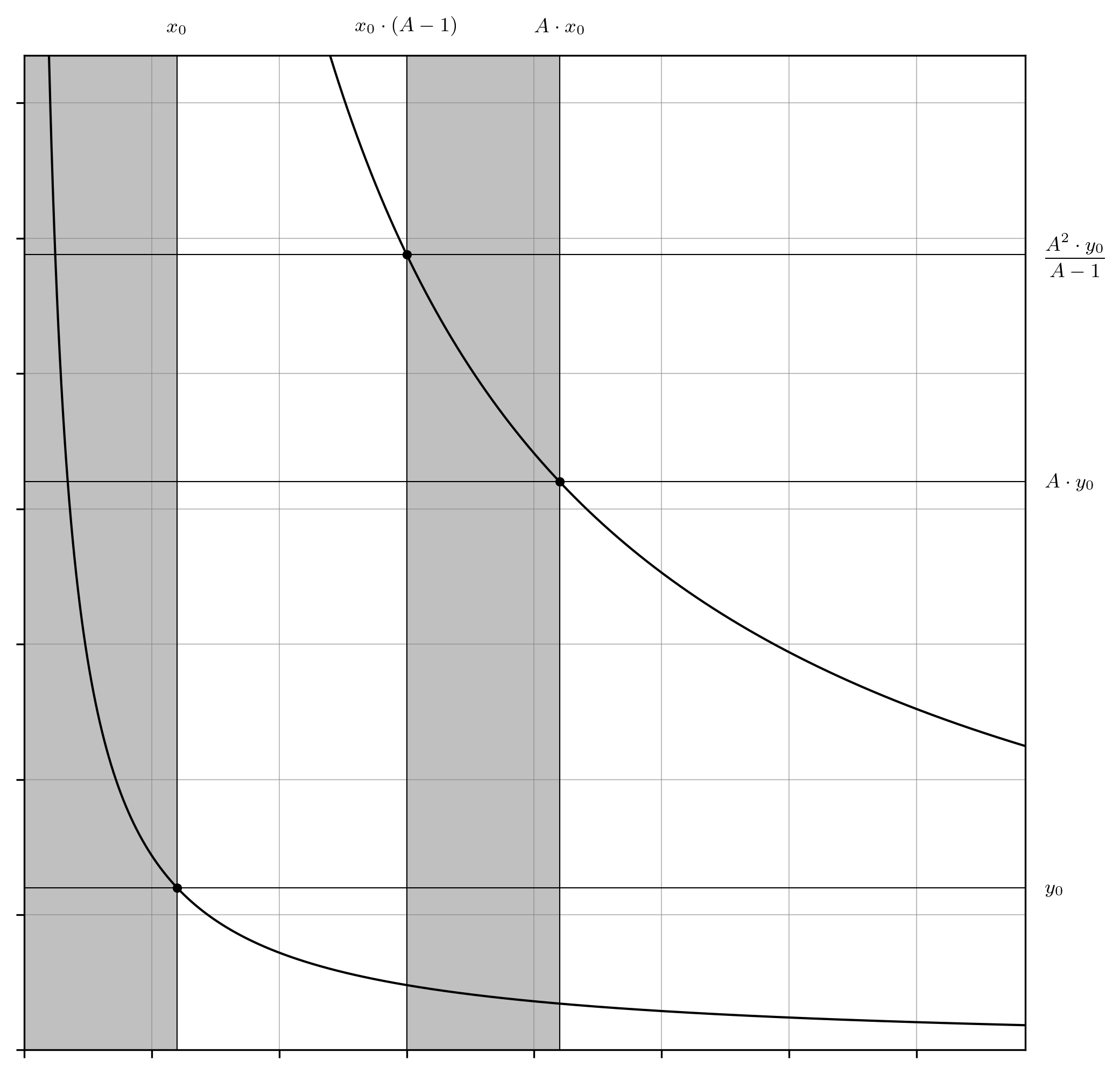}
    \captionsetup{
        justification=raggedright,
        singlelinecheck=false,
        font=small,
        labelfont=bf,
        labelsep=quad,
        format=plain
    }
    \caption{The $\min\left( x_{\text{v}} \right)$ and $\max\left( y_{\text{v}} \right)$ coordinates (Equations \ref{eq30} and \ref{eq31}). The correspondence between the virtual token balance and the real token balance is depicted with shaded areas for reference.}
    \label{fig8}
\end{figure}

The derivation is straightforward. The virtual curve begins with $A \cdot x_{0}$ virtual $x_{\text{v}}$ tokens and will become depleted when $x_{0}$ tokens have been traded out of it (Equations \ref{eq30}) (Figure \ref{fig8}). Let this be the minimum virtual x-coordinate, $\min \left( x_{\text{v}} \right)$; which is necessarily coupled with the maximum virtual y-coordinate, $\max \left( y_{\text{v}} \right)$. The identity of $\max \left( y_{\text{v}} \right)$ can be found by substitution of $\min \left( x_{\text{v}} \right)$ into Equation \ref{eq21} (Equation \ref{eq31}).

\begin{equation} \label{eq30}
\min\left( x_{\text{v}} \right) = A \cdot x_{0} - x_{0} = x_{0} \cdot \left( A - 1 \right)
\end{equation}

\begin{flalign}
& \text{\renewcommand{\arraystretch}{0.66}
    \begin{tabular}{@{}c@{}}
    \scriptsize from \\
    \scriptsize (\ref{eq21})\\\scriptsize (\ref{eq30})
  \end{tabular}} 
  & 
  \max\left( y_{\text{v}} \right) = \displaystyle \frac{A^{2} \cdot x_{0} \cdot y_{0}}{\min\left( x_{\text{v}} \right)} = \displaystyle \frac{A^{2} \cdot y_{0}}{A - 1}
  &  
  \label{eq31} 
  &
\end{flalign}

\begin{figure}[ht]
    \centering
    \includegraphics[width=\textwidth]{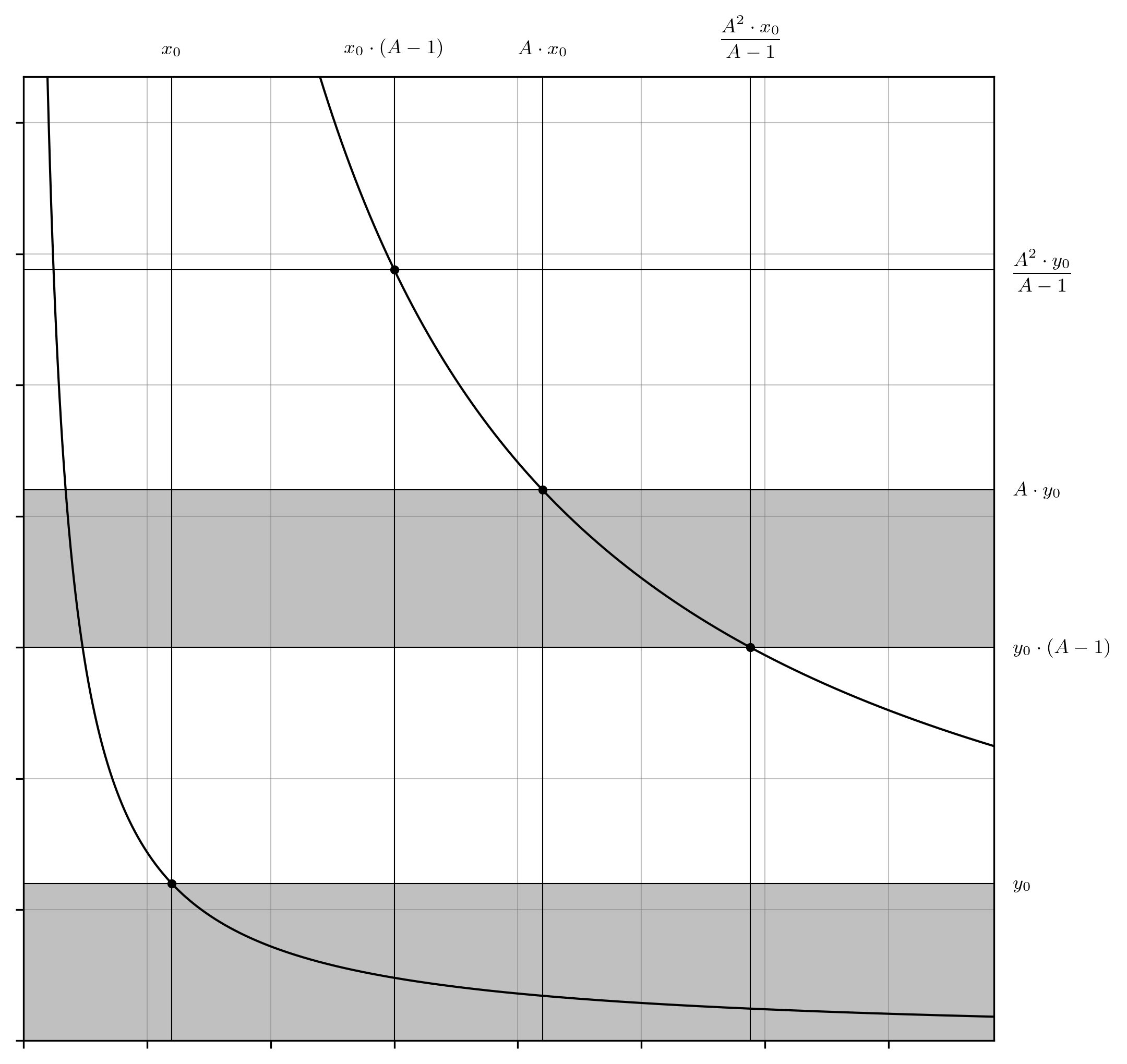}
    \captionsetup{
        justification=raggedright,
        singlelinecheck=false,
        font=small,
        labelfont=bf,
        labelsep=quad,
        format=plain
    }
    \caption{The $\max\left( x_{\text{v}} \right)$ and $\min\left( y_{\text{v}} \right)$ coordinates (Equations \ref{eq32} and \ref{eq33}). The correspondence between the virtual token balance and the real token balance is depicted with shaded areas for reference.}
    \label{fig9}
\end{figure}

This process is repeated to discover the identity of the other boundary, with coordinates at $\max\left( x_{\text{v}} \right)$ and $\min\left( y_{\text{v}} \right)$ (Figure \ref{fig9}). The virtual curve begins with $A \cdot y_{0}$ virtual $y$ tokens and is depleted when $y_{0}$ tokens have been traded out of it (Equation \ref{eq32}). The identity of $\max\left( x_{\text{v}} \right)$ can be found by substitution of $\min\left( y_{\text{v}} \right)$ into Equation \ref{eq22} (Equation \ref{eq33}).

\begin{equation} \label{eq32}
\min\left( y_{\text{v}} \right) = A \cdot y_{0} - y_{0} = y_{0} \cdot \left( A - 1 \right)
\end{equation}

\begin{flalign}
& \text{\renewcommand{\arraystretch}{0.66}
    \begin{tabular}{@{}c@{}}
    \scriptsize from \\
    \scriptsize (\ref{eq22})\\\scriptsize (\ref{eq30})
  \end{tabular}} 
  & 
  \max\left( x_{\text{v}} \right) = \displaystyle \frac{A^{2} \cdot x_{0} \cdot y_{0}}{\min\left( y_{\text{v}} \right)} = \displaystyle \frac{A^{2} \cdot x_{0}}{A - 1}
  &  
  \label{eq33} 
  &
\end{flalign}

\begin{figure}[ht]
    \centering
    \includegraphics[width=\textwidth]{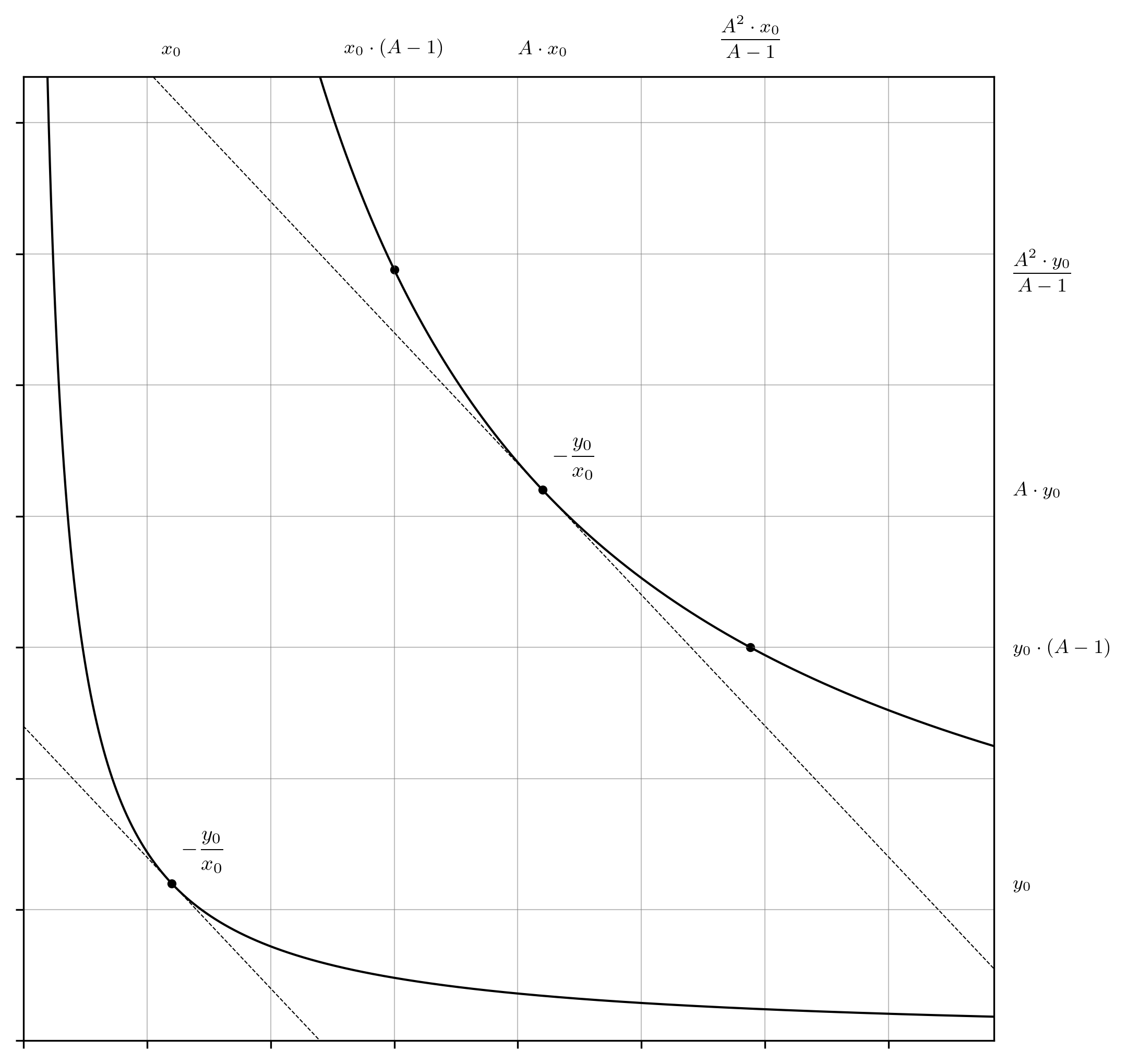}
    \captionsetup{
        justification=raggedright,
        singlelinecheck=false,
        font=small,
        labelfont=bf,
        labelsep=quad,
        format=plain
    }
    \caption{It is trivial that the gradient of the curves at the points $\left( x_{0}, y_{0} \right)$ on the reference curve and $\left( A \cdot x_{0}, A \cdot y_{0} \right)$ on the virtual curve are equal (Equations \ref{eq15} and \ref{eq34}).}
    \label{fig10}
\end{figure}

Turning our attention now to the exchange rates implied at these points of interest. It is trivial that the gradient of the curves at the points $\left( x_{0}, y_{0} \right)$ on the reference curve and $\left( A \cdot x_{0}, A \cdot y_{0} \right)$ on the virtual curve are equal (Equation \ref{eq34}); this identity is proved by substitution of these values into Equation \ref{eq15} for the reference curve, or any of Equations \ref{eq15}, \ref{eq28}, or \ref{eq29} for the virtual curve (Figure \ref{fig10}).

\begin{flalign}
& \text{\renewcommand{\arraystretch}{0.66}
    \begin{tabular}{@{}c@{}}
    \scriptsize from \\
    \scriptsize (\ref{eq15})\\\scriptsize (\ref{eq28})\\\scriptsize (\ref{eq29})
  \end{tabular}} 
  & 
  \displaystyle \frac{\partial y_{\text{v}}}{\partial x_{\text{v}}} = - \displaystyle \frac{{A \cdot y_{0}}}{{A \cdot x_{0}}} = - \displaystyle \frac{\left( A \cdot y_{0} \right)^{2}}{A^{2} \cdot x_{0} \cdot y_{0}} = - \displaystyle \frac{A^{2} \cdot x_{0} \cdot y_{0}}{\left( A \cdot x_{0} \right)^{2}} = - \displaystyle \frac{y_{0}}{x_{0}}
  &  
  \label{eq34} 
  &
\end{flalign}

\begin{figure}[ht]
    \centering
    \includegraphics[width=\textwidth]{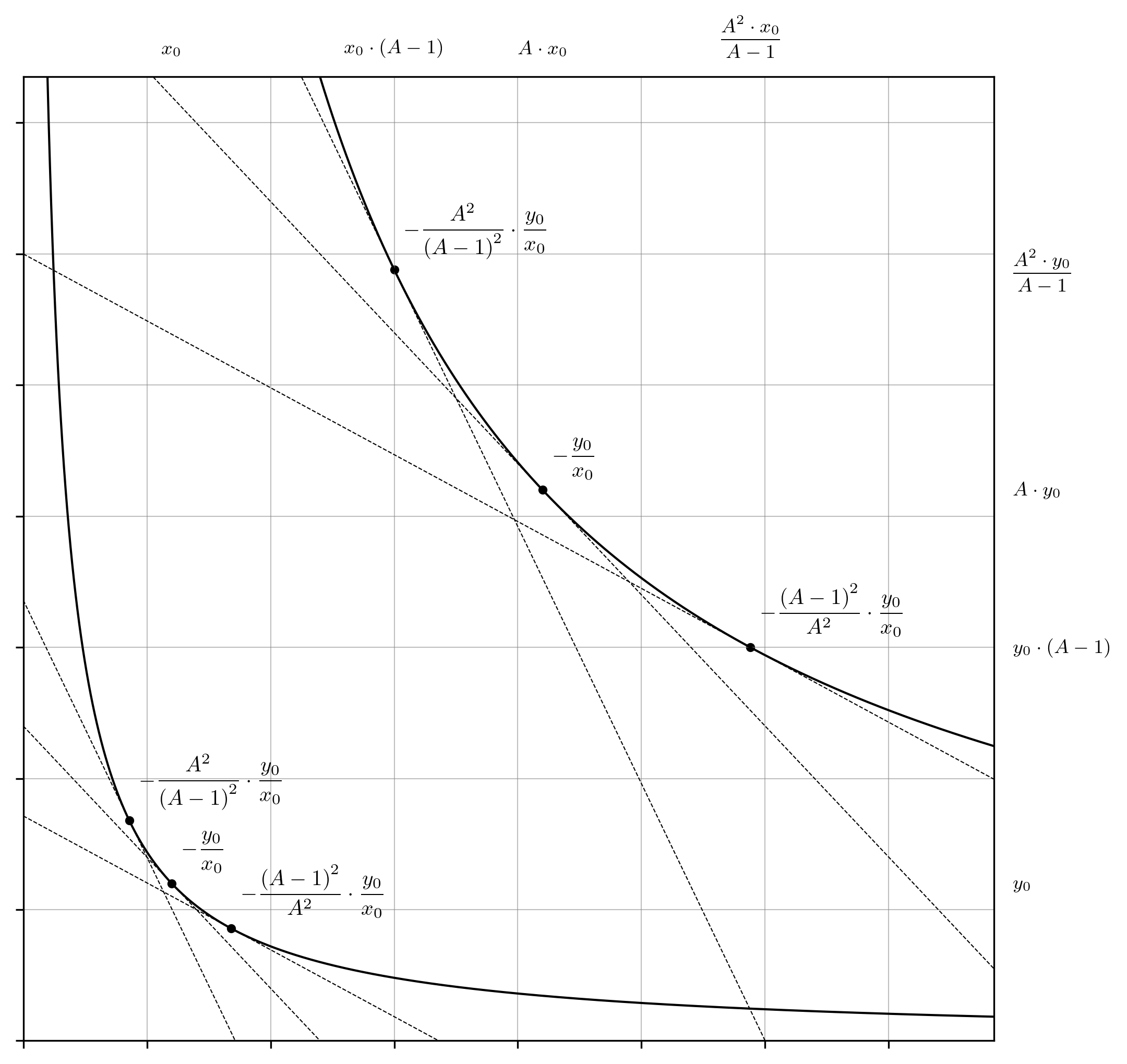}
    \captionsetup{
        justification=raggedright,
        singlelinecheck=false,
        font=small,
        labelfont=bf,
        labelsep=quad,
        format=plain
    }
    \caption{The slope of tangent lines drawn at the points $\left(\min \left( x_{\text{v}} \right), \max \left( y_{\text{v}} \right) \right)$ and $\left(\max \left( x_{\text{v}} \right), \min \left( y_{\text{v}} \right) \right)$ on the virtual curve are commensurate with its price boundaries. Tangent lines are drawn on the reference curve parallel to those on the virtual curve; the algebraic identities of the points where these parallel tangents kiss the reference curve are elucidated later in this exercise (Equations \ref{eq37}, \ref{eq39} and \ref{eq41}).}
    \label{fig11}
\end{figure}

The price boundaries represented by $\left(\min \left( x_{\text{v}} \right), \max \left( y_{\text{v}} \right) \right)$ and $\left(\max \left( x_{\text{v}} \right), \min \left( y_{\text{v}} \right) \right)$ can be defined via a similar process; substituting their identities into any of Equations \ref{eq15}, \ref{eq28}, or \ref{eq29} yields the same result (Equations \ref{eq35} and \ref{eq36}). Note that since the price curves implied by both the virtual and reference curves each encompass the same range, $-\infty < \partial y_{\text{v}} / \partial x_{\text{v}} < 0, \forall x_{\text{v}}, y_{\text{v}} \in \mathbb{R} ^ {+}$ and $-\infty < \partial y / \partial x < 0, \forall x, y \in \mathbb{R} ^ {+}$, there must exist points on the reference curve where evaluation of the first derivative of $x \cdot y = x_{0} \cdot y_{0}$ at these points yields the same marginal prices as the boundaries of the virtual curve. Geometrically, these are the unique points where tangent lines drawn on the reference curve are parallel to the tangent lines drawn at the boundaries of the virtual curve (Figure \ref{fig11}). 

\begin{flalign}
& \text{\renewcommand{\arraystretch}{0.66}
    \begin{tabular}{@{}c@{}}
    \scriptsize from \\
    \scriptsize (\ref{eq30})\\\scriptsize (\ref{eq31})
  \end{tabular}} 
  & 
  \displaystyle \frac{\partial y_{\text{v}}}{\partial x_{\text{v}}} = - \displaystyle \frac{\max\left( y_{\text{v}} \right)}{\min\left( x_{\text{v}} \right)} = - \displaystyle \frac{\max^{2}\left( y_{\text{v}} \right)}{A^{2} \cdot x_{0} \cdot y_{0}} = - \displaystyle \frac{A^{2} \cdot x_{0} \cdot y_{0}}{\min^{2}\left( x_{\text{v}} \right)} = - \displaystyle \frac{A^{2}}{\left( A - 1 \right)^{2}} \cdot \displaystyle \frac{y_{0}}{x_{0}}
  &  
  \label{eq35} 
  &
\end{flalign}

\begin{flalign}
& \text{\renewcommand{\arraystretch}{0.66}
    \begin{tabular}{@{}c@{}}
    \scriptsize from \\
    \scriptsize (\ref{eq32})\\\scriptsize (\ref{eq33})
  \end{tabular}} 
  & 
  \displaystyle \frac{\partial y_{\text{v}}}{\partial x_{\text{v}}} = - \displaystyle \frac{\min\left( y_{\text{v}} \right)}{\max\left( x_{\text{v}} \right)} = - \displaystyle \frac{\min^{2}\left( y_{\text{v}} \right)}{A^{2} \cdot x_{0} \cdot y_{0}} = - \displaystyle \frac{A^{2} \cdot x_{0} \cdot y_{0}}{\max^{2}\left( x_{\text{v}} \right)} = - \displaystyle \frac{\left( A - 1 \right)^{2}}{A^{2}} \cdot \displaystyle \frac{y_{0}}{x_{0}}
  &  
  \label{eq36} 
  &
\end{flalign}

Owing to their significance, let the marginal prices at the points $\left(\min \left( x_{\text{v}} \right), \max \left( y_{\text{v}} \right) \right)$ and $\left(\max \left( x_{\text{v}} \right), \min \left( y_{\text{v}} \right) \right)$ be represented by the variable names $P_{\text{high}}$, $P_{0}$, and $P_{\text{low}}$, respectively. These variables are defined here to be the absolute value of the marginal exchange rates at these points as this renders some of the analysis performed later in this document a little less complicated (Equations \ref{eq37}, \ref{eq38} and \ref{eq41}). However, one ignores the implicit negative sign of the slopes on these curves at their own peril. To save face, this negation is stated plainly in Equations \ref{eq38}, \ref{eq40} and \ref{eq42}. 

\begin{flalign}
& \text{\renewcommand{\arraystretch}{0.66}
    \begin{tabular}{@{}c@{}}
    \scriptsize from \\
    \scriptsize (\ref{eq35})
  \end{tabular}} 
  & 
  P_{\text{high}} = \displaystyle \frac{A^{2}}{\left( A - 1 \right)^{2}} \cdot \displaystyle \frac{y_{0}}{x_{0}};\ \sqrt{P_{\text{high}}} = \displaystyle \frac{A}{A - 1} \cdot \displaystyle \frac{\sqrt{y_{0}}}{\sqrt{x_{0}}}
  &  
  \label{eq37} 
  &
\end{flalign}

\begin{flalign}
& \text{\renewcommand{\arraystretch}{0.66}
    \begin{tabular}{@{}c@{}}
    \scriptsize from \\
    \scriptsize (\ref{eq35})\\\scriptsize (\ref{eq37})
  \end{tabular}} 
  & 
  \left. \ \displaystyle \frac{\partial y_{\text{v}}}{\partial x_{\text{v}}} \right|_{\min\left( x_{\text{v}} \right)} = - P_{\text{high}}
  &  
  \label{eq38} 
  &
\end{flalign}

\begin{flalign}
& \text{\renewcommand{\arraystretch}{0.66}
    \begin{tabular}{@{}c@{}}
    \scriptsize from \\
    \scriptsize (\ref{eq34})
  \end{tabular}} 
  & 
  P_{0} = \displaystyle \frac{y_{0}}{x_{0}}; \sqrt{P_{0}} = \displaystyle \frac{\sqrt{y_{0}}}{\sqrt{x_{0}}}
  &  
  \label{eq39} 
  &
\end{flalign}

\begin{flalign}
& \text{\renewcommand{\arraystretch}{0.66}
    \begin{tabular}{@{}c@{}}
    \scriptsize from \\
    \scriptsize (\ref{eq34})\\\scriptsize (\ref{eq39})
  \end{tabular}} 
  & 
  \left. \ \displaystyle \frac{\partial y_{\text{v}}}{\partial x_{\text{v}}} \right|_{A \cdot x_{0}} = - P_{0}
  &  
  \label{eq40} 
  &
\end{flalign}

\begin{flalign}
& \text{\renewcommand{\arraystretch}{0.66}
    \begin{tabular}{@{}c@{}}
    \scriptsize from \\
    \scriptsize (\ref{eq36})
  \end{tabular}} 
  & 
  P_{\text{low}} = \displaystyle \frac{\left( A - 1 \right)^{2}}{A^{2}} \cdot \displaystyle \frac{y_{0}}{x_{0}};\ \sqrt{P_{\text{low}}} = \displaystyle \frac{A - 1}{A} \cdot \displaystyle \frac{\sqrt{y_{0}}}{\sqrt{x_{0}}}
  &  
  \label{eq41} 
  &
\end{flalign}

\begin{flalign}
& \text{\renewcommand{\arraystretch}{0.66}
    \begin{tabular}{@{}c@{}}
    \scriptsize from \\
    \scriptsize (\ref{eq36})\\\scriptsize (\ref{eq41})
  \end{tabular}} 
  & 
  \left. \ \displaystyle \frac{\partial y_{\text{v}}}{\partial x_{\text{v}}} \right|_{\max\left( x_{\text{v}} \right)} = - P_{\text{low}}
  &  
  \label{eq42} 
  &
\end{flalign}

There are a few significant capstone identities that can be found at this juncture. Firstly, the geometric mean of $P_{\text{high}}$ and $P_{\text{low}}$ is equal to $P_{0}$ (Equations \ref{eq43} and \ref{eq44}). Similarly, the geometric means of $\max(x_{\text{v}})$ and $\min(x_{\text{v}})$, and $\max(y_{\text{v}})$ and $\min(y_{\text{v}})$ are equal to $A \cdot x_{0}$ and $A \cdot y_{0}$, respectively (which are the coordinates at which the first derivative evaluates to $P_{0}$) (Equations \ref{eq43}, \ref{eq44}, \ref{eq45}, \ref{eq46}, \ref{eq47} and \ref{eq48}). These are expected results, and can be understood as a consequence of the Mean Value Theorem\footnote{wikipedia.org/wiki/Mean\_value\_theorem} applied to the implicit curve and price curve over the closed interval defined by the bounds elucidated above. This is even more evident in the quotients of $\max(y_{\text{v}})$ and $\max(x_{\text{v}})$, and $\min(y_{\text{v}})$ and $\min(x_{\text{v}})$, which evaluate to $y_{0} / x_{0}$ and is equal to $P_{0}$ (Equations \ref{eq49} and \ref{eq50}).

\begin{flalign}
& \text{\renewcommand{\arraystretch}{0.66}
    \begin{tabular}{@{}c@{}}
    \scriptsize from \\
    \scriptsize (\ref{eq37})\\\scriptsize (\ref{eq41})
  \end{tabular}} 
  & 
  P_{\text{high}} \cdot P_{\text{low}} = \displaystyle \frac{\left( A - 1 \right)^{2}}{A^{2}} \cdot \displaystyle \frac{A^{2}}{\left( A - 1 \right)^{2}} \cdot \displaystyle \frac{y_{0}}{x_{0}} \cdot \displaystyle \frac{y_{0}}{x_{0}} = \displaystyle \frac{y_{0}^{2}}{x_{0}^{2}}
  &  
  \label{eq43} 
  &
\end{flalign}

\begin{flalign}
& \text{\renewcommand{\arraystretch}{0.66}
    \begin{tabular}{@{}c@{}}
    \scriptsize from \\
    \scriptsize (\ref{eq39})\\\scriptsize (\ref{eq43})
  \end{tabular}} 
  & 
  \sqrt{P_{\text{high}}} \cdot \sqrt{P_{\text{low}}} = P_{0} = \displaystyle \frac{y_{0}}{x_{0}}
  &  
  \label{eq44} 
  &
\end{flalign}

\begin{flalign}
& \text{\renewcommand{\arraystretch}{0.66}
    \begin{tabular}{@{}c@{}}
    \scriptsize from \\
    \scriptsize (\ref{eq30})\\\scriptsize (\ref{eq33})
  \end{tabular}} 
  & 
  \max\left( x_{\text{v}} \right) \cdot \min\left( x_{\text{v}} \right) = \displaystyle \frac{A^{2} \cdot x_{0}}{A - 1} \cdot x_{0} \cdot \left( A - 1 \right) = A^{2} \cdot x_{0}^{2}
  &  
  \label{eq45} 
  &
\end{flalign}

\begin{flalign}
& \text{\renewcommand{\arraystretch}{0.66}
    \begin{tabular}{@{}c@{}}
    \scriptsize from \\
    \scriptsize (\ref{eq45})
  \end{tabular}} 
  & 
  \sqrt{\max\left( x_{\text{v}} \right) \cdot}\sqrt{\min\left( x_{\text{v}} \right)} = A \cdot x_{0}
  &  
  \label{eq46} 
  &
\end{flalign}

\begin{flalign}
& \text{\renewcommand{\arraystretch}{0.66}
    \begin{tabular}{@{}c@{}}
    \scriptsize from \\
    \scriptsize (\ref{eq31})\\\scriptsize (\ref{eq32})
  \end{tabular}} 
  & 
  \max\left( y_{\text{v}} \right) \cdot \min\left( y_{\text{v}} \right) = \displaystyle \frac{A^{2} \cdot y_{0}}{A - 1} \cdot y_{0} \cdot \left( A - 1 \right) = A^{2} \cdot y_{0}^{2}
  &  
  \label{eq47} 
  &
\end{flalign}

\begin{flalign}
& \text{\renewcommand{\arraystretch}{0.66}
    \begin{tabular}{@{}c@{}}
    \scriptsize from \\
    \scriptsize (\ref{eq47})
  \end{tabular}} 
  & 
  \sqrt{\max\left( y_{\text{v}} \right) \cdot}\sqrt{\min\left( y_{\text{v}} \right)} = A \cdot y_{0}
  &  
  \label{eq48} 
  &
\end{flalign}

\begin{flalign}
& \text{\renewcommand{\arraystretch}{0.66}
    \begin{tabular}{@{}c@{}}
    \scriptsize from \\
    \scriptsize (\ref{eq30})\\\scriptsize (\ref{eq33})
  \end{tabular}} 
  & 
  \displaystyle \frac{\max\left( y_{\text{v}} \right)}{\max\left( x_{\text{v}} \right)} = \displaystyle \frac{\displaystyle \frac{A^{2} \cdot y_{0}}{A - 1}}{\displaystyle \frac{A^{2} \cdot x_{0}}{A - 1}} = \displaystyle \frac{y_{0}}{x_{0}} = P_{0}
  &  
  \label{eq49} 
  &
\end{flalign}

\begin{flalign}
& \text{\renewcommand{\arraystretch}{0.66}
    \begin{tabular}{@{}c@{}}
    \scriptsize from \\
    \scriptsize (\ref{eq31})\\\scriptsize (\ref{eq32})
  \end{tabular}} 
  & 
  \displaystyle \frac{\min\left( y_{\text{v}} \right)}{\min\left( x_{\text{v}} \right)} = \displaystyle \frac{y_{0} \cdot \left( A - 1 \right)}{x_{0} \cdot \left( A - 1 \right)} = \displaystyle \frac{y_{0}}{x_{0}} = P_{0}
  &  
  \label{eq50} 
  &
\end{flalign}

Secondly, and less trivially, the quotient of $P_{\text{high}}$ and $P_{\text{low}}$ can be expressed entirely as a function of $A$ (Equations \ref{eq51}, \ref{eq52} and \ref{eq53}). This result is commonly referred to as the “capital efficiency” identity without context or explanation but can now be understood through the lens of a smaller liquidity pool emulating a larger liquidity pool $A$ times larger than itself. 

\begin{flalign}
& \text{\renewcommand{\arraystretch}{0.66}
    \begin{tabular}{@{}c@{}}
    \scriptsize from \\
    \scriptsize (\ref{eq37})\\\scriptsize (\ref{eq41})
  \end{tabular}} 
  & 
  \displaystyle \frac{P_{\text{high}}}{P_{\text{low}}} = \displaystyle \frac{\displaystyle \frac{A^{2}}{\left( A - 1 \right)^{2}} \cdot \displaystyle \frac{y_{0}}{x_{0}}}{\displaystyle \frac{\left( A - 1 \right)^{2}}{A^{2}} \cdot \displaystyle \frac{y_{0}}{x_{0}}} = \displaystyle \frac{A^{4}}{\left( A - 1 \right)^{4}}
  &  
  \label{eq51} 
  &
\end{flalign}

\begin{flalign}
& \text{\renewcommand{\arraystretch}{0.66}
    \begin{tabular}{@{}c@{}}
    \scriptsize from \\
    \scriptsize (\ref{eq51})
  \end{tabular}} 
  & 
  \displaystyle \frac{\sqrt[4]{P_{\text{high}}}}{\sqrt[4]{P_{\text{low}}}} = \displaystyle \frac{A}{A - 1}
  &  
  \label{eq52} 
  &
\end{flalign}

\begin{flalign}
& \text{\renewcommand{\arraystretch}{0.66}
    \begin{tabular}{@{}c@{}}
    \scriptsize from \\
    \scriptsize (\ref{eq52})
  \end{tabular}} 
  & 
  A = \displaystyle \frac{1}{1 - \displaystyle \frac{\sqrt[4]{P_{\text{low}}}}{\sqrt[4]{P_{\text{high}}}}} = \displaystyle \frac{\sqrt[4]{P_{\text{high}}}}{\sqrt[4]{P_{\text{high}}} - \sqrt[4]{P_{\text{low}}}}
  &  
  \label{eq53} 
  &
\end{flalign}

The so-called “capital efficiency” term receives a lot of attention, but it is ultimately a red herring. The more important observation is that the quotients of $P_{\text{high}}$ and $P_{0}$, and $P_{0}$ and $P_{\text{low}}$ are equal to each other and to some other constant, also expressed in terms of $A$ (Equations \ref{eq54}, \ref{eq55} and \ref{eq56}). For now, let this constant be the symbol $C$; its identity has a deep connection to the underlying hyperbolic trigonometry demonstrated later in this exercise. The constant $C$ is also obtainable from the quotients of $\max \left( x_{\text{v}} \right)$ and $\min \left( x_{\text{v}} \right)$, and $\max \left( y_{\text{v}} \right)$ and $\min \left( y_{\text{v}} \right)$ (Equations \ref{eq57} and \ref{eq58}).

\begin{flalign}
& \text{\renewcommand{\arraystretch}{0.66}
    \begin{tabular}{@{}c@{}}
    \scriptsize from \\
    \scriptsize (\ref{eq37})\\\scriptsize (\ref{eq39})
  \end{tabular}} 
  & 
  \displaystyle \frac{P_{\text{high}}}{P_{0}} = \displaystyle \frac{\displaystyle \frac{A^{2}}{\left( A - 1 \right)^{2}} \cdot \displaystyle \frac{y_{0}}{x_{0}}}{\displaystyle \frac{y_{0}}{x_{0}}} = \displaystyle \frac{A^{2}}{\left( A - 1 \right)^{2}} = C
  &  
  \label{eq54} 
  &
\end{flalign}

\begin{flalign}
& \text{\renewcommand{\arraystretch}{0.66}
    \begin{tabular}{@{}c@{}}
    \scriptsize from \\
    \scriptsize (\ref{eq39})\\\scriptsize (\ref{eq41})
  \end{tabular}} 
  & 
  \displaystyle \frac{P_{0}}{P_{\text{low}}} = \displaystyle \frac{\displaystyle \frac{y_{0}}{x_{0}}}{\displaystyle \frac{\left( A - 1 \right)^{2}}{A^{2}} \cdot \displaystyle \frac{y_{0}}{x_{0}}} = \displaystyle \frac{A^{2}}{\left( A - 1 \right)^{2}} = C
  &  
  \label{eq55} 
  &
\end{flalign}

\begin{flalign}
& \text{\renewcommand{\arraystretch}{0.66}
    \begin{tabular}{@{}c@{}}
    \scriptsize from \\
    \scriptsize (\ref{eq51})\\\scriptsize (\ref{eq54})\\\scriptsize (\ref{eq55})
  \end{tabular}} 
  & 
  \displaystyle \frac{P_{\text{high}}}{P_{\text{low}}} = \displaystyle \frac{A^{4}}{\left( A - 1 \right)^{4}} = C^{2};\ \displaystyle \frac{\sqrt{P_{\text{high}}}}{\sqrt{P_{\text{low}}}} = \displaystyle \frac{A^{2}}{\left( A - 1 \right)^{2}} = C
  &  
  \label{eq56} 
  &
\end{flalign}

\begin{flalign}
& \text{\renewcommand{\arraystretch}{0.66}
    \begin{tabular}{@{}c@{}}
    \scriptsize from \\
    \scriptsize (\ref{eq37})\\\scriptsize (\ref{eq39})
  \end{tabular}} 
  & 
  \displaystyle \frac{\max\left( x_{\text{v}} \right)}{\min\left( x_{\text{v}} \right)} = \displaystyle \frac{\displaystyle \frac{A^{2} \cdot x_{0}}{A - 1}}{x_{0} \cdot \left( A - 1 \right)} = \displaystyle \frac{A^{2}}{\left( A - 1 \right)^{2}} = C
  &  
  \label{eq57} 
  &
\end{flalign}

\begin{flalign}
& \text{\renewcommand{\arraystretch}{0.66}
    \begin{tabular}{@{}c@{}}
    \scriptsize from \\
    \scriptsize (\ref{eq39})\\\scriptsize (\ref{eq41})
  \end{tabular}} 
  & 
  \displaystyle \frac{\max\left( y_{\text{v}} \right)}{\min\left( y_{\text{v}} \right)} = \displaystyle \frac{\displaystyle \frac{A^{2} \cdot y_{0}}{A - 1}}{y_{0} \cdot \left( A - 1 \right)} = \displaystyle \frac{A^{2}}{\left( A - 1 \right)^{2}} = C
  &  
  \label{eq58} 
  &
\end{flalign}

Turning our attention now to the tangents drawn on the reference curve, parallel to those on the virtual curve. There are two convenient methods to elucidate the algebraic identities of the points on the reference curve where the marginal prices are equal to the price boundaries of the virtual curve. The naïve method (which still works) is to simply reverse the amplification by taking the quotients of $\min \left( x_{\text{v}} \right)$, $\max \left( y_{\text{v}} \right)$, $\min \left( y_{\text{v}} \right)$, $\max \left( x_{\text{v}} \right)$, and the amplification constant $A$ (Equations \ref{eq59}, \ref{eq60}, \ref{eq61} and \ref{eq62}). 

\begin{flalign}
& \text{\renewcommand{\arraystretch}{0.66}
    \begin{tabular}{@{}c@{}}
    \scriptsize from \\
    \scriptsize (\ref{eq30})
  \end{tabular}} 
  & 
  \min \left( x \right) = \displaystyle \frac{\min\left( x_{\text{v}} \right)}{A} = \displaystyle \frac{x_{0} \cdot \left( A - 1 \right)}{A}
  &  
  \label{eq59} 
  &
\end{flalign}

\begin{flalign}
& \text{\renewcommand{\arraystretch}{0.66}
    \begin{tabular}{@{}c@{}}
    \scriptsize from \\
    \scriptsize (\ref{eq31})
  \end{tabular}} 
  & 
  \max \left( y \right) = \displaystyle \frac{\max\left( y_{\text{v}} \right)}{A} = \displaystyle \frac{\displaystyle \frac{A^{2} \cdot y_{0}}{A - 1}}{A} = \displaystyle \frac{A \cdot y_{0}}{A - 1}
  &  
  \label{eq60} 
  &
\end{flalign}

\begin{flalign}
& \text{\renewcommand{\arraystretch}{0.66}
    \begin{tabular}{@{}c@{}}
    \scriptsize from \\
    \scriptsize (\ref{eq32})
  \end{tabular}} 
  & 
  \max \left( x \right) = \displaystyle \frac{\max\left( x_{\text{v}} \right)}{A} = \displaystyle \frac{\displaystyle \frac{A^{2} \cdot x_{0}}{A - 1}}{A} = \displaystyle \frac{A \cdot x_{0}}{A - 1}
  &  
  \label{eq61} 
  &
\end{flalign}

\begin{flalign}
& \text{\renewcommand{\arraystretch}{0.66}
    \begin{tabular}{@{}c@{}}
    \scriptsize from \\
    \scriptsize (\ref{eq33})
  \end{tabular}} 
  & 
  \min \left( y \right) = \displaystyle \frac{\min\left( y_{\text{v}} \right)}{A} = \displaystyle \frac{y_{0} \cdot \left( A - 1 \right)}{A}
  &  
  \label{eq62} 
  &
\end{flalign}

The other method is by way of the previously derived marginal rate expressions, Equations \ref{eq16} and \ref{eq17}, after substituting the appropriate marginal rate identity from Equations \ref{eq35} and \ref{eq36}. This method is demonstrated only for $\min \left( x \right)$ below (Equations \ref{eq63} and \ref{eq64}) but can be used to confirm all identities in Equations \ref{eq59}, \ref{eq60}, \ref{eq61} and \ref{eq62}, adding some rigor to the amplification reversal method used above. 

\begin{flalign}
& \text{\renewcommand{\arraystretch}{0.66}
    \begin{tabular}{@{}c@{}}
    \scriptsize from \\
    \scriptsize (\ref{eq17})\\\scriptsize (\ref{eq35})
  \end{tabular}} 
  & 
  \displaystyle \frac{\partial y}{\partial x} = - \displaystyle \frac{x_{0} \cdot y_{0}}{\min^{2}\ (x)\ } = - \displaystyle \frac{A^{2}}{\left( A - 1 \right)^{2}} \cdot \displaystyle \frac{y_{0}}{x_{0}}
  &  
  \label{eq63} 
  &
\end{flalign}

\begin{flalign}
& \text{\renewcommand{\arraystretch}{0.66}
    \begin{tabular}{@{}c@{}}
    \scriptsize from \\
    \scriptsize (\ref{eq63})
  \end{tabular}} 
  & 
  \displaystyle \frac{1}{\min^{2}\ (x)\ } = \displaystyle \frac{A^{2}}{\left( A - 1 \right)^{2}} \cdot \displaystyle \frac{1}{x_{0}^{2}} \Rightarrow \min \left( x \right) = \displaystyle \frac{x_{0} \cdot \left( A - 1 \right)}{A}
  &  
  \label{eq64} 
  &
\end{flalign}

\begin{figure}[ht]
    \centering
    \includegraphics[width=\textwidth]{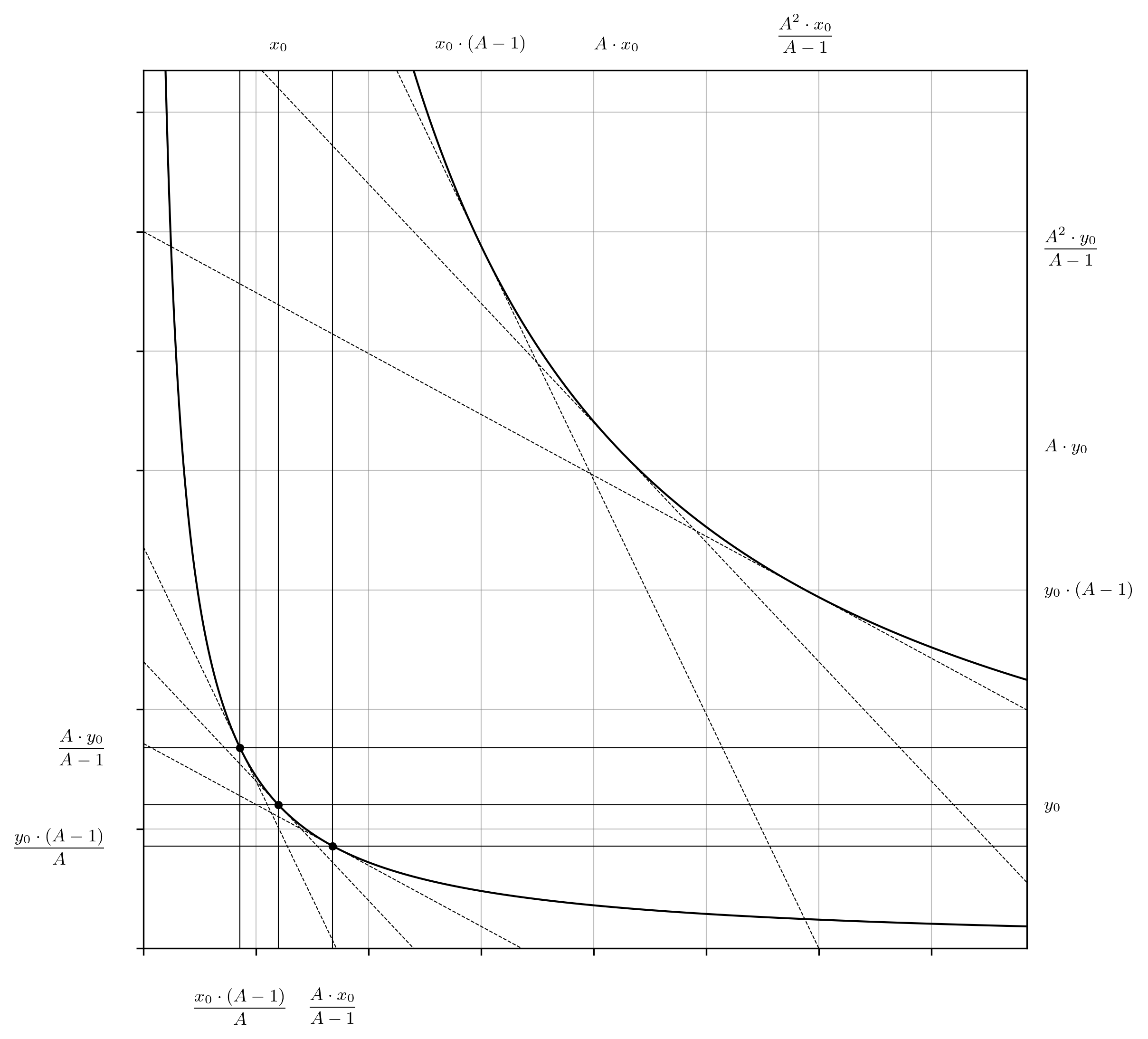}
    \captionsetup{
        justification=raggedright,
        singlelinecheck=false,
        font=small,
        labelfont=bf,
        labelsep=quad,
        format=plain
    }
    \caption{The algebraic identities of the points where the lines tangent to the reference curve and parallel to the tangent lines at the price boundaries of the virtual curve are now elucidated (Equations \ref{eq59}, \ref{eq60}, \ref{eq61} and \ref{eq62}).}
    \label{fig12}
\end{figure}

These points have been appended to the evolving plot in Figure \ref{fig12}. For completeness, the shadow of the Mean Value Theorem can be observed here, too. The geometric means of $\min(x)$ and $\max(x)$, and $\max(y)$ and $\min(y)$ are equal to $x_{0}$ and $y_{0}$, respectively (which are also the coordinates at which the first derivative evaluates to $P_{0}$) (Equations \ref{eq65}, \ref{eq66}, \ref{eq67} and \ref{eq68}). The quotients of $\max(y)$ and $\max(x)$, and $\min(y)$ and $\min(x)$ also evaluate to $y_{0}/x_{0}$ and are equal to $P_{0}$ (Equations \ref{eq69} and \ref{eq70}). Further foreshadowing the hyperbolic trigonometry discussion, the quotients of $\max(x)$ and $\min(x)$, and $\max(y)$ and $\min(y)$ also yield the previously observed constant, $C$ (Equations \ref{eq71} and \ref{eq72}).

\begin{flalign}
& \text{\renewcommand{\arraystretch}{0.66}
    \begin{tabular}{@{}c@{}}
    \scriptsize from \\
    \scriptsize (\ref{eq59})\\\scriptsize (\ref{eq61})
  \end{tabular}} 
  & 
  \min \left( x \right) \cdot \max \left( x \right) = \displaystyle \frac{x_{0} \cdot \left( A - 1 \right)}{A} \cdot \displaystyle \frac{A \cdot x_{0}}{A - 1} = x_{0}^{2}
  &  
  \label{eq65} 
  &
\end{flalign}

\begin{flalign}
& \text{\renewcommand{\arraystretch}{0.66}
    \begin{tabular}{@{}c@{}}
    \scriptsize from \\
    \scriptsize (\ref{eq65})
  \end{tabular}} 
  & 
  \sqrt{\min \left( x \right)} \cdot \sqrt{\max \left( x \right)} = x_{0}
  &  
  \label{eq66} 
  &
\end{flalign}

\begin{flalign}
& \text{\renewcommand{\arraystretch}{0.66}
    \begin{tabular}{@{}c@{}}
    \scriptsize from \\
    \scriptsize (\ref{eq60})\\\scriptsize (\ref{eq62})
  \end{tabular}} 
  & 
  \max \left( y \right) \cdot \min \left( y \right) = \displaystyle \frac{A \cdot y_{0}}{A - 1} \cdot \displaystyle \frac{y_{0} \cdot \left( A - 1 \right)}{A} = y_{0}^{2}
  &  
  \label{eq67} 
  &
\end{flalign}

\begin{flalign}
& \text{\renewcommand{\arraystretch}{0.66}
    \begin{tabular}{@{}c@{}}
    \scriptsize from \\
    \scriptsize (\ref{eq67})
  \end{tabular}} 
  & 
  \sqrt{\max \left( y \right)} \cdot \sqrt{\min \left( y \right)} = y_{0}
  &  
  \label{eq68} 
  &
\end{flalign}

\begin{flalign}
& \text{\renewcommand{\arraystretch}{0.66}
    \begin{tabular}{@{}c@{}}
    \scriptsize from \\
    \scriptsize (\ref{eq60})\\\scriptsize (\ref{eq61})
  \end{tabular}} 
  & 
  \displaystyle \frac{\max \left( y \right)}{\max \left( x \right)} = \displaystyle \frac{\displaystyle \frac{A \cdot y_{0}}{A - 1}}{\displaystyle \frac{A \cdot x_{0}}{A - 1}} = \displaystyle \frac{y_{0}}{x_{0}} = P_{0}
  &  
  \label{eq69} 
  &
\end{flalign}

\begin{flalign}
& \text{\renewcommand{\arraystretch}{0.66}
    \begin{tabular}{@{}c@{}}
    \scriptsize from \\
    \scriptsize (\ref{eq59})\\\scriptsize (\ref{eq62})
  \end{tabular}} 
  & 
  \displaystyle \frac{\min \left( y \right)}{\min \left( x \right)} = \displaystyle \frac{\displaystyle \frac{y_{0} \cdot \left( A - 1 \right)}{A}}{\displaystyle \frac{x_{0} \cdot \left( A - 1 \right)}{A}} = \displaystyle \frac{y_{0}}{x_{0}} = P_{0}
  &  
  \label{eq70} 
  &
\end{flalign}

\begin{flalign}
& \text{\renewcommand{\arraystretch}{0.66}
    \begin{tabular}{@{}c@{}}
    \scriptsize from \\
    \scriptsize (\ref{eq59})\\\scriptsize (\ref{eq61})
  \end{tabular}} 
  & 
  \displaystyle \frac{\max \left( x \right)}{\min \left( x \right)} = \displaystyle \frac{\displaystyle \frac{A \cdot x_{0}}{A - 1}}{\displaystyle \frac{x_{0} \cdot \left( A - 1 \right)}{A}} = \displaystyle \frac{A^{2}}{\left( A - 1 \right)^{2}} = C
  &  
  \label{eq71} 
  &
\end{flalign}

\begin{flalign}
& \text{\renewcommand{\arraystretch}{0.66}
    \begin{tabular}{@{}c@{}}
    \scriptsize from \\
    \scriptsize (\ref{eq60})\\\scriptsize (\ref{eq62})
  \end{tabular}} 
  & 
  \displaystyle \frac{\max \left( y \right)}{\min \left( y \right)} = \displaystyle \frac{\displaystyle \frac{A \cdot y_{0}}{A - 1}}{\displaystyle \frac{y_{0} \cdot \left( A - 1 \right)}{A}} = \displaystyle \frac{A^{2}}{\left( A - 1 \right)^{2}} = C
  &  
  \label{eq72} 
  &
\end{flalign}

\begin{figure}[ht]
    \centering
    \includegraphics[width=\textwidth]{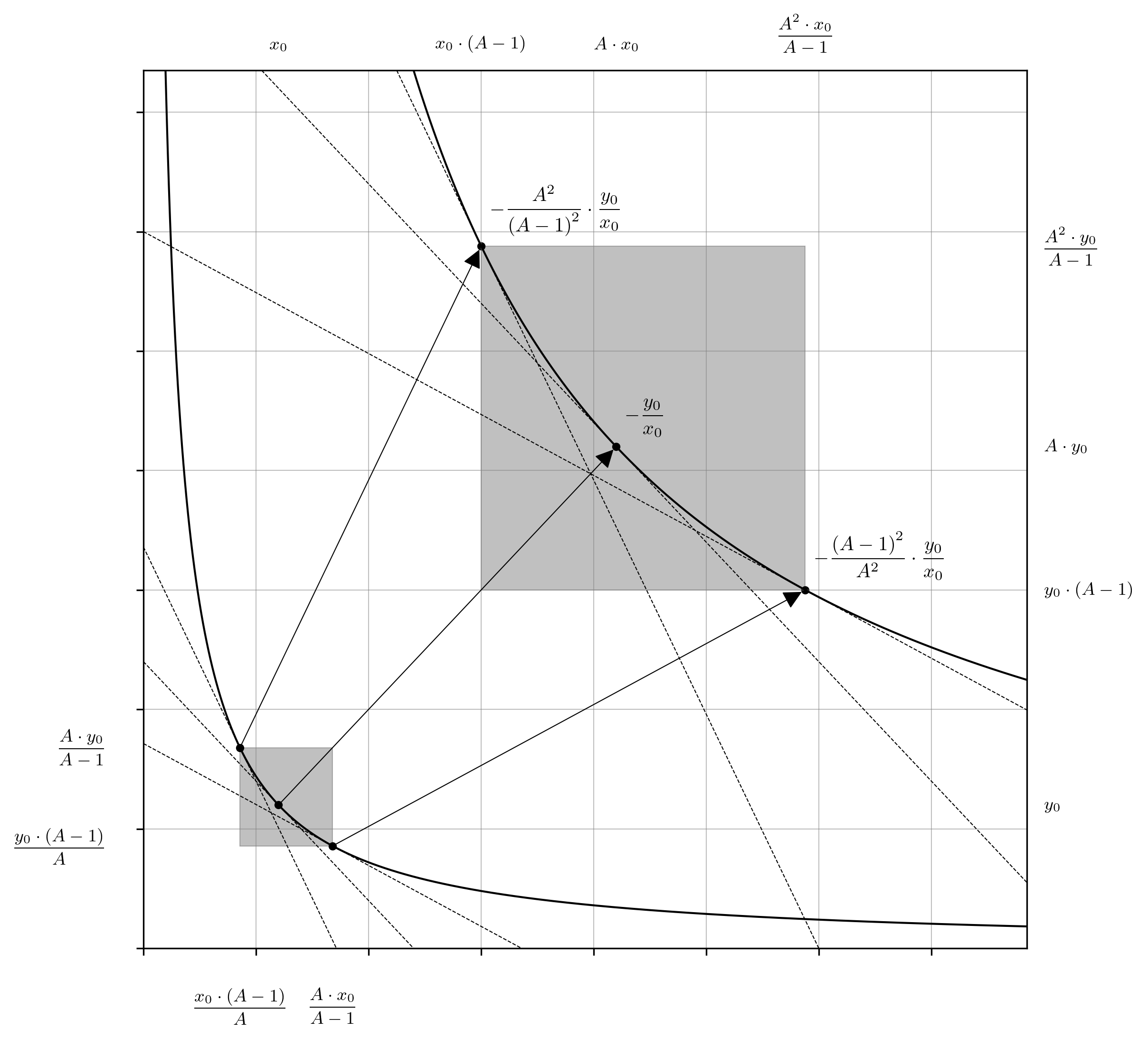}
    \captionsetup{
        justification=raggedright,
        singlelinecheck=false,
        font=small,
        labelfont=bf,
        labelsep=quad,
        format=plain
    }
    \caption{The amplification process is depicted with corresponding shaded areas of the reference and virtual curves, respectively. Points where the first derivative of each curve evaluates to the same result are shown as white dots, and the mapping of these points is depicted with white arrows.}
    \label{fig13}
\end{figure}

A heuristic understanding of the token balance virtualization can now be attained. A section of the standard $x \cdot y = x_{0} \cdot y_{0}$ implicit curve, defined by a geometric center with the reference coordinates $\left( x_{0}, y_{0} \right)$, and relative distances from this reference point defined by the amplification constant $A$, is mapped to the implicit curve of $x_{\text{v}} \cdot y_{\text{v}} = A^{2} \cdot x_{0} \cdot y_{0}$. Shaded areas and arrows which depict the mapping between the two curves have been added to the developing plot to visualize the overall process (Figure \ref{fig13}).

\begin{figure}[ht]
    \centering
    \includegraphics[width=\textwidth]{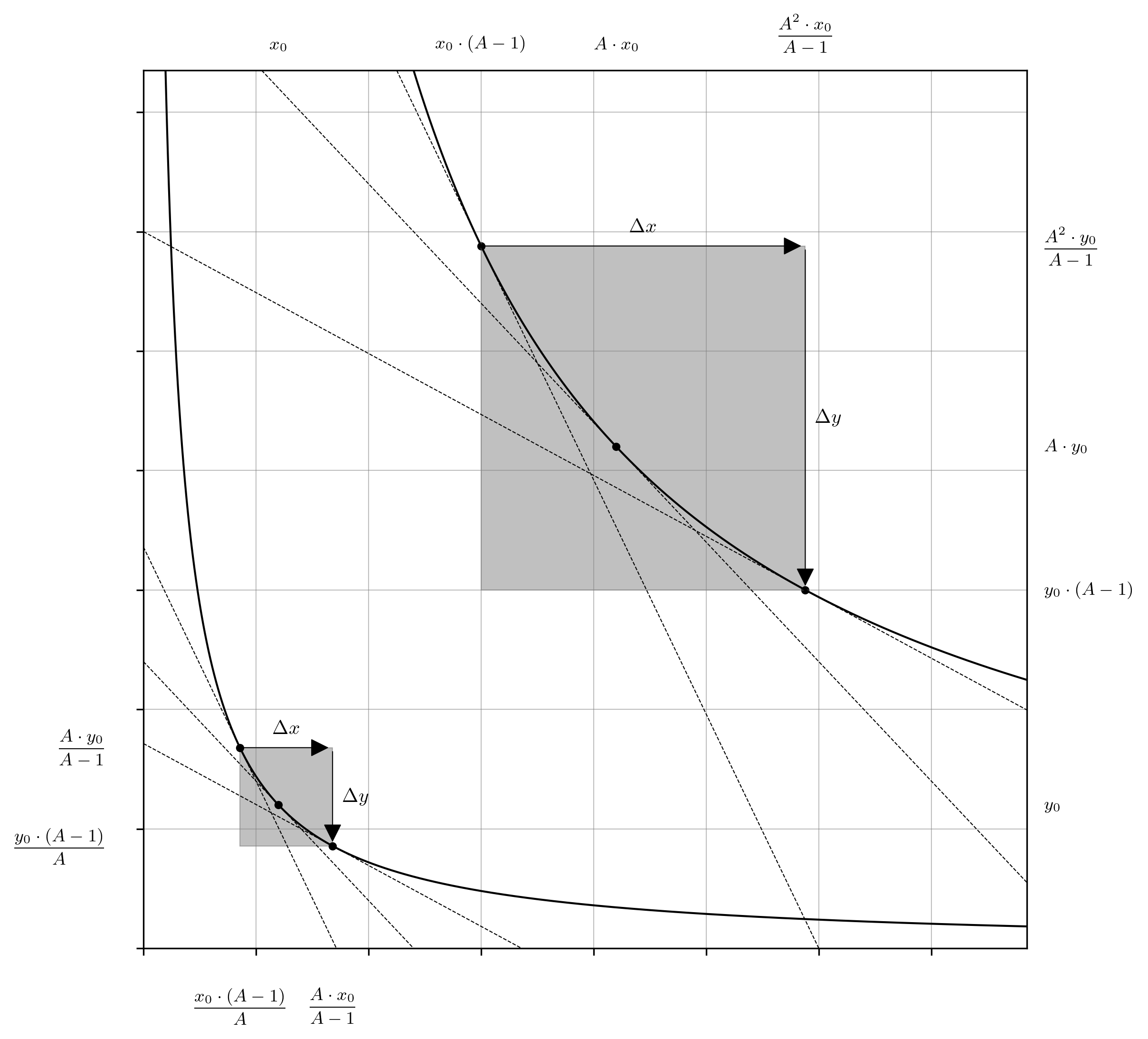}
    \captionsetup{
        justification=raggedright,
        singlelinecheck=false,
        font=small,
        labelfont=bf,
        labelsep=quad,
        format=plain
    }
    \caption{Traversal upon the rectangular hyperbolas $x \cdot y = x_{0} \cdot y_{0}$ and $x_{\text{v}} \cdot y_{\text{v}} = A^{2} \cdot x_{0} \cdot y_{0}$ (Equations \ref{eq2} and \ref{eq20}), representing a token swap against the reference and amplified liquidity pools, where $\mathrm{\Delta}x > 0$ and $\mathrm{\Delta}y < 0$. The marginal rates of exchange before and after the swap are identical. The ratio of the $\mathrm{\Delta}x$ and $\mathrm{\Delta}y$ arrow lengths for each curve are also identical, and therefore the overall rate of exchange, $\mathrm{\Delta}y / \mathrm{\Delta}x$ is equal in both cases.}
    \label{fig14}
\end{figure}

\begin{figure}[ht]
    \centering
    \includegraphics[width=\textwidth]{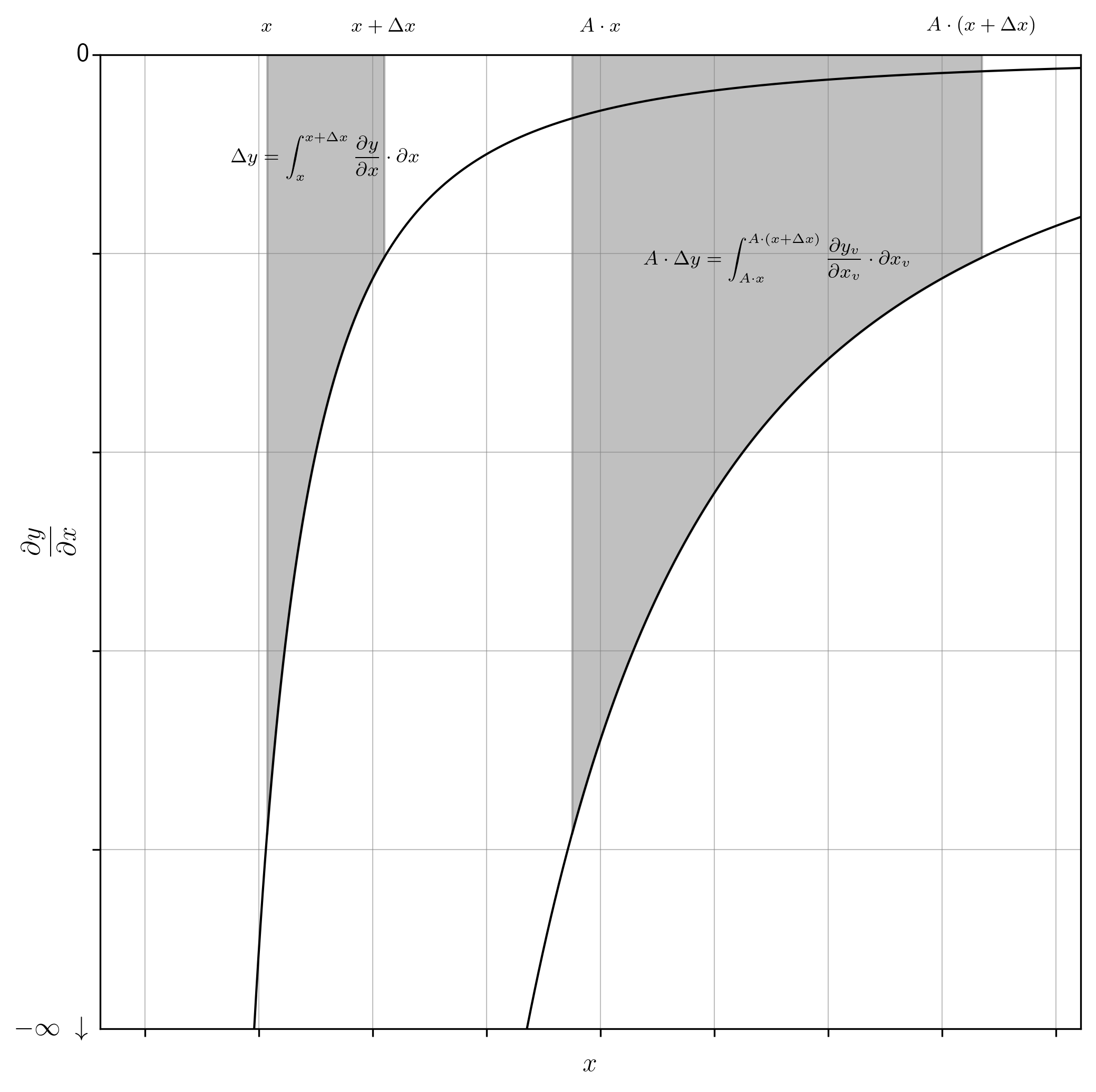}
    \captionsetup{
        justification=raggedright,
        singlelinecheck=false,
        font=small,
        labelfont=bf,
        labelsep=quad,
        format=plain
    }
    \caption{The integration above $\partial y / \partial x = - x_{0} \cdot y_{0} / x^{2}$ and $\partial y_{\text{v}} / \partial x_{\text{v}} = - A^{2} \cdot x_{0} \cdot y_{0} / x_{\text{v}}^{2}$ (Equations \ref{eq17} and \ref{eq29}) over the intervals $x \rightarrow x + \mathrm{\Delta}x$ and $A \cdot x \rightarrow A \cdot \left( x + \mathrm{\Delta}x\right)$ (i.e. $x_{\text{v}} \rightarrow x_{\text{v}} + \mathrm{\Delta}x$) representing a token swap against both the reference and amplified liquidity pools, where $\mathrm{\Delta}x > 0$ and $\mathrm{\Delta}y < 0$. Note that that the integration areas for the reference and virtual curves are $\mathrm{\Delta}y$ and $A \cdot \mathrm{\Delta}y$, respectively. By inference, any token swap against the virtual curve can be calculated from the reference curve by multiplying the output by the amplification constant.}
    \label{fig15}
\end{figure}

The capacity of the virtual curve to absorb higher trading volumes compared to the reference curve can also be interrogated by considering a token swap from one price bound to the other (Figures \ref{fig14} and \ref{fig15}). Note that such an action will execute at the same effective price in both cases. While the marginal rates before and after the exchange (and therefore the overall exchange rate) are identical in both the reference and virtual curves, the trade amounts are significantly greater in the virtual curve. The improved trade volume can be inspected visually from the increased arrow lengths in Figure \ref{fig14}, and the integrated area in Figure \ref{fig15}.

While outside of the focus of the present discussion, the financial ramification of this process is worthy of consideration. The token swaps depicted in Figures \ref{fig14} and \ref{fig15} obscure the fact that the virtual curve has traded its entire reserve of one token for the other token, whereas the more conservative reference curve has only traded a small proportion of its reserves. This can have dramatic consequences\footnote{Loesch, S.; Hindman, N.; Richardson, M. B.; and Welch, N. Impermanent Loss in Uniswap v3. arxiv.org/abs/2111.09192, 2021.} on the portfolio value represented by the liquidity position of the virtual curve. In lieu of a more thorough investigation, the Medium articles\footnote{medium.com/auditless/impermanent-loss-in-uniswap-v3-6c7161d3b445}\textsuperscript{,}\footnote{medium.com/coinmonks/top-5-mysterious-liquidity-providers-in-uniswap-v3-and-what-we-can-learn-from-them-1894bd27096f} authored by Peteris Erins and Ivan Vakhmyanin do an excellent job of elaborating this topic material further. 

\subsection{The Bancor v2 Real Curve}\label{subsec3.2}

\begin{figure}[ht]
    \centering
    \includegraphics[width=\textwidth]{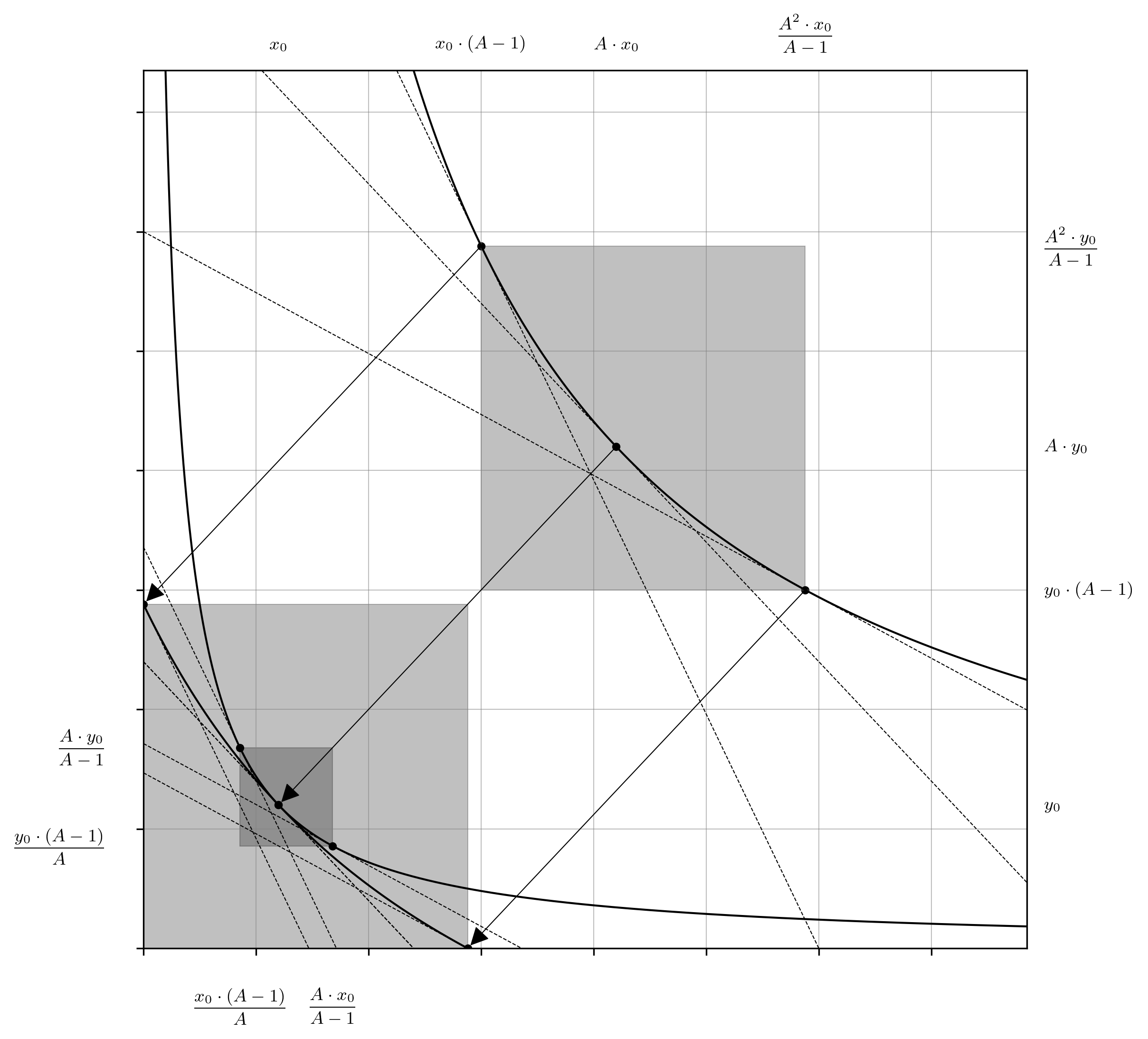}
    \captionsetup{
        justification=raggedright,
        singlelinecheck=false,
        font=small,
        labelfont=bf,
        labelsep=quad,
        format=plain
    }
    \caption{Construction of the real concentrated liquidity curve is depicted as a map of the virtual curve back to the x- and y-axes.}
    \label{fig16}
\end{figure}

\begin{figure}[ht]
    \centering
    \includegraphics[width=\textwidth]{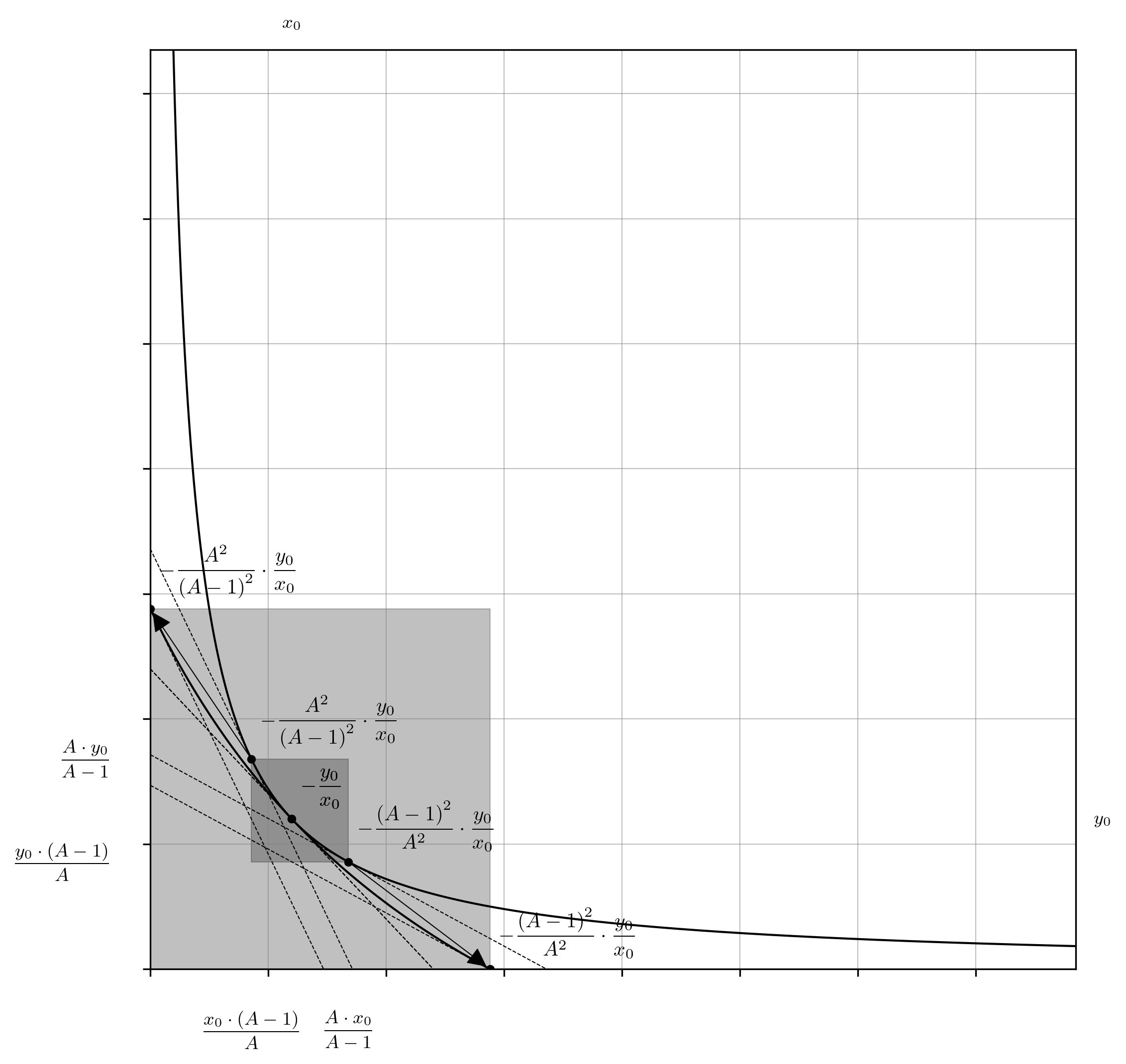}
    \captionsetup{
        justification=raggedright,
        singlelinecheck=false,
        font=small,
        labelfont=bf,
        labelsep=quad,
        format=plain
    }
    \caption{Construction of the real concentrated liquidity curve is depicted as a map of the reference curve section to new coordinates such that the previously defined price boundaries occur at the x- and y-intercepts, the geometric center remains stationary, and the marginal exchange rate at its coordinates remains unchanged. }
    \label{fig17}
\end{figure}

The objective of this part of the exercise is to translate the virtual invariant into a form where there is no recourse to virtual token balances. That is, to derive the real concentrated liquidity invariant which behaves identically to the virtual curve, but which correctly reports its own true token balances. This can be interpreted visually as a second mapping process which drives the boundaries of the virtual curve back to the x- and y-axes (Figure \ref{fig16}). Alternatively (and equivalently), this construction can also be interpreted as a direct mapping of the previously identified points and prices on the reference curve back to the x- and y-axes such that the geometric center remains stationary, and such that the marginal price at this point remains unchanged during the transformation (Figure \ref{fig17}).

Thankfully, this part is relatively easy. Introduction of horizontal, $H$, and vertical, $V$, shift parameters into the virtual invariant equation (Equation \ref{eq20}) yields Equation \ref{eq73}. These shifts are equal to the gap between the x- and y-axes, and the shaded areas representing the bounds of the virtual curve in Figure \ref{fig16}. Therefore, the horizontal shift is equal to $\min \left( x_{\text{v}} \right)$, and the vertical shift is equal to $\min \left( y_{\text{v}} \right)$; substitution of the algebraic identities for these terms (Equations \ref{eq30} and \ref{eq32}) into $H$ and $V$ in Equation \ref{eq73} yields the real concentrated liquidity invariant (Equation \ref{eq74}).

\begin{flalign}
& \text{\renewcommand{\arraystretch}{0.66}
    \begin{tabular}{@{}c@{}}
    \scriptsize from \\
    \scriptsize (\ref{eq20})
  \end{tabular}} 
  & 
  (x + H) \cdot (y + V) = A^{2} \cdot x_{0} \cdot y_{0}
  &  
  \label{eq73} 
  &
\end{flalign}

\begin{flalign}
& \text{\renewcommand{\arraystretch}{0.66}
    \begin{tabular}{@{}c@{}}
    \scriptsize from \\
    \scriptsize (\ref{eq30})\\\scriptsize (\ref{eq32})\\\scriptsize (\ref{eq73})
  \end{tabular}} 
  & 
  \left( x + x_{0} \cdot \left( A - 1 \right) \right) \cdot \left( y + y_{0} \cdot \left( A - 1 \right) \right) = A^{2} \cdot x_{0} \cdot y_{0}
  &  
  \label{eq74} 
  &
\end{flalign}

Algebraic manipulation of the real invariant (Equation \ref{eq74}) to derive the token swap identities, $\mathrm{\Delta}x$ and $\mathrm{\Delta}y$, and the marginal rate equation is performed as follows. First, the expressions isolating $x$ and $y$ are derived (Equations \ref{eq75} and \ref{eq76}).

\begin{flalign}
& \text{\renewcommand{\arraystretch}{0.66}
    \begin{tabular}{@{}c@{}}
    \scriptsize from \\
    \scriptsize (\ref{eq74})
  \end{tabular}} 
  & 
  x = \displaystyle \frac{A^{2} \cdot x_{0} \cdot y_{0}}{y + y_{0} \cdot \left( A - 1 \right)} - x_{0} \cdot \left( A - 1 \right)
  &  
  \label{eq75} 
  &
\end{flalign}

\begin{flalign}
& \text{\renewcommand{\arraystretch}{0.66}
    \begin{tabular}{@{}c@{}}
    \scriptsize from \\
    \scriptsize (\ref{eq74})
  \end{tabular}} 
  & 
  y = \displaystyle \frac{A^{2} \cdot x_{0} \cdot y_{0}}{x + x_{0} \cdot \left( A - 1 \right)} - y_{0} \cdot \left( A - 1 \right)
  &  
  \label{eq76} 
  &
\end{flalign}

The token swap equations are constructed as before; the $x$ and $y$ terms in Equation \ref{eq74} or Equations \ref{eq75} and \ref{eq76} are substituted for $x + \mathrm{\Delta}x$ and $y + \mathrm{\Delta}y$ (Equations \ref{eq77}, \ref{eq78} and \ref{eq79}), then rearranged to make either $\mathrm{\Delta}x$ or $\mathrm{\Delta}y$ the subject. 

\begin{flalign}
& \text{\renewcommand{\arraystretch}{0.66}
    \begin{tabular}{@{}c@{}}
    \scriptsize from \\
    \scriptsize (\ref{eq74})
  \end{tabular}} 
  & 
  \left( x + \mathrm{\Delta}x + x_{0} \cdot \left( A - 1 \right) \right) \cdot \left( y + \mathrm{\Delta}y + y_{0} \cdot \left( A - 1 \right) \right) = A^{2} \cdot x_{0} \cdot y_{0}
  &  
  \label{eq77} 
  &
\end{flalign}

\begin{flalign}
& \text{\renewcommand{\arraystretch}{0.66}
    \begin{tabular}{@{}c@{}}
    \scriptsize from \\
    \scriptsize (\ref{eq75})\\\scriptsize (\ref{eq77})
  \end{tabular}} 
  & 
  \mathrm{\Delta}x = \displaystyle \frac{A^{2} \cdot x_{0} \cdot y_{0}}{y + \mathrm{\Delta}y + y_{0} \cdot \left( A - 1 \right)} - x_{0} \cdot \left( A - 1 \right) - x
  &  
  \label{eq78} 
  &
\end{flalign}

\begin{flalign}
& \text{\renewcommand{\arraystretch}{0.66}
    \begin{tabular}{@{}c@{}}
    \scriptsize from \\
    \scriptsize (\ref{eq76})\\\scriptsize (\ref{eq77})
  \end{tabular}} 
  & 
  \mathrm{\Delta}y = \displaystyle \frac{A^{2} \cdot x_{0} \cdot y_{0}}{x + \mathrm{\Delta}x + x_{0} \cdot \left( A - 1 \right)} - y_{0} \cdot \left( A - 1 \right) - y
  &  
  \label{eq79} 
  &
\end{flalign}

To reduce the number of dimensions, the identities for $x$ and $y$ in Equations \ref{eq75} and \ref{eq76} are substituted into Equations \ref{eq78} and \ref{eq79}, respectively (Equations \ref{eq80} and \ref{eq82}). Combining the minuends and subtrahends into a single fraction yields Equations \ref{eq81} and \ref{eq83}.

\begin{flalign}
& \text{\renewcommand{\arraystretch}{0.66}
    \begin{tabular}{@{}c@{}}
    \scriptsize from \\
    \scriptsize (\ref{eq75})\\\scriptsize (\ref{eq78})
  \end{tabular}} 
  & 
  \mathrm{\Delta}x = \displaystyle \frac{A^{2} \cdot x_{0} \cdot y_{0}}{y + \mathrm{\Delta}y + y_{0} \cdot \left( A - 1 \right)} - \displaystyle \frac{A^{2} \cdot x_{0} \cdot y_{0}}{y + y_{0} \cdot \left( A - 1 \right)}
  &  
  \label{eq80} 
  &
\end{flalign}

\begin{flalign}
& \text{\renewcommand{\arraystretch}{0.66}
    \begin{tabular}{@{}c@{}}
    \scriptsize from \\
    \scriptsize (\ref{eq80})
  \end{tabular}} 
  & 
  \mathrm{\Delta}x = - \displaystyle \frac{\mathrm{\Delta}y \cdot A^{2} \cdot x_{0} \cdot y_{0}}{\left( y + y_{0} \cdot \left( A - 1 \right) \right) \cdot \left( y + \mathrm{\Delta}y + y_{0} \cdot \left( A - 1 \right) \right)}
  &  
  \label{eq81} 
  &
\end{flalign}

\begin{flalign}
& \text{\renewcommand{\arraystretch}{0.66}
    \begin{tabular}{@{}c@{}}
    \scriptsize from \\
    \scriptsize (\ref{eq76})\\\scriptsize (\ref{eq79})
  \end{tabular}} 
  & 
  \mathrm{\Delta}y = \displaystyle \frac{A^{2} \cdot x_{0} \cdot y_{0}}{x + \mathrm{\Delta}x + x_{0} \cdot \left( A - 1 \right)} - \displaystyle \frac{A^{2} \cdot x_{0} \cdot y_{0}}{x + x_{0} \cdot \left( A - 1 \right)}
  &  
  \label{eq82} 
  &
\end{flalign}

\begin{flalign}
& \text{\renewcommand{\arraystretch}{0.66}
    \begin{tabular}{@{}c@{}}
    \scriptsize from \\
    \scriptsize (\ref{eq82})
  \end{tabular}} 
  & 
  \mathrm{\Delta}y = - \displaystyle \frac{\mathrm{\Delta}x \cdot A^{2} \cdot x_{0} \cdot y_{0}}{\left( x + x_{0} \cdot \left( A - 1 \right) \right) \cdot \left( x + \mathrm{\Delta}x + x_{0} \cdot \left( A - 1 \right) \right)}
  &  
  \label{eq83} 
  &
\end{flalign}

As done previously for Equations \ref{eq10} and \ref{eq11}, the invariant term $A^{2} \cdot x_{0} \cdot y_{0}$ in Equations \ref{eq81} and \ref{eq83} can be substituted for its identity, $\left( x + x_{0} \cdot \left( A - 1\right) \right) \cdot \left( y + y_{0} \cdot \left( A - 1\right) \right)$ (i.e. the LHS of Equation \ref{eq74}). This process yields the familiar two-dimensional swap equations which depend on both the $x$ and $y$ coordinates (Equations \ref{eq338} and \ref{eq339}). 

\begin{flalign}
& \text{\renewcommand{\arraystretch}{0.66}
    \begin{tabular}{@{}c@{}}
    \scriptsize from \\
    \scriptsize (\ref{eq81})
  \end{tabular}} 
  & 
  \mathrm{\Delta}x = - \displaystyle \frac{\mathrm{\Delta}y \cdot \left( x + x_{0} \cdot \left( A - 1\right) \right)}{ y + \mathrm{\Delta}y + y_{0} \cdot \left( A - 1\right) }  
  &  
  \label{eq338} 
  &
\end{flalign}

\begin{flalign}
& \text{\renewcommand{\arraystretch}{0.66}
    \begin{tabular}{@{}c@{}}
    \scriptsize from \\
    \scriptsize (\ref{eq83})
  \end{tabular}} 
  & 
  \mathrm{\Delta}y = - \displaystyle \frac{\mathrm{\Delta}x \cdot \left( y + y_{0} \cdot \left( A - 1\right) \right)}{ x + \mathrm{\Delta}x + x_{0} \cdot \left( A - 1\right) }  
  &  
  \label{eq339} 
  &
\end{flalign}

Again, the marginal price equations can be derived in several ways. Repeating the process demonstrated in Equations \ref{eq12}, \ref{eq13}, \ref{eq14} and \ref{eq15}, rearrangement of Equations \ref{eq81} and \ref{eq83} to get the \textit{effective rate of exchange}, followed by determination of the limit as the denominator goes to zero gives the \textit{instantaneous rate of exchange}. It should be apparent that direct differentiation of Equations \ref{eq75} and \ref{eq76} via the quotient rule yields the same results. 

\begin{flalign}
& \text{\renewcommand{\arraystretch}{0.66}
    \begin{tabular}{@{}c@{}}
    \scriptsize from \\
    \scriptsize (\ref{eq81})
  \end{tabular}} 
  & 
  \displaystyle \frac{\mathrm{\Delta}x}{\mathrm{\Delta}y} = - \displaystyle \frac{A^{2} \cdot x_{0} \cdot y_{0}}{\left( y + y_{0} \cdot \left( A - 1 \right) \right) \cdot \left( y + \mathrm{\Delta}y + y_{0} \cdot \left( A - 1 \right) \right)}
  &  
  \label{eq84} 
  &
\end{flalign}

\begin{flalign}
& \text{\renewcommand{\arraystretch}{0.66}
    \begin{tabular}{@{}c@{}}
    \scriptsize from \\
    \scriptsize (\ref{eq84})
  \end{tabular}} 
  & 
  \displaystyle \frac{\partial x}{\partial y} = \lim_{\mathrm{\Delta}y \rightarrow 0}\displaystyle \frac{\mathrm{\Delta}x}{\mathrm{\Delta}y} = - \displaystyle \frac{A^{2} \cdot x_{0} \cdot y_{0}}{\left( y + y_{0} \cdot \left( A - 1 \right) \right)^{2}};\ \displaystyle \frac{\partial y}{\partial x} = - \displaystyle \frac{\left( y + y_{0} \cdot \left( A - 1 \right) \right)^{2}}{A^{2} \cdot x_{0} \cdot y_{0}}
  &  
  \label{eq85} 
  &
\end{flalign}

\begin{flalign}
& \text{\renewcommand{\arraystretch}{0.66}
    \begin{tabular}{@{}c@{}}
    \scriptsize from \\
    \scriptsize (\ref{eq83})
  \end{tabular}} 
  & 
  \displaystyle \frac{\mathrm{\Delta}y}{\mathrm{\Delta}x} = - \displaystyle \frac{A^{2} \cdot x_{0} \cdot y_{0}}{\left( x + x_{0} \cdot \left( A - 1 \right) \right) \cdot \left( x + \mathrm{\Delta}x + x_{0} \cdot \left( A - 1 \right) \right)}
  &  
  \label{eq86} 
  &
\end{flalign}

\begin{flalign}
& \text{\renewcommand{\arraystretch}{0.66}
    \begin{tabular}{@{}c@{}}
    \scriptsize from \\
    \scriptsize (\ref{eq86})
  \end{tabular}} 
  & 
  \displaystyle \frac{\partial y}{\partial x} = \lim_{\mathrm{\Delta}x \rightarrow 0}\displaystyle \frac{\mathrm{\Delta}y}{\mathrm{\Delta}x} = - \displaystyle \frac{A^{2} \cdot x_{0} \cdot y_{0}}{\left( x + x_{0} \cdot \left( A - 1 \right) \right)^{2}};\ \displaystyle \frac{\partial x}{\partial y} = - \displaystyle \frac{\left( x + x_{0} \cdot \left( A - 1 \right) \right)^{2}}{A^{2} \cdot x_{0} \cdot y_{0}}
  &  
  \label{eq87} 
  &
\end{flalign}

Lastly, the marginal rate equations can also be expressed in terms of both token balances via substitution of the $x$ and $y$ terms in Equations \ref{eq13} and \ref{eq15} with their horizontally- or vertically-shifted transformations (Equation \ref{eq88}). If that seems like a leap, the truth of this identity can also be proved by substituting the $A^{2} \cdot x_{0} \cdot y_{0}$ term in Equations \ref{eq85} and \ref{eq87} for its identity, the LHS of Equation \ref{eq74}.

\begin{flalign}
& \text{\renewcommand{\arraystretch}{0.66}
    \begin{tabular}{@{}c@{}}
    \scriptsize from \\
    \scriptsize (\ref{eq74})\\\scriptsize (\ref{eq85})\\\scriptsize (\ref{eq87})
  \end{tabular}} 
  & 
  \displaystyle \frac{\partial x}{\partial y} = - \displaystyle \frac{x + x_{0} \cdot (A - 1)}{y + y_{0} \cdot (A - 1)};\ \displaystyle \frac{\partial y}{\partial x} = - \displaystyle \frac{y + y_{0} \cdot (A - 1)}{x + x_{0} \cdot (A - 1)}
  &  
  \label{eq88} 
  &
\end{flalign}

The token swap equations can now be independently derived, and further understood, from the continuous summation over the price curves. Explicit integration of Equations \ref{eq85} and \ref{eq87} over the interval representing the number of tokens being swapped yields results identical to Equations \ref{eq80}, \ref{eq81}, \ref{eq82} and \ref{eq83} (Equations \ref{eq89} and \ref{eq90}). The integration above $\partial y / \partial x = - A^{2} \cdot x_{0} \cdot y_{0} / \left( x + x_{0} \cdot \left( A -1 \right)^{2} \right)$ over the interval $x \rightarrow x + \mathrm{\Delta}x$ representing a token swap is depicted in due course.

\begin{flalign}
& \text{\renewcommand{\arraystretch}{0.66}
    \begin{tabular}{@{}c@{}}
    \scriptsize from \\
    \scriptsize (\ref{eq80})\\\scriptsize (\ref{eq81})\\\scriptsize (\ref{eq85})
  \end{tabular}} 
  & 
  \mathrm{\Delta}x = - \int_{y}^{y + \mathrm{\Delta}y}{\displaystyle \frac{A^{2} \cdot x_{0} \cdot y_{0}}{\left( y + y_{0} \cdot \left( A - 1 \right) \right)^{2}}} \cdot \partial y = \left\lbrack \displaystyle \frac{A^{2} \cdot x_{0} \cdot y_{0}}{y + y_{0} \cdot (A - 1)} \right\rbrack_{y}^{y + \mathrm{\Delta}y} \Rightarrow \mathbf{Eqns.}\ \mathbf{86\ }and\mathbf{\ 87}
  &  
  \label{eq89} 
  &
\end{flalign}

\begin{flalign}
& \text{\renewcommand{\arraystretch}{0.66}
    \begin{tabular}{@{}c@{}}
    \scriptsize from \\
    \scriptsize (\ref{eq82})\\\scriptsize (\ref{eq83})\\\scriptsize (\ref{eq87})
  \end{tabular}} 
  & 
  \mathrm{\Delta}y = - \int_{x}^{x + \mathrm{\Delta}x}{\displaystyle \frac{A^{2} \cdot x_{0} \cdot y_{0}}{\left( x + x_{0} \cdot \left( A - 1 \right) \right)^{2}}} \cdot \partial x = \left\lbrack \displaystyle \frac{A^{2} \cdot x_{0} \cdot y_{0}}{x + x_{0} \cdot (A - 1)} \right\rbrack_{x}^{x + \mathrm{\Delta}x} \Rightarrow \mathbf{Eqns.}\ \mathbf{88\ }and\mathbf{\ 89}
  &  
  \label{eq90} 
  &
\end{flalign}

Be reminded that since $x$ and $y$ are dependent on each other, trying to obtain the token swap equation from direct integration of Equation \ref{eq88} requires separation of the $x$ and $y$ variables first (Equations \ref{eq340} and \ref{eq341}). Evaluation of the definite integrals over $x \rightarrow x + \mathrm{\Delta}x$ and $y \rightarrow y + \mathrm{\Delta}y$ (Equations \ref{eq342}, \ref{eq343} and \ref{eq344}) followed by rearrangement of the resulting expression to isolate either $\mathrm{\Delta}x$ or $\mathrm{\Delta}y$ gives the familiar two-dimensional swap equations (Equations \ref{eq345} and \ref{eq346}), which are redundant with those previously obtained (Equations \ref{eq338} and \ref{eq339}).  

\begin{flalign}
& \text{\renewcommand{\arraystretch}{0.66}
    \begin{tabular}{@{}c@{}}
    \scriptsize from \\
    \scriptsize (\ref{eq88})
  \end{tabular}} 
  & 
  \displaystyle \frac{1}{y + y_{0} \cdot \left( A - 1\right)} \cdot \partial{y} = - \displaystyle \frac{1}{x + x_{0} \cdot \left( A - 1\right)} \cdot \partial{x} 
  &  
  \label{eq340} 
  &
\end{flalign}

\begin{flalign}
& \text{\renewcommand{\arraystretch}{0.66}
    \begin{tabular}{@{}c@{}}
    \scriptsize from \\
    \scriptsize (\ref{eq340})
  \end{tabular}} 
  & 
  \int_{y}^{y + \mathrm{\Delta}y} \displaystyle \frac{1}{y + y_{0} \cdot \left( A - 1\right)} \cdot \partial y = - \int_{x}^{x + \mathrm{\Delta}x} \displaystyle \frac{1}{x + x_{0} \cdot \left( A - 1\right)} \cdot \partial x 
  &  
  \label{eq341} 
  &
\end{flalign}

\begin{flalign}
& \text{\renewcommand{\arraystretch}{0.66}
    \begin{tabular}{@{}c@{}}
    \scriptsize from \\
    \scriptsize (\ref{eq341})
  \end{tabular}} 
  & 
  \left\lbrack \ln \left( y + y_{0} \cdot \left( A - 1\right) \right) \right\rbrack_{y}^{y + \mathrm{\Delta}y} = - \left\lbrack \ln \left( x + x_{0} \cdot \left( A - 1\right) \right) \right\rbrack_{x}^{x + \mathrm{\Delta}x}  
  &  
  \label{eq342} 
  &
\end{flalign}

\begin{flalign}
& \text{\renewcommand{\arraystretch}{0.66}
    \begin{tabular}{@{}c@{}}
    \scriptsize from \\
    \scriptsize (\ref{eq342})
  \end{tabular}} 
  & 
  \ln \left( \displaystyle \frac{y + \mathrm{\Delta}y + y_{0} \cdot \left( A - 1\right)}{y + y_{0} \cdot \left( A - 1\right)} \right) = \ln \left( \displaystyle \frac{x + x_{0} \cdot \left( A - 1\right)}{x + \mathrm{\Delta}x + x_{0} \cdot \left( A - 1\right)} \right)  
  &  
  \label{eq343} 
  &
\end{flalign}

\begin{flalign}
& \text{\renewcommand{\arraystretch}{0.66}
    \begin{tabular}{@{}c@{}}
    \scriptsize from \\
    \scriptsize (\ref{eq343})
  \end{tabular}} 
  & 
  \displaystyle \frac{y + \mathrm{\Delta}y + y_{0} \cdot \left( A - 1\right)}{y + y_{0} \cdot \left( A - 1\right)} = \displaystyle \frac{x + x_{0} \cdot \left( A - 1\right)}{x + \mathrm{\Delta}x + x_{0} \cdot \left( A - 1\right)}  
  &  
  \label{eq344} 
  &
\end{flalign}

\begin{flalign}
& \text{\renewcommand{\arraystretch}{0.66}
    \begin{tabular}{@{}c@{}}
    \scriptsize from \\
    \scriptsize (\ref{eq338})\\\scriptsize (\ref{eq344})
  \end{tabular}} 
  & 
  \mathrm{\Delta}x = - \left( x + x_{0} \cdot \left( A - 1\right) \right) \cdot \left( 1 - \displaystyle \frac{y + y_{0} \cdot \left( A - 1\right)}{ y + \mathrm{\Delta}y + y_{0} \cdot \left( A - 1\right)} \right) \Rightarrow \mathbf{Eqn.\ 90} 
  &  
  \label{eq345} 
  &
\end{flalign}

\begin{flalign}
& \text{\renewcommand{\arraystretch}{0.66}
    \begin{tabular}{@{}c@{}}
    \scriptsize from \\
    \scriptsize (\ref{eq339})\\\scriptsize (\ref{eq344})
  \end{tabular}} 
  & 
  \mathrm{\Delta}y = - \left( y + y_{0} \cdot \left( A - 1\right) \right) \cdot \left( 1 - \displaystyle \frac{x + x_{0} \cdot \left( A - 1\right)}{ x + \mathrm{\Delta}x + x_{0} \cdot \left( A - 1\right)} \right) \Rightarrow \mathbf{Eqn.\ 91} 
  &  
  \label{eq346} 
  &
\end{flalign}

\begin{figure}[ht]
    \centering
    \includegraphics[width=\textwidth]{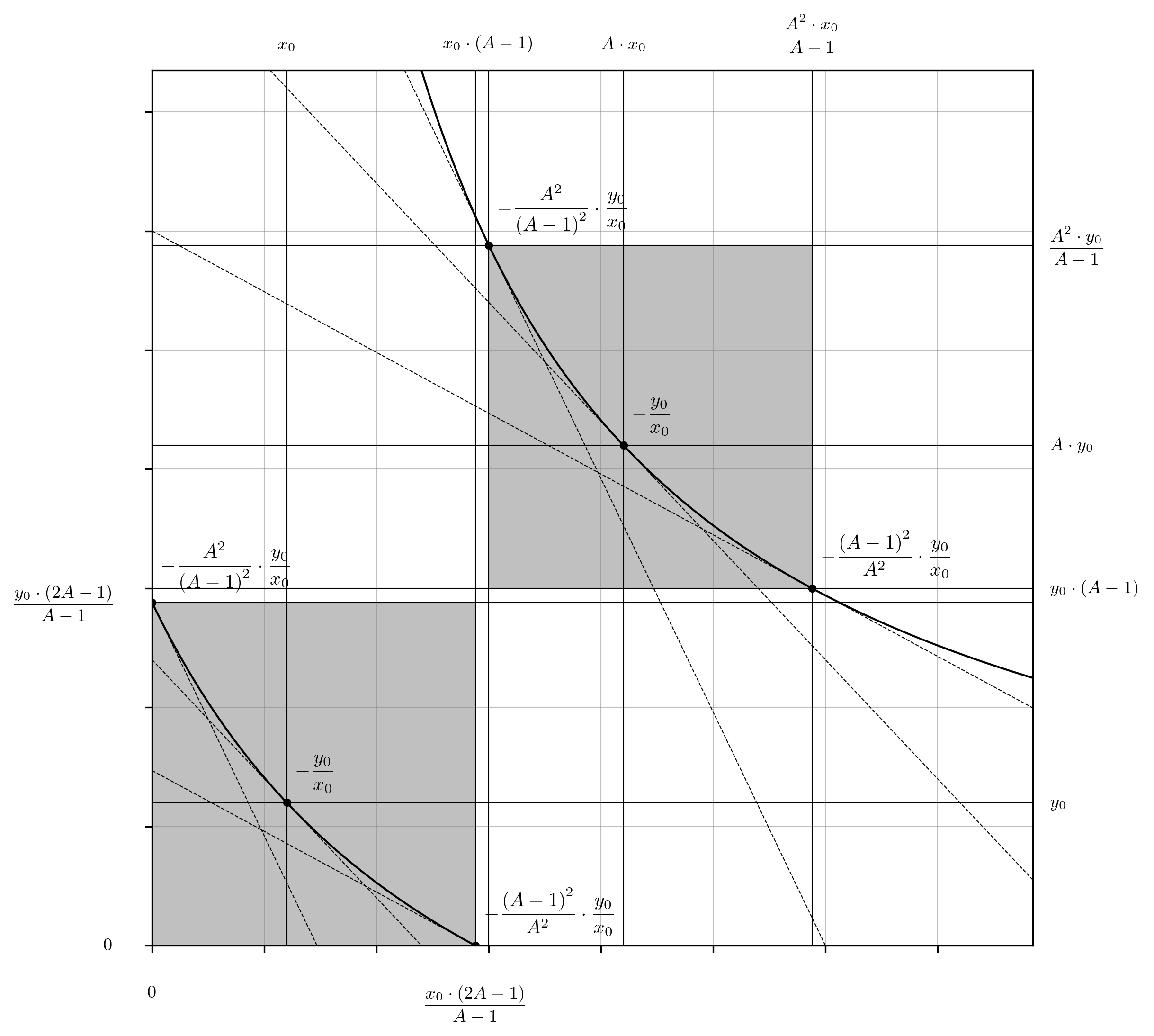}
    \captionsetup{
        justification=raggedright,
        singlelinecheck=false,
        font=small,
        labelfont=bf,
        labelsep=quad,
        format=plain
    }
    \caption{The x- and y-intercepts of the real curve (Equations \ref{eq91} and \ref{eq92}) are annotated, which completes its characterization. The real and virtual curves are depicted with the capstone algebraic identities elucidated thus far, including the value of the derivative evaluated at these points, illustrated with annotated tangent lines to the curve.}
    \label{fig18}
\end{figure}

Turning our attention back to the plot, the intercepts of the real curve (i.e. the points where it cuts either the x- or y-axes) can be determined easily (Figure \ref{fig18}). There are two convenient methods to achieve this. Substitution of $y$ and $x$ for zero in Equations \ref{eq75} and \ref{eq76}, respectively, services the same result as taking the difference between $\max \left( x_{\text{v}} \right)$ and $\min \left( x_{\text{v}} \right)$, and $\max \left( y_{\text{v}} \right)$ and $\min \left( y_{\text{v}} \right)$, respectively (Equations \ref{eq91} and \ref{eq92}). The latter approach is more geometrically intuitive given the construction thus far, but to each their own. 

\begin{flalign}
& \text{\renewcommand{\arraystretch}{0.66}
    \begin{tabular}{@{}c@{}}
    \scriptsize from \\
    \scriptsize (\ref{eq30})\\\scriptsize (\ref{eq33})\\\scriptsize (\ref{eq75})
  \end{tabular}} 
  & 
  x_{\text{int}} = \max\left( x_{\text{v}} \right) - \min\left( x_{\text{v}} \right) = \displaystyle \frac{A^{2} \cdot x_{0}}{A - 1} - x_{0} \cdot \left( A - 1 \right) = \displaystyle \frac{x_{0} \cdot (2A - 1)}{A - 1}
  &  
  \label{eq91} 
  &
\end{flalign}

\begin{flalign}
& \text{\renewcommand{\arraystretch}{0.66}
    \begin{tabular}{@{}c@{}}
    \scriptsize from \\
    \scriptsize (\ref{eq31})\\\scriptsize (\ref{eq32})\\\scriptsize (\ref{eq76})
  \end{tabular}} 
  & 
  y_{\text{int}} = \max\left( y_{\text{v}} \right) - \min\left( y_{\text{v}} \right) = \displaystyle \frac{A^{2} \cdot y_{0}}{A - 1} - y_{0} \cdot \left( A - 1 \right) = \displaystyle \frac{y_{0} \cdot (2A - 1)}{A - 1}
  &  
  \label{eq92} 
  &
\end{flalign}

\begin{figure}[ht]
    \centering
    \includegraphics[width=\textwidth]{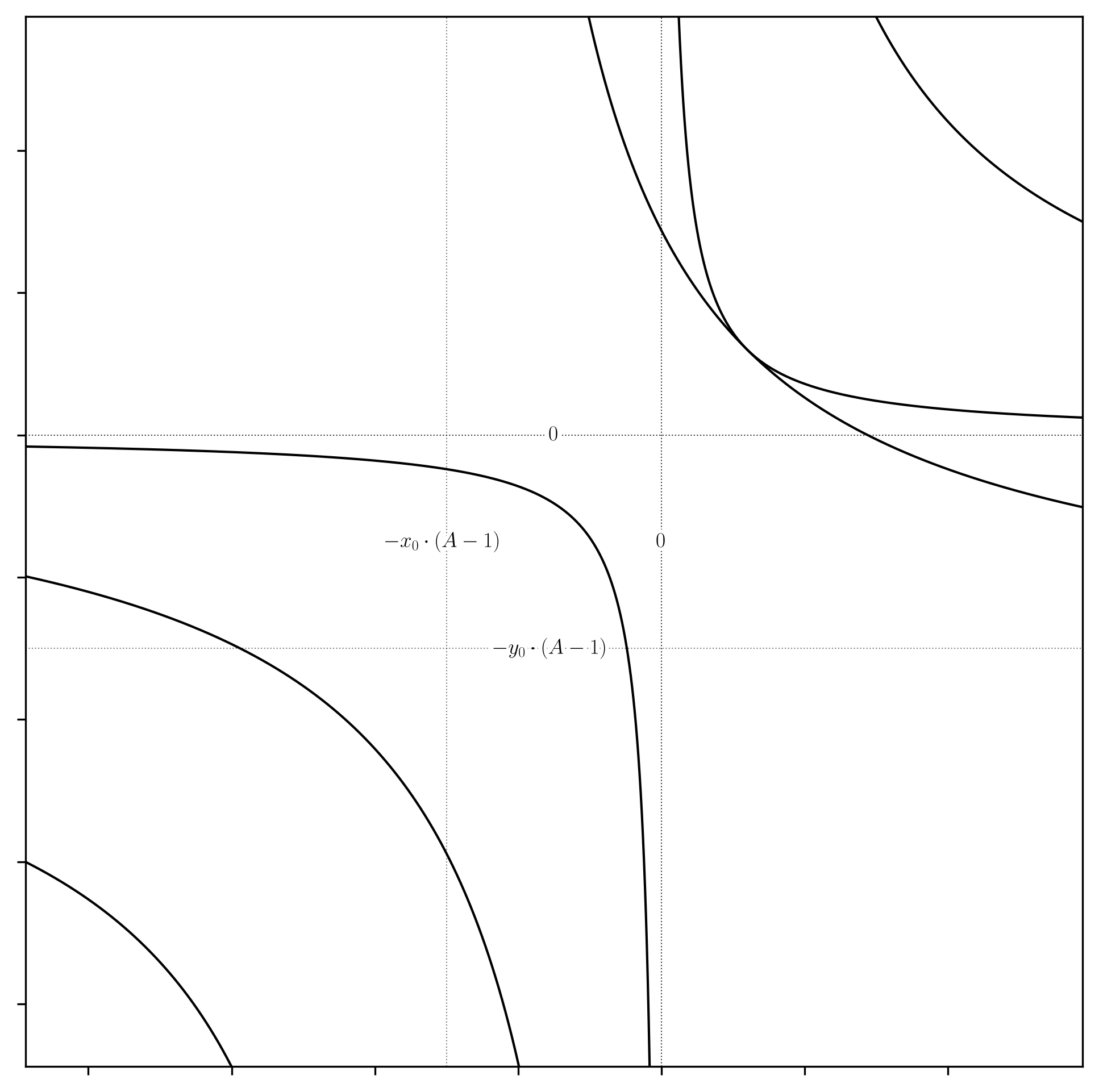}
    \captionsetup{
        justification=raggedright,
        singlelinecheck=false,
        font=small,
        labelfont=bf,
        labelsep=quad,
        format=plain
    }
    \caption{The horizontal and vertical asymptotes of the virtual (Equations \ref{eq93} and \ref{eq94}) and reference curves, and the real curve are depicted with dotted lines, and their coordinates are annotated appropriately.}
    \label{fig19}
\end{figure}

Since the real curve is shifted wholesale relative to the origin, so are its horizontal and vertical asymptotes. Again, the geometric intuition is to simply subtract from the origin the magnitude of the shift in the x- and y-dimensions, whereas the analytic approach is to substitute the y- or x-component of Equations \ref{eq75} and \ref{eq76} as required to set the denominator of the curve's equation to zero (Equations \ref{eq93} and \ref{eq94}). In either case you get the same result for about the same effort (Figure \ref{fig19}).

\begin{flalign}
& \text{\renewcommand{\arraystretch}{0.66}
    \begin{tabular}{@{}c@{}}
    \scriptsize from \\
    \scriptsize (\ref{eq30})\\\scriptsize (\ref{eq76})
  \end{tabular}} 
  & 
  x_{\text{asym}} = - x_{0} \cdot \left( A - 1 \right)
  &  
  \label{eq93} 
  &
\end{flalign}

\begin{flalign}
& \text{\renewcommand{\arraystretch}{0.66}
    \begin{tabular}{@{}c@{}}
    \scriptsize from \\
    \scriptsize (\ref{eq32})\\\scriptsize (\ref{eq75})
  \end{tabular}} 
  & 
  y_{\text{asym}} = - y_{0} \cdot \left( A - 1 \right)
  &  
  \label{eq94} 
  &
\end{flalign}

The quotient properties of $x_{\text{int}}$, $y_{\text{int}}$, $x_{\text{asym}}$ and $y_{\text{asym}}$ are preserved with respect to $P_{0}$ (Equations \ref{eq95} and \ref{eq96}), analogous to that previously observed for the virtual and reference curves (Equations \ref{eq49}, \ref{eq50}, \ref{eq69} and \ref{eq70}).

\begin{flalign}
& \text{\renewcommand{\arraystretch}{0.66}
    \begin{tabular}{@{}c@{}}
    \scriptsize from \\
    \scriptsize (\ref{eq91})\\\scriptsize (\ref{eq92})
  \end{tabular}} 
  & 
  \displaystyle \frac{y_{\text{int}}}{x_{\text{int}}} = \displaystyle \frac{\displaystyle \frac{y_{0} \cdot (2A - 1)}{A - 1}}{\displaystyle \frac{x_{0} \cdot (2A - 1)}{A - 1}} = \displaystyle \frac{y_{0}}{x_{0}} = P_{0}
  &  
  \label{eq95} 
  &
\end{flalign}

\begin{flalign}
& \text{\renewcommand{\arraystretch}{0.66}
    \begin{tabular}{@{}c@{}}
    \scriptsize from \\
    \scriptsize (\ref{eq93})\\\scriptsize (\ref{eq94})
  \end{tabular}} 
  & 
  \displaystyle \frac{y_{\text{asym}}}{x_{\text{asym}}} = \displaystyle \frac{- y_{0} \cdot \left( A - 1 \right)}{- x_{0} \cdot \left( A - 1 \right)} = \displaystyle \frac{y_{0}}{x_{0}} = P_{0}
  &  
  \label{eq96} 
  &
\end{flalign}

The geometric means of the significant x- and y-coordinates is also preserved, but the perspective needs to be generalized to be their distances from the asymptotes, instead of implicitly their distance relative to the origin (which was the case previously). While many of the manipulations in this document may be [justly] criticized for their mathematical redundancy, this one is especially so (Equations \ref{eq97}, \ref{eq98}, \ref{eq99} and \ref{eq100}). Measuring the distance of the intercepts and the origin from the asymptotes is literally the same as reversing the shift of the real curve back to the virtual curve, making these identities and the process of acquiring them identical to Equations \ref{eq45}, \ref{eq46}, \ref{eq47} and \ref{eq48}. The mysterious constant $C$ shows up again here for the same reason (Equations \ref{eq101} and \ref{eq102}). This part of the exercise is not entirely without merit, though; these symbolic representations provide a convenient handle for further algebraic abstraction when reparametrizing the curve, as will be discussed shortly.

\begin{flalign}
& \text{\renewcommand{\arraystretch}{0.66}
    \begin{tabular}{@{}c@{}}
    \scriptsize from \\
    \scriptsize (\ref{eq82})\\\scriptsize (\ref{eq84})
  \end{tabular}} 
  & 
  \left( x_{\text{int}} - x_{\text{asym}} \right) \cdot \left( 0 - x_{\text{asym}} \right) = \max\left( x_{\text{v}} \right) \cdot \min\left( x_{\text{v}} \right) = A^{2} \cdot x_{0}^{2}
  &  
  \label{eq97} 
  &
\end{flalign}

\begin{flalign}
& \text{\renewcommand{\arraystretch}{0.66}
    \begin{tabular}{@{}c@{}}
    \scriptsize from \\
    \scriptsize (\ref{eq97})
  \end{tabular}} 
  & 
  \sqrt{\left( x_{\text{int}} - x_{\text{asym}} \right)} \cdot \sqrt{\left( 0 - x_{\text{asym}} \right)} = \sqrt{\max\left( x_{\text{v}} \right)} \cdot \sqrt{\min\left( x_{\text{v}} \right)} = A \cdot x_{0}
  &  
  \label{eq98} 
  &
\end{flalign}

\begin{flalign}
& \text{\renewcommand{\arraystretch}{0.66}
    \begin{tabular}{@{}c@{}}
    \scriptsize from \\
    \scriptsize (\ref{eq83})\\\scriptsize (\ref{eq85})
  \end{tabular}} 
  & 
  \left( y_{\text{int}} - y_{\text{asym}} \right) \cdot \left( 0 - y_{\text{asym}} \right) = \max\left( y_{\text{v}} \right) \cdot \min\left( y_{\text{v}} \right) = A^{2} \cdot y_{0}^{2}
  &  
  \label{eq99} 
  &
\end{flalign}

\begin{flalign}
& \text{\renewcommand{\arraystretch}{0.66}
    \begin{tabular}{@{}c@{}}
    \scriptsize from \\
    \scriptsize (\ref{eq99})
  \end{tabular}} 
  & 
  \sqrt{\left( y_{\text{int}} - y_{\text{asym}} \right)} \cdot \sqrt{\left( 0 - y_{\text{asym}} \right)} = \sqrt{\max\left( y_{\text{v}} \right)} \cdot \sqrt{\min\left( y_{\text{v}} \right)} = A \cdot y_{0}
  &  
  \label{eq100} 
  &
\end{flalign}

\begin{flalign}
& \text{\renewcommand{\arraystretch}{0.66}
    \begin{tabular}{@{}c@{}}
    \scriptsize from \\
    \scriptsize (\ref{eq82})\\\scriptsize (\ref{eq84})
  \end{tabular}} 
  & 
  \displaystyle \frac{\left( x_{\text{int}} - x_{\text{asym}} \right)}{\left( 0 - x_{\text{asym}} \right)} = \displaystyle \frac{\max\left( x_{\text{v}} \right)}{\min\left( x_{\text{v}} \right)} = \displaystyle \frac{A^{2}}{\left( A - 1 \right)^{2}} = C
  &  
  \label{eq101} 
  &
\end{flalign}

\begin{flalign}
& \text{\renewcommand{\arraystretch}{0.66}
    \begin{tabular}{@{}c@{}}
    \scriptsize from \\
    \scriptsize (\ref{eq83})\\\scriptsize (\ref{eq85})
  \end{tabular}} 
  & 
  \displaystyle \frac{\left( y_{\text{int}} - y_{\text{asym}} \right)}{\left( 0 - y_{\text{asym}} \right)} = \displaystyle \frac{\max\left( y_{\text{v}} \right)}{\min\left( y_{\text{v}} \right)} = \displaystyle \frac{A^{2}}{\left( A - 1 \right)^{2}} = C
  &  
  \label{eq102} 
  &
\end{flalign}

\begin{figure}[ht]
    \centering
    \includegraphics[width=\textwidth]{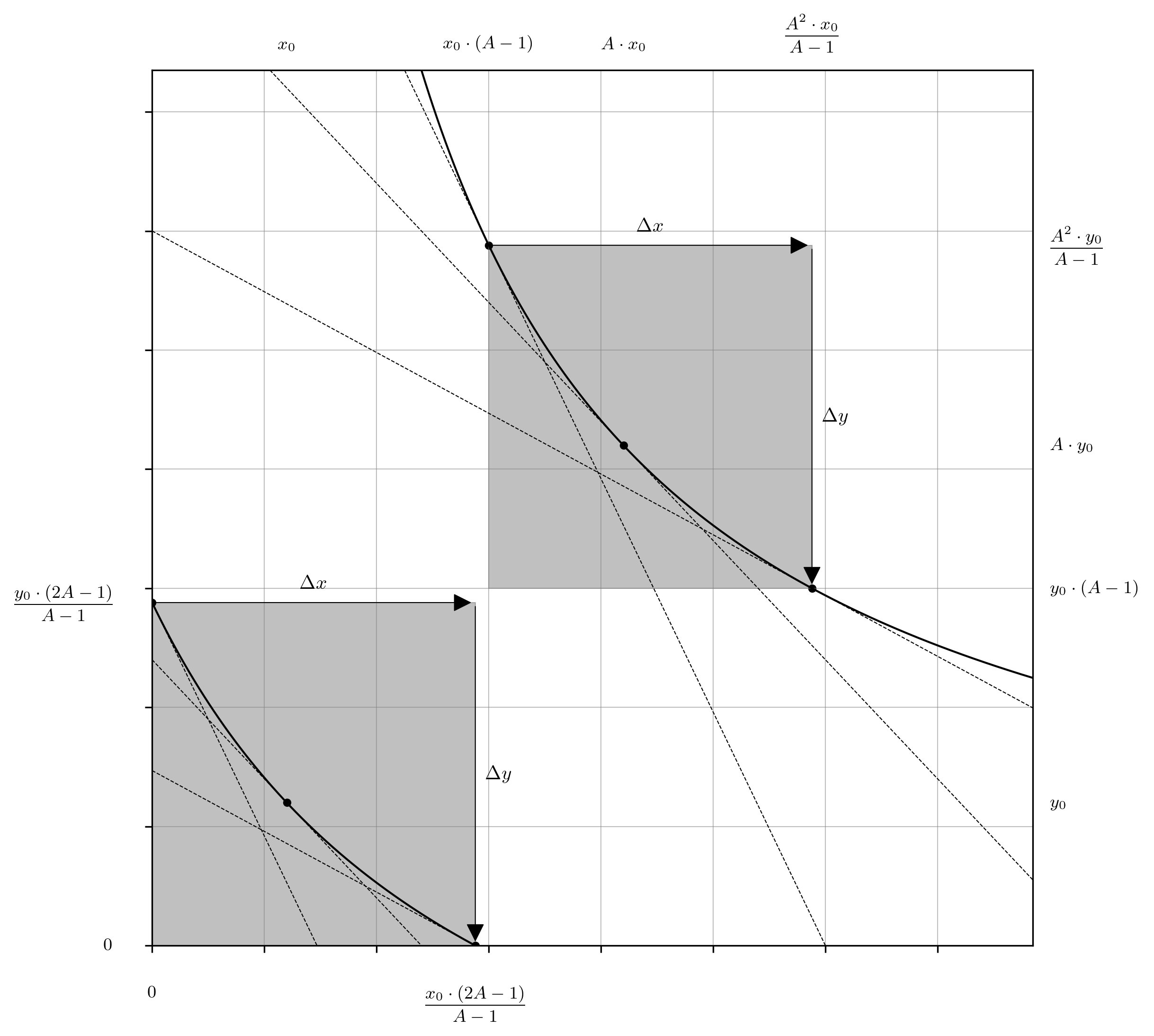}
    \captionsetup{
        justification=raggedright,
        singlelinecheck=false,
        font=small,
        labelfont=bf,
        labelsep=quad,
        format=plain
    }
    \caption{Traversal upon the rectangular hyperbolas $x_{\text{v}} \cdot y_{\text{v}} = A^{2} \cdot x_{0} \cdot y_{0}$ and $\left(x + x_{0} \cdot \left( A - 1 \right) \right) \cdot \left(y + y_{0} \cdot \left( A - 1 \right) \right) = A^{2} \cdot x_{0} \cdot y_{0}$ (Equations \ref{eq20} and \ref{eq74}), representing a token swap against the virtual and real depictions of the amplified liquidity pools, where $\mathrm{\Delta}x > 0$ and $\mathrm{\Delta}y < 0$. The swap quantities $\mathrm{\Delta}x$ and $\mathrm{\Delta}y$, and marginal rates of exchange before and after the swap are identical. Therefore, with respect to the outcome of the exchange the difference between the two curve implementations is nothing.}
    \label{fig20}
\end{figure}

Tying up loose ends, the same simulated token swap performed in Figures \ref{fig14} and \ref{fig15} is repeated for the real curve and compared with the virtual curve in Figures \ref{fig19} and \ref{fig20}. The important thing to note is that the change in the coordinates conventions makes zero difference with respect to the outcome of the swap. The magnitudes of the $\mathrm{\Delta}x$ and $\mathrm{\Delta}y$ arrows in Figure \ref{fig19}, and the integrated area in Figure \ref{fig20} are identical; only the position on the plane where the algorithm is performed has changed, which has no effect on either the marginal or effective rates of exchange. Therefore, they are equivalent in every way that matters to us. 

\begin{figure}[ht]
    \centering
    \includegraphics[width=\textwidth]{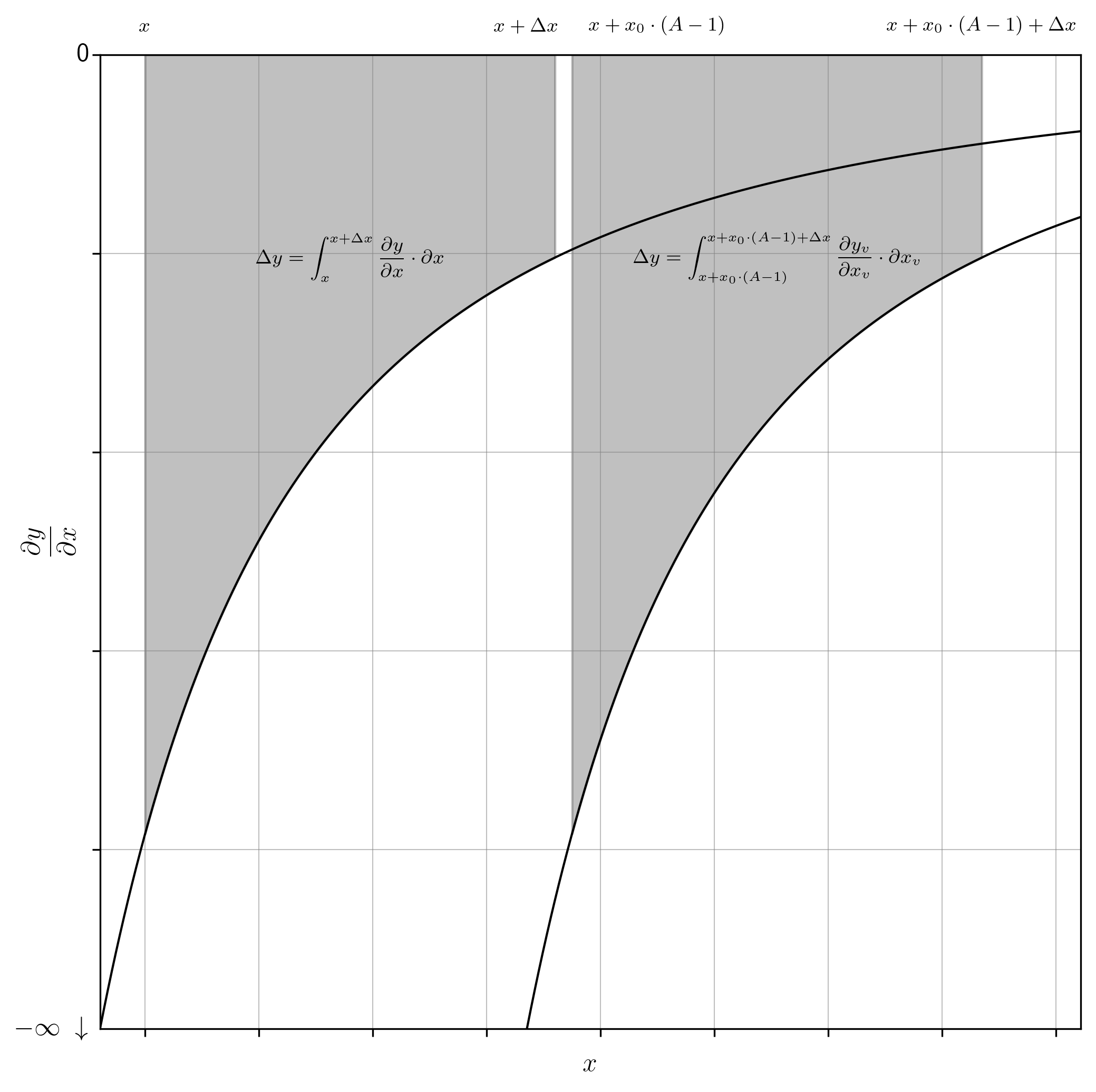}
    \captionsetup{
        justification=raggedright,
        singlelinecheck=false,
        font=small,
        labelfont=bf,
        labelsep=quad,
        format=plain
    }
    \caption{The integration above $\partial y_{\text{v}} / \partial x_{\text{v}} = - A^{2} \cdot x_{0} \cdot y_{0} / x_{\text{v}}^{2}$ and $\partial y / \partial x = - A^{2} \cdot x_{0} \cdot y_{0} / \left(x + x_{0} \cdot \left( A - 1 \right) \right)^{2}$ (Equations \ref{eq29} and \ref{eq87}) over the intervals $x \rightarrow x + \mathrm{\Delta}x$ and $x + x_{0} \cdot \left( A - 1 \right) \rightarrow x + x_{0} \cdot \left( A - 1 \right) + \mathrm{\Delta}x$ representing a token swap against both the real and virtual curves, where $\mathrm{\Delta}x > 0$ and $\mathrm{\Delta}y < 0$. Apart from the shift along the x-axis, these are identical in every other aspect. Note that whereas the relationship between the reference and virtual price curves in Figure \ref{fig15} was multiplicative ($x$ and $\mathrm{\Delta}x$, versus $A \cdot x$ and $A \cdot \left( x + \mathrm{\Delta}x\right)$, the relationship between the real and virtual price curves depicted here is additive ($x$ and $\mathrm{\Delta}x$ versus $x + x_{0} \cdot \left( A - 1 \right)$ and $x + x_{0} \cdot \left( A - 1 \right) + \mathrm{\Delta}x$).}
    \label{fig21}
\end{figure}

\section{Other “Unnatural” Reparameterizations of the Concentrated Liquidity Invariant Function}\label{sec4}

\subsection{The Uniswap v3 Real Curve}\label{subsec4.1}

The most common reparameterization of the real concentrated liquidity invariant (Equation \ref{eq74}) is that popularized by Uniswap v3 in its whitepaper.\footnote{Adams, H.; Zinmeister, N.; Salem, M.; Keefer, R.; Robinson, D. Uniswap v3 Core uniswap.org/whitepaper-v3.pdf} To construct this form, first obtain the identities of $\left( A - 1 \right)$ in terms of the square roots of the price bounds via Equations \ref{eq37} and \ref{eq41} (Equations \ref{eq103}, \ref{eq104}, \ref{eq105} and \ref{eq106}). Then, substitute the obtained $\left( A - 1 \right)$ identities into the horizontal and vertical shift components of Equation \ref{eq74} as appropriate (Equation \ref{eq107}); the substitution pattern must be such that the coefficient of the shift ($x_{0}$ or $y_{0}$) simplifies to its square root ($\sqrt{x_{0}}$ or $\sqrt{y_{0}}$) as shown in Equation \ref{eq108}. Substitution of the constant term $A \cdot \sqrt{x_{0}} \cdot \sqrt{y_{0}}$ for $L$ (Equation \ref{eq109}) yields the Uniswap v3 real curve equation (Equation \ref{eq110}).

\begin{flalign}
& \text{\renewcommand{\arraystretch}{0.66}
    \begin{tabular}{@{}c@{}}
    \scriptsize from \\
    \scriptsize (\ref{eq37})
  \end{tabular}} 
  & 
  \sqrt{P_{\text{high}}} = \displaystyle \frac{A}{A - 1} \cdot \displaystyle \frac{\sqrt{y_{0}}}{\sqrt{x_{0}}}
  &  
  \label{eq103} 
  &
\end{flalign}

\begin{flalign}
& \text{\renewcommand{\arraystretch}{0.66}
    \begin{tabular}{@{}c@{}}
    \scriptsize from \\
    \scriptsize (\ref{eq103})
  \end{tabular}} 
  & 
  A - 1 = \displaystyle \frac{A}{\sqrt{P_{\text{high}}}} \cdot \displaystyle \frac{\sqrt{y_{0}}}{\sqrt{x_{0}}}
  &  
  \label{eq104} 
  &
\end{flalign}

\begin{flalign}
& \text{\renewcommand{\arraystretch}{0.66}
    \begin{tabular}{@{}c@{}}
    \scriptsize from \\
    \scriptsize (\ref{eq41})
  \end{tabular}} 
  & 
  \sqrt{P_{\text{low}}} = \displaystyle \frac{A - 1}{A} \cdot \displaystyle \frac{\sqrt{y_{0}}}{\sqrt{x_{0}}}
  &  
  \label{eq105} 
  &
\end{flalign}

\begin{flalign}
& \text{\renewcommand{\arraystretch}{0.66}
    \begin{tabular}{@{}c@{}}
    \scriptsize from \\
    \scriptsize (\ref{eq105})
  \end{tabular}} 
  & 
  A - 1 = A \cdot \sqrt{P_{\text{low}}} \cdot \displaystyle \frac{\sqrt{x_{0}}}{\sqrt{y_{0}}}
  &  
  \label{eq106} 
  &
\end{flalign}

\begin{flalign}
& \text{\renewcommand{\arraystretch}{0.66}
    \begin{tabular}{@{}c@{}}
    \scriptsize from \\
    \scriptsize (\ref{eq74})\\\scriptsize (\ref{eq104})\\\scriptsize (\ref{eq106})
  \end{tabular}} 
  & 
  \left( x + x_{0} \cdot \left( \displaystyle \frac{A}{\sqrt{P_{\text{high}}}} \cdot \displaystyle \frac{\sqrt{y_{0}}}{\sqrt{x_{0}}} \right) \right) \cdot \left( y + y_{0} \cdot \left( A \cdot \sqrt{P_{\text{low}}} \cdot \displaystyle \frac{\sqrt{x_{0}}}{\sqrt{y_{0}}} \right) \right) = A^{2} \cdot x_{0} \cdot y_{0}
  &  
  \label{eq107} 
  &
\end{flalign}

\begin{flalign}
& \text{\renewcommand{\arraystretch}{0.66}
    \begin{tabular}{@{}c@{}}
    \scriptsize from \\
    \scriptsize (\ref{eq104})
  \end{tabular}} 
  & 
  \left( x + \displaystyle \frac{A \cdot \sqrt{y_{0}} \cdot \sqrt{x_{0}}}{\sqrt{P_{\text{high}}}} \right) \cdot \left( y + A \cdot \sqrt{y_{0}} \cdot \sqrt{x_{0}} \cdot \sqrt{P_{\text{low}}} \right) = A^{2} \cdot x_{0} \cdot y_{0}
  &  
  \label{eq108} 
  &
\end{flalign}

\begin{flalign}
& \text{\renewcommand{\arraystretch}{0.66}
    \begin{tabular}{@{}c@{}}
    \scriptsize from \\
    \scriptsize (\ref{eq108})
  \end{tabular}} 
  & 
  A \cdot \sqrt{y_{0}} \cdot \sqrt{x_{0}} = L
  &  
  \label{eq109} 
  &
\end{flalign}

\begin{flalign}
& \text{\renewcommand{\arraystretch}{0.66}
    \begin{tabular}{@{}c@{}}
    \scriptsize from \\
    \scriptsize (\ref{eq108})\\\scriptsize (\ref{eq109})
  \end{tabular}} 
  & 
  \left( x + \displaystyle \frac{L}{\sqrt{P_{\text{high}}}} \right) \cdot \left( y + L \cdot \sqrt{P_{\text{low}}} \right) = L^{2}
  &  
  \label{eq110} 
  &
\end{flalign}

This process affords a new expression for the same mathematical object; fundamentally, Equations \ref{eq74} and \ref{eq110} are one and the same thing. However, this does not mean that they are equally useful given their context in smart contract implementations. For example, Equation \ref{eq110} conveniently lists its own price bounds in the invariant itself but requires one to perform a little extra work to deduce its reference coordinates and relative scaling. On the other hand, Equation \ref{eq74} reports its own reference coordinates and relative scaling at the expense of the price bounds. From an analytical perspective the difference is meaningless, but from an implementation perspective the difference could be important. In the case of Uniswap v3, the price bounds are commensurate with an overall smart contract architecture decision that requires these parameters (i.e. the tick boundaries) to be available and reliable. While on-the-fly calculations can recreate these parameters in situ, these operations add more computational overhead and rounding errors which may present gas efficiency issues, or even smart contract security concerns. Therefore, the expression used (Equation \ref{eq110}) reflects the overall systems design and serves as an abstraction of how to interpret and use the data one is likely to find stored there. 

This is an important discussion point, because the difference between well and poorly designed systems can come down to the choice of how to parameterize the equations that describe it, even if those descriptions are analytically redundant. In the case of Uniswap v3, Equation \ref{eq110} is in some sense the \textit{correct parameterization for that system}; however, it is a dangerous idea to consider it a catch all that should be used by any concentrated liquidity system. 

\begin{figure}[ht]
    \centering
    \includegraphics[width=\textwidth]{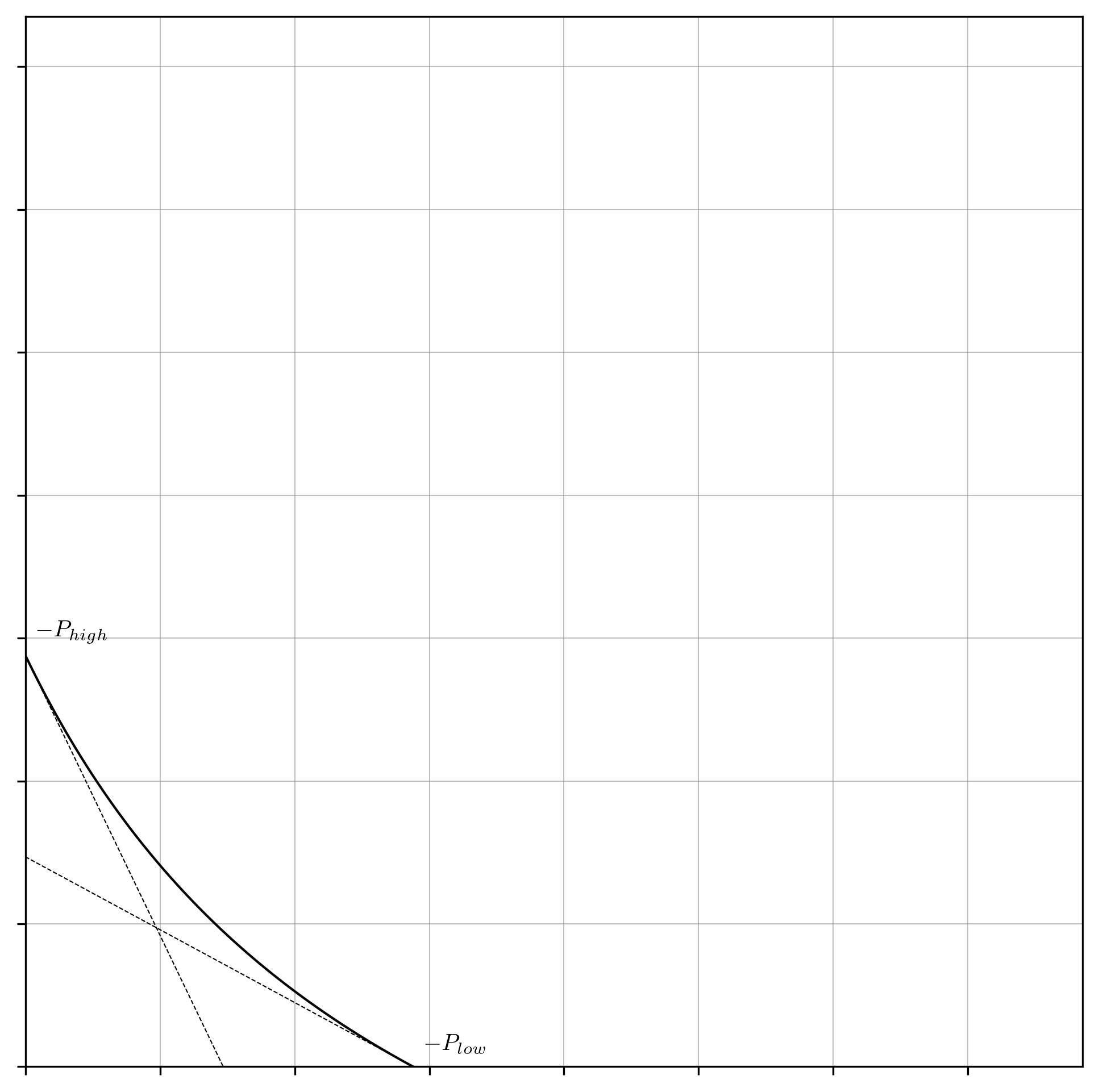}
    \captionsetup{
        justification=raggedright,
        singlelinecheck=false,
        font=small,
        labelfont=bf,
        labelsep=quad,
        format=plain
    }
    \caption{Initial Uniswap v3 real curve. Due to its parameterisation and prior to performing any additional analysis, only the slopes of the tangent lines at the intercepts are known ($-P_{\text{high}}$ and $-P_{\text{low}}$).}
    \label{fig22}
\end{figure}

The Uniswap v3 real curve (Equation \ref{eq110}) is plotted in Figure \ref{fig22} and annotated with the slopes of the tangent lines at the intercepts, which is the only information presently available at this stage of the analysis. 

Returning to the objective of this part of the exercise, we will now fully elucidate the reparametrized version of the model beginning from the Uniswap v3 invariant function (Equation \ref{eq110}), characterise the real, virtual, and reference curves, and re-contextualize them in terms of the general theory as outlined in the previous sections. Manipulation of the Uniswap v3 invariant (Equation \ref{eq110}) is performed using the same methods as before. First, the expressions isolating $x$ and $y$ are derived (Equations \ref{eq111} and \ref{eq112}).

\begin{flalign}
& \text{\renewcommand{\arraystretch}{0.66}
    \begin{tabular}{@{}c@{}}
    \scriptsize from \\
    \scriptsize (\ref{eq110})
  \end{tabular}} 
  & 
  x = \displaystyle \frac{L^{2}}{\left( y + L \cdot \sqrt{P_{\text{low}}} \right)} - \displaystyle \frac{L}{\sqrt{P_{\text{high}}}}
  &  
  \label{eq111} 
  &
\end{flalign}

\begin{flalign}
& \text{\renewcommand{\arraystretch}{0.66}
    \begin{tabular}{@{}c@{}}
    \scriptsize from \\
    \scriptsize (\ref{eq110})
  \end{tabular}} 
  & 
  y = \displaystyle \frac{L^{2}}{x + \displaystyle \frac{L}{\sqrt{P_{\text{high}}}}} - L \cdot \sqrt{P_{\text{low}}}
  &  
  \label{eq112} 
  &
\end{flalign}

Then, the token swap equations are constructed as normal; the $x$ and $y$ terms in Equation \ref{eq110} or Equations \ref{eq111} and \ref{eq112} are substituted for $x + \mathrm{\Delta}x$ and $y + \mathrm{\Delta}y$ then rearranged to make either $\mathrm{\Delta}x$ or $\mathrm{\Delta}y$ the subject (Equations \ref{eq113}, \ref{eq114} and \ref{eq115}). 

\begin{flalign}
& \text{\renewcommand{\arraystretch}{0.66}
    \begin{tabular}{@{}c@{}}
    \scriptsize from \\
    \scriptsize (\ref{eq110})
  \end{tabular}} 
  & 
  \left( x + \mathrm{\Delta}x + \displaystyle \frac{L}{\sqrt{P_{\text{high}}}} \right) \cdot \left( y + \mathrm{\Delta}y + L \cdot \sqrt{P_{\text{low}}} \right) = L^{2}
  &  
  \label{eq113} 
  &
\end{flalign}

\begin{flalign}
& \text{\renewcommand{\arraystretch}{0.66}
    \begin{tabular}{@{}c@{}}
    \scriptsize from \\
    \scriptsize (\ref{eq111})\\\scriptsize (\ref{eq113})
  \end{tabular}} 
  & 
  \mathrm{\Delta}x = \displaystyle \frac{L^{2}}{\left( y + \ \mathrm{\Delta}y + L \cdot \sqrt{P_{\text{low}}} \right)} - \displaystyle \frac{L}{\sqrt{P_{\text{high}}}} - x
  &  
  \label{eq114} 
  &
\end{flalign}

\begin{flalign}
& \text{\renewcommand{\arraystretch}{0.66}
    \begin{tabular}{@{}c@{}}
    \scriptsize from \\
    \scriptsize (\ref{eq112})\\\scriptsize (\ref{eq113})
  \end{tabular}} 
  & 
  \mathrm{\Delta}y = \displaystyle \frac{L^{2}}{x + \mathrm{\Delta}x + \displaystyle \frac{L}{\sqrt{P_{\text{high}}}}} - L \cdot \sqrt{P_{\text{low}}} - y
  &  
  \label{eq115} 
  &
\end{flalign}

Dimension reduction is achieved via the same process as previously demonstrated; the identities for $x$ and $y$ in Equations \ref{eq111} and \ref{eq112} are substituted into Equations \ref{eq114} and \ref{eq115}, respectively (Equations \ref{eq116} and \ref{eq118}). Simplifying these expressions into single fractions yields Equations \ref{eq117} and \ref{eq119}.

\begin{flalign}
& \text{\renewcommand{\arraystretch}{0.66}
    \begin{tabular}{@{}c@{}}
    \scriptsize from \\
    \scriptsize (\ref{eq111})\\\scriptsize (\ref{eq114})
  \end{tabular}} 
  & 
  \mathrm{\Delta}x = \displaystyle \frac{L^{2}}{\left( y + \ \mathrm{\Delta}y + L \cdot \sqrt{P_{\text{low}}} \right)} - \displaystyle \frac{L}{\sqrt{P_{\text{high}}}} - \displaystyle \frac{L^{2}}{\left( y + L \cdot \sqrt{P_{\text{low}}} \right)} + \displaystyle \frac{L}{\sqrt{P_{\text{high}}}}
  &  
  \label{eq116} 
  &
\end{flalign}

\begin{flalign}
& \text{\renewcommand{\arraystretch}{0.66}
    \begin{tabular}{@{}c@{}}
    \scriptsize from \\
    \scriptsize (\ref{eq113})
  \end{tabular}} 
  & 
  \mathrm{\Delta}x = - \displaystyle \frac{\mathrm{\Delta}y \cdot L^{2}}{\left( y + L \cdot \sqrt{P_{\text{low}}} \right) \cdot \left( y + \mathrm{\Delta}y + L \cdot \sqrt{P_{\text{low}}} \right)}
  &  
  \label{eq117} 
  &
\end{flalign}

\begin{flalign}
& \text{\renewcommand{\arraystretch}{0.66}
    \begin{tabular}{@{}c@{}}
    \scriptsize from \\
    \scriptsize (\ref{eq112})\\\scriptsize (\ref{eq115})
  \end{tabular}} 
  & 
  \mathrm{\Delta}y = \displaystyle \frac{L^{2}}{x + \mathrm{\Delta}x + \displaystyle \frac{L}{\sqrt{P_{\text{high}}}}} - L \cdot \sqrt{P_{\text{low}}} - \displaystyle \frac{L^{2}}{x + \displaystyle \frac{L}{\sqrt{P_{\text{high}}}}} + L \cdot \sqrt{P_{\text{low}}}
  &  
  \label{eq118} 
  &
\end{flalign}

\begin{flalign}
& \text{\renewcommand{\arraystretch}{0.66}
    \begin{tabular}{@{}c@{}}
    \scriptsize from \\
    \scriptsize (\ref{eq118})
  \end{tabular}} 
  & 
  \mathrm{\Delta}y = - \displaystyle \frac{\mathrm{\Delta}x \cdot L^{2}}{\left( x + \displaystyle \frac{L}{\sqrt{P_{\text{high}}}} \right) \cdot \left( x + \mathrm{\Delta}x + \displaystyle \frac{L}{\sqrt{P_{\text{high}}}} \right)}
  &  
  \label{eq119} 
  &
\end{flalign}

Note that Equations \ref{eq117} and \ref{eq119} can also be obtained directly from their predecessors, Equations \ref{eq81} and \ref{eq83}, by substituting the scaling term, $A^{2} \cdot x_{0} \cdot y_{0}$, and the horizontal and vertical shift terms, $x_{0} \cdot \left( A - 1 \right)$ and $y_{0} \cdot \left( A - 1 \right)$, for their implicit identities, $L^{2}$, $L / \sqrt{P_{\text{high}}}$, and $L \cdot \sqrt{P_{\text{low}}}$, respectively (Equations \ref{eq103}, \ref{eq104}, \ref{eq105}, \ref{eq106}, \ref{eq107}, \ref{eq108}, \ref{eq109} and \ref{eq110}). This substitution pattern can also be used to obtain the effective and marginal price equations, but for consistency with the previous work, these identities will be elaborated below according to the previously established methods. 

Repeating the process demonstrated in Equations \ref{eq12}, \ref{eq13}, \ref{eq14}, \ref{eq15} and Equations \ref{eq84}, \ref{eq85}, \ref{eq86}, \ref{eq87}, rearrangement of Equations \ref{eq114}, and \ref{eq116} to get the \textit{effective rate of exchange}, followed by determination of the limit as the denominator goes to zero gives the \textit{instantaneous rate of exchange} (i.e. the marginal price)(Equations \ref{eq120}, \ref{eq121}, \ref{eq122}, \ref{eq123}). 

\begin{flalign}
& \text{\renewcommand{\arraystretch}{0.66}
    \begin{tabular}{@{}c@{}}
    \scriptsize from \\
    \scriptsize (\ref{eq117})
  \end{tabular}} 
  & 
  \displaystyle \frac{\mathrm{\Delta}x}{\mathrm{\Delta}y} = - \displaystyle \frac{L^{2}}{\left( y + L \cdot \sqrt{P_{\text{low}}} \right) \cdot \left( y + \mathrm{\Delta}y + L \cdot \sqrt{P_{\text{low}}} \right)}
  &  
  \label{eq120} 
  &
\end{flalign}

\begin{flalign}
& \text{\renewcommand{\arraystretch}{0.66}
    \begin{tabular}{@{}c@{}}
    \scriptsize from \\
    \scriptsize (\ref{eq120})
  \end{tabular}} 
  & 
  \displaystyle \frac{\partial x}{\partial y} = \lim_{\mathrm{\Delta}y \rightarrow 0}\displaystyle \frac{\mathrm{\Delta}x}{\mathrm{\Delta}y} = - \displaystyle \frac{L^{2}}{\left( y + L \cdot \sqrt{P_{\text{low}}} \right)^{2}};\ \displaystyle \frac{\partial y}{\partial x} = - \displaystyle \frac{\left( y + L \cdot \sqrt{P_{\text{low}}} \right)^{2}}{L^{2}}
  &  
  \label{eq121} 
  &
\end{flalign}

\begin{flalign}
& \text{\renewcommand{\arraystretch}{0.66}
    \begin{tabular}{@{}c@{}}
    \scriptsize from \\
    \scriptsize (\ref{eq119})
  \end{tabular}} 
  & 
  \displaystyle \frac{\mathrm{\Delta}y}{\mathrm{\Delta}x} = - \displaystyle \frac{L^{2}}{\left( x + \displaystyle \frac{L}{\sqrt{P_{\text{high}}}} \right) \cdot \left( x + \mathrm{\Delta}x + \displaystyle \frac{L}{\sqrt{P_{\text{high}}}} \right)}
  &  
  \label{eq122} 
  &
\end{flalign}

\begin{flalign}
& \text{\renewcommand{\arraystretch}{0.66}
    \begin{tabular}{@{}c@{}}
    \scriptsize from \\
    \scriptsize (\ref{eq122})
  \end{tabular}} 
  & 
  \displaystyle \frac{\partial y}{\partial x} = \lim_{\mathrm{\Delta}x \rightarrow 0}\displaystyle \frac{\mathrm{\Delta}y}{\mathrm{\Delta}x} = - \displaystyle \frac{L^{2}}{\left( x + \displaystyle \frac{L}{\sqrt{P_{\text{high}}}} \right)^{2}};\ \displaystyle \frac{\partial x}{\partial y} = - \displaystyle \frac{\left( x + \displaystyle \frac{L}{\sqrt{P_{\text{high}}}} \right)^{2}}{L^{2}}
  &  
  \label{eq123} 
  &
\end{flalign}

The Uniswap v3 marginal rate equations can also be expressed in terms of both token balances via substitution of the $x$ and $y$ terms in Equations \ref{eq13}, and \ref{eq15} with their horizontally or vertically shifted transformations (Equation \ref{eq124}), or via substitution of the identity of $L^{2}$ from the LHS of Equation \ref{eq110} into Equations \ref{eq121}, and \ref{eq123}. Simplifying the fractions in Equation \ref{eq124} yields Equation \ref{eq125}.

\begin{flalign}
& \text{\renewcommand{\arraystretch}{0.66}
    \begin{tabular}{@{}c@{}}
    \scriptsize from \\
    \scriptsize (\ref{eq110})\\\scriptsize (\ref{eq121})\\\scriptsize (\ref{eq123})
  \end{tabular}} 
  & 
  \displaystyle \frac{\partial x}{\partial y} = - \displaystyle \frac{x + \displaystyle \frac{L}{\sqrt{P_{\text{high}}}}}{y + L \cdot \sqrt{P_{\text{low}}}};\ \displaystyle \frac{\partial y}{\partial x} = - \displaystyle \frac{y + L \cdot \sqrt{P_{\text{low}}}}{x + \displaystyle \frac{L}{\sqrt{P_{\text{high}}}}}
  &  
  \label{eq124} 
  &
\end{flalign}

\begin{flalign}
& \text{\renewcommand{\arraystretch}{0.66}
    \begin{tabular}{@{}c@{}}
    \scriptsize from \\
    \scriptsize (\ref{eq124})
  \end{tabular}} 
  & 
  \displaystyle \frac{\partial x}{\partial y} = - \displaystyle \frac{x \cdot \sqrt{P_{\text{high}}} + L}{\sqrt{P_{\text{high}}} \cdot \left( y + L \cdot \sqrt{P_{\text{low}}} \right)};\ \displaystyle \frac{\partial y}{\partial x} = - \displaystyle \frac{\sqrt{P_{\text{high}}} \cdot \left( y + L \cdot \sqrt{P_{\text{low}}} \right)}{x \cdot \sqrt{P_{\text{high}}} + L}
  &  
  \label{eq125} 
  &
\end{flalign}

Continuous summation over the price curves (Equations \ref{eq121} and \ref{eq123}) over the interval representing the number of tokens being swapped yields results identical to Equations \ref{eq116}, \ref{eq117}, \ref{eq118} and \ref{eq119}. Again, do not be tempted to ignore the dependence of $x$ and $y$ on each other and attempt an integration of Equations \ref{eq124} and \ref{eq125}, without separating these variables first (not shown). 

\begin{flalign}
& \text{\renewcommand{\arraystretch}{0.66}
    \begin{tabular}{@{}c@{}}
    \scriptsize from \\
    \scriptsize (\ref{eq116})\\\scriptsize (\ref{eq117})\\\scriptsize (\ref{eq121})
  \end{tabular}} 
  & 
  \mathrm{\Delta}x = - \int_{y}^{y + \mathrm{\Delta}y}{\displaystyle \frac{L^{2}}{\left( y + L \cdot \sqrt{P_{\text{low}}} \right)^{2}}} \cdot \partial y = \left\lbrack \displaystyle \frac{L^{2}}{y + L \cdot \sqrt{P_{\text{low}}}} \right\rbrack_{y}^{y + \mathrm{\Delta}y} \Rightarrow \mathbf{Eqns.}\ \mathbf{131\ }and\mathbf{\ 132}
  &  
  \label{eq126} 
  &
\end{flalign}

\begin{flalign}
& \text{\renewcommand{\arraystretch}{0.66}
    \begin{tabular}{@{}c@{}}
    \scriptsize from \\
    \scriptsize (\ref{eq118})\\\scriptsize (\ref{eq119})\\\scriptsize (\ref{eq123})
  \end{tabular}} 
  & 
  \mathrm{\Delta}y = - \int_{x}^{x + \mathrm{\Delta}x}{\displaystyle \frac{L^{2}}{\left( x + \displaystyle \frac{L}{\sqrt{P_{\text{high}}}} \right)^{2}}} \cdot \partial x = \left\lbrack \displaystyle \frac{L^{2}}{x + \displaystyle \frac{L}{\sqrt{P_{\text{high}}}}} \right\rbrack_{x}^{x + \mathrm{\Delta}x} \Rightarrow \mathbf{Eqns.}\ \mathbf{133\ }and\mathbf{\ 134}
  &  
  \label{eq127} 
  &
\end{flalign}

\begin{figure}[ht]
    \centering
    \includegraphics[width=\textwidth]{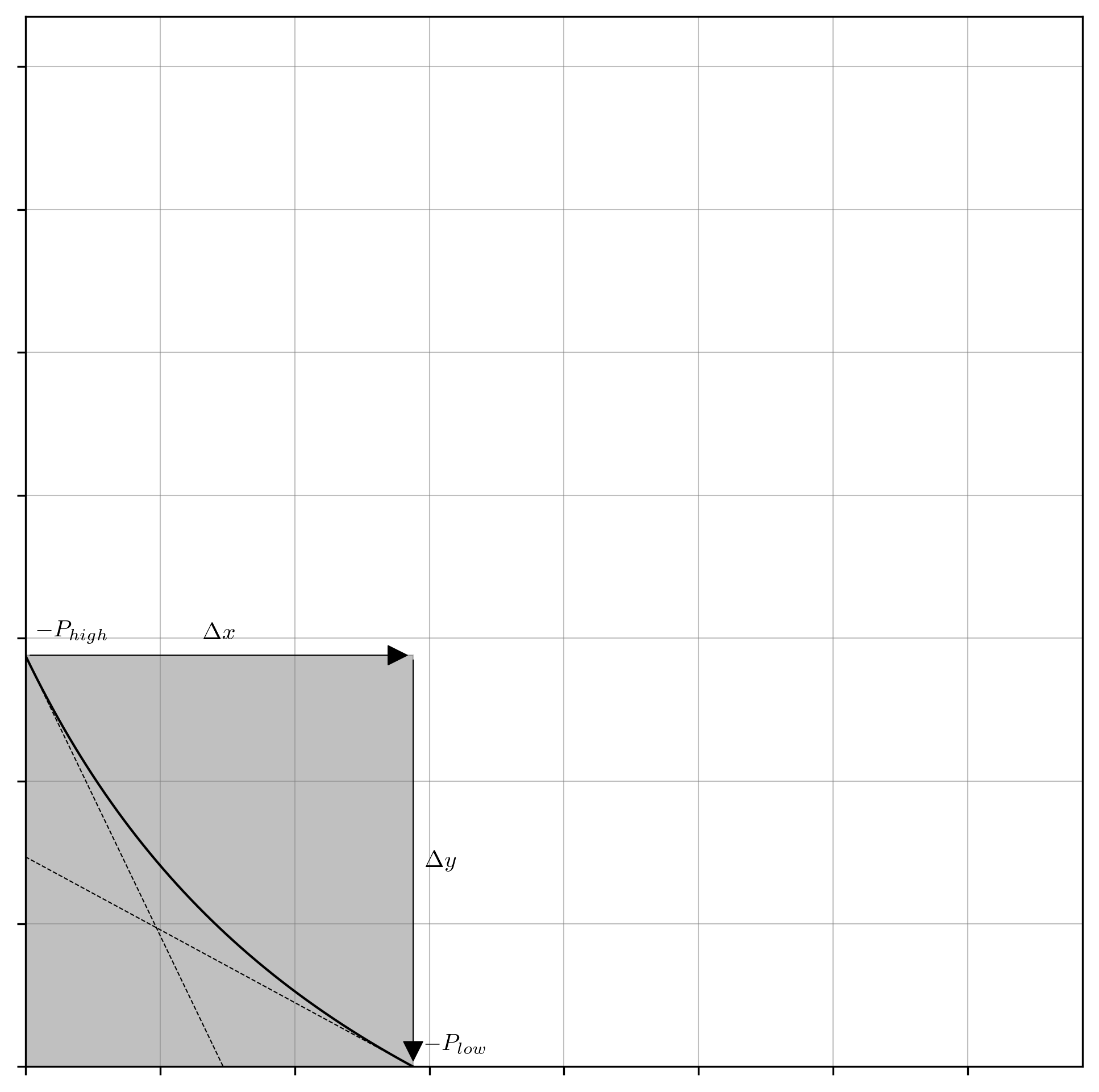}
    \captionsetup{
        justification=raggedright,
        singlelinecheck=false,
        font=small,
        labelfont=bf,
        labelsep=quad,
        format=plain
    }
    \caption{Traversal upon the rectangular hyperbola $\left( x + L / \sqrt{P_{\text{high}}} \right) \cdot \left( y + L \cdot \sqrt{P_{\text{low}}} \right) = L^{2}$ (Equation \ref{eq110}), representing a token swap against the Uniswap v3 real curve, where $\mathrm{\Delta}x > 0$ and $\mathrm{\Delta}y < 0$.}
    \label{fig23}
\end{figure}

\begin{figure}[ht]
    \centering
    \includegraphics[width=\textwidth]{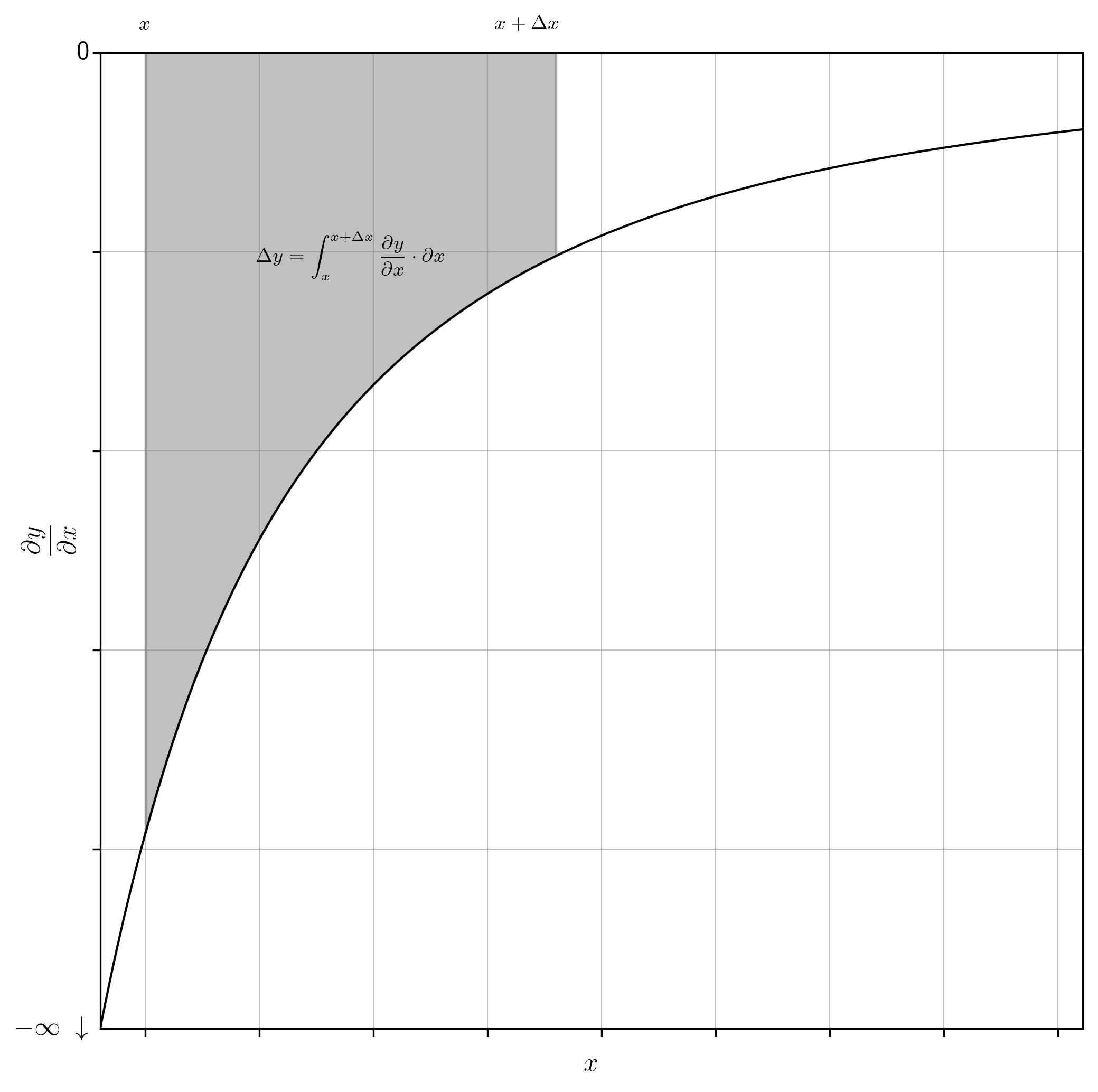}
    \captionsetup{
        justification=raggedright,
        singlelinecheck=false,
        font=small,
        labelfont=bf,
        labelsep=quad,
        format=plain
    }
    \caption{The integration above $\partial y / \partial x = - L^{2} / \left( x + L / \sqrt{P_{\text{high}}} \right)^{2}$ (Equation \ref{eq123}) over the interval $x \rightarrow x + \mathrm{\Delta}x$ representing a token swap against the Uniswap v3 real curve, where $\mathrm{\Delta}x > 0$ and $\mathrm{\Delta}y < 0$.}
    \label{fig24}
\end{figure}

Both the direct token swap by traversal upon the bonding curve $\left( x + L / \sqrt{P_{\text{high}}} \right) \cdot \left( y + L \cdot \sqrt{P_{\text{low}}} \right) = L^{2}$ and the integration above $\partial y / \partial x = - L^{2} / \left( x + L / \sqrt{P_{\text{high}}} \right)^{2}$ over the interval $x \rightarrow x + \mathrm{\Delta}x$ are depicted in Figures \ref{fig23} and \ref{fig24}.

\begin{figure}[ht]
    \centering
    \includegraphics[width=\textwidth]{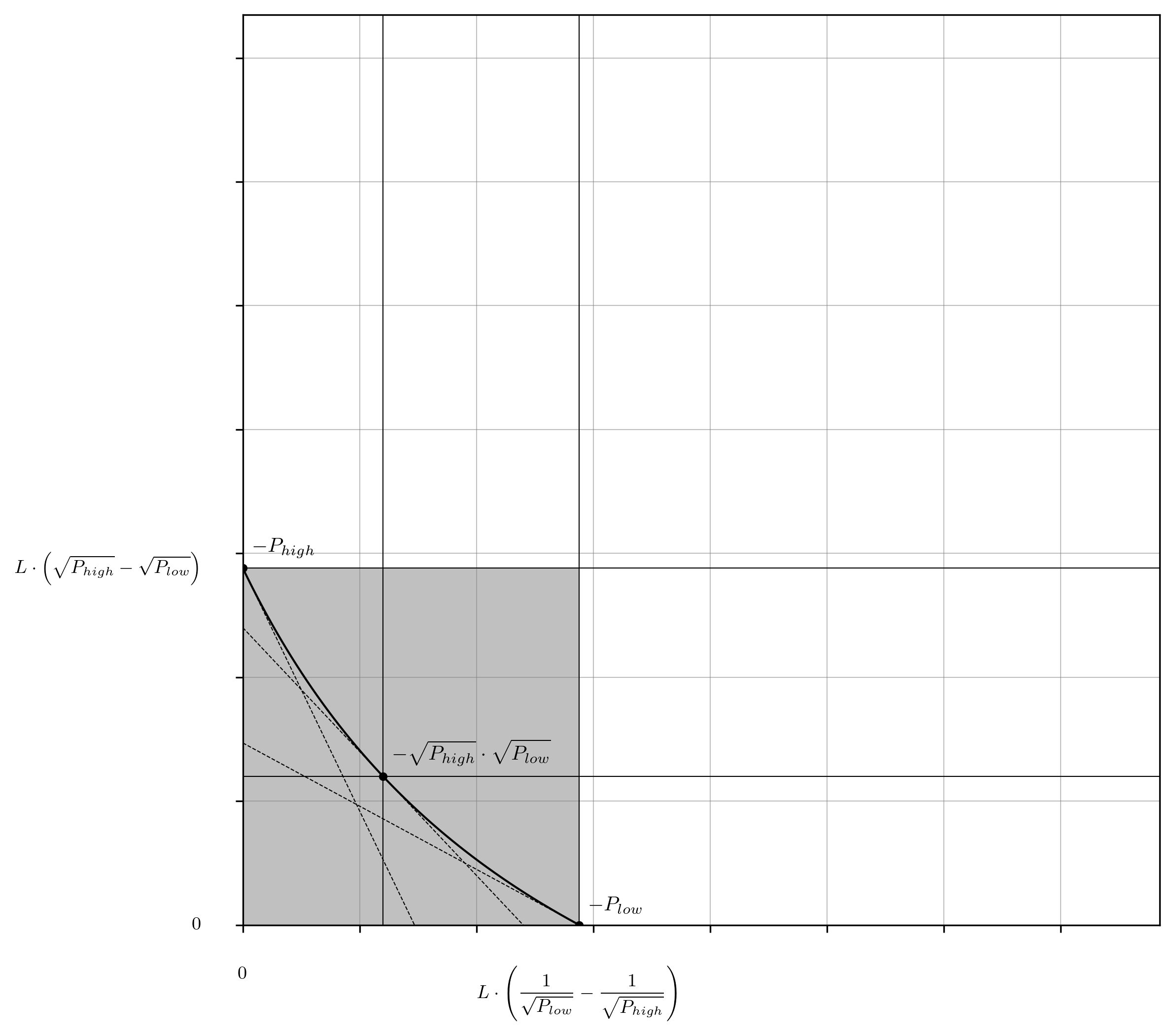}
    \captionsetup{
        justification=raggedright,
        singlelinecheck=false,
        font=small,
        labelfont=bf,
        labelsep=quad,
        format=plain
    }
    \caption{The x- and y-intercepts of the Uniswap v3 real curve are annotated on the appropriate axes (Equations \ref{eq128}, \ref{eq129} and \ref{eq130}).}
    \label{fig25}
\end{figure}

\begin{figure}[ht]
    \centering
    \includegraphics[width=\textwidth]{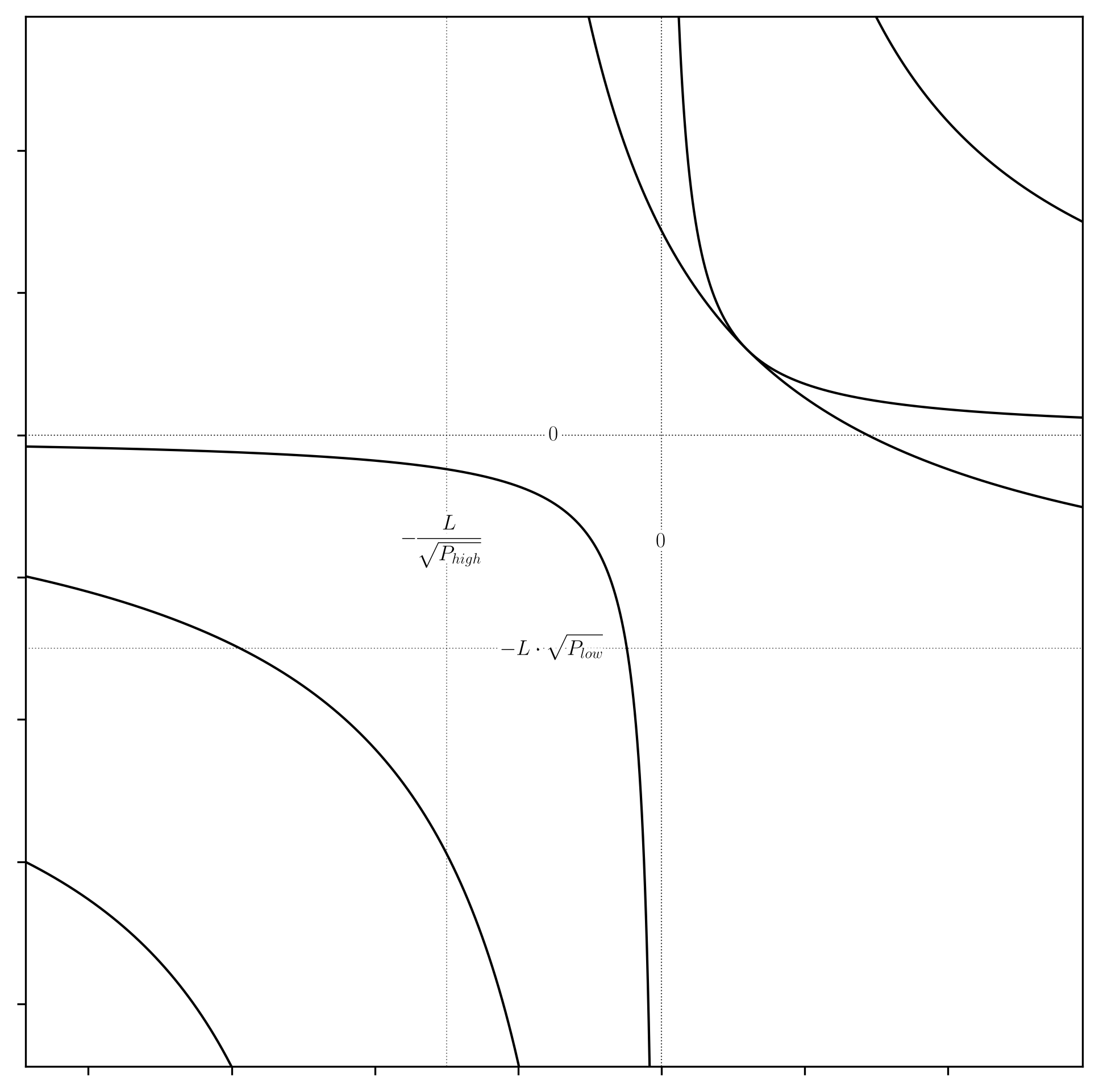}
    \captionsetup{
        justification=raggedright,
        singlelinecheck=false,
        font=small,
        labelfont=bf,
        labelsep=quad,
        format=plain
    }
    \caption{The horizontal and vertical asymptotes of the Uniswap v3 virtual and reference curves, and its real curve (Equations \ref{eq132} and \ref{eq133}), are depicted with dotted lines and their coordinates are annotated appropriately. While we have not yet approached the characterization of the reference and virtual curves, it is trivial that their asymptotes are at zero.}
    \label{fig26}
\end{figure}

The x- and y-intercepts under the Uniswap v3 parameterization are afforded by direct substitution of $y$ and $x$ for zero in Equations \ref{eq111}, and \ref{eq112} (Equations \ref{eq128}, \ref{eq129}, and \ref{eq130}) (Figure \ref{fig25}). As expected, the quotient of $y_{\text{int}}$ and $x_{\text{int}}$ is equal to $P_{0}$ as previously observed in Equation \ref{eq95} (Equation \ref{eq131}). The x- and y-asymptotes can again be derived geometrically or analytically (Figure \ref{fig26}). Since the Uniswap v3 curve makes use of the same horizontal and vertical shift parameters as previously discussed, albeit expressed in terms of $L$ and either $\sqrt{P_{\text{high}}}$ or $\sqrt{P_{\text{low}}}$, the same geometric intuition can be applied here. Simply subtract from the origin these shift terms to obtain the asymptotes. Alternatively, substitute the y- or x-component of Equations \ref{eq111} and \ref{eq112} as required to set the denominator of the curve's equation to zero (Equations \ref{eq132} and \ref{eq133}). Again, as seen previously (Equation \ref{eq96}), the quotient of the y- and x-asymptotes is equal to $P_{0}$ (Equation \ref{eq134}). 

\begin{flalign}
& \text{\renewcommand{\arraystretch}{0.66}
    \begin{tabular}{@{}c@{}}
    \scriptsize from \\
    \scriptsize (\ref{eq111})
  \end{tabular}} 
  & 
  x_{\text{int}} = \displaystyle \frac{L}{\sqrt{P_{\text{low}}}} - \displaystyle \frac{L}{\sqrt{P_{\text{high}}}} = \displaystyle \frac{L}{\sqrt{P_{\text{high}}} \cdot \sqrt{P_{\text{low}}}} \cdot \left( \sqrt{P_{\text{high}}} - \sqrt{P_{\text{low}}} \right)
  &  
  \label{eq128} 
  &
\end{flalign}

\begin{flalign}
& \text{\renewcommand{\arraystretch}{0.66}
    \begin{tabular}{@{}c@{}}
    \scriptsize from \\
    \scriptsize (\ref{eq128})
  \end{tabular}} 
  & 
  x_{\text{int}} = \displaystyle \frac{L}{P_{0}} \cdot \left( \sqrt{P_{\text{high}}} - \sqrt{P_{\text{low}}} \right)
  &  
  \label{eq129} 
  &
\end{flalign}

\begin{flalign}
& \text{\renewcommand{\arraystretch}{0.66}
    \begin{tabular}{@{}c@{}}
    \scriptsize from \\
    \scriptsize (\ref{eq112})
  \end{tabular}} 
  & 
  y_{\text{int}} = L \cdot \sqrt{P_{\text{high}}} - L \cdot \sqrt{P_{\text{low}}} = L \cdot \left( \sqrt{P_{\text{high}}} - \sqrt{P_{\text{low}}} \right)
  &  
  \label{eq130} 
  &
\end{flalign}

\begin{flalign}
& \text{\renewcommand{\arraystretch}{0.66}
    \begin{tabular}{@{}c@{}}
    \scriptsize from \\
    \scriptsize (\ref{eq128})\\\scriptsize (\ref{eq129}) \\\scriptsize (\ref{eq130})
  \end{tabular}} 
  & 
  \displaystyle \frac{y_{\text{int}}}{x_{\text{int}}} = \sqrt{P_{\text{high}}} \cdot \sqrt{P_{\text{low}}} = P_{0}
  &  
  \label{eq131} 
  &
\end{flalign}

\begin{flalign}
& \text{\renewcommand{\arraystretch}{0.66}
    \begin{tabular}{@{}c@{}}
    \scriptsize from \\
    \scriptsize (\ref{eq112})
  \end{tabular}} 
  & 
  x_{\text{asym}} = - \displaystyle \frac{L}{\sqrt{P_{\text{high}}}}
  &  
  \label{eq132} 
  &
\end{flalign}

\begin{flalign}
& \text{\renewcommand{\arraystretch}{0.66}
    \begin{tabular}{@{}c@{}}
    \scriptsize from \\
    \scriptsize (\ref{eq111})
  \end{tabular}} 
  & 
  y_{\text{asym}} = - L \cdot \sqrt{P_{\text{low}}}
  &  
  \label{eq133} 
  &
\end{flalign}

\begin{flalign}
& \text{\renewcommand{\arraystretch}{0.66}
    \begin{tabular}{@{}c@{}}
    \scriptsize from \\
    \scriptsize (\ref{eq132})\\\scriptsize (\ref{eq133})
  \end{tabular}} 
  & 
  \displaystyle \frac{y_{\text{asym}}}{x_{\text{asym}}} = \sqrt{P_{\text{high}}} \cdot \sqrt{P_{\text{low}}} = P_{0}
  &  
  \label{eq134} 
  &
\end{flalign}

The remainder of this section is not strictly necessary – the algebra does not lie; however, as we are now several layers of abstraction above the original construction, this cross-referencing also helps to re-contextualize the derivation and connect back to some of the establishing theory. 

The truth of the x- and y-intercept identities, $x_{\text{int}}$ and $y_{\text{int}}$, derived in Equations \ref{eq128}, \ref{eq129} and  \ref{eq130} can be verified with reference to work already completed. Since Equations \ref{eq91} and \ref{eq129} refer to the same object (i.e. the x-intercept of the real curve), we can assert their equivalence (Equation \ref{eq147}). Substitution of the $P_{0}$ and $L$ terms via their definitions (Equations \ref{eq39} and \ref{eq109}) results in Equation \ref{eq148}, which after handling the radicals yields a new relationship between $A$ and the quotient of the difference of the square roots of the price bounds, $\sqrt{P_{\text{high}}}$ and $\sqrt{P_{\text{low}}}$, and the square root of their geometric mean, $\sqrt{P_{0}}$ (Equation \ref{eq149}). Similar algebraic manipulations of Equations \ref{eq92} and \ref{eq130} (i.e. the y-intercepts) produce the same result (Equations \ref{eq150}, \ref{eq151} and \ref{eq152}). The independent verification of these expressions can then be obtained directly from Equations \ref{eq37}, \ref{eq39} and \ref{eq41}, which affords additional proof that these parameterizations are indeed describing the same object (Equations \ref{eq153} and \ref{eq154}).

\begin{flalign}
& \text{\renewcommand{\arraystretch}{0.66}
    \begin{tabular}{@{}c@{}}
    \scriptsize from \\
    \scriptsize (\ref{eq92})\\\scriptsize (\ref{eq129})
  \end{tabular}} 
  & 
  \displaystyle \frac{x_{0} \cdot (2A - 1)}{A - 1} = \displaystyle \frac{L}{P_{0}} \cdot \left( \sqrt{P_{\text{high}}} - \sqrt{P_{\text{low}}} \right)
  &  
  \label{eq147} 
  &
\end{flalign}

\begin{flalign}
& \text{\renewcommand{\arraystretch}{0.66}
    \begin{tabular}{@{}c@{}}
    \scriptsize from \\
    \scriptsize (\ref{eq39})\\\scriptsize (\ref{eq109})\\\scriptsize (\ref{eq147})
  \end{tabular}} 
  & 
  \displaystyle \frac{x_{0} \cdot (2A - 1)}{A - 1} = \displaystyle \frac{A \cdot \sqrt{x_{0}} \cdot \sqrt{y_{0}}}{\displaystyle \frac{y_{0}}{x_{0}}} \cdot \left( \sqrt{P_{\text{high}}} - \sqrt{P_{\text{low}}} \right)
  &  
  \label{eq148} 
  &
\end{flalign}

\begin{flalign}
& \text{\renewcommand{\arraystretch}{0.66}
    \begin{tabular}{@{}c@{}}
    \scriptsize from \\
    \scriptsize (\ref{eq39})\\\scriptsize (\ref{eq148})
  \end{tabular}} 
  & 
  \displaystyle \frac{2A - 1}{A \cdot \left( A - 1 \right)} = \displaystyle \frac{\sqrt{x_{0}}}{\sqrt{y_{0}}} \cdot \left( \sqrt{P_{\text{high}}} - \sqrt{P_{\text{low}}} \right) = \displaystyle \frac{\sqrt{P_{\text{high}}} - \sqrt{P_{\text{low}}}}{\sqrt{P_{0}}}
  &  
  \label{eq149} 
  &
\end{flalign}

\begin{flalign}
& \text{\renewcommand{\arraystretch}{0.66}
    \begin{tabular}{@{}c@{}}
    \scriptsize from \\
    \scriptsize (\ref{eq92})\\\scriptsize (\ref{eq130})
  \end{tabular}} 
  & 
  \displaystyle \frac{y_{0} \cdot (2A - 1)}{A - 1} = L \cdot \left( \sqrt{P_{\text{high}}} - \sqrt{P_{\text{low}}} \right)
  &  
  \label{eq150} 
  &
\end{flalign}

\begin{flalign}
& \text{\renewcommand{\arraystretch}{0.66}
    \begin{tabular}{@{}c@{}}
    \scriptsize from \\
    \scriptsize (\ref{eq109})\\\scriptsize (\ref{eq150})
  \end{tabular}} 
  & 
  \displaystyle \frac{y_{0} \cdot (2A - 1)}{A - 1} = A \cdot \sqrt{x_{0}} \cdot \sqrt{y_{0}} \cdot \left( \sqrt{P_{\text{high}}} - \sqrt{P_{\text{low}}} \right)
  &  
  \label{eq151} 
  &
\end{flalign}

\begin{flalign}
& \text{\renewcommand{\arraystretch}{0.66}
    \begin{tabular}{@{}c@{}}
    \scriptsize from \\
    \scriptsize (\ref{eq39})\\\scriptsize (\ref{eq151})
  \end{tabular}} 
  & 
  \displaystyle \frac{(2A - 1)}{A \cdot \left( A - 1 \right)} = \displaystyle \frac{\sqrt{x_{0}}}{\sqrt{y_{0}}} \cdot \left( \sqrt{P_{\text{high}}} - \sqrt{P_{\text{low}}} \right) = \displaystyle \frac{\sqrt{P_{\text{high}}} - \sqrt{P_{\text{low}}}}{\sqrt{P_{0}}}
  &  
  \label{eq152} 
  &
\end{flalign}

\begin{flalign}
& \text{\renewcommand{\arraystretch}{0.66}
    \begin{tabular}{@{}c@{}}
    \scriptsize from \\
    \scriptsize (\ref{eq37})\\\scriptsize (\ref{eq39})\\\scriptsize (\ref{eq41})
  \end{tabular}} 
  & 
  \sqrt{P_{\text{high}}} - \sqrt{P_{\text{low}}} = \displaystyle \frac{A}{A - 1} \cdot \displaystyle \frac{\sqrt{y_{0}}}{\sqrt{x_{0}}} - \displaystyle \frac{A - 1}{A} \cdot \displaystyle \frac{\sqrt{y_{0}}}{\sqrt{x_{0}}} = \displaystyle \frac{(2A - 1)}{A \cdot \left( A - 1 \right)} \cdot \displaystyle \frac{\sqrt{y_{0}}}{\sqrt{x_{0}}}
  &  
  \label{eq153} 
  &
\end{flalign}

\begin{flalign}
& \text{\renewcommand{\arraystretch}{0.66}
    \begin{tabular}{@{}c@{}}
    \scriptsize from \\
    \scriptsize (\ref{eq39})\\\scriptsize (\ref{eq153})
  \end{tabular}} 
  & 
  \displaystyle \frac{\sqrt{P_{\text{high}}} - \sqrt{P_{\text{low}}}}{\sqrt{P_{0}}} = \displaystyle \frac{(2A - 1)}{A \cdot \left( A - 1 \right)}
  &  
  \label{eq154} 
  &
\end{flalign}

Further confirmation is possible via analysis of the x- and y-asymptotes, via Equations \ref{eq93} and \ref{eq132}, and \ref{eq94} and \ref{eq133}, respectively, which produce the identities previously obtained in Equations \ref{eq37} and \ref{eq41} for $P_{\text{high}}$ and $P_{\text{low}}$ (Equations \ref{eq155} and \ref{eq156}). 

\begin{flalign}
& \text{\renewcommand{\arraystretch}{0.66}
    \begin{tabular}{@{}c@{}}
    \scriptsize from \\
    \scriptsize (\ref{eq93})\\\scriptsize (\ref{eq109})\\\scriptsize (\ref{eq132})
  \end{tabular}} 
  & 
  x_{0} \cdot \left( A - 1 \right) = \displaystyle \frac{L}{\sqrt{P_{\text{high}}}} \Rightarrow \displaystyle \frac{A^{2}}{\left( A - 1 \right)^{2}} \cdot \displaystyle \frac{y_{0}}{x_{0}} = P_{\text{high}}
  &  
  \label{eq155} 
  &
\end{flalign}

\begin{flalign}
& \text{\renewcommand{\arraystretch}{0.66}
    \begin{tabular}{@{}c@{}}
    \scriptsize from \\
    \scriptsize (\ref{eq94})\\\scriptsize (\ref{eq109})\\\scriptsize (\ref{eq133})
  \end{tabular}} 
  & 
  y_{0} \cdot \left( A - 1 \right) = L \cdot \sqrt{P_{\text{low}}} \Rightarrow \displaystyle \frac{\left( A - 1 \right)^{2}}{A^{2}} \cdot \displaystyle \frac{y_{0}}{x_{0}} = P_{\text{low}}
  &  
  \label{eq156} 
  &
\end{flalign}

\subsection{The Uniswap v3 Virtual Curve}\label{subsec4.2}

The objective of this next section is to derive the same identities as was done for Equations \ref{eq97}, \ref{eq98}, \ref{eq99}, \ref{eq100}, \ref{eq101} and \ref{eq102}, then prove these identities by way of substitution with previously obtained relationships. 

\begin{figure}[ht]
    \centering
    \includegraphics[width=\textwidth]{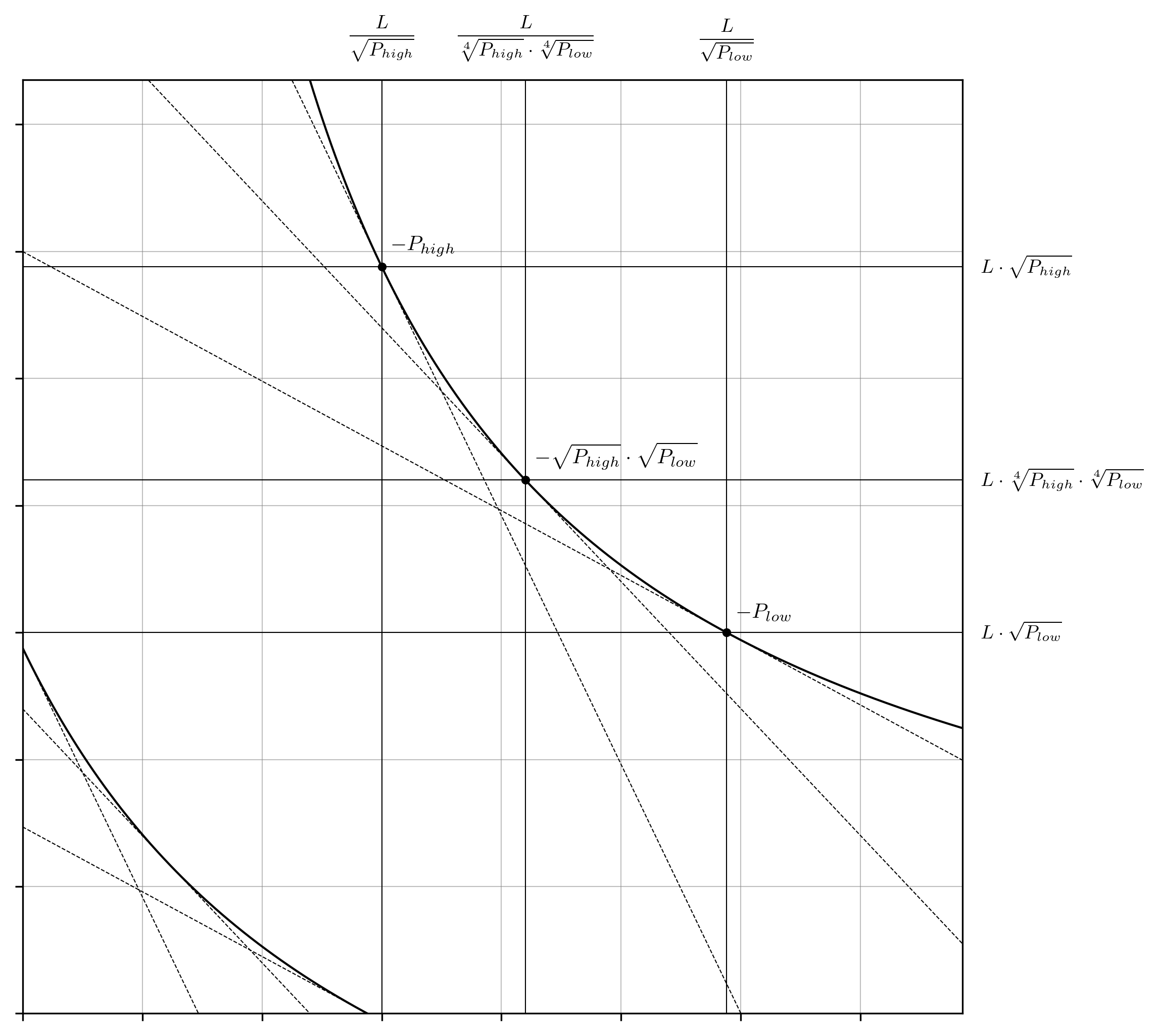}
    \captionsetup{
        justification=raggedright,
        singlelinecheck=false,
        font=small,
        labelfont=bf,
        labelsep=quad,
        format=plain
    }
    \caption{The completed characterisation of the Uniswap v3 virtual curve, with the coordinates at its price bounds and geometric mean of the price bounds annotated as appropriate (Equations \ref{eq136}, \ref{eq137}, \ref{eq141} and \ref{eq142}).}
    \label{fig27}
\end{figure}

Recall that the geometric means of the x- and y-bounds of the virtual curve have a correspondence with the amplified coordinates of the reference curve (to which its size is being compared). To perform this process for the Uniwap v3 reparametrized real curve, as before, we are comparing the bounds with respect to the asymptotes of the shifted curve, which is equivalent to reversing the shift before measuring these geometric means (Equations \ref{eq135}, \ref{eq136}, \ref{eq137}, \ref{eq138}, \ref{eq139}, \ref{eq140}, \ref{eq141}, \ref{eq142}, \ref{eq143} and  \ref{eq144}), thus recreating the virtual curve coordinates from the corresponding real curve coordinates (Figure \ref{fig27}). Consistent with prior work, the quotient of the maximum and minimum bounds in both dimensions are equal to the [yet] unexplained mystery constant, $C$ (Equations \ref{eq145} and \ref{eq146}). However, unlike last time it is now presented in terms of $\sqrt{P_{\text{high}}}$ and $\sqrt{P_{\text{low}}}$, instead of $A$. This identity has already been proven via Equation \ref{eq56}. 

\begin{flalign}
& \text{\renewcommand{\arraystretch}{0.66}
    \begin{tabular}{@{}c@{}}
    \scriptsize from \\
    \scriptsize (\ref{eq128})\\\scriptsize (\ref{eq132})
  \end{tabular}} 
  & 
  \max\left( x_{\text{v}} \right) = \left( x_{\text{int}} - x_{\text{asym}} \right) = \displaystyle \frac{L}{\sqrt{P_{\text{low}}}} - \displaystyle \frac{L}{\sqrt{P_{\text{high}}}} + \displaystyle \frac{L}{\sqrt{P_{\text{high}}}}
  &  
  \label{eq135} 
  &
\end{flalign}

\begin{flalign}
& \text{\renewcommand{\arraystretch}{0.66}
    \begin{tabular}{@{}c@{}}
    \scriptsize from \\
    \scriptsize (\ref{eq135})
  \end{tabular}} 
  & 
  \max\left( x_{\text{v}} \right) = \displaystyle \frac{L}{\sqrt{P_{\text{low}}}}
  &  
  \label{eq136} 
  &
\end{flalign}

\begin{flalign}
& \text{\renewcommand{\arraystretch}{0.66}
    \begin{tabular}{@{}c@{}}
    \scriptsize from \\
    \scriptsize (\ref{eq132})
  \end{tabular}} 
  & 
  \min\left( x_{\text{v}} \right) = \left( 0 - x_{\text{asym}} \right) = \displaystyle \frac{L}{\sqrt{P_{\text{high}}}}
  &  
  \label{eq137} 
  &
\end{flalign}

\begin{flalign}
& \text{\renewcommand{\arraystretch}{0.66}
    \begin{tabular}{@{}c@{}}
    \scriptsize from \\
    \scriptsize (\ref{eq136})\\\scriptsize (\ref{eq137})
  \end{tabular}} 
  & 
  \max\left( x_{\text{v}} \right) \cdot \min\left( x_{\text{v}} \right) = \displaystyle \frac{L^{2}}{\sqrt{P_{\text{high}}} \cdot \sqrt{P_{\text{low}}}} = \displaystyle \frac{L^{2}}{P_{0}}
  &  
  \label{eq138} 
  &
\end{flalign}

\begin{flalign}
& \text{\renewcommand{\arraystretch}{0.66}
    \begin{tabular}{@{}c@{}}
    \scriptsize from \\
    \scriptsize (\ref{eq138})
  \end{tabular}} 
  & 
  \sqrt{\max\left( x_{\text{v}} \right)} \cdot \sqrt{\min\left( x_{\text{v}} \right)} = \displaystyle \frac{L}{\sqrt[4]{P_{\text{high}}} \cdot \sqrt[4]{P_{\text{low}}}} = \displaystyle \frac{L}{\sqrt{P_{0}}}
  &  
  \label{eq139} 
  &
\end{flalign}

\begin{flalign}
& \text{\renewcommand{\arraystretch}{0.66}
    \begin{tabular}{@{}c@{}}
    \scriptsize from \\
    \scriptsize (\ref{eq130})\\\scriptsize (\ref{eq133})
  \end{tabular}} 
  & 
  \max\left( y_{\text{v}} \right) = \left( y_{\text{int}} - y_{\text{asym}} \right) = L \cdot \left( \sqrt{P_{\text{high}}} - \sqrt{P_{\text{low}}} \right) + \ L \cdot \sqrt{P_{\text{low}}}
  &  
  \label{eq140} 
  &
\end{flalign}

\begin{flalign}
& \text{\renewcommand{\arraystretch}{0.66}
    \begin{tabular}{@{}c@{}}
    \scriptsize from \\
    \scriptsize (\ref{eq140})
  \end{tabular}} 
  & 
  \max\left( y_{\text{v}} \right) = L \cdot \sqrt{P_{\text{high}}}
  &  
  \label{eq141} 
  &
\end{flalign}

\begin{flalign}
& \text{\renewcommand{\arraystretch}{0.66}
    \begin{tabular}{@{}c@{}}
    \scriptsize from \\
    \scriptsize (\ref{eq133})
  \end{tabular}} 
  & 
  \min\left( y_{\text{v}} \right) = \left( 0 - y_{\text{asym}} \right) = L \cdot \sqrt{P_{\text{low}}}
  &  
  \label{eq142} 
  &
\end{flalign}

\begin{flalign}
& \text{\renewcommand{\arraystretch}{0.66}
    \begin{tabular}{@{}c@{}}
    \scriptsize from \\
    \scriptsize (\ref{eq141})\\\scriptsize (\ref{eq142})
  \end{tabular}} 
  & 
  \max\left( y_{\text{v}} \right) \cdot \min\left( y_{\text{v}} \right) = L^{2} \cdot \sqrt{P_{\text{high}}} \cdot \sqrt{P_{\text{low}}} = L^{2} \cdot P_{0}
  &  
  \label{eq143} 
  &
\end{flalign}

\begin{flalign}
& \text{\renewcommand{\arraystretch}{0.66}
    \begin{tabular}{@{}c@{}}
    \scriptsize from \\
    \scriptsize (\ref{eq143})
  \end{tabular}} 
  & 
  \sqrt{\max\left( y_{\text{v}} \right)} \cdot \sqrt{\min\left( y_{\text{v}} \right)} = L \cdot \sqrt[4]{P_{\text{high}}} \cdot \sqrt[4]{P_{\text{low}}} = L \cdot \sqrt{P_{0}}
  &  
  \label{eq144} 
  &
\end{flalign}

\begin{flalign}
& \text{\renewcommand{\arraystretch}{0.66}
    \begin{tabular}{@{}c@{}}
    \scriptsize from \\
    \scriptsize (\ref{eq136})\\\scriptsize (\ref{eq137})
  \end{tabular}} 
  & 
  \displaystyle \frac{\max\left( x_{\text{v}} \right)}{\min\left( x_{\text{v}} \right)} = \displaystyle \frac{\sqrt{P_{\text{high}}}}{\sqrt{P_{\text{low}}}} = C
  &  
  \label{eq145} 
  &
\end{flalign}

\begin{flalign}
& \text{\renewcommand{\arraystretch}{0.66}
    \begin{tabular}{@{}c@{}}
    \scriptsize from \\
    \scriptsize (\ref{eq141})\\\scriptsize (\ref{eq142})
  \end{tabular}} 
  & 
  \displaystyle \frac{\max\left( y_{\text{v}} \right)}{\min\left( y_{\text{v}} \right)} = \displaystyle \frac{\sqrt{P_{\text{high}}}}{\sqrt{P_{\text{low}}}} = C
  &  
  \label{eq146} 
  &
\end{flalign}

Finally, validation of the geometric means of $\max \left( x_{\text{v}}\right)$ and $\min \left( x_{\text{v}}\right)$, and $\max \left( y_{\text{v}}\right)$ and $\min \left( y_{\text{v}}\right)$ (i.e. the x- and y-bounds of the implied virtual curve) (Equations \ref{eq139} and \ref{eq144}) can be found by asserting their equivalence to the identities previously attained (Equations \ref{eq98} and \ref{eq100}), and then deriving the identity of $P_{0}$ from this relationship (Equations \ref{eq157} and \ref{eq158}). 

\begin{flalign}
& \text{\renewcommand{\arraystretch}{0.66}
    \begin{tabular}{@{}c@{}}
    \scriptsize from \\
    \scriptsize (\ref{eq98})\\\scriptsize (\ref{eq109})\\\scriptsize (\ref{eq139})
  \end{tabular}} 
  & 
  A \cdot x_{0} = \displaystyle \frac{L}{\sqrt{P_{0}}} \Rightarrow \displaystyle \frac{y_{0}}{x_{0}} = P_{0}
  &  
  \label{eq157} 
  &
\end{flalign}

\begin{flalign}
& \text{\renewcommand{\arraystretch}{0.66}
    \begin{tabular}{@{}c@{}}
    \scriptsize from \\
    \scriptsize (\ref{eq100})\\\scriptsize (\ref{eq109})\\\scriptsize (\ref{eq144})
  \end{tabular}} 
  & 
  A \cdot y_{0} = L \cdot \sqrt{P_{0}} \Rightarrow \displaystyle \frac{y_{0}}{x_{0}} = P_{0}
  &  
  \label{eq158} 
  &
\end{flalign}

To complete the characterisation of the real curve and begin reconstructing the reparametrized reference curve, recall that the point at the coordinates ($x_{0}$, $y_{0}$) has two useful properties: the derivative of the reference curve evaluated at this point is equal to $P_{0}$, and its position relative to the point on the virtual curve with the same derivative is uniquely determined by the same horizontal and vertical shift parameters as the real curve (none of the other points on the reference curve align with the virtual curve after performing the shift). Moreover, this is the one unique point that is shared by both the reference and the real curve, and their derivative evaluated at this point is also equal. 

\begin{figure}[ht]
    \centering
    \includegraphics[width=\textwidth]{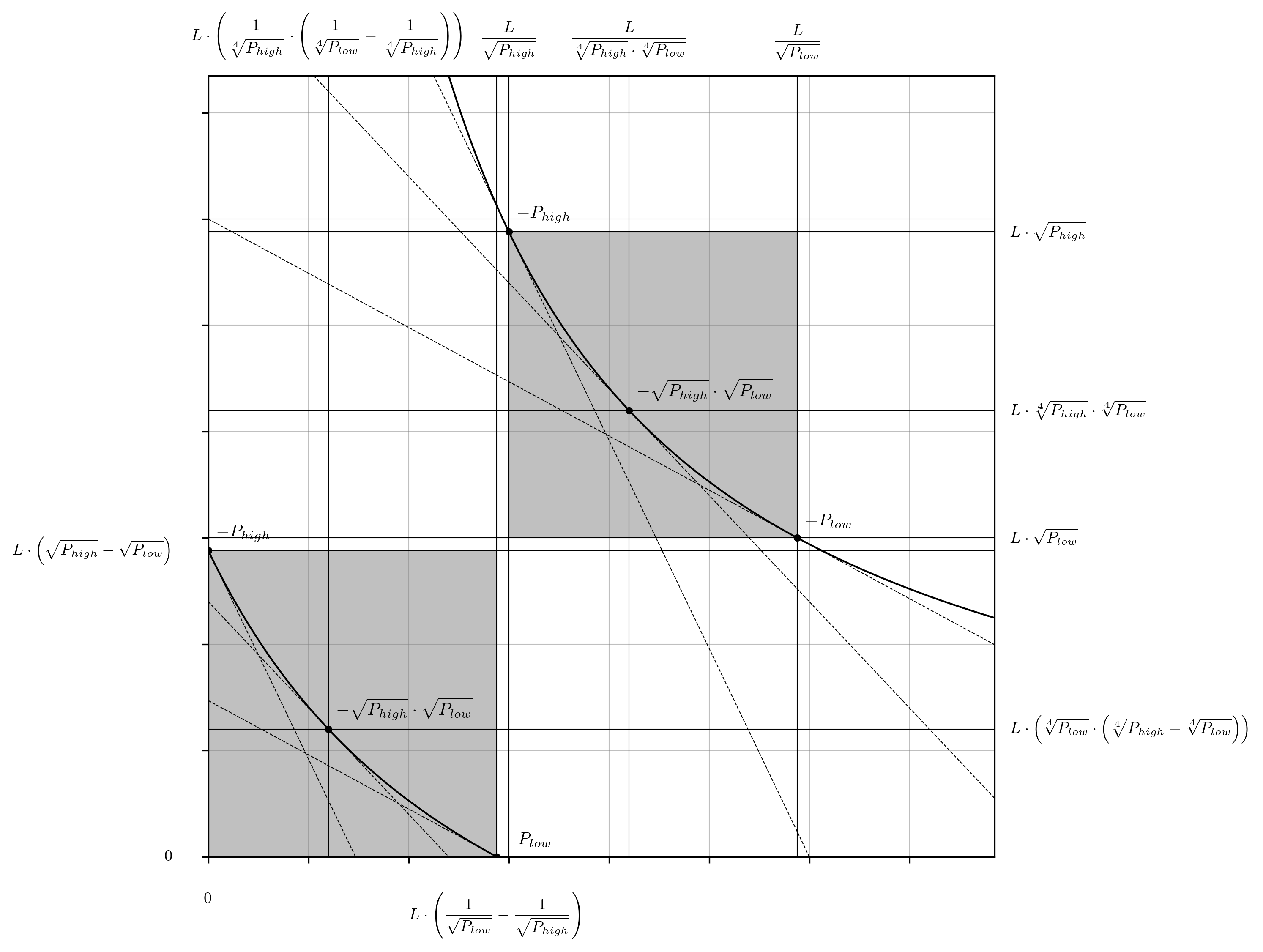}
    \captionsetup{
        justification=raggedright,
        singlelinecheck=false,
        font=small,
        labelfont=bf,
        labelsep=quad,
        format=plain
    }
    \caption{Completed characterizations of the Uniswap v3 real and virtual curves are depicted and annotated appropriately (Equations \ref{eq128}, \ref{eq130}, \ref{eq136}, \ref{eq137}, \ref{eq139}, \ref{eq141}, \ref{eq142}, \ref{eq144}, \ref{eq160}, \ref{eq162}).}
    \label{fig28}
\end{figure}

To obtain the identities of $x_{0}$ and $y_{0}$, the shift parameter (equivalent to $x_{\text{asym}}$ and $y_{\text{asym}}$) is applied as appropriate to the corresponding coordinates on the reference curve (i.e. the geometric means of $\min \left( x_{\text{v}}\right)$ and $\max \left( x_{\text{v}}\right)$, and $\min \left( y_{\text{v}}\right)$ and $\max \left( y_{\text{v}}\right)$) (Equations \ref{eq159} and \ref{eq161}). In both cases, the expressions can be homogenized by distributing a lone radical into the square of its fourth root before factoring (Equations \ref{eq160} and \ref{eq162}). There are two different representations apiece for the $x_{0}$ and $y_{0}$ identities included in Equations \ref{eq160} and \ref{eq162}, respectively. One of these is a “better fit” for the real curve and the other for the reference curve, and both are used as appropriate when annotating the plots. With these coordinates, the characterization of the real curve is now complete (Figure \ref{fig28}).

\begin{flalign}
& \text{\renewcommand{\arraystretch}{0.66}
    \begin{tabular}{@{}c@{}}
    \scriptsize from \\
    \scriptsize (\ref{eq132})\\\scriptsize (\ref{eq139})
  \end{tabular}} 
  & 
  x_{0} = \sqrt{\max\left( x_{\text{v}} \right)} \cdot \sqrt{\min\left( x_{\text{v}} \right)} + x_{\text{asym}} = \displaystyle \frac{L}{\sqrt[4]{P_{\text{high}}} \cdot \sqrt[4]{P_{\text{low}}}} - \displaystyle \frac{L}{\sqrt{P_{\text{high}}}}
  &  
  \label{eq159} 
  &
\end{flalign}

\begin{flalign}
& \text{\renewcommand{\arraystretch}{0.66}
    \begin{tabular}{@{}c@{}}
    \scriptsize from \\
    \scriptsize (\ref{eq159})
  \end{tabular}} 
  & 
  x_{0} = L \cdot \left( \displaystyle \frac{1}{\sqrt[4]{P_{\text{high}}}} \cdot \left( \displaystyle \frac{1}{\sqrt[4]{P_{\text{low}}}} - \displaystyle \frac{1}{\sqrt[4]{P_{\text{high}}}} \right) \right) = \displaystyle \frac{L}{\sqrt[4]{P_{\text{high}}} \cdot \sqrt[4]{P_{\text{low}}}} \cdot \left( 1 - \displaystyle \frac{\sqrt[4]{P_{\text{low}}}}{\sqrt[4]{P_{\text{high}}}} \right)
  &  
  \label{eq160} 
  &
\end{flalign}

\begin{flalign}
& \text{\renewcommand{\arraystretch}{0.66}
    \begin{tabular}{@{}c@{}}
    \scriptsize from \\
    \scriptsize (\ref{eq133})\\\scriptsize (\ref{eq144})
  \end{tabular}} 
  & 
  y_{0} = \sqrt{\max\left( y_{\text{v}} \right)} \cdot \sqrt{\min\left( y_{\text{v}} \right)} + y_{\text{asym}} = L \cdot \sqrt[4]{P_{\text{high}}} \cdot \sqrt[4]{P_{\text{low}}} - L \cdot \sqrt{P_{\text{low}}}
  &  
  \label{eq161} 
  &
\end{flalign}

\begin{flalign}
& \text{\renewcommand{\arraystretch}{0.66}
    \begin{tabular}{@{}c@{}}
    \scriptsize from \\
    \scriptsize (\ref{eq161})
  \end{tabular}} 
  & 
  y_{0} = L \cdot \left( \sqrt[4]{P_{\text{low}}} \cdot \left( \sqrt[4]{P_{\text{high}}} - \sqrt[4]{P_{\text{low}}} \right) \right) = L \cdot \sqrt[4]{P_{\text{high}}} \cdot \sqrt[4]{P_{\text{low}}} \cdot \left( 1 - \displaystyle \frac{\sqrt[4]{P_{\text{low}}}}{\sqrt[4]{P_{\text{high}}}} \right)
  &  
  \label{eq162} 
  &
\end{flalign}

\begin{figure}[ht]
    \centering
    \includegraphics[width=\textwidth]{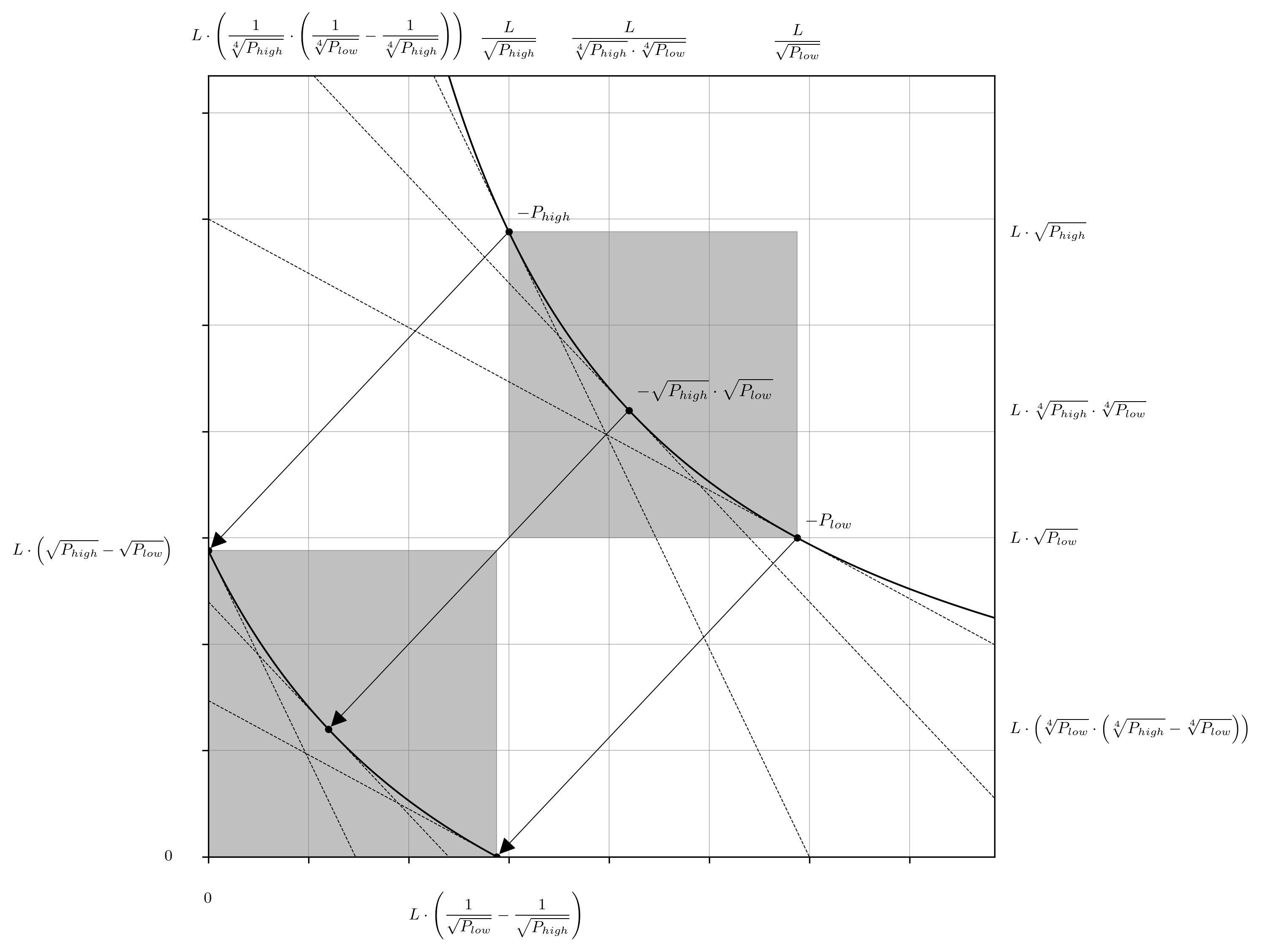}
    \captionsetup{
        justification=raggedright,
        singlelinecheck=false,
        font=small,
        labelfont=bf,
        labelsep=quad,
        format=plain
    }
    \caption{Reconstruction of the Uniswap v3 real curve is depicted as a map of the virtual curve back to the x- and y-axes.}
    \label{fig29}
\end{figure}

In contrast to the prior construction, which began with a reference curve followed by its virtual amplification and then horizontal and vertical shifts to arrive at the real curve, here, we have derived the virtual curve beginning from the shifted real curve. While the process is backwards, the result is the same, and it is still appropriate to define the real curve in terms of the virtual curve and the horizontal and vertical shifts (Figure \ref{fig29}). 

Recall that the amplified virtual curve is just a regular rectangular hyperbola. The only important piece of information is its scaling constant, previously defined as $A^{2} \cdot x_{0} \cdot y_{0}$; none of these terms are native parameters in this case. However, the same scaling constant can be obtained directly from inspection of the RHS of Equation \ref{eq110}, or from the product of any set of coordinates along the curve, for which we have already elucidated three unique pairs. It does not matter which you use, the result will be the same (Equation \ref{eq163}). After that, the elaboration of the marginal price and token swap equations is performed as was done for Equations \ref{eq20}, \ref{eq21}, \ref{eq22}, \ref{eq23}, \ref{eq24}, \ref{eq25}, \ref{eq26}, \ref{eq27}, \ref{eq28}, and \ref{eq29} (Equations \ref{eq163}, \ref{eq164}, \ref{eq165}, \ref{eq166}, \ref{eq167}, \ref{eq168}, \ref{eq169}, and \ref{eq170}).

\begin{flalign}
& \text{\renewcommand{\arraystretch}{0.66}
    \begin{tabular}{@{}c@{}}
    \scriptsize from \\
    \scriptsize (\ref{eq2})
  \end{tabular}} 
  & 
  x_{\text{v}} \cdot y_{\text{v}} = \displaystyle \frac{L}{\sqrt[4]{P_{\text{high}}} \cdot \sqrt[4]{P_{\text{low}}}} \cdot L \cdot \sqrt[4]{P_{\text{high}}} \cdot \sqrt[4]{P_{\text{low}}} = L^{2}
  &  
  \label{eq163} 
  &
\end{flalign}

\begin{flalign}
& \text{\renewcommand{\arraystretch}{0.66}
    \begin{tabular}{@{}c@{}}
    \scriptsize from \\
    \scriptsize (\ref{eq8})
  \end{tabular}} 
  & 
  y_{\text{v}} = \displaystyle \frac{L^{2}}{x_{\text{v}}}
  &  
  \label{eq164} 
  &
\end{flalign}

\begin{flalign}
& \text{\renewcommand{\arraystretch}{0.66}
    \begin{tabular}{@{}c@{}}
    \scriptsize from \\
    \scriptsize (\ref{eq6})
  \end{tabular}} 
  & 
  x_{\text{v}} = \displaystyle \frac{L^{2}}{y_{\text{v}}}
  &  
  \label{eq165} 
  &
\end{flalign}

\begin{flalign}
& \text{\renewcommand{\arraystretch}{0.66}
    \begin{tabular}{@{}c@{}}
    \scriptsize from \\
    \scriptsize (\ref{eq3})
  \end{tabular}} 
  & 
  \left( x_{\text{v}} + \mathrm{\Delta}x \right) \cdot \left( y_{\text{v}} + \mathrm{\Delta}y \right) = L^{2}
  &  
  \label{eq166} 
  &
\end{flalign}

\begin{flalign}
& \text{\renewcommand{\arraystretch}{0.66}
    \begin{tabular}{@{}c@{}}
    \scriptsize from \\
    \scriptsize (\ref{eq4})\\\scriptsize (\ref{eq7})
  \end{tabular}} 
  & 
  \mathrm{\Delta}x = \displaystyle \frac{L^{2}}{y_{\text{v}} + \mathrm{\Delta}y} - x_{\text{v}} = - \displaystyle \frac{\mathrm{\Delta}y \cdot L^{2}}{y_{\text{v}} \cdot \left( y_{\text{v}} + \mathrm{\Delta}y \right)}
  &  
  \label{eq167} 
  &
\end{flalign}

\begin{flalign}
& \text{\renewcommand{\arraystretch}{0.66}
    \begin{tabular}{@{}c@{}}
    \scriptsize from \\
    \scriptsize (\ref{eq5})\\\scriptsize (\ref{eq9})
  \end{tabular}} 
  & 
  \mathrm{\Delta}y = \displaystyle \frac{A^{2} \cdot L^{2}}{x_{\text{v}} + \mathrm{\Delta}x} - y_{\text{v}} = - \displaystyle \frac{\mathrm{\Delta}x \cdot L^{2}}{x_{\text{v}} \cdot \left( x_{\text{v}} + \mathrm{\Delta}x \right)}
  &  
  \label{eq168} 
  &
\end{flalign}

\begin{flalign}
& \text{\renewcommand{\arraystretch}{0.66}
    \begin{tabular}{@{}c@{}}
    \scriptsize from \\
    \scriptsize (\ref{eq16})
  \end{tabular}} 
  & 
  \displaystyle \frac{\partial x_{\text{v}}}{\partial y_{\text{v}}} = - \displaystyle \frac{L^{2}}{y_{\text{v}}^{2}};\ \ \ \displaystyle \frac{\partial y_{\text{v}}}{\partial x_{\text{v}}} = - \displaystyle \frac{y_{\text{v}}^{2}}{L^{2}}
  &  
  \label{eq169} 
  &
\end{flalign}

\begin{flalign}
& \text{\renewcommand{\arraystretch}{0.66}
    \begin{tabular}{@{}c@{}}
    \scriptsize from \\
    \scriptsize (\ref{eq17})
  \end{tabular}} 
  & 
  \displaystyle \frac{\partial y_{\text{v}}}{\partial x_{\text{v}}} = - \displaystyle \frac{L^{2}}{x_{\text{v}}^{2}};\ \ \ \displaystyle \frac{\partial x_{\text{v}}}{\partial y_{\text{v}}} = - \displaystyle \frac{x_{\text{v}}^{2}}{L^{2}}
  &  
  \label{eq170} 
  &
\end{flalign}

\begin{figure}[ht]
    \centering
    \includegraphics[width=\textwidth]{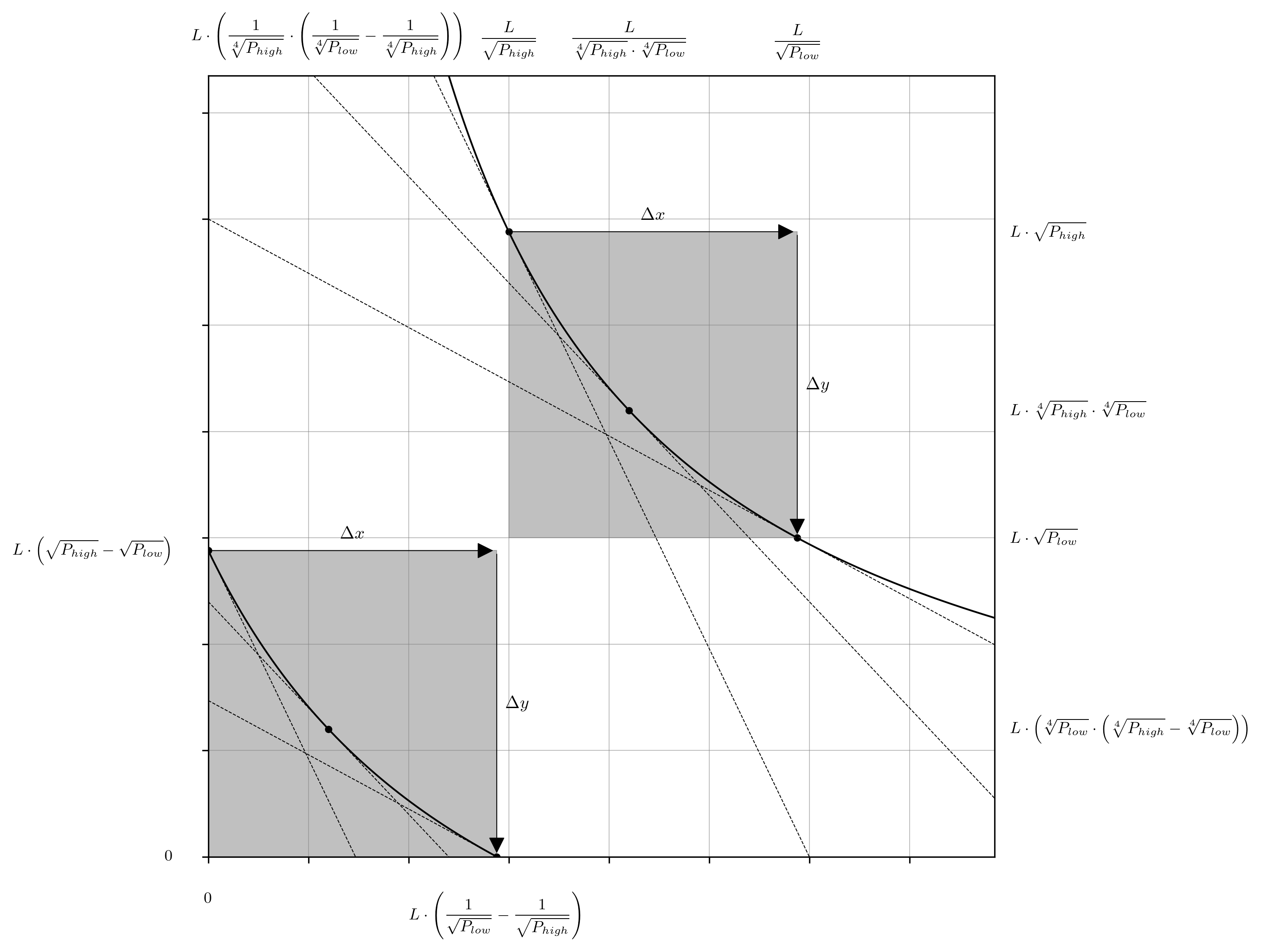}
    \captionsetup{
        justification=raggedright,
        singlelinecheck=false,
        font=small,
        labelfont=bf,
        labelsep=quad,
        format=plain
    }
    \caption{Traversal upon the rectangular hyperbolas $\left( x + L / \sqrt{P_{\text{high}}}\right) \cdot \left( y + L \cdot \sqrt{P_{\text{low}}}\right) = L^{2}$, and $x_{\text{v}} \cdot y_{\text{v}} = L^{2}$ (Equations \ref{eq110} and \ref{eq163}), representing a token swap against the Uniswap v3 real and virtual curves, where $\mathrm{\Delta}x > 0$ and $\mathrm{\Delta}y < 0$. The swap quantities $\mathrm{\Delta}x$ and $\mathrm{\Delta}y$, and marginal rates of exchange before and after the swap are identical. Therefore, with respect to the outcome of the exchange the difference between the two curve implementations is zero.}
    \label{fig30}
\end{figure}

\begin{figure}[ht]
    \centering
    \includegraphics[width=\textwidth]{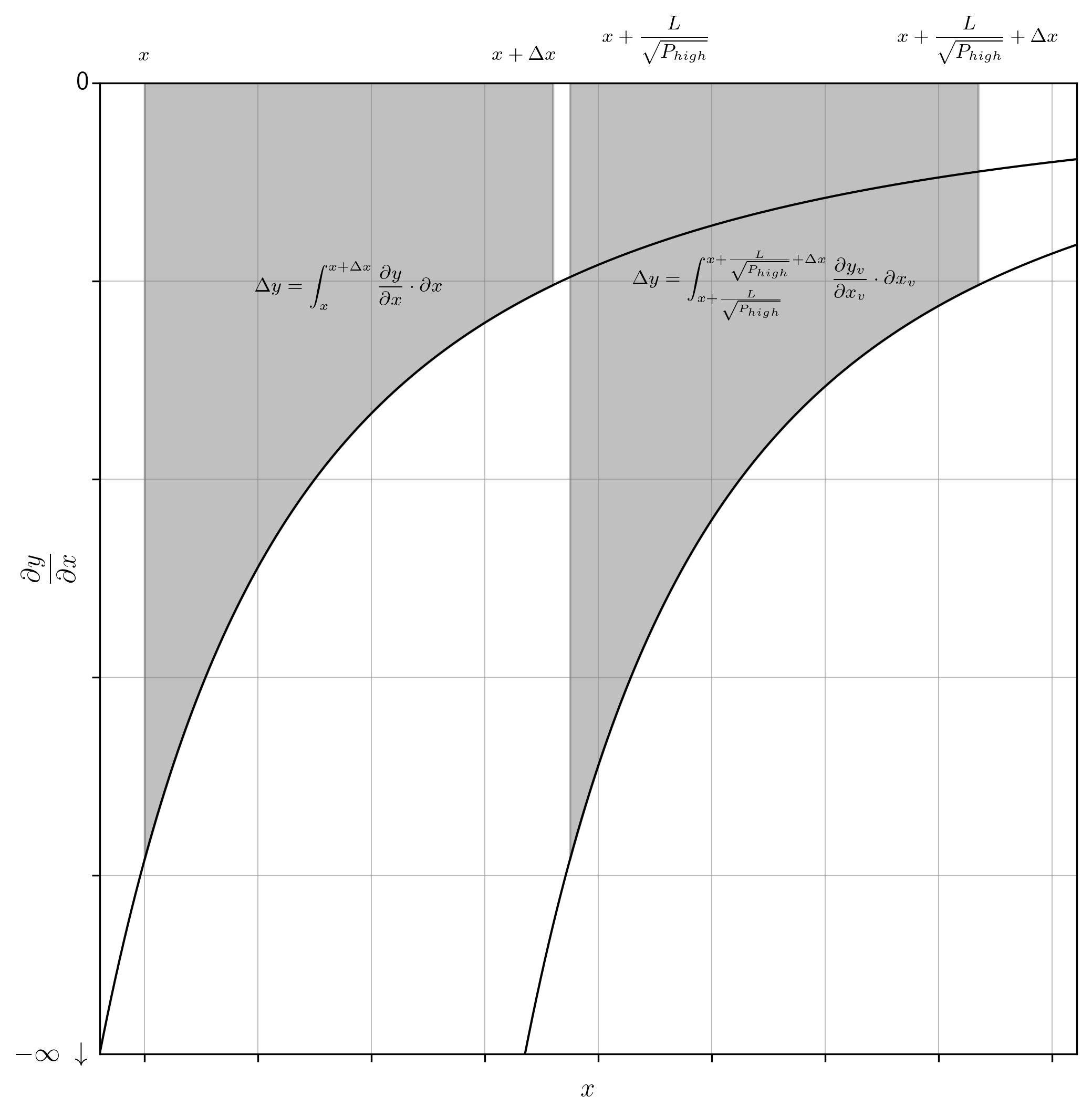}
    \captionsetup{
        justification=raggedright,
        singlelinecheck=false,
        font=small,
        labelfont=bf,
        labelsep=quad,
        format=plain
    }
    \caption{The integration above $\partial y / \partial x = - L^{2} / \left( x + L / \sqrt{P_{\text{high}}} \right)^{2}$ and $\partial y_{\text{v}} / \partial x_{\text{v}} = - L^{2} / x_{\text{v}}^{2}$ (Equations \ref{eq123} and \ref{eq170}) over the intervals $x \rightarrow x + \mathrm{\Delta}x$ and $x + L / \sqrt{P_{\text{high}}} \rightarrow x + L / \sqrt{P_{\text{high}}} + \mathrm{\Delta}x$ (i.e. $x_{\text{v}} \rightarrow x_{\text{v}} + \mathrm{\Delta}x$) representing a token swap against both the real and virtual curves, where $\mathrm{\Delta}x > 0$ and $\mathrm{\Delta}y < 0$. Apart from the shift along the x-axis, these are identical in every other aspect. Note that the relationship between the real and virtual price curves is additive ($x$ and  $\mathrm{\Delta}x$ versus $x + L / \sqrt{P_{\text{high}}}$ and 
    $x + L / \sqrt{P_{\text{high}}} + \mathrm{\Delta}x$).}
    \label{fig31}
\end{figure}

The indifference of the real and virtual curves with respect the calculated swap quantities is again evident in the analysis (Figures \ref{fig30}, \ref{fig31}). Translocation upon the integrated forms (i.e. the bonding curves, Figure \ref{fig30}) and integration above their implied price curves (Figure \ref{fig31}) yields the same token amounts, $\mathrm{\Delta}x$ and $\mathrm{\Delta}y$. The only difference is the frame of reference, which is shifted by ($-x_{\text{asym}}$, $-y_{\text{asym}}$) in the virtual curve compared to the real curve, which has no effect on either the marginal rate, or the effective rate of exchange. Since the $L$ term is exposed directly on the Uniswap v3 smart contracts, legacy infrastructure that was developed to integrate with prior iterations of the canonical $x \cdot y = k$ on-chain liquidity system can be retrofit with relatively little effort. Therefore, the virtual curve is in some aspect the preferred context through which third party software observes and interacts with the Uniswap v3 system, which was likely designed and implemented with $x \cdot y = k$ as a core assumption. It would be naïve to assume that this in an accident. The parameterization is deliberately self-explanatory and allows for legacy software to be quickly patched with conditional arguments regarding the price boundaries, while allowing the core swap functions to remain in-tact.

\subsection{The Uniswap v3 Reference Curve}\label{subsec4.3}

To begin constructing the Uniswap v3 reference curve, first recall that the product of $x_{0}$ and $y_{0}$ is used as its scaling constant (Equation \ref{eq2}). In other words, $x_{0} \cdot y_{0}$ is equl to $k$ in the proverbial “constant product” equation, $x \cdot y = k$. With the identities of $x_{0}$ and $y_{0}$ in-hand, this scaling constant can be derived from the product of the RHS of Equations \ref{eq160} and \ref{eq162} (Equations \ref{eq171} and \ref{eq172}). While there are many compelling algebraic rearrangements for this identity, the form in Equation \ref{eq172} was chosen for its ease of use in the derivations to follow, and as a call-back to the so-called “capital efficiency” term previously isolated in Equation \ref{eq53}. 

\begin{flalign}
& \text{\renewcommand{\arraystretch}{0.66}
    \begin{tabular}{@{}c@{}}
    \scriptsize from \\
    \scriptsize (\ref{eq2})\\\scriptsize (\ref{eq160})\\\scriptsize (\ref{eq162})
  \end{tabular}} 
  & 
  x \cdot y = L \cdot \left( \frac{1}{\sqrt[4]{P_{\text{high}}}} \cdot \left( \frac{1}{\sqrt[4]{P_{\text{low}}}} - \frac{1}{\sqrt[4]{P_{\text{high}}}} \right) \right) \cdot L \cdot \left( \sqrt[4]{P_{\text{low}}} \cdot \left( \sqrt[4]{P_{\text{high}}} - \sqrt[4]{P_{\text{low}}} \right) \right)
  &  
  \label{eq171} 
  &
\end{flalign}

\begin{flalign}
& \text{\renewcommand{\arraystretch}{0.66}
    \begin{tabular}{@{}c@{}}
    \scriptsize from \\
    \scriptsize (\ref{eq171})
  \end{tabular}} 
  & 
  x \cdot y = L^{2} \cdot \left( 1 - 2 \cdot \displaystyle \frac{\sqrt[4]{P_{\text{low}}}}{\sqrt[4]{P_{\text{high}}}} + \left( \displaystyle \frac{\sqrt[4]{P_{\text{low}}}}{\sqrt[4]{P_{\text{high}}}} \right)^{2} \right) = L^{2} \cdot \left( 1 - \displaystyle \frac{\sqrt[4]{P_{\text{low}}}}{\sqrt[4]{P_{\text{high}}}} \right)^{2}
  &  
  \label{eq172} 
  &
\end{flalign}

\begin{figure}[ht]
    \centering
    \includegraphics[width=\textwidth]{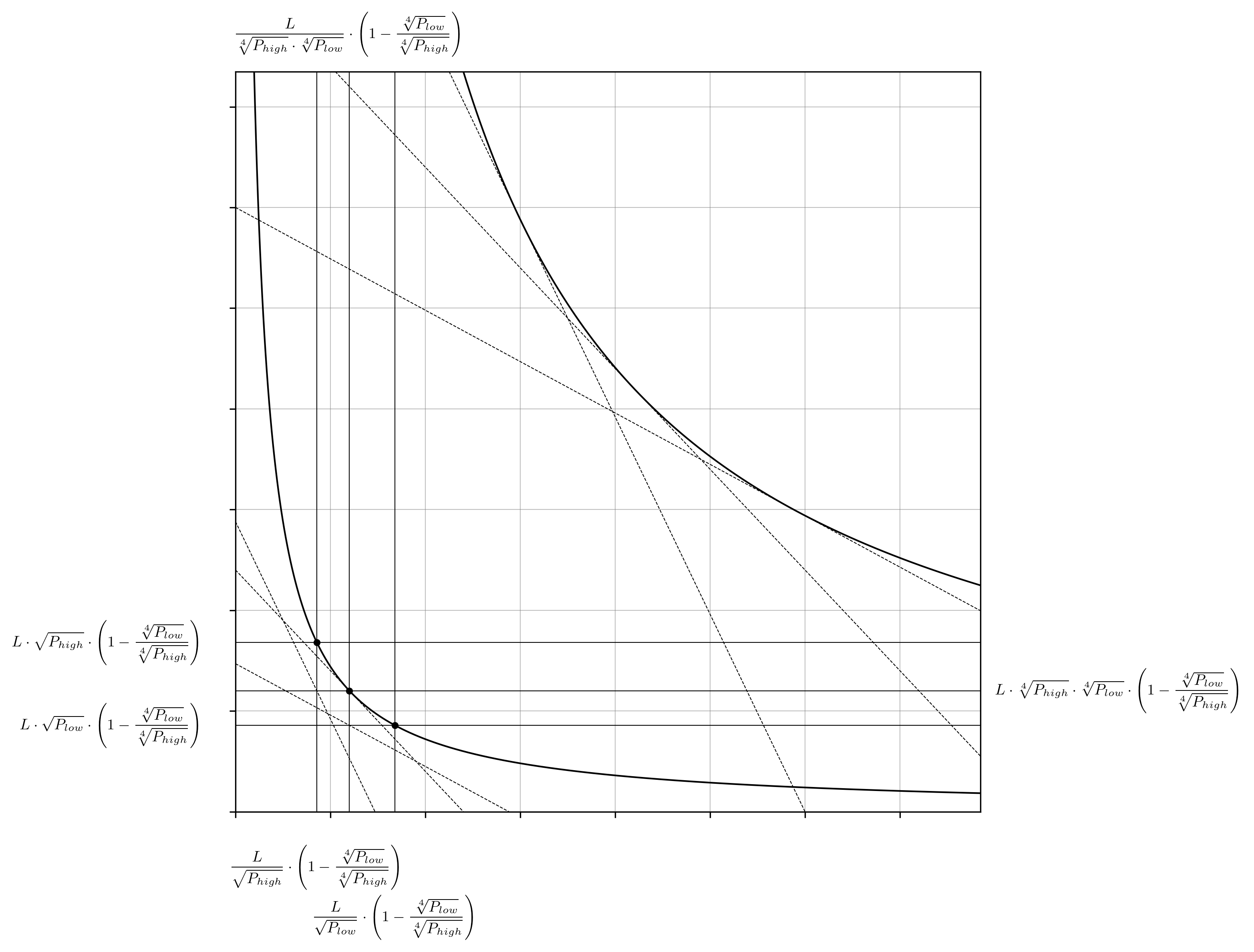}
    \captionsetup{
        justification=raggedright,
        singlelinecheck=false,
        font=small,
        labelfont=bf,
        labelsep=quad,
        format=plain
    }
    \caption{The algebraic identities of the points where the lines tangent to the Uniswap v3 reference curve and parallel to the tangent lines at the price boundaries of the virtual curve are now elucidated (Equations \ref{eq160}, \ref{eq162}, \ref{eq174}, \ref{eq176}, \ref{eq178}, \ref{eq180}).}
    \label{fig32}
\end{figure}

With the reference curve’s scaling constant now defined in terms of $L$, $P_{\text{high}}$ and $P_{\text{low}}$, the coordinates corresponding to the price bounds, $\min(x)$, $\max(x)$, $\min(y)$, $\max(y)$, can also be expressed with the same parameters. In an earlier demonstration, this process was made trivial by “reversing” the effective amplification of the virtual curve relative to the reference curve by taking the quotients of $\min(x_{\text{v}})$, $\max(y_{\text{v}})$, $\max(x_{\text{v}})$, and $\min(y_{\text{v}})$, and the amplification constant $A$ (Equations \ref{eq59}, \ref{eq60}, \ref{eq61}, \ref{eq62}). While this process is still possible via the definition of $A$ in terms of $P_{\text{high}}$ and $P_{\text{low}}$ (Equation \ref{eq53}), it no longer has the same intuitive benefit with respect to the construction process. This is a good excuse to explore the more analytical approach. The process is identical to that demonstrated when attempting to add some rigor to the amplification reversal method (Equations \ref{eq63} and \ref{eq64}); the derivative of the reference curve (Equations \ref{eq16} and \ref{eq17}) is forced to either $P_{\text{high}}$ or $P_{\text{low}}$ (native parameters in this case), and the $x$ or $y$ variable is substituted for $\min(x)$, $\max(x)$, $\min(y)$, or $\max(y)$, as appropriate. Then the coordinate of interest is obtained via trivial rearrangements of the resulting expressions (Equations \ref{eq173}, \ref{eq174}, \ref{eq175}, \ref{eq176}, \ref{eq177}, \ref{eq178}, \ref{eq179} and \ref{eq180}) (Figure \ref{fig32}).

\begin{flalign}
& \text{\renewcommand{\arraystretch}{0.66}
    \begin{tabular}{@{}c@{}}
    \scriptsize from \\
    \scriptsize (\ref{eq17})\\\scriptsize (\ref{eq162})
  \end{tabular}} 
  & 
  \displaystyle \frac{\partial y}{\partial x} = - P_{\text{high}} = - \displaystyle \frac{L^{2} \cdot \left( 1 - \displaystyle \frac{\sqrt[4]{P_{\text{low}}}}{\sqrt[4]{P_{\text{high}}}} \right)^{2}}{\min^{2}(x)}
  &  
  \label{eq173} 
  &
\end{flalign}

\begin{flalign}
& \text{\renewcommand{\arraystretch}{0.66}
    \begin{tabular}{@{}c@{}}
    \scriptsize from \\
    \scriptsize (\ref{eq173})
  \end{tabular}} 
  & 
  \min \left( x \right) = \displaystyle \frac{L}{\sqrt{P_{\text{high}}}} \cdot \left( 1 - \displaystyle \frac{\sqrt[4]{P_{\text{low}}}}{\sqrt[4]{P_{\text{high}}}} \right)
  &  
  \label{eq174} 
  &
\end{flalign}

\begin{flalign}
& \text{\renewcommand{\arraystretch}{0.66}
    \begin{tabular}{@{}c@{}}
    \scriptsize from \\
    \scriptsize (\ref{eq17})\\\scriptsize (\ref{eq162})
  \end{tabular}} 
  & 
  \displaystyle \frac{\partial y}{\partial x} = - P_{\text{low}} = - \displaystyle \frac{L^{2} \cdot \left( 1 - \displaystyle \frac{\sqrt[4]{P_{\text{low}}}}{\sqrt[4]{P_{\text{high}}}} \right)^{2}}{\max^{2}(x)}
  &  
  \label{eq175} 
  &
\end{flalign}

\begin{flalign}
& \text{\renewcommand{\arraystretch}{0.66}
    \begin{tabular}{@{}c@{}}
    \scriptsize from \\
    \scriptsize (\ref{eq175})
  \end{tabular}} 
  & 
  \max \left( x \right) = \displaystyle \frac{L}{\sqrt{P_{\text{low}}}} \cdot \left( 1 - \displaystyle \frac{\sqrt[4]{P_{\text{low}}}}{\sqrt[4]{P_{\text{high}}}} \right)
  &  
  \label{eq176} 
  &
\end{flalign}

\begin{flalign}
& \text{\renewcommand{\arraystretch}{0.66}
    \begin{tabular}{@{}c@{}}
    \scriptsize from \\
    \scriptsize (\ref{eq16})\\\scriptsize (\ref{eq162})
  \end{tabular}} 
  & 
  \displaystyle \frac{\partial y}{\partial x} = - P_{\text{low}} = - \displaystyle \frac{\min^{2}(y)}{L^{2} \cdot \left( 1 - \displaystyle \frac{\sqrt[4]{P_{\text{low}}}}{\sqrt[4]{P_{\text{high}}}} \right)^{2}}
  &  
  \label{eq177} 
  &
\end{flalign}

\begin{flalign}
& \text{\renewcommand{\arraystretch}{0.66}
    \begin{tabular}{@{}c@{}}
    \scriptsize from \\
    \scriptsize (\ref{eq177})
  \end{tabular}} 
  & 
  \min \left( y \right) = L \cdot \sqrt{P_{\text{low}}} \cdot \left( 1 - \displaystyle \frac{\sqrt[4]{P_{\text{low}}}}{\sqrt[4]{P_{\text{high}}}} \right)
  &  
  \label{eq178} 
  &
\end{flalign}

\begin{flalign}
& \text{\renewcommand{\arraystretch}{0.66}
    \begin{tabular}{@{}c@{}}
    \scriptsize from \\
    \scriptsize (\ref{eq16})\\\scriptsize (\ref{eq162})
  \end{tabular}} 
  & 
  \displaystyle \frac{\partial y}{\partial x} = - P_{\text{high}} = - \displaystyle \frac{{\max^{2}}(y)}{L^{2} \cdot \left( 1 - \displaystyle \frac{\sqrt[4]{P_{\text{low}}}}{\sqrt[4]{P_{\text{high}}}} \right)^{2}}
  &  
  \label{eq179} 
  &
\end{flalign}

\begin{flalign}
& \text{\renewcommand{\arraystretch}{0.66}
    \begin{tabular}{@{}c@{}}
    \scriptsize from \\
    \scriptsize (\ref{eq179})
  \end{tabular}} 
  & 
  \max \left( y \right) = L \cdot \sqrt{P_{\text{high}}} \cdot \left( 1 - \displaystyle \frac{\sqrt[4]{P_{\text{low}}}}{\sqrt[4]{P_{\text{high}}}} \right)
  &  
  \label{eq180} 
  &
\end{flalign}

\begin{figure}[ht]
    \centering
    \includegraphics[width=\textwidth]{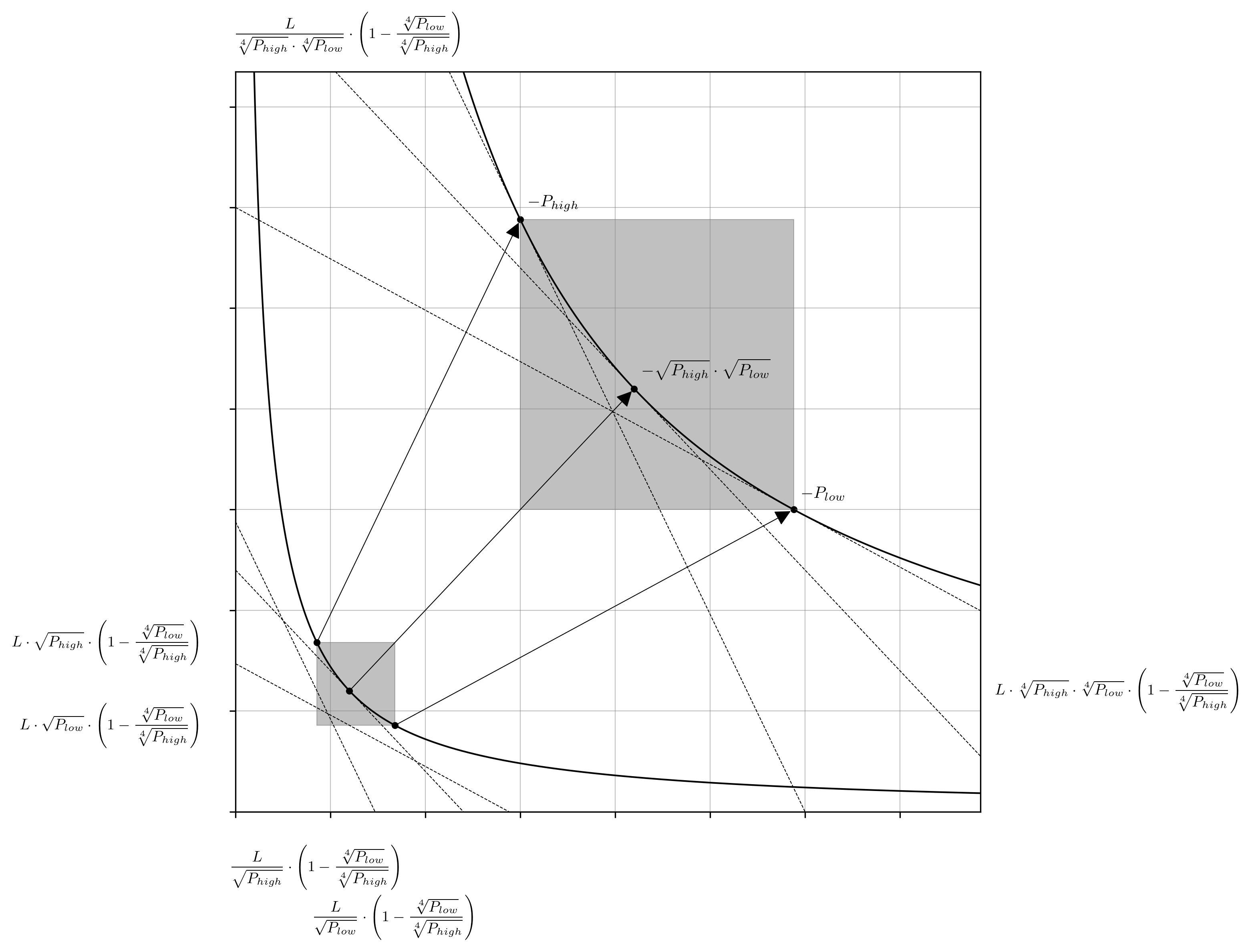}
    \captionsetup{
        justification=raggedright,
        singlelinecheck=false,
        font=small,
        labelfont=bf,
        labelsep=quad,
        format=plain
    }
    \caption{The amplification process is depicted with corresponding shaded areas of the Uniswap v3 reference and virtual curves, respectively. Points where the first derivative of each curve evaluates to the same result are shown as white dots, and the mapping of these points is depicted with white arrows.}
    \label{fig33}
\end{figure}

As noted above, the analysis of the Uniswap v3 reparameterization and its real, virtual, and reference curves has been necessarily performed in the reverse order compared to the seminal work (Figures \ref{fig1} to \ref{fig21}). Regardless, the liquidity amplification heuristic applies all the same, and can be represented as before with arrows connecting pairs of points on the reference and virtual curves, where the derivatives evaluated at these points are equal (Figures \ref{fig33}).

\begin{figure}[ht]
    \centering
    \includegraphics[width=\textwidth]{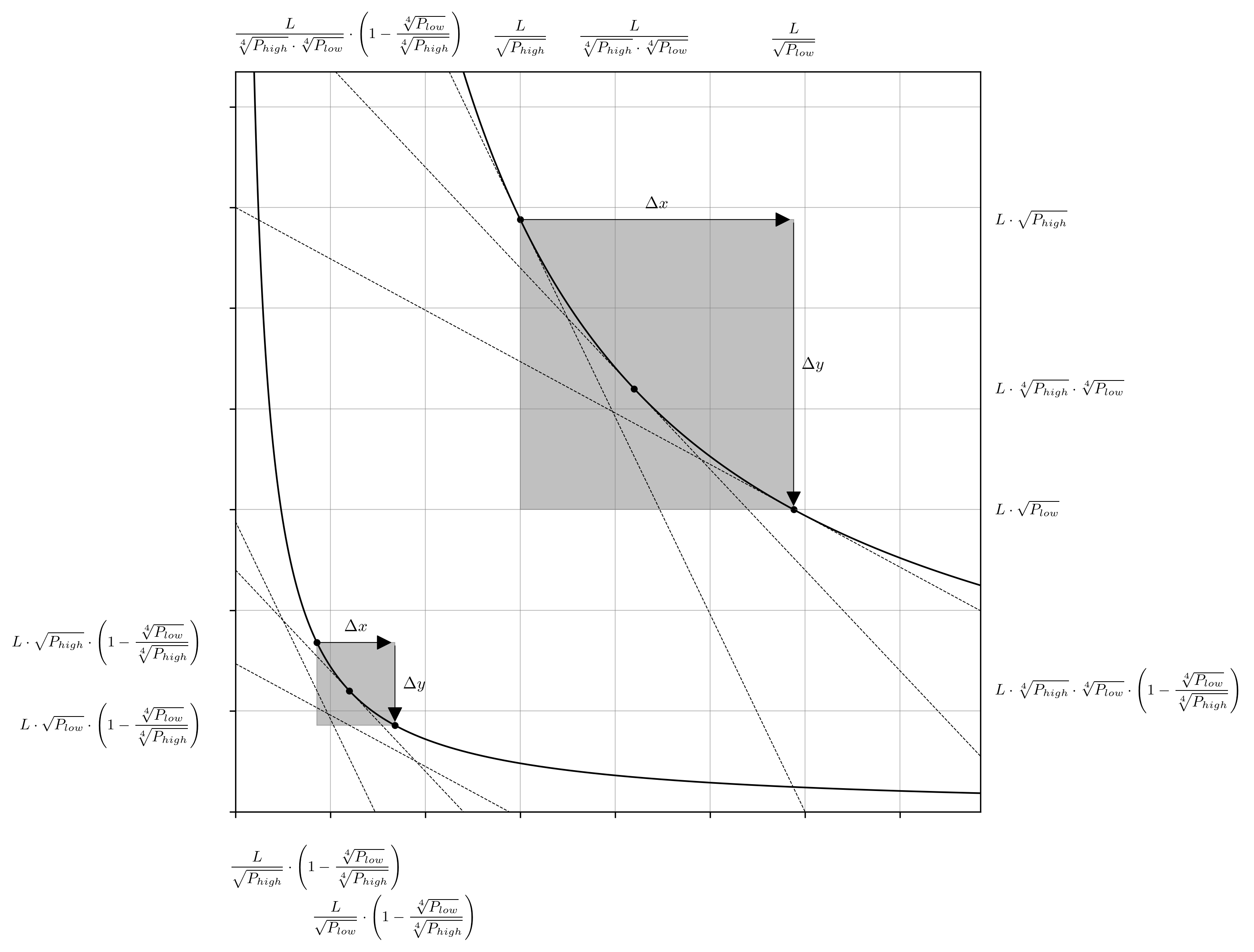}
    \captionsetup{
        justification=raggedright,
        singlelinecheck=false,
        font=small,
        labelfont=bf,
        labelsep=quad,
        format=plain
    }
    \caption{Traversal upon the rectangular hyperbolas $x_{\text{v}} \cdot y_{\text{v}} = L^{2}$ and $x \cdot y = L^{2} \cdot \left(1 - \sqrt[4]{P_{\text{low}}} / \sqrt[4]{P_{\text{high}}} \right)^{2}$ (Equations \ref{eq163} and \ref{eq172}), representing a token swap against the reference and amplified liquidity pools, where $\mathrm{\Delta}x > 0$ and $\mathrm{\Delta}y < 0$. The marginal rates of exchange before and after the swap are identical. The ratio of the $\mathrm{\Delta}x$ and $\mathrm{\Delta}y$ arrow lengths for each curve are also identical, and therefore the overall rate of exchange, $\mathrm{\Delta}y / \mathrm{\Delta}x$ is equal in both cases.}
    \label{fig34}
\end{figure}

\begin{figure}[ht]
    \centering
    \includegraphics[width=\textwidth]{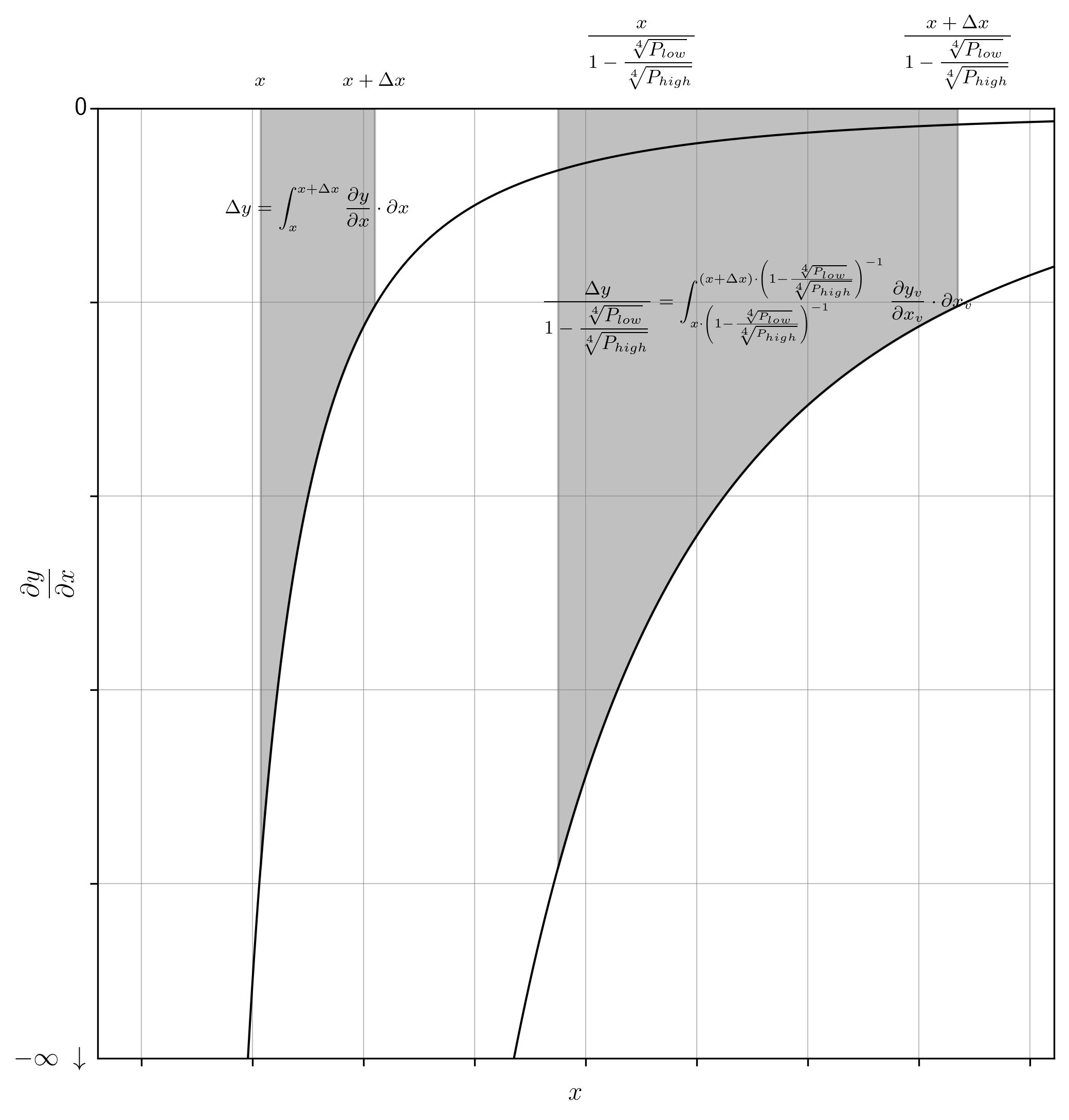}
    \captionsetup{
        justification=raggedright,
        singlelinecheck=false,
        font=small,
        labelfont=bf,
        labelsep=quad,
        format=plain
    }
    \caption{The integration above $\partial y / \partial x = - L^{2} \cdot \left( 1 - \sqrt[4]{P_{\text{low}}} / \sqrt[4]{P_{\text{high}}} \right)^{2} / x^{2}$ (i.e. $\partial / \partial x$ Equation \ref{eq172}) and $\partial y / \partial x = - L^{2} / x^{2}$ (Equation \ref{eq170}) over the intervals $x \rightarrow x + \mathrm{\Delta}x$ and $x / \left( 1 - \sqrt[4]{P_{\text{low}}} / \sqrt[4]{P_{\text{high}}} \right) \rightarrow \left(x + \mathrm{\Delta}x \right) / \left( 1 - \sqrt[4]{P_{\text{low}}} / \sqrt[4]{P_{\text{high}}} \right)$ (i.e. $x_{\text{v}} \rightarrow x_{\text{v}} + \mathrm{\Delta}x$) representing a token swap against both the reference and amplified liquidity pools, where $\mathrm{\Delta}x > 0$ and $\mathrm{\Delta}y < 0$. Note that that the integration areas for the reference and virtual curves are $\mathrm{\Delta}y$ and $\mathrm{\Delta}y / \left( 1 - \sqrt[4]{P_{\text{low}}} / \sqrt[4]{P_{\text{high}}} \right)$, respectively. By inference, any token swap against the virtual curve can be calculated from the reference curve by taking its quotient versus the poorly named “capital efficiency” term (Equation \ref{eq53}).}
    \label{fig35}
\end{figure}

The relative trade volumes supported by the reference and virtual curves over the same price interval can now be compared (Figures \ref{fig34} to \ref{fig35}). As before, note that while the trade action executes from and terminates at the same marginal price values and achieves the same overall effective exchange rate, the trade amounts are significantly greater in the virtual curve. The increased trade volume can be inspected visually from the increased arrow lengths in Figure \ref{fig34}, and the expansion observed for the integrated area in Figure \ref{fig35}.

This concludes our examination of the Uniswap v3 reparameterization. 

\subsection{The Carbon DeFi Real Curve}\label{subsec4.4}

Bancor’s peer-to-peer trading product, \textit{Carbon DeFi}, also uses a unique, “unnatural” reparameterization of the seminal concentrated liquidity invariant informed by the protocol’s specific requirements. Unlike its predecessors, Carbon Defi’s purpose is not to re-create Bancor’s original vision for automatic liquidity systems vis-à-vis continuous pricing methods that attempt to follow the market, but rather to support the arbitrary price quoting behaviours exhibited by market participants. The particulars of the Carbon DeFi design are beyond the scope of this document and deserve a dedicated discourse. However, the necessary context is simple enough to state ad hoc. 

In short, Carbon DeFi system must be able to adjust the curve’s scaling parameter reliably and accurately as required under certain circumstances to maintain price quote boundaries as dictated by its users. This is potentially a sensitive process. Naïvely shoe-horning either of the implementations previously discussed for Bancor v2 or Uniswap v3 is an inelegant and unnecessarily computationally intensive choice at best, and risks exposing an exploit resulting from precision loss during fixed-point operations at worst. Therefore, Carbon DeFi requires the scaling parameter of the curve to be expressed exclusively in terms of one token balance, which can be adjusted with reference to the same without performing any calculations at all, thus observing best practices in defensive programming. Just as the Uniswap v3 reparameterization is the correct form for the system it represents, the Carbon DeFi reparameterization is correct for is own systems, but \textit{both refer to the same underlying mathematical object}.

As an aside, Carbon DeFi’s curves are also unusual because they refer to only one token balance, rather than two. This has no immediate ramifications for the analyses we are about to perform but is worth raising for its own sake and to foreshadow a future publication which will handle these details more thoroughly. The only detail needed is that, by convention, Carbon DeFi treats the y-axis as belonging to the only token its bonding curve describes. Therefore, the derivation of the Carbon DeFi real curve begins from its most conspicuous identity, the y-intercept. 

\begin{figure}[ht]
    \centering
    \includegraphics[width=\textwidth]{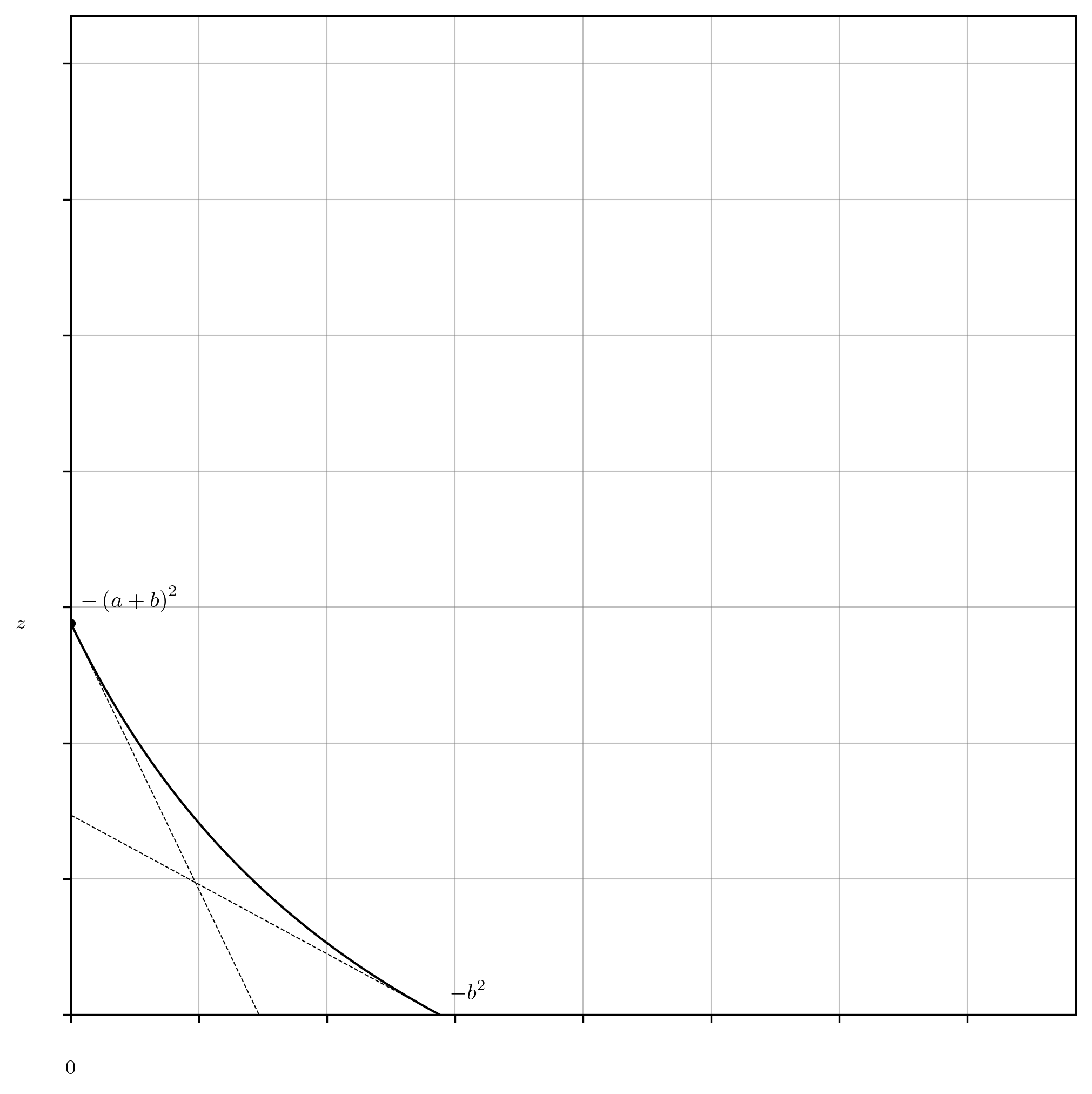}
    \captionsetup{
        justification=raggedright,
        singlelinecheck=false,
        font=small,
        labelfont=bf,
        labelsep=quad,
        format=plain
    }
    \caption{Initial Carbon DeFi real curve. Due to its parameterization and prior to performing any additional analysis, only the slopes of the tangent lines at the intercepts $-P_{\text{high}} = - \left( a + b \right)^{2}$ and $-P_{\text{low}} = - b^{2}$, and the y-intercept itself, $y_{\text{int}} = z$, are known.}
    \label{fig36}
\end{figure}

Since Carbon DeFi is a price quoting protocol, the price boundaries represented by $P_{\text{high}}$ and $P_{\text{low}}$ are obvious choices to encapsulate natively. The objective then becomes to substitute the $L$ term in Equation \ref{eq110} with a redundant definition expressed in terms of the y-intercept, $y_{\text{int}}$, and the price boundaries, $P_{\text{high}}$ and $P_{\text{low}}$. This can be accomplished easily with reference to the work already done. The y-intercept was previously defined in terms of $L$, $P_{\text{high}}$, and $P_{\text{low}}$ in Equation \ref{eq130}; rearrangement of this identity to isolate $L$ (Equation \ref{eq181}), followed by substitution into Equation \ref{eq110} yields the reparametrized Carbon DeFi real curve invariant (Equation \ref{eq182}). Refactoring the expression into a more presentable phenotype is done by defining the parameters $a$, $b$, and $z$, which are equal to $\sqrt{P_{\text{high}}} - \sqrt{P_{\text{low}}}$, $\sqrt{P_{\text{low}}}$, and the y-intercept, $y_{\text{int}}$, of the real curve, respectively (Equations \ref{eq183}, \ref{eq184} and  \ref{eq185}). Substitution of these definitions into Equation \ref{eq182} yields Equation \ref{eq186}, and rearrangement yields Equations \ref{eq187} and \ref{eq188}. Note that the differences between Equations \ref{eq182}, \ref{eq186}, \ref{eq187} and \ref{eq188} are only cosmetic. These features are depicted in Figure \ref{fig36}.

\begin{flalign}
& \text{\renewcommand{\arraystretch}{0.66}
    \begin{tabular}{@{}c@{}}
    \scriptsize from \\
    \scriptsize (\ref{eq130})
  \end{tabular}} 
  & 
  L = \displaystyle \frac{y_{\text{int}}}{\sqrt{P_{\text{high}}} - \sqrt{P_{\text{low}}}}
  &  
  \label{eq181} 
  &
\end{flalign}

\begin{flalign}
& \text{\renewcommand{\arraystretch}{0.66}
    \begin{tabular}{@{}c@{}}
    \scriptsize from \\
    \scriptsize (\ref{eq110})\\\scriptsize (\ref{eq181})
  \end{tabular}} 
  & 
  \left( x + \displaystyle \frac{y_{\text{int}}}{\sqrt{P_{\text{high}}} \cdot \left( \sqrt{P_{\text{high}}} - \sqrt{P_{\text{low}}} \right)} \right) \cdot \left( y + \displaystyle \frac{\sqrt{P_{\text{low}}} \cdot y_{\text{int}}}{\sqrt{P_{\text{high}}} - \sqrt{P_{\text{low}}}} \right) = \displaystyle \frac{{y_{\text{int}}}^{2}}{\left( \sqrt{P_{\text{high}}} - \sqrt{P_{\text{low}}} \right)^{2}}
  &  
  \label{eq182} 
  &
\end{flalign}

\begin{equation} \label{eq183}
a = \sqrt{P_{\text{high}}} - \sqrt{P_{\text{low}}}
\end{equation}

\begin{equation} \label{eq184}
b = \sqrt{P_{\text{low}}}
\end{equation}

\begin{equation} \label{eq185}
z = y_{\text{int}}
\end{equation}

\begin{flalign}
& \text{\renewcommand{\arraystretch}{0.66}
    \begin{tabular}{@{}c@{}}
    \scriptsize from \\
    \scriptsize (\ref{eq182})\\\scriptsize (\ref{eq183})\\\scriptsize (\ref{eq184})\\\scriptsize (\ref{eq185})
  \end{tabular}} 
  & 
  \left( x + \displaystyle \frac{z}{a \cdot (a + b)} \right) \cdot \left( y + \displaystyle \frac{z \cdot b}{a} \right) = \displaystyle \frac{z^{2}}{a^{2}}
  &  
  \label{eq186} 
  &
\end{flalign}

\begin{flalign}
& \text{\renewcommand{\arraystretch}{0.66}
    \begin{tabular}{@{}c@{}}
    \scriptsize from \\
    \scriptsize (\ref{eq186})
  \end{tabular}} 
  & 
  y \cdot \left( z + x \cdot a \cdot (a + b) \right) = z \cdot \left( z - x \cdot b \cdot (a + b) \right)
  &  
  \label{eq187} 
  &
\end{flalign}

\begin{flalign}
& \text{\renewcommand{\arraystretch}{0.66}
    \begin{tabular}{@{}c@{}}
    \scriptsize from \\
    \scriptsize (\ref{eq186})
  \end{tabular}} 
  & 
  \displaystyle \frac{y}{z} = \displaystyle \frac{z - x \cdot b \cdot (a + b)}{z + x \cdot a \cdot (a + b)}
  &  
  \label{eq188} 
  &
\end{flalign}

It might be helpful to consider how this arrangement can be expressed in more familiar terms. The expression $a \cdot \left( a + b \right)$ is redundant with $P_{\text{high}} - P_{0}$, and $b \cdot \left( a + b \right)$ is redundant with $P_{0}$ (Equations \ref{eq189}, \ref{eq190} and \ref{eq191}). Substitution of these relationships and the identity $z = y_{\text{int}}$ into Equation \ref{eq188} yields Equation \ref{eq191}. 

\begin{flalign}
& \text{\renewcommand{\arraystretch}{0.66}
    \begin{tabular}{@{}c@{}}
    \scriptsize from \\
    \scriptsize (\ref{eq183})\\\scriptsize (\ref{eq184})
  \end{tabular}} 
  & 
  a \cdot (a + b) = \left( \sqrt{P_{\text{high}}} - \sqrt{P_{\text{low}}} \right) \cdot \left( \sqrt{P_{\text{high}}} - \sqrt{P_{\text{low}}} + \sqrt{P_{\text{low}}} \right) = P_{\text{high}} - P_{0}
  &  
  \label{eq189} 
  &
\end{flalign}

\begin{flalign}
& \text{\renewcommand{\arraystretch}{0.66}
    \begin{tabular}{@{}c@{}}
    \scriptsize from \\
    \scriptsize (\ref{eq183})\\\scriptsize (\ref{eq184})
  \end{tabular}} 
  & 
  b \cdot (a + b) = \sqrt{P_{\text{low}}} \cdot \left( \sqrt{P_{\text{high}}} - \sqrt{P_{\text{low}}} + \sqrt{P_{\text{low}}} \right) = P_{0}
  &  
  \label{eq190} 
  &
\end{flalign}

\begin{flalign}
& \text{\renewcommand{\arraystretch}{0.66}
    \begin{tabular}{@{}c@{}}
    \scriptsize from \\
    \scriptsize (\ref{eq186})\\\scriptsize (\ref{eq187})\\\scriptsize (\ref{eq188})
  \end{tabular}} 
  & 
  \displaystyle \frac{y}{y_{\text{int}}} = \displaystyle \frac{y_{\text{int}} - x \cdot P_{0}}{y_{\text{int}} + x \cdot \left( P_{\text{high}} - P_{0} \right)} = \mathbf{Eqn}\mathbf{.\ }\mathbf{194}
  &  
  \label{eq191} 
  &
\end{flalign}

As was performed for the Uniswap v3 reparameterization, we will now begin the process of completely elucidating the Carbon DeFi model commencing from its real curve description (Equations \ref{eq186}, \ref{eq187} and  \ref{eq188}). The objectives are the same as before – to characterise the real, virtual, and reference curves, and recontextualize them in terms of the general theory. First, manipulation of the Carbon DeFi invariant (Equation \ref{eq186}) to isolate the $x$ and $y$ terms yields Equations \ref{eq192} and \ref{eq193}.

\begin{flalign}
& \text{\renewcommand{\arraystretch}{0.66}
    \begin{tabular}{@{}c@{}}
    \scriptsize from \\
    \scriptsize (\ref{eq186})
  \end{tabular}} 
  & 
  x = \displaystyle \frac{z^{2}}{a^{2} \cdot \left( y + \displaystyle \frac{b \cdot z}{a} \right)} - \displaystyle \frac{z}{a \cdot (a + b)}
  &  
  \label{eq192} 
  &
\end{flalign}

\begin{flalign}
& \text{\renewcommand{\arraystretch}{0.66}
    \begin{tabular}{@{}c@{}}
    \scriptsize from \\
    \scriptsize (\ref{eq186})
  \end{tabular}} 
  & 
  y = \displaystyle \frac{z^{2}}{a^{2} \cdot \left( x + \displaystyle \frac{z}{a \cdot (a + b)} \right)} - \displaystyle \frac{b \cdot z}{a}
  &  
  \label{eq193} 
  &
\end{flalign}

The token swap equations are obtained via now familiar methods. The $x$ and $y$ terms in Equation \ref{eq186}, or Equations \ref{eq192} and \ref{eq193} are substituted for $x + \mathrm{\Delta}x$ and $y + \mathrm{\Delta}y$ then rearranged to make either $\mathrm{\Delta}x$ or $\mathrm{\Delta}y$ the subject (Equations \ref{eq194}, \ref{eq195} and \ref{eq196}).
 
\begin{flalign}
& \text{\renewcommand{\arraystretch}{0.66}
    \begin{tabular}{@{}c@{}}
    \scriptsize from \\
    \scriptsize (\ref{eq186})
  \end{tabular}} 
  & 
  \left( x + \mathrm{\Delta}x + \displaystyle \frac{z}{a \cdot (a + b)} \right) \cdot \left( y + \mathrm{\Delta}y + \displaystyle \frac{b \cdot z}{a} \right) = \displaystyle \frac{z^{2}}{a^{2}}
  &  
  \label{eq194} 
  &
\end{flalign}

\begin{flalign}
& \text{\renewcommand{\arraystretch}{0.66}
    \begin{tabular}{@{}c@{}}
    \scriptsize from \\
    \scriptsize (\ref{eq192})\\\scriptsize (\ref{eq194})
  \end{tabular}} 
  & 
  \mathrm{\Delta}x = \displaystyle \frac{z^{2}}{a^{2} \cdot \left( y + \mathrm{\Delta}y + \displaystyle \frac{b \cdot z}{a} \right)} - \displaystyle \frac{z}{a \cdot (a + b)} - x
  &  
  \label{eq195} 
  &
\end{flalign}

\begin{flalign}
& \text{\renewcommand{\arraystretch}{0.66}
    \begin{tabular}{@{}c@{}}
    \scriptsize from \\
    \scriptsize (\ref{eq193})\\\scriptsize (\ref{eq194})
  \end{tabular}} 
  & 
  \mathrm{\Delta}y = \displaystyle \frac{z^{2}}{a^{2} \cdot \left( x + \mathrm{\Delta}x + \displaystyle \frac{z}{a \cdot (a + b)} \right)} - \displaystyle \frac{b \cdot z}{a} - y
  &  
  \label{eq196} 
  &
\end{flalign}

Substituting the identities for $x$ and $y$ in Equations \ref{eq192} and \ref{eq193} into Equations \ref{eq195} and \ref{eq196} (Equations \ref{eq197} and \ref{eq199}), then simplifying, results in the single-dimension swap equations (Equations \ref{eq198} and \ref{eq200}).

\begin{flalign}
& \text{\renewcommand{\arraystretch}{0.66}
    \begin{tabular}{@{}c@{}}
    \scriptsize from \\
    \scriptsize (\ref{eq192})\\\scriptsize (\ref{eq195})
  \end{tabular}} 
  & 
  \mathrm{\Delta}x = \displaystyle \frac{z^{2}}{a^{2} \cdot \left( y + \mathrm{\Delta}y + \displaystyle \frac{b \cdot z}{a} \right)} - \displaystyle \frac{z}{a \cdot (a + b)} - \displaystyle \frac{z^{2}}{a^{2} \cdot \left( y + \displaystyle \frac{b \cdot z}{a} \right)} + \displaystyle \frac{z}{a \cdot (a + b)}
  &  
  \label{eq197} 
  &
\end{flalign}

\begin{flalign}
& \text{\renewcommand{\arraystretch}{0.66}
    \begin{tabular}{@{}c@{}}
    \scriptsize from \\
    \scriptsize (\ref{eq197})
  \end{tabular}} 
  & 
  \mathrm{\Delta}x = - \displaystyle \frac{\mathrm{\Delta}y \cdot z^{2}}{(a \cdot y + b \cdot z) \cdot \left( a \cdot (y + \mathrm{\Delta}y) + b \cdot z \right)}
  &  
  \label{eq198} 
  &
\end{flalign}

\begin{flalign}
& \text{\renewcommand{\arraystretch}{0.66}
    \begin{tabular}{@{}c@{}}
    \scriptsize from \\
    \scriptsize (\ref{eq193})\\\scriptsize (\ref{eq196})
  \end{tabular}} 
  & 
  \mathrm{\Delta}y = \displaystyle \frac{z^{2}}{a^{2} \cdot \left( x + \mathrm{\Delta}x + \displaystyle \frac{z}{a \cdot (a + b)} \right)} - \displaystyle \frac{b \cdot z}{a} - \displaystyle \frac{z^{2}}{a^{2} \cdot \left( x + \displaystyle \frac{z}{a \cdot (a + b)} \right)} + \displaystyle \frac{b \cdot z}{a}
  &  
  \label{eq199} 
  &
\end{flalign}

\begin{flalign}
& \text{\renewcommand{\arraystretch}{0.66}
    \begin{tabular}{@{}c@{}}
    \scriptsize from \\
    \scriptsize (\ref{eq199})
  \end{tabular}} 
  & 
  \mathrm{\Delta}y = - \displaystyle \frac{\mathrm{\Delta}x \cdot z^{2} \cdot (a + b)^{2}}{\left( x \cdot a \cdot (a + b) + z \right) \cdot \left( (x + \mathrm{\Delta}x) \cdot a \cdot (a + b) + z \right)}
  &  
  \label{eq200} 
  &
\end{flalign}

The swap formulas Equations \ref{eq198} and \ref{eq200} can also be obtained directly from their seminal counterparts, Equations \ref{eq81} and \ref{eq83}, by substituting the scaling term, $A^{2} \cdot x_{0} \cdot y_{0}$, and the horizontal and vertical shift terms, $x_{0} \cdot \left( A - 1 \right)$ and $y_{0} \cdot \left( A - 1 \right)$, for their reparameterized forms $z^{2}$, $z / \left( a \cdot \left( a + b \right) \right)$, and $\left( b \cdot z \right) / a$, respectively (Equations \ref{eq181}, \ref{eq182}, \ref{eq183}, \ref{eq184}, \ref{eq185}, \ref{eq186}, \ref{eq187} and \ref{eq188}). This approach is similarly well-suited to obtaining the effective and marginal rate equations, but as before, these identities will be elaborated according to the previously established methods. Repeating the process demonstrated in Equations \ref{eq12}, \ref{eq13}, \ref{eq14} and \ref{eq15}, Equations \ref{eq84}, \ref{eq85}, \ref{eq86} and \ref{eq87}, and Equations \ref{eq120}, \ref{eq121}, \ref{eq122} and \ref{eq123}, rearrangement of Equations \ref{eq198} and \ref{eq200} gives the \textit{effective rate of exchange} (Equations \ref{eq201} and \ref{eq203}), followed by determination of the limit as the denominator goes to zero gives the \textit{instantaneous rate of exchange} (i.e. the marginal price) (Equations \ref{eq202} and \ref{eq204}). 

\begin{flalign}
& \text{\renewcommand{\arraystretch}{0.66}
    \begin{tabular}{@{}c@{}}
    \scriptsize from \\
    \scriptsize (\ref{eq196})
  \end{tabular}} 
  & 
  \displaystyle \frac{\mathrm{\Delta}x}{\mathrm{\Delta}y} = - \displaystyle \frac{z^{2}}{(a \cdot y + b \cdot z) \cdot \left( a \cdot (y + \mathrm{\Delta}y) + b \cdot z \right)}
  &  
  \label{eq201} 
  &
\end{flalign}

\begin{flalign}
& \text{\renewcommand{\arraystretch}{0.66}
    \begin{tabular}{@{}c@{}}
    \scriptsize from \\
    \scriptsize (\ref{eq199})
  \end{tabular}} 
  & 
  \displaystyle \frac{\partial x}{\partial y} = \lim_{\mathrm{\Delta}y \rightarrow 0}\displaystyle \frac{\mathrm{\Delta}x}{\mathrm{\Delta}y} = - \displaystyle \frac{z^{2}}{(a \cdot y + b \cdot z)^{2}};\ \displaystyle \frac{\partial y}{\partial x} = - \displaystyle \frac{(a \cdot y + b \cdot z)^{2}}{z^{2}}
  &  
  \label{eq202} 
  &
\end{flalign}

\begin{flalign}
& \text{\renewcommand{\arraystretch}{0.66}
    \begin{tabular}{@{}c@{}}
    \scriptsize from \\
    \scriptsize (\ref{eq198})
  \end{tabular}} 
  & 
  \displaystyle \frac{\mathrm{\Delta}y}{\mathrm{\Delta}x} = - \displaystyle \frac{z^{2} \cdot (a + b)^{2}}{\left( x \cdot a \cdot (a + b) + z \right) \cdot \left( (x + \mathrm{\Delta}x) \cdot a \cdot (a + b) + z \right)}
  &  
  \label{eq203} 
  &
\end{flalign}

\begin{flalign}
& \text{\renewcommand{\arraystretch}{0.66}
    \begin{tabular}{@{}c@{}}
    \scriptsize from \\
    \scriptsize (\ref{eq201})
  \end{tabular}} 
  & 
  \displaystyle \frac{\partial y}{\partial x} = \lim_{\mathrm{\Delta}x \rightarrow 0}\displaystyle \frac{\mathrm{\Delta}y}{\mathrm{\Delta}x} = - \displaystyle \frac{z^{2} \cdot (a + b)^{2}}{\left( x \cdot a \cdot (a + b) + z \right)^{2}};\ \displaystyle \frac{\partial x}{\partial y} = - \displaystyle \frac{\left( x \cdot a \cdot (a + b) + z \right)^{2}}{z^{2} \cdot (a + b)^{2}}
  &  
  \label{eq204} 
  &
\end{flalign}

The Carbon DeFi marginal rate equations can also be expressed in terms of both the x- and y-dimensions (remember, for Carbon DeFi only the y-dimension is a token balance) via substitution of the $x$ and $y$ terms in Equations \ref{eq13} and \ref{eq15} with their horizontally or vertically shifted transformations (Equation \ref{eq205}). Simplifying the fractions in Equation \ref{eq205} yields Equation \ref{eq206}.

\begin{flalign}
& \text{\renewcommand{\arraystretch}{0.66}
    \begin{tabular}{@{}c@{}}
    \scriptsize from \\
    \scriptsize (\ref{eq110})\\\scriptsize (\ref{eq121})\\\scriptsize (\ref{eq123})
  \end{tabular}} 
  & 
  \displaystyle \frac{\partial x}{\partial y} = - \displaystyle \frac{x + \displaystyle \frac{z}{a \cdot (a + b)}}{y + \displaystyle \frac{b \cdot z}{a}};\ \displaystyle \frac{\partial y}{\partial x} = - \displaystyle \frac{y + \displaystyle \frac{b \cdot z}{a}}{x + \displaystyle \frac{z}{a \cdot (a + b)}}
  &  
  \label{eq205} 
  &
\end{flalign}

\begin{flalign}
& \text{\renewcommand{\arraystretch}{0.66}
    \begin{tabular}{@{}c@{}}
    \scriptsize from \\
    \scriptsize (\ref{eq124})
  \end{tabular}} 
  & 
  \displaystyle \frac{\partial x}{\partial y} = - \displaystyle \frac{x \cdot a \cdot (a + b) + z}{(a + b)(a \cdot y + b \cdot z)};\ \displaystyle \frac{\partial y}{\partial x} = - \displaystyle \frac{(a + b)(a \cdot y + b \cdot z)}{x \cdot a \cdot (a + b) + z}
  &  
  \label{eq206} 
  &
\end{flalign}

Continuous summation over the price curves (Equations \ref{eq202} and \ref{eq204}) over the interval representing the number of tokens being swapped yields results identical to Equations \ref{eq197}, \ref{eq198}, \ref{eq199} and \ref{eq200}. Be reminded that $x$ and $y$ are functions of each other, so it is a mistake to attempt direct integration of Equation \ref{eq206} without separating these variables first (not shown).

\begin{flalign}
& \text{\renewcommand{\arraystretch}{0.66}
    \begin{tabular}{@{}c@{}}
    \scriptsize from \\
    \scriptsize (\ref{eq197})\\\scriptsize (\ref{eq198})\\\scriptsize (\ref{eq202})
  \end{tabular}} 
  & 
  \mathrm{\Delta}x = - \int_{y}^{y + \mathrm{\Delta}y}{\displaystyle \frac{z^{2}}{(a \cdot y + b \cdot z)^{2}};\ } \cdot \partial y = \left\lbrack \displaystyle \frac{z^{2}}{a \cdot y + b \cdot z} \right\rbrack_{y}^{y + \mathrm{\Delta}y} \Rightarrow \mathbf{Eqns.}\ \mathbf{212\ }and\mathbf{\ 213}
  &  
  \label{eq207} 
  &
\end{flalign}

\begin{flalign}
& \text{\renewcommand{\arraystretch}{0.66}
    \begin{tabular}{@{}c@{}}
    \scriptsize from \\
    \scriptsize (\ref{eq199})\\\scriptsize (\ref{eq200})\\\scriptsize (\ref{eq204})
  \end{tabular}} 
  & 
  \mathrm{\Delta}y = - \int_{x}^{x + \mathrm{\Delta}x}{\displaystyle \frac{z^{2} \cdot (a + b)^{2}}{\left( x \cdot a \cdot (a + b) + z \right)^{2}}} \cdot \partial x = \left\lbrack \displaystyle \frac{z^{2} \cdot (a + b)^{2}}{x \cdot a \cdot (a + b) + z} \right\rbrack_{x}^{x + \mathrm{\Delta}x} \Rightarrow \mathbf{Eqns.}\ \mathbf{214\ }and\mathbf{\ 215}
  &  
  \label{eq208} 
  &
\end{flalign}

\begin{figure}[ht]
    \centering
    \includegraphics[width=\textwidth]{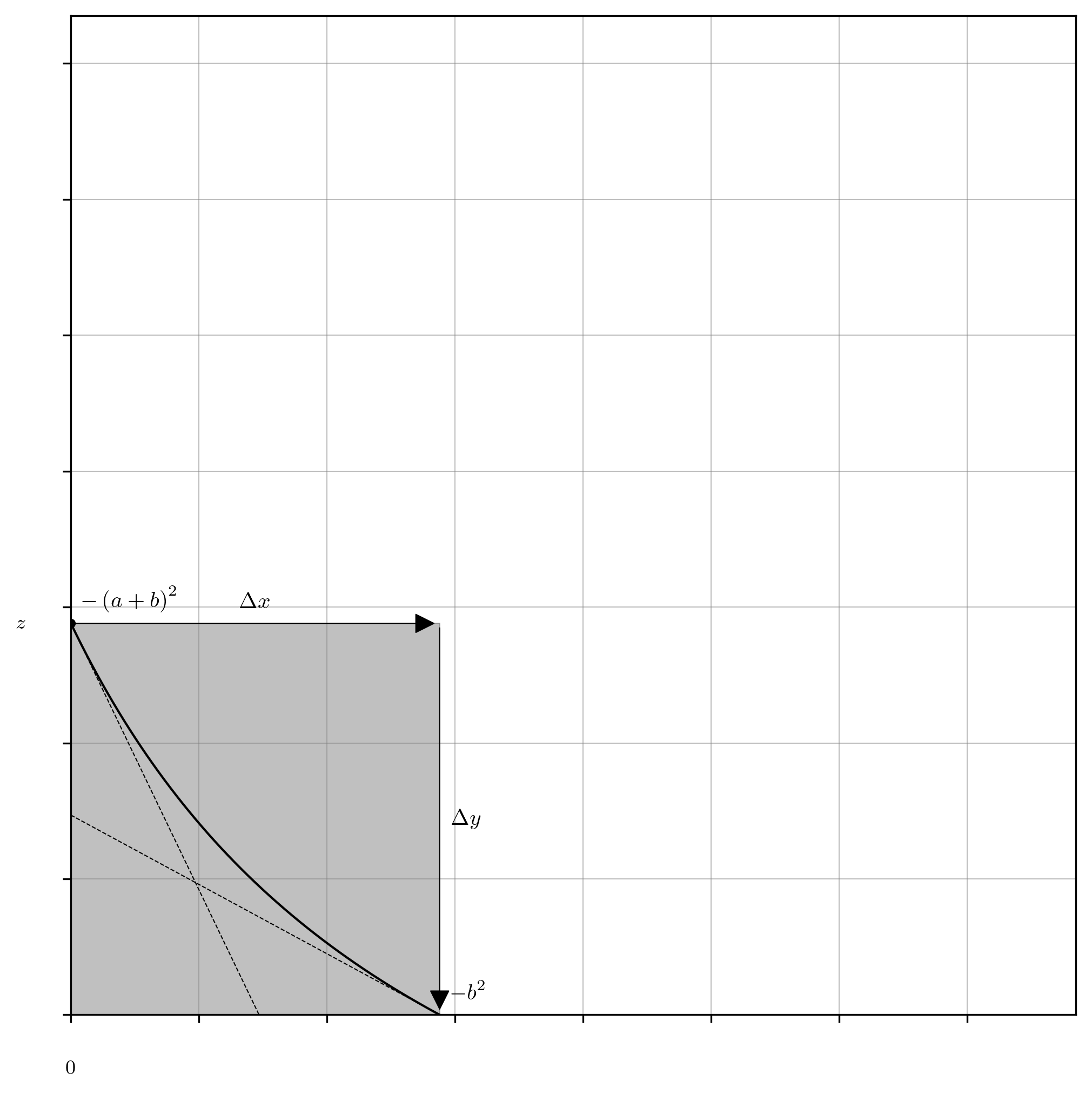}
    \captionsetup{
        justification=raggedright,
        singlelinecheck=false,
        font=small,
        labelfont=bf,
        labelsep=quad,
        format=plain
    }
    \caption{Traversal upon the rectangular hyperbola $y \cdot \left( z + x \cdot a \cdot \left( a + b \right) \right) = z \cdot \left( z - x \cdot b \cdot \left( a + b \right) \right)$ (Equation \ref{eq187}), representing a token swap against the Carbon DeFi real curve, where $\mathrm{\Delta}x > 0$ and $\mathrm{\Delta}y < 0$.}
    \label{fig37}
\end{figure}

\begin{figure}[ht]
    \centering
    \includegraphics[width=\textwidth]{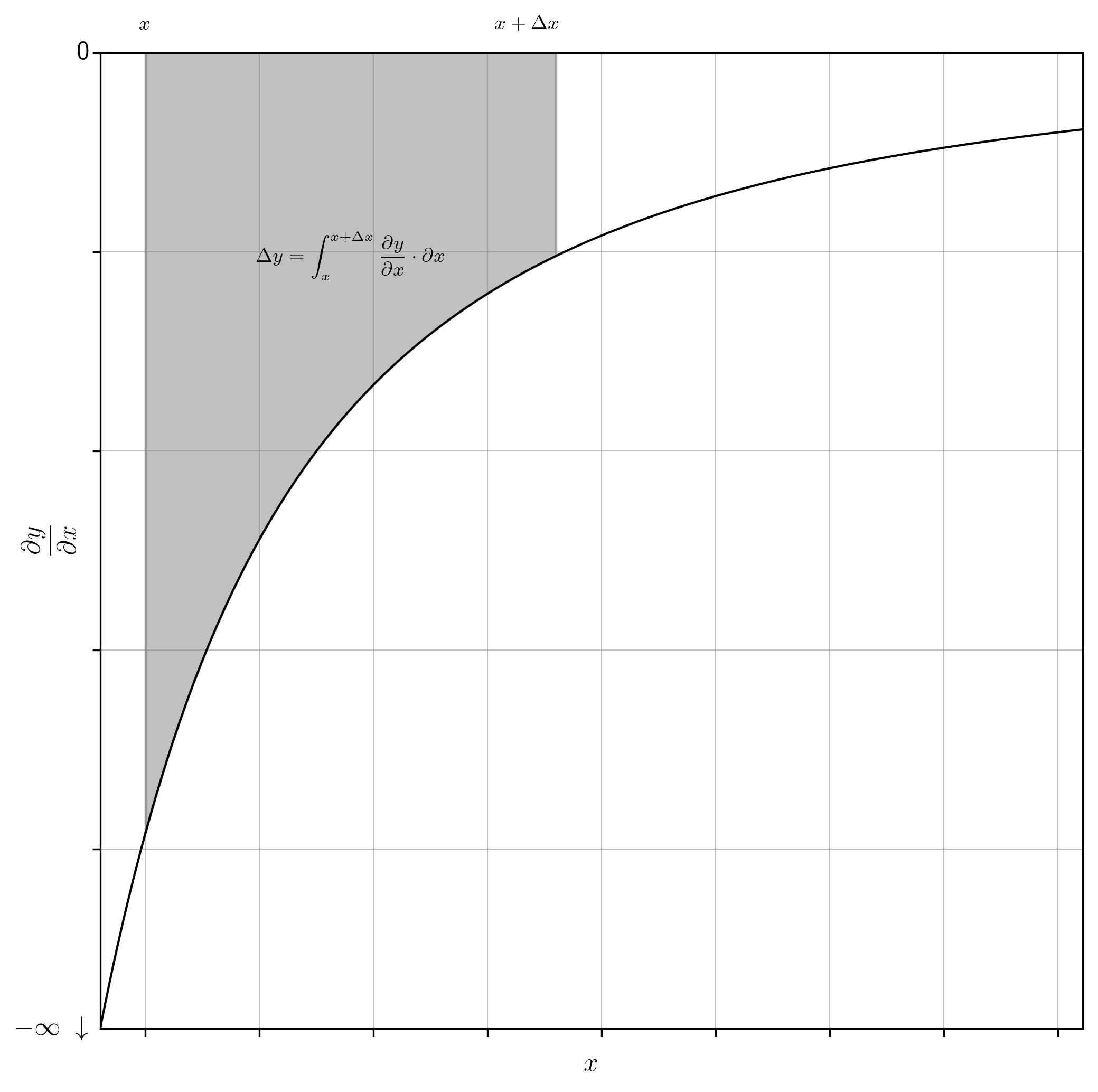}
    \captionsetup{
        justification=raggedright,
        singlelinecheck=false,
        font=small,
        labelfont=bf,
        labelsep=quad,
        format=plain
    }
    \caption{The integration above $\partial y / \partial x = - z^{2} \cdot \left( a + b \right)^{2} / \left( x \cdot a \cdot \left( a + b \right) + z \right)^{2}$ (Equation \ref{eq204}) over the interval $x \rightarrow x + \mathrm{\Delta}x$ representing a token swap against the Carbon DeFi real curve,where $\mathrm{\Delta}x > 0$ and $\mathrm{\Delta}y < 0$.}
    \label{fig38}
\end{figure}

Both the direct token swap by traversal upon the bonding curve $y \cdot \left( z + x \cdot a \cdot \left( a + b \right) \right) = z \cdot \left( z - x \cdot b \cdot \left( a + b \right) \right)$ (Equation \ref{eq187}) and the integration above $\partial y / \partial x = - z^{2} \cdot \left( a + b \right)^{2} / \left( x \cdot a \cdot \left( a + b \right) + z \right)^{2}$ (Equation \ref{eq204}) over the interval $x \rightarrow x + \mathrm{\Delta}x$ are depicted in Figures \ref{fig37} and \ref{fig38}.

\begin{figure}[ht]
    \centering
    \includegraphics[width=\textwidth]{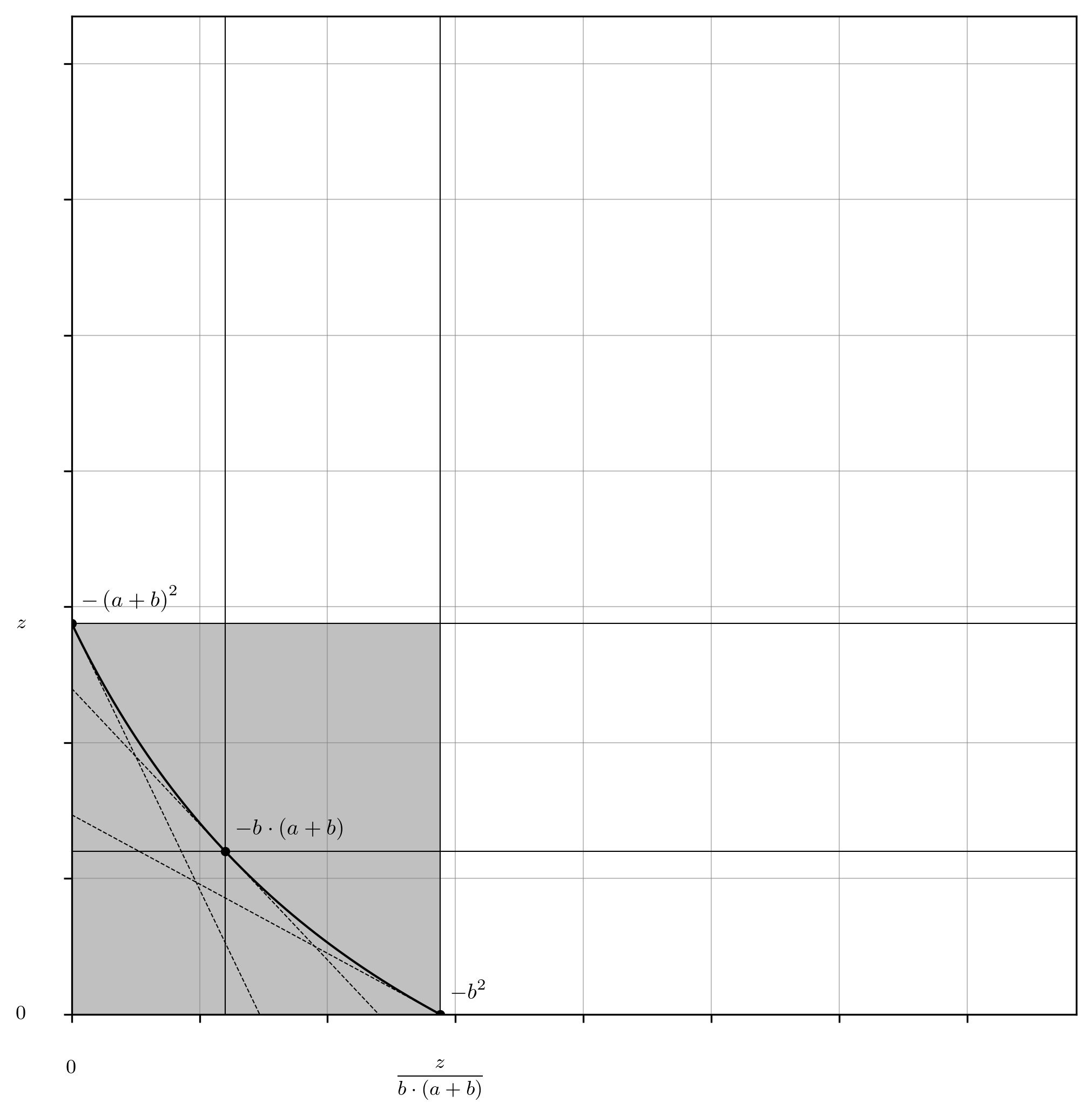}
    \captionsetup{
        justification=raggedright,
        singlelinecheck=false,
        font=small,
        labelfont=bf,
        labelsep=quad,
        format=plain
    }
    \caption{The x- and y-intercepts of the Carbon DeFi real curve are annotated on the appropriate axes (Equations \ref{eq185} and \ref{eq209}), and the algebraic identity for the first derivative of the curve evaluated at these points (via Equations \ref{eq183} and  \ref{eq184}), and their geometric mean (Equation \ref{eq190}), are illustrated with annotated tangent lines to the curve.}
    \label{fig39}
\end{figure}

\begin{figure}[ht]
    \centering
    \includegraphics[width=\textwidth]{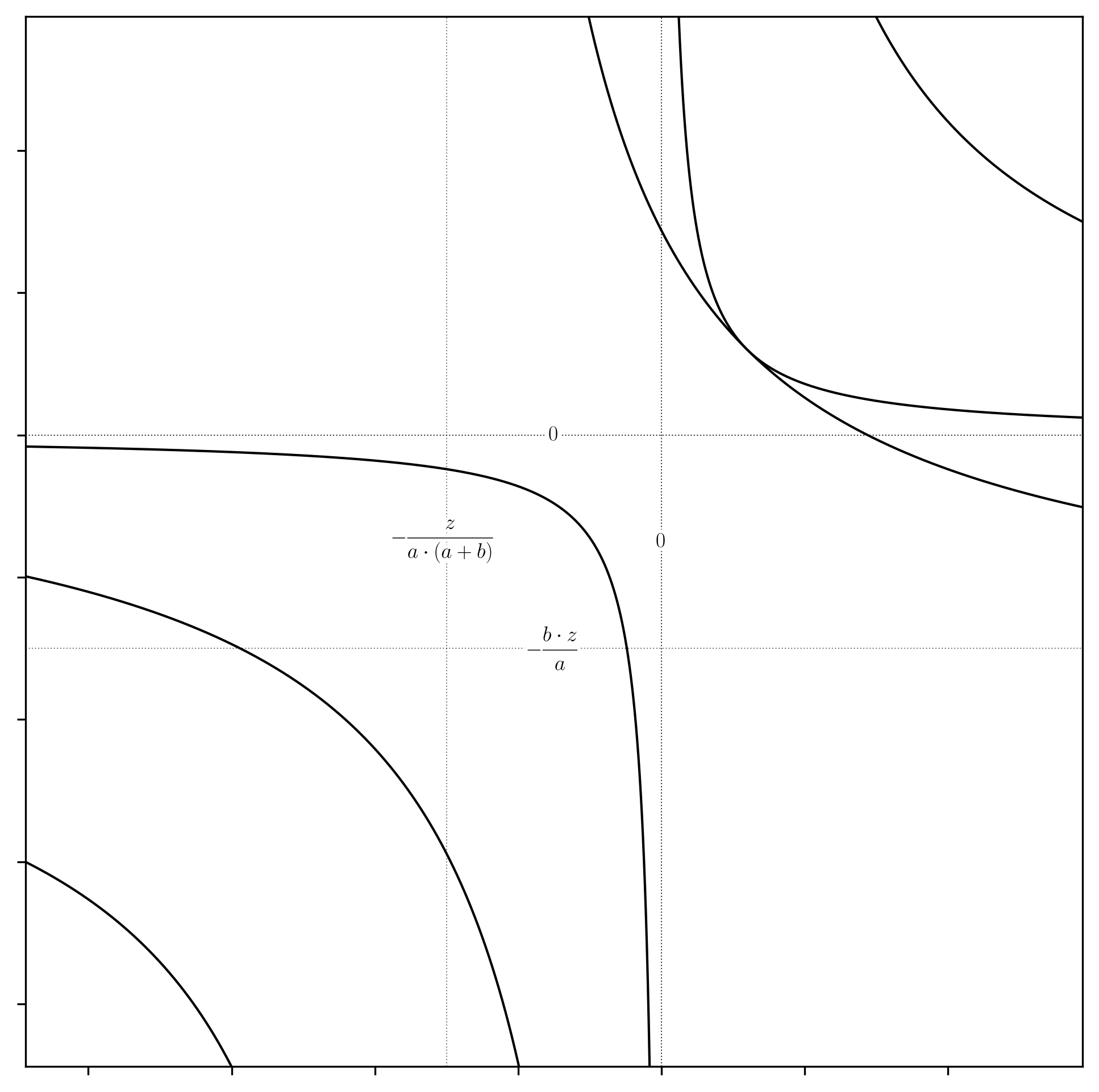}
    \captionsetup{
        justification=raggedright,
        singlelinecheck=false,
        font=small,
        labelfont=bf,
        labelsep=quad,
        format=plain
    }
    \caption{The horizontal and vertical asymptotes of the Carbon DeFi virtual and reference curves, and its real curve (Equations \ref{eq211} and \ref{eq212}), are depicted with dotted lines and their coordinates are annotated appropriately. While we have not yet approached the characterization of the reference and virtual curves, it is trivial that their asymptotes are at zero.}
    \label{fig40}
\end{figure}

The y-intercept of the Carbon DeFi invariant is one of its parameters, $z$ (Equation \ref{eq185}), and the x-intercept is obtained by direct substitution of $y$ for zero in Equation \ref{eq192} (Equation \ref{eq209}) (Figure \ref{fig39}). As anticipated, the quotient of $y_{\text{int}}$ and $x_{\text{int}}$ is equal to $P_{0}$ as previously observed in Equations \ref{eq95} and \ref{eq131} (Equation \ref{eq210}). The x- and y-asymptotes can be obtained as demonstrated before, either geometrically (subtract from the origin the implied horizontal and vertical shift parameters in Equation \ref{eq186}), or analytically (substitute the y- or x-component required to set the denominator of Equations \ref{eq192} and \ref{eq193} to zero) (Equations \ref{eq211} and \ref{eq212}) (Figure \ref{fig40}). Consistent with prior observations (Equations \ref{eq96} and \ref{eq132}), the quotient of the y- and x-asymptotes is equal to $P_{0}$ (Equation \ref{eq213}). 

\begin{flalign}
& \text{\renewcommand{\arraystretch}{0.66}
    \begin{tabular}{@{}c@{}}
    \scriptsize from \\
    \scriptsize (\ref{eq192})
  \end{tabular}} 
  & 
  x_{\text{int}} = \displaystyle \frac{z}{a \cdot b} - \displaystyle \frac{z}{a \cdot (a + b)} = \displaystyle \frac{z}{b \cdot (a + b)}
  &  
  \label{eq209} 
  &
\end{flalign}

\begin{flalign}
& \text{\renewcommand{\arraystretch}{0.66}
    \begin{tabular}{@{}c@{}}
    \scriptsize from \\
    \scriptsize (\ref{eq185})\\\scriptsize (\ref{eq190})\\\scriptsize (\ref{eq209})
  \end{tabular}} 
  & 
  \displaystyle \frac{y_{\text{int}}}{x_{\text{int}}} = b \cdot (a + b) = P_{0}
  &  
  \label{eq210} 
  &
\end{flalign}

\begin{flalign}
& \text{\renewcommand{\arraystretch}{0.66}
    \begin{tabular}{@{}c@{}}
    \scriptsize from \\
    \scriptsize (\ref{eq193})
  \end{tabular}} 
  & 
  x_{\text{asym}} = - \displaystyle \frac{z}{a \cdot (a + b)}
  &  
  \label{eq211} 
  &
\end{flalign}

\begin{flalign}
& \text{\renewcommand{\arraystretch}{0.66}
    \begin{tabular}{@{}c@{}}
    \scriptsize from \\
    \scriptsize (\ref{eq191})
  \end{tabular}} 
  & 
  y_{\text{asym}} = - \displaystyle \frac{b \cdot z}{a}
  &  
  \label{eq212} 
  &
\end{flalign}

\begin{flalign}
& \text{\renewcommand{\arraystretch}{0.66}
    \begin{tabular}{@{}c@{}}
    \scriptsize from \\
    \scriptsize (\ref{eq209})\\\scriptsize (\ref{eq210})
  \end{tabular}} 
  & 
  \displaystyle \frac{y_{\text{asym}}}{x_{\text{asym}}} = b \cdot (a + b) = P_{0}
  &  
  \label{eq213} 
  &
\end{flalign}

\subsection{The Carbon DeFi Virtual Curve}\label{subsec4.5}

\begin{figure}[ht]
    \centering
    \includegraphics[width=\textwidth]{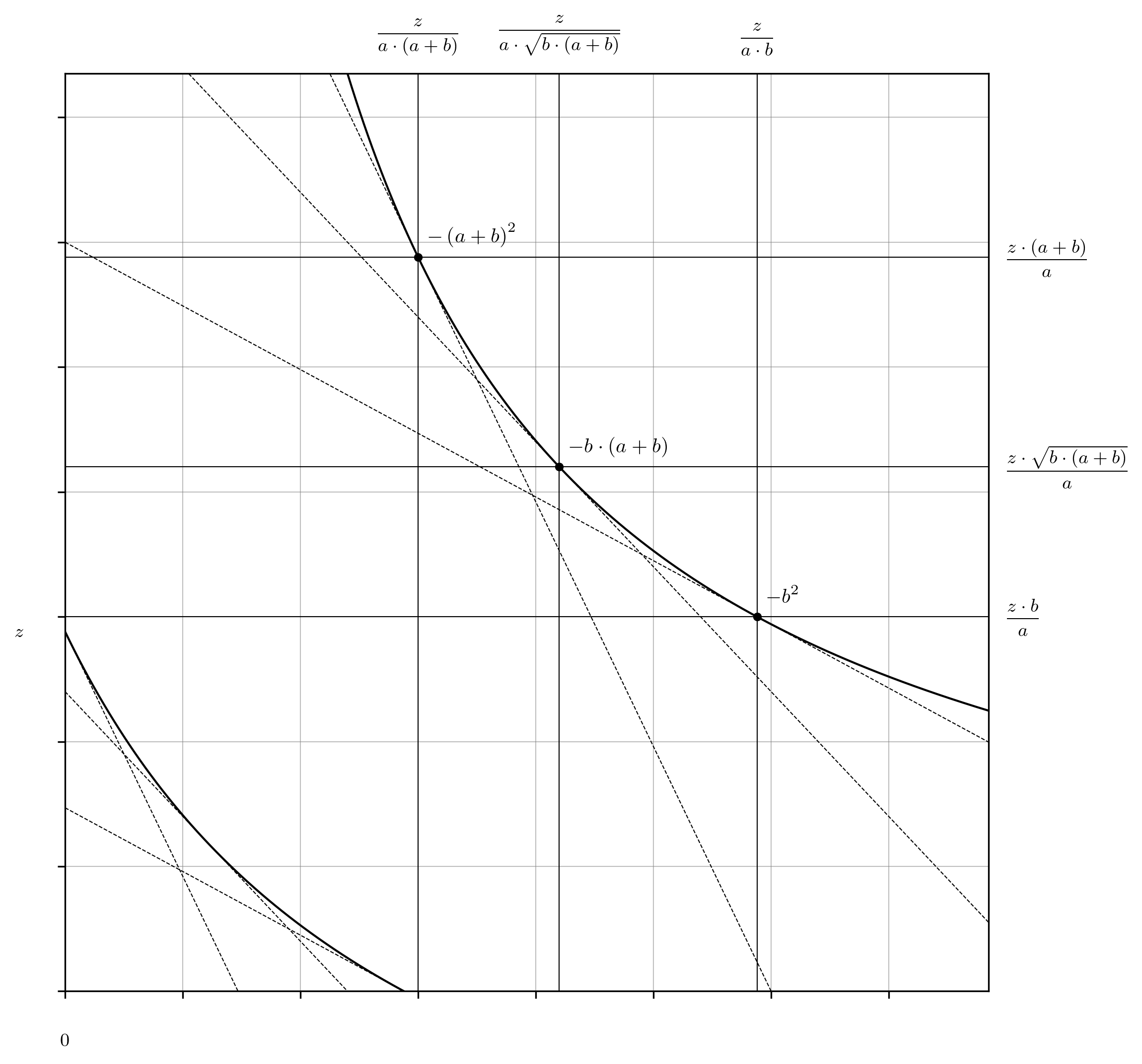}
    \captionsetup{
        justification=raggedright,
        singlelinecheck=false,
        font=small,
        labelfont=bf,
        labelsep=quad,
        format=plain
    }
    \caption{The completed characterisation of the Carbon DeFi virtual curve, with the coordinates at its price bounds and geometric mean of the price bounds annotated as appropriate (Equations \ref{eq214}, \ref{eq215}, \ref{eq217}, \ref{eq218}, \ref{eq219} and \ref{eq221}).}
    \label{fig41}
\end{figure}

The process of characterizing the Carbon DeFi virtual curve is essentially identical to that demonstrated for the Bancor v2 and Uniswap v3 virtual curves (Equations \ref{eq97}, \ref{eq98}, \ref{eq99}, and \ref{eq100}; and \ref{eq135}, \ref{eq136}, \ref{eq137}, \ref{eq138}, \ref{eq139}, \ref{eq140}, \ref{eq141}, \ref{eq142}, \ref{eq143}, and \ref{eq144}). First, the $\max(x_{\text{v}})$, $\min(x_{\text{v}})$, $\max(y_{\text{v}})$, and $\min(y_{\text{v}})$ coordinates are obtained by reversing the horizontal and vertical shifts implied by the real curve description (Equations \ref{eq214}, \ref{eq215}, \ref{eq218} and \ref{eq219}), then locating “center” by evaluating the geometric means of the virtual curve boundaries (Equations \ref{eq217} and \ref{eq221}) (Figure \ref{fig41}). Lastly, the mystery constant $C$ is confirmed by taking the quotient of the maximum and minimum bounds in both dimensions (Equations \ref{eq222} and \ref{eq223}).

\begin{flalign}
& \text{\renewcommand{\arraystretch}{0.66}
    \begin{tabular}{@{}c@{}}
    \scriptsize from \\
    \scriptsize (\ref{eq209})\\\scriptsize (\ref{eq211})
  \end{tabular}} 
  & 
  \max\left( x_{\text{v}} \right) = \left( x_{\text{int}} - x_{\text{asym}} \right) = \displaystyle \frac{z}{b \cdot (a + b)} + \displaystyle \frac{z}{a \cdot (a + b)} = \displaystyle \frac{z}{a \cdot b}
  &  
  \label{eq214} 
  &
\end{flalign}

\begin{flalign}
& \text{\renewcommand{\arraystretch}{0.66}
    \begin{tabular}{@{}c@{}}
    \scriptsize from \\
    \scriptsize (\ref{eq211})
  \end{tabular}} 
  & 
  \min\left( x_{\text{v}} \right) = \left( 0 - x_{\text{asym}} \right) = \displaystyle \frac{z}{a \cdot (a + b)}
  &  
  \label{eq215} 
  &
\end{flalign}

\begin{flalign}
& \text{\renewcommand{\arraystretch}{0.66}
    \begin{tabular}{@{}c@{}}
    \scriptsize from \\
    \scriptsize (\ref{eq214})\\\scriptsize (\ref{eq215})
  \end{tabular}} 
  & 
  \max\left( x_{\text{v}} \right) \cdot \min\left( x_{\text{v}} \right) = \displaystyle \frac{z}{a \cdot b} \cdot \displaystyle \frac{z}{a \cdot (a + b)} = \displaystyle \frac{z^{2}}{a^{2} \cdot b \cdot (a + b)}
  &  
  \label{eq216} 
  &
\end{flalign}

\begin{flalign}
& \text{\renewcommand{\arraystretch}{0.66}
    \begin{tabular}{@{}c@{}}
    \scriptsize from \\
    \scriptsize (\ref{eq216})
  \end{tabular}} 
  & 
  \sqrt{\max\left( x_{\text{v}} \right)} \cdot \sqrt{\min\left( x_{\text{v}} \right)} = \displaystyle \frac{z}{a \cdot \sqrt{b \cdot (a + b)}}
  &  
  \label{eq217} 
  &
\end{flalign}

\begin{flalign}
& \text{\renewcommand{\arraystretch}{0.66}
    \begin{tabular}{@{}c@{}}
    \scriptsize from \\
    \scriptsize (\ref{eq185})\\\scriptsize (\ref{eq212})
  \end{tabular}} 
  & 
  \max\left( y_{\text{v}} \right) = \left( y_{\text{int}} - y_{\text{asym}} \right) = z + \displaystyle \frac{b \cdot z}{a} = \displaystyle \frac{z \cdot (a + b)}{a}
  &  
  \label{eq218} 
  &
\end{flalign}

\begin{flalign}
& \text{\renewcommand{\arraystretch}{0.66}
    \begin{tabular}{@{}c@{}}
    \scriptsize from \\
    \scriptsize (\ref{eq212})
  \end{tabular}} 
  & 
  \min\left( y_{\text{v}} \right) = \left( 0 - y_{\text{asym}} \right) = \displaystyle \frac{b \cdot z}{a}
  &  
  \label{eq219} 
  &
\end{flalign}

\begin{flalign}
& \text{\renewcommand{\arraystretch}{0.66}
    \begin{tabular}{@{}c@{}}
    \scriptsize from \\
    \scriptsize (\ref{eq218})\\\scriptsize (\ref{eq219})
  \end{tabular}} 
  & 
  \max\left( y_{\text{v}} \right) \cdot \min\left( y_{\text{v}} \right) = \displaystyle \frac{z \cdot (a + b)}{a} \cdot \displaystyle \frac{b \cdot z}{a} = \displaystyle \frac{z^{2} \cdot b \cdot (a + b)}{a^{2}}
  &  
  \label{eq220} 
  &
\end{flalign}

\begin{flalign}
& \text{\renewcommand{\arraystretch}{0.66}
    \begin{tabular}{@{}c@{}}
    \scriptsize from \\
    \scriptsize (\ref{eq220})
  \end{tabular}} 
  & 
  \sqrt{\max\left( y_{\text{v}} \right)} \cdot \sqrt{\min\left( y_{\text{v}} \right)} = \displaystyle \frac{z \cdot \sqrt{b \cdot (a + b)}}{a}
  &  
  \label{eq221} 
  &
\end{flalign}

\begin{flalign}
& \text{\renewcommand{\arraystretch}{0.66}
    \begin{tabular}{@{}c@{}}
    \scriptsize from \\
    \scriptsize (\ref{eq214})\\\scriptsize (\ref{eq215})
  \end{tabular}} 
  & 
  \displaystyle \frac{\max\left( x_{\text{v}} \right)}{\min\left( x_{\text{v}} \right)} = \displaystyle \frac{\displaystyle \frac{z}{a \cdot b}}{\displaystyle \frac{z}{a \cdot (a + b)}} = \displaystyle \frac{a + b}{b} = \displaystyle \frac{\sqrt{P_{\text{high}}}}{\sqrt{P_{\text{low}}}} = C
  &  
  \label{eq222} 
  &
\end{flalign}

\begin{flalign}
& \text{\renewcommand{\arraystretch}{0.66}
    \begin{tabular}{@{}c@{}}
    \scriptsize from \\
    \scriptsize (\ref{eq218})\\\scriptsize (\ref{eq219})
  \end{tabular}} 
  & 
  \displaystyle \frac{\max\left( y_{\text{v}} \right)}{\min\left( y_{\text{v}} \right)} = \displaystyle \frac{\displaystyle \frac{z \cdot (a + b)}{a}}{\displaystyle \frac{b \cdot z}{a}} = \displaystyle \frac{a + b}{b} = \displaystyle \frac{\sqrt{P_{\text{high}}}}{\sqrt{P_{\text{low}}}} = C
  &  
  \label{eq223} 
  &
\end{flalign}

To complete the characterization of the real curve (and begin characterizing the reference curve), the identities of $x_{0}$ and $y_{0}$ must be acquired by compensating for the shift parameter (equivalent to $x_{\text{asym}}$ and $y_{\text{asym}}$) with respect to the “center” coordinates of the virtual curve (i.e., the geometric means of $\min(x_{\text{v}})$ and $\max(x_{\text{v}})$, and $\min(y_{\text{v}})$ and $\max(y_{\text{v}})$). Reversing the shift (Equations \ref{eq224} and \ref{eq226}) then simplifying (Equations \ref{eq225} and \ref{eq227}) yields the identities of $x_{0}$ and $y_{0}$ in terms of $a$, $b$, and $z$. In both cases, the simplified forms, as well as a seemingly unnecessary expansion of $b \cdot (a + b)$ into the square of its square root, are presented. The simplified versions are innocuous with respect to the annotation of the real curve; however, the expanded version reveals a continued pattern with respect to the reference curve capstone identities. Both are used in their appropriate contexts.

\begin{figure}[ht]
    \centering
    \includegraphics[width=\textwidth]{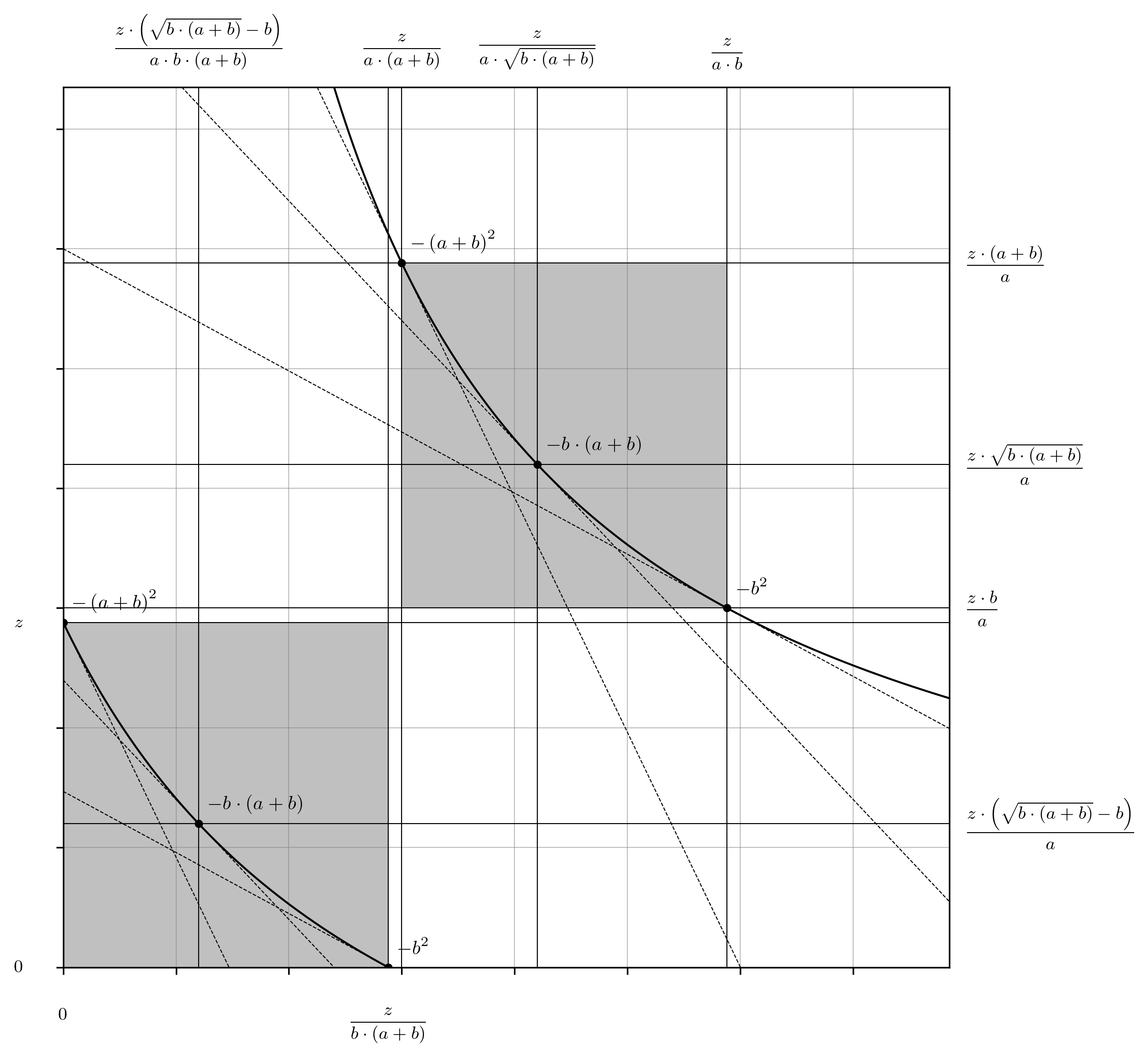}
    \captionsetup{
        justification=raggedright,
        singlelinecheck=false,
        font=small,
        labelfont=bf,
        labelsep=quad,
        format=plain
    }
    \caption{Completed characterizations of the Carbon DeFi real and virtual curves are depicted and annotated appropriately (Equations \ref{eq185}, \ref{eq209}, \ref{eq214}, \ref{eq215}, \ref{eq217}, \ref{eq218}, \ref{eq219}, \ref{eq221}, \ref{eq225} and \ref{eq227}).}
    \label{fig42}
\end{figure}

With these coordinates, the characterization of the real curve is now complete (Figure \ref{fig42}).

\begin{flalign}
& \text{\renewcommand{\arraystretch}{0.66}
    \begin{tabular}{@{}c@{}}
    \scriptsize from \\
    \scriptsize (\ref{eq211})\\\scriptsize (\ref{eq217})
  \end{tabular}} 
  & 
  x_{0} = \sqrt{\max\left( x_{\text{v}} \right)} \cdot \sqrt{\min\left( x_{\text{v}} \right)} + x_{\text{asym}} = \displaystyle \frac{z}{a \cdot \sqrt{b \cdot (a + b)}} - \displaystyle \frac{z}{a \cdot (a + b)}
  &  
  \label{eq224} 
  &
\end{flalign}

\begin{flalign}
& \text{\renewcommand{\arraystretch}{0.66}
    \begin{tabular}{@{}c@{}}
    \scriptsize from \\
    \scriptsize (\ref{eq224})
  \end{tabular}} 
  & 
  x_{0} = \displaystyle \frac{z \cdot \left( \sqrt{b \cdot (a + b)} - b \right)}{a \cdot b \cdot (a + b)} = \displaystyle \frac{z}{a \cdot \sqrt{b \cdot (a + b)}} \cdot \displaystyle \frac{\sqrt{b \cdot (a + b)} - b}{\sqrt{b \cdot (a + b)}}
  &  
  \label{eq225} 
  &
\end{flalign}

\begin{flalign}
& \text{\renewcommand{\arraystretch}{0.66}
    \begin{tabular}{@{}c@{}}
    \scriptsize from \\
    \scriptsize (\ref{eq212})\\\scriptsize (\ref{eq221})
  \end{tabular}} 
  & 
  y_{0} = \sqrt{\max\left( y_{\text{v}} \right)} \cdot \sqrt{\min\left( y_{\text{v}} \right)} + y_{\text{asym}} = \displaystyle \frac{z \cdot \sqrt{b \cdot (a + b)}}{a} - \displaystyle \frac{b \cdot z}{a}
  &  
  \label{eq226} 
  &
\end{flalign}

\begin{flalign}
& \text{\renewcommand{\arraystretch}{0.66}
    \begin{tabular}{@{}c@{}}
    \scriptsize from \\
    \scriptsize (\ref{eq226})
  \end{tabular}} 
  & 
  y_{0} = \displaystyle \frac{z \cdot \left( \sqrt{b \cdot (a + b)} - b \right)}{a} = \displaystyle \frac{z \cdot \sqrt{b \cdot (a + b)}}{a} \cdot \displaystyle \frac{\sqrt{b \cdot (a + b)} - b}{\sqrt{b \cdot (a + b)}}
  &  
  \label{eq227} 
  &
\end{flalign}

\begin{figure}[ht]
    \centering
    \includegraphics[width=\textwidth]{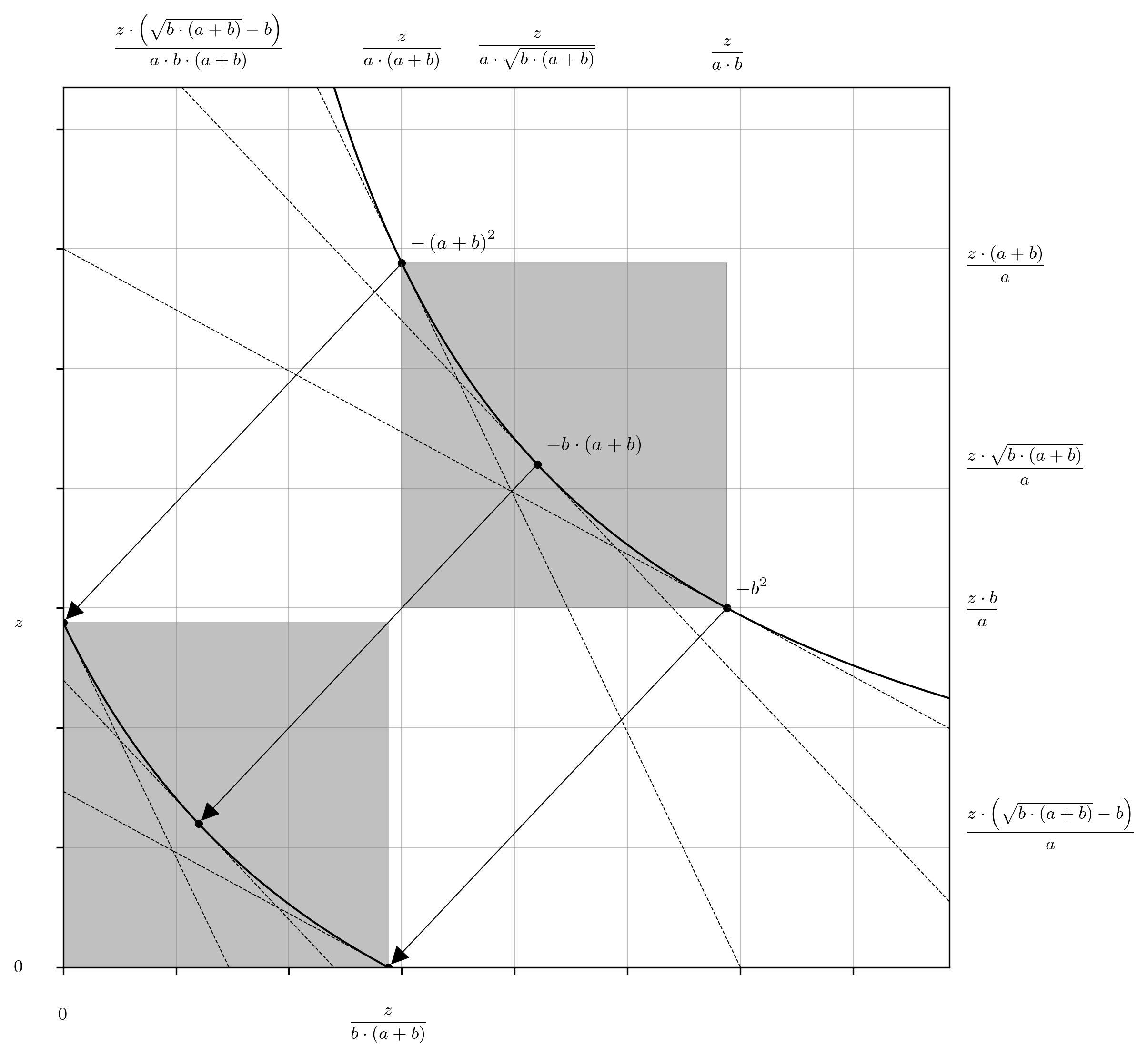}
    \captionsetup{
        justification=raggedright,
        singlelinecheck=false,
        font=small,
        labelfont=bf,
        labelsep=quad,
        format=plain
    }
    \caption{Reconstruction of the Carbon DeFi real curve is depicted as a map of the virtual curve back to the x- and y-axes.}
    \label{fig43}
\end{figure}

Similar to the Uniswap v3 curve constructions and in contrast to the Bancor v2 constructions, we have derived the virtual curve beginning from the shifted real curve. Regardless, it is still appropriate to define the real curve in terms of the virtual curve and the horizontal and vertical shifts (Figure \ref{fig43}). 

To complete the virtual curve characterization, recall that it is just a regular rectangular hyperbola. Its scaling constant can be obtained directly from inspection of the RHS of Equation \ref{eq186}, or from the product of any set of coordinates along the curve, (e.g., $\min(x_{\text{v}}) \cdot \max(y_{\text{v}})$, $\max(x_{\text{v}}) \cdot \min(y_{\text{v}})$, or $\sqrt{\max(x_{\text{v}})} \cdot \sqrt{\min(x_{\text{v}})} \cdot \sqrt{\max(y_{\text{v}})} \cdot \sqrt{\min(y_{\text{v}})}$) (Equation \ref{eq228}). Then, elaboration of the marginal price and token swap equations is performed as was done for Equations \ref{eq20}, \ref{eq21}, \ref{eq22}, \ref{eq23}, \ref{eq24}, \ref{eq25}, \ref{eq26}, \ref{eq27}, \ref{eq28} and \ref{eq29} and 
Equations \ref{eq163}, \ref{eq164}, \ref{eq165}, \ref{eq166}, \ref{eq167}, \ref{eq168}, \ref{eq169}, and \ref{eq170} (Equations \ref{eq228}, \ref{eq229}, \ref{eq230}, \ref{eq231}, \ref{eq232}, \ref{eq233}, \ref{eq234}, and \ref{eq235}).

\begin{flalign}
& \text{\renewcommand{\arraystretch}{0.66}
    \begin{tabular}{@{}c@{}}
    \scriptsize from \\
    \scriptsize (\ref{eq2})
  \end{tabular}} 
  & 
  x_{\text{v}} \cdot y_{\text{v}} = \displaystyle \frac{z}{a \cdot \sqrt{b \cdot (a + b)}} \cdot \displaystyle \frac{z \cdot \sqrt{b \cdot (a + b)}}{a} = \displaystyle \frac{z^{2}}{a^{2}}
  &  
  \label{eq228} 
  &
\end{flalign}

\begin{flalign}
& \text{\renewcommand{\arraystretch}{0.66}
    \begin{tabular}{@{}c@{}}
    \scriptsize from \\
    \scriptsize (\ref{eq8})
  \end{tabular}} 
  & 
  y_{\text{v}} = \displaystyle \frac{z^{2}}{{a^{2} \cdot x}_{\text{v}}}
  &  
  \label{eq229} 
  &
\end{flalign}

\begin{flalign}
& \text{\renewcommand{\arraystretch}{0.66}
    \begin{tabular}{@{}c@{}}
    \scriptsize from \\
    \scriptsize (\ref{eq6})
  \end{tabular}} 
  & 
  x_{\text{v}} = \displaystyle \frac{z^{2}}{{a^{2} \cdot y}_{\text{v}}}
  &  
  \label{eq230} 
  &
\end{flalign}

\begin{flalign}
& \text{\renewcommand{\arraystretch}{0.66}
    \begin{tabular}{@{}c@{}}
    \scriptsize from \\
    \scriptsize (\ref{eq3})
  \end{tabular}} 
  & 
  \left( x_{\text{v}} + \mathrm{\Delta}x \right) \cdot \left( y_{\text{v}} + \mathrm{\Delta}y \right) = \displaystyle \frac{z^{2}}{a^{2}}
  &  
  \label{eq231} 
  &
\end{flalign}

\begin{flalign}
& \text{\renewcommand{\arraystretch}{0.66}
    \begin{tabular}{@{}c@{}}
    \scriptsize from \\
    \scriptsize (\ref{eq4})\\\scriptsize (\ref{eq7})
  \end{tabular}} 
  & 
  \mathrm{\Delta}x = \displaystyle \frac{z^{2}}{a^{2} \cdot \left( y_{\text{v}} + \mathrm{\Delta}y \right)} - x_{\text{v}} = - \displaystyle \frac{\mathrm{\Delta}y \cdot z^{2}}{a^{2} \cdot \left( y_{\text{v}} \cdot \left( y_{\text{v}} + \mathrm{\Delta}y \right) \right)}
  &  
  \label{eq232} 
  &
\end{flalign}

\begin{flalign}
& \text{\renewcommand{\arraystretch}{0.66}
    \begin{tabular}{@{}c@{}}
    \scriptsize from \\
    \scriptsize (\ref{eq5})\\\scriptsize (\ref{eq9})
  \end{tabular}} 
  & 
  \mathrm{\Delta}y = \displaystyle \frac{A^{2} \cdot z^{2}}{a^{2} \cdot \left( x_{\text{v}} + \mathrm{\Delta}x \right)} - y_{\text{v}} = - \displaystyle \frac{\mathrm{\Delta}x \cdot z^{2}}{a^{2} \cdot \left( x_{\text{v}} \cdot \left( x_{\text{v}} + \mathrm{\Delta}x \right) \right)}
  &  
  \label{eq233} 
  &
\end{flalign}

\begin{flalign}
& \text{\renewcommand{\arraystretch}{0.66}
    \begin{tabular}{@{}c@{}}
    \scriptsize from \\
    \scriptsize (\ref{eq16})
  \end{tabular}} 
  & 
  \displaystyle \frac{\partial x_{\text{v}}}{\partial y_{\text{v}}} = - \displaystyle \frac{z^{2}}{{a^{2} \cdot y}_{\text{v}}^{2}};\ \ \ \displaystyle \frac{\partial y_{\text{v}}}{\partial x_{\text{v}}} = - \displaystyle \frac{{a^{2} \cdot y}_{\text{v}}^{2}}{z^{2}}
  &  
  \label{eq234} 
  &
\end{flalign}

\begin{flalign}
& \text{\renewcommand{\arraystretch}{0.66}
    \begin{tabular}{@{}c@{}}
    \scriptsize from \\
    \scriptsize (\ref{eq17})
  \end{tabular}} 
  & 
  \displaystyle \frac{\partial y_{\text{v}}}{\partial x_{\text{v}}} = - \displaystyle \frac{z^{2}}{{a^{2} \cdot x}_{\text{v}}^{2}};\ \ \ \displaystyle \frac{\partial x_{\text{v}}}{\partial y_{\text{v}}} = - \displaystyle \frac{{a^{2} \cdot x}_{\text{v}}^{2}}{z^{2}}
  &  
  \label{eq235} 
  &
\end{flalign}

\begin{figure}[ht]
    \centering
    \includegraphics[width=\textwidth]{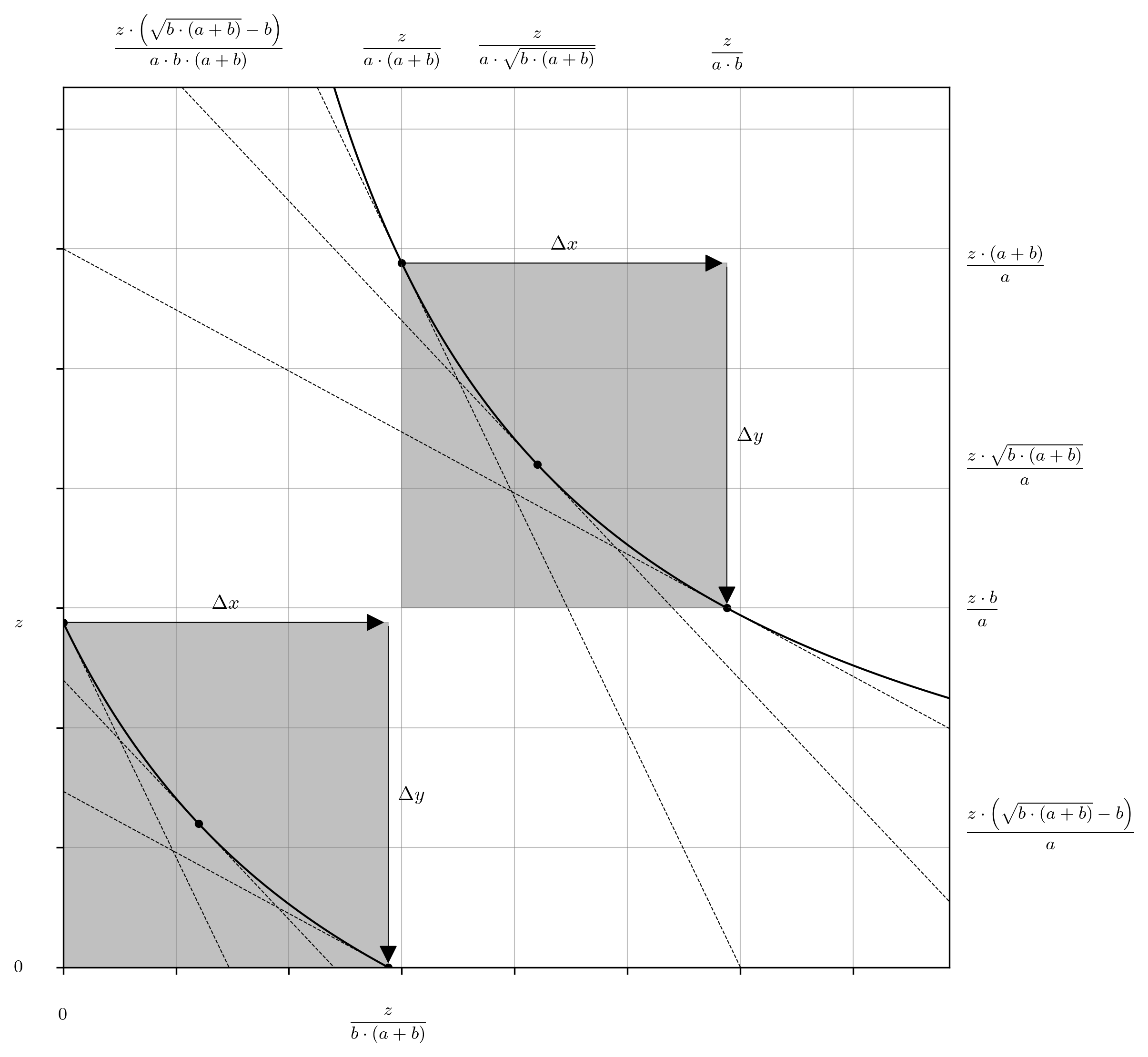}
    \captionsetup{
        justification=raggedright,
        singlelinecheck=false,
        font=small,
        labelfont=bf,
        labelsep=quad,
        format=plain
    }
    \caption{Traversal upon the rectangular hyperbolas $y \cdot \left( z + x \cdot a \cdot \left( a + b \right) \right) = z \cdot \left( z - x \cdot b \cdot \left( a + b \right) \right)$ and $x_{\text{v}} \cdot y_{\text{v}} = z^{2} / a^{2}$ (Equations \ref{eq187} and \ref{eq228}), representing a token swap against the Carbon DeFi real and virtual curves, where $\mathrm{\Delta}x > 0$ and $\mathrm{\Delta}y < 0$. The swap quantities $\mathrm{\Delta}x$ and $\mathrm{\Delta}y$, and marginal rates of exchange before and after the swap are identical. Therefore, with respect to the outcome of the exchange the difference between the two curve implementations is zero.}
    \label{fig44}
\end{figure}

\begin{figure}[ht]
    \centering
    \includegraphics[width=\textwidth]{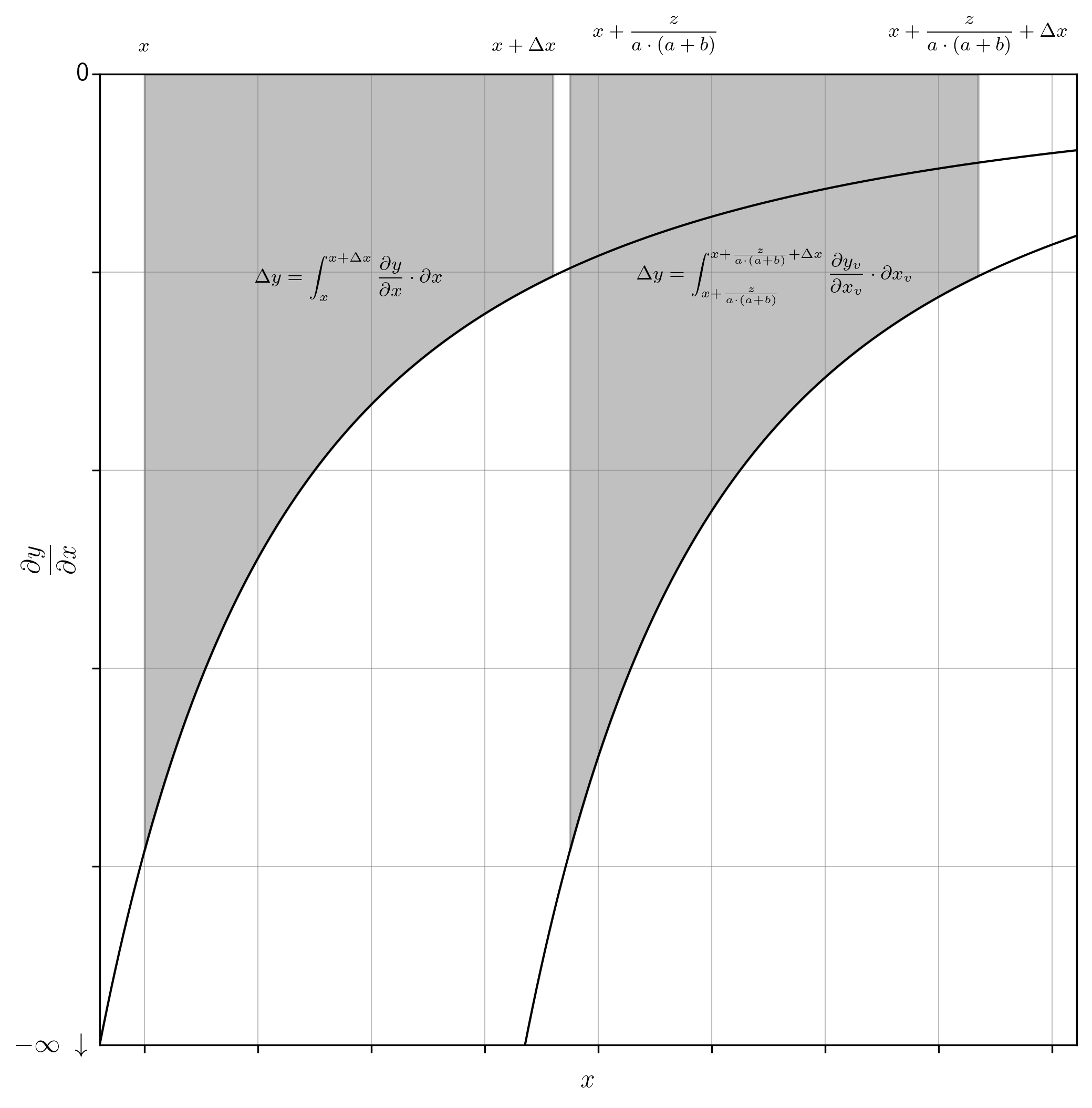}
    \captionsetup{
        justification=raggedright,
        singlelinecheck=false,
        font=small,
        labelfont=bf,
        labelsep=quad,
        format=plain
    }
    \caption{The integration above $\partial y / \partial x = - z^{2} \cdot \left( a + b \right)^{2} / \left( x \cdot a \cdot \left( a + b \right) + z \right)^{2}$ and $\partial y_{\text{v}} / \partial x_{\text{v}} = - z^{2} / \left( a \cdot x_{\text{v}} \right)^{2}$  (Equations \ref{eq204} and \ref{eq235}) over the intervals $x \rightarrow x + \mathrm{\Delta}x$ and $x + z / \left( a \cdot \left( a + b \right) \right) \rightarrow x + z / \left( a \cdot \left( a + b \right) \right) + \mathrm{\Delta}x$ (i.e. $x_{\text{v}} \rightarrow x_{\text{v}} + \mathrm{\Delta}x$) representing a token swap against both the real and virtual curves, where $\mathrm{\Delta}x > 0$ and $\mathrm{\Delta}y < 0$. Apart from the shift along the x-axis, these are identical in every other aspect. Note that the relationship between the real and virtual price curves is additive ($x$ and $\mathrm{\Delta}x$ versus $x + z / \left( a \cdot \left( a + b \right) \right)$ and $x + z / \left( a \cdot \left( a + b \right) \right) + \mathrm{\Delta}x$).}
    \label{fig45}
\end{figure}

The indifference of the real and virtual curves with respect the calculated swap quantities is again evident in the analysis (Figures \ref{fig44} and \ref{fig45}). Translocation upon the integrated forms (i.e. the bonding curves, Figure \ref{fig44}) and integration above their implied price curves (Figure \ref{fig45}) yields the same token amounts, $\mathrm{\Delta}x$ and $\mathrm{\Delta}y$. The only difference is the frame of reference, which is shifted by ($-x_{\text{asym}}$, $-y_{\text{asym}}$) in the virtual curve compared to the real curve, which has no effect on either the marginal rate, or the effective rate of exchange. As with Uniswap v3, Carbon Defi exposes the terms necessary to adapt legacy software that is dependent on the $x \cdot y = k$ assumption; simply read $z$ and $a$ from the smart contracts and $k$ becomes the square of their quotient, $k = \left( z / a\right)^{2}$.

\subsection{The Carbon DeFi Reference Curve}\label{subsec4.6}

The Carbon DeFi reference curve is constructed according to the general methods established above. The product of $x_{0}$ and $y_{0}$ is equal to its scaling constant (i.e. $k$ in the proverbial “constant product” equation, $x \cdot y = k$), and can be obtained from the RHS of Equations \ref{eq225} and \ref{eq227} (Equations \ref{eq236} and \ref{eq237}). The form the expression takes in Equation \ref{eq237} is to make the re-scaling of the reference curve with respect to the virtual curve obvious, which also allows for the original amplification term, $A$, to be extracted and represented in terms of $a$, $b$, and $z$ (Equations \ref{eq53}, \ref{eq237} and \ref{eq238}). 

\begin{flalign}
& \text{\renewcommand{\arraystretch}{0.66}
    \begin{tabular}{@{}c@{}}
    \scriptsize from \\
    \scriptsize (\ref{eq2})\\\scriptsize (\ref{eq225})\\\scriptsize (\ref{eq227})
  \end{tabular}} 
  & 
  x \cdot y = \displaystyle \frac{z \cdot \left( \sqrt{b \cdot (a + b)} - b \right)}{a \cdot b \cdot (a + b)} \cdot \displaystyle \frac{z \cdot \left( \sqrt{b \cdot (a + b)} - b \right)}{a}
  &  
  \label{eq236} 
  &
\end{flalign}

\begin{flalign}
& \text{\renewcommand{\arraystretch}{0.66}
    \begin{tabular}{@{}c@{}}
    \scriptsize from \\
    \scriptsize (\ref{eq236})
  \end{tabular}} 
  & 
  x \cdot y = \displaystyle \frac{z^{2} \cdot \left( \sqrt{b \cdot (a + b)} - b \right)^{2}}{a^{2} \cdot b \cdot (a + b)} = \displaystyle \frac{z^{2}}{a^{2}} \cdot \left( \displaystyle \frac{\sqrt{b \cdot (a + b)} - b}{\sqrt{b \cdot (a + b)}} \right)^{2}
  &  
  \label{eq237} 
  &
\end{flalign}

\begin{flalign}
& \text{\renewcommand{\arraystretch}{0.66}
    \begin{tabular}{@{}c@{}}
    \scriptsize from \\
    \scriptsize (\ref{eq53})\\\scriptsize (\ref{eq184})\\\scriptsize (\ref{eq190})\\\scriptsize (\ref{eq237})
  \end{tabular}} 
  & 
  \displaystyle \frac{\sqrt{b \cdot (a + b)}}{\sqrt{b \cdot (a + b)} - b} = \displaystyle \frac{\sqrt{P_{0}}}{\sqrt{P_{0}} - \sqrt{P_{\text{low}}}} \Rightarrow \displaystyle \frac{\sqrt[4]{P_{\text{high}}}}{\sqrt[4]{P_{\text{high}}} - \sqrt[4]{P_{\text{low}}}} = A
  &  
  \label{eq238} 
  &
\end{flalign}

\begin{figure}[ht]
    \centering
    \includegraphics[width=\textwidth]{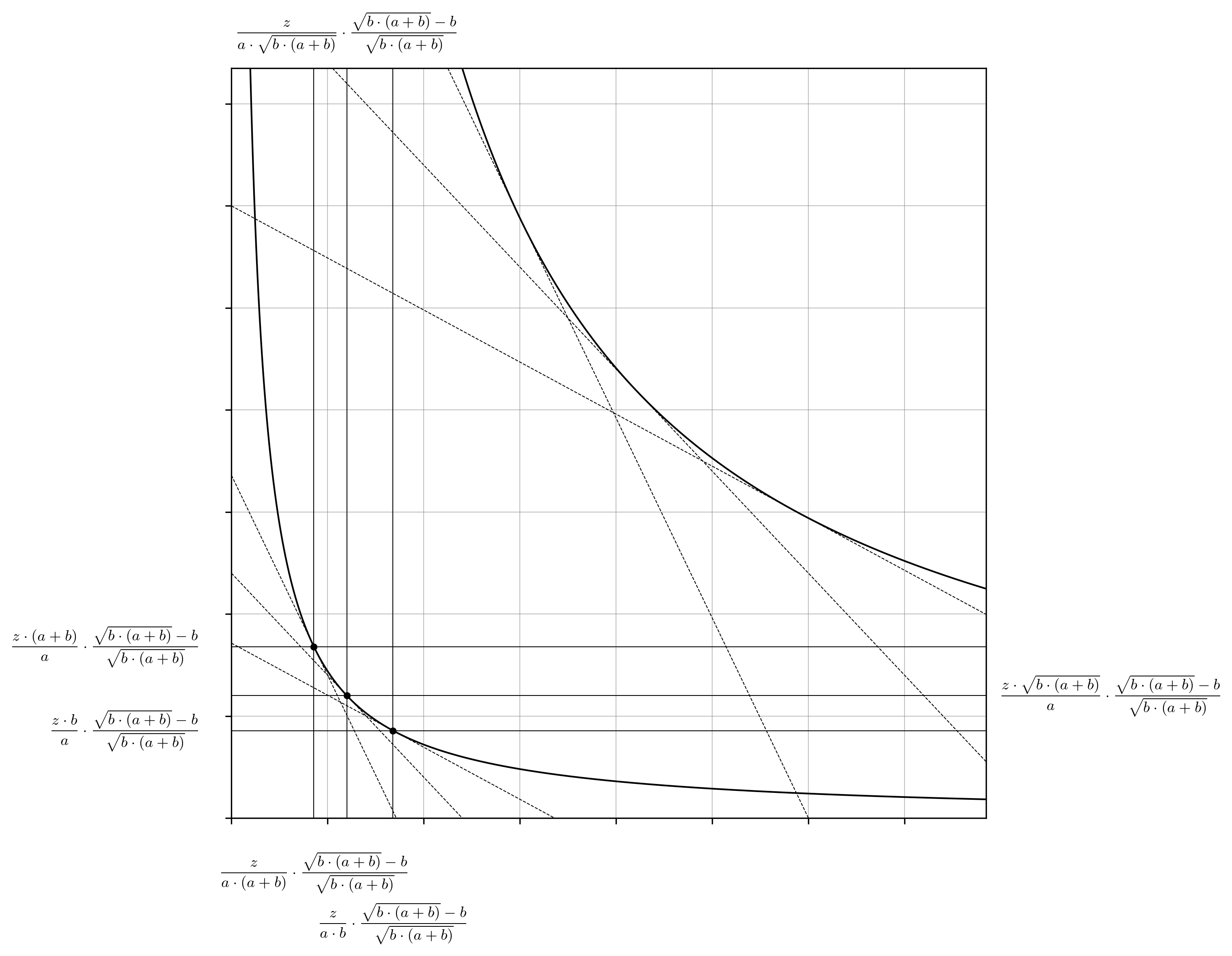}
    \captionsetup{
        justification=raggedright,
        singlelinecheck=false,
        font=small,
        labelfont=bf,
        labelsep=quad,
        format=plain
    }
    \caption{The algebraic identities of the points where the lines tangent to the Carbon DeFi reference curve and parallel to the tangent lines at the price boundaries of the virtual curve are now elucidated (Equations \ref{eq225}, \ref{eq227}, \ref{eq240}, \ref{eq242}, \ref{eq244} and \ref{eq246}).}
    \label{fig46}
\end{figure}

With the reference curve’s scaling constant now defined in terms of $a$, $b$, and $z$, the coordinates corresponding to the price bounds, $\min(x)$, $\max(x)$, $\min(y)$, and $\max(y)$, can also be expressed with the same parameters. As was done for the Uniswap v3 case, we will approach this analytically. First, the derivative of the reference curve (Equations \ref{eq16} and \ref{eq17}) is forced to either $-P_{\text{high}}$ or $-P_{\text{low}}$; in the Carbon DeFi parameterization, $P_{\text{low}} = b^{2}$ and $P_{\text{high}} = \left( a + b \right)^{2}$. Then, the $x$ or $y$ variable is substituted for $\min(x)$, $\max(x)$, $\min(y)$, or $\max(y)$, as appropriate, and the coordinate of interest is obtained via trivial rearrangements of the resulting expressions (Equations \ref{eq239}, \ref{eq240}, \ref{eq241}, \ref{eq242}, \ref{eq243}, \ref{eq244}, \ref{eq245}, and \ref{eq246}) (Figure \ref{fig46}).

\begin{flalign}
& \text{\renewcommand{\arraystretch}{0.66}
    \begin{tabular}{@{}c@{}}
    \scriptsize from \\
    \scriptsize (\ref{eq17})\\\scriptsize (\ref{eq237})
  \end{tabular}} 
  & 
  \displaystyle \frac{\partial y}{\partial x} = - (a + b)^{2} = - \displaystyle \frac{\displaystyle \frac{z^{2}}{a^{2}} \cdot \left( \displaystyle \frac{\sqrt{b \cdot (a + b)} - b}{\sqrt{b \cdot (a + b)}} \right)^{2}}{\min^{2}(x)}
  &  
  \label{eq239} 
  &
\end{flalign}

\begin{flalign}
& \text{\renewcommand{\arraystretch}{0.66}
    \begin{tabular}{@{}c@{}}
    \scriptsize from \\
    \scriptsize (\ref{eq239})
  \end{tabular}} 
  & 
  \min \left( x \right) = \displaystyle \frac{z}{a \cdot (a + b)} \cdot \left( \displaystyle \frac{\sqrt{b \cdot (a + b)} - b}{\sqrt{b \cdot (a + b)}} \right)
  &  
  \label{eq240} 
  &
\end{flalign}

\begin{flalign}
& \text{\renewcommand{\arraystretch}{0.66}
    \begin{tabular}{@{}c@{}}
    \scriptsize from \\
    \scriptsize (\ref{eq17})\\\scriptsize (\ref{eq237})
  \end{tabular}} 
  & 
  \displaystyle \frac{\partial y}{\partial x} = - b^{2} = - \displaystyle \frac{\displaystyle \frac{z^{2}}{a^{2}} \cdot \left( \displaystyle \frac{\sqrt{b \cdot (a + b)} - b}{\sqrt{b \cdot (a + b)}} \right)^{2}}{\max^{2}(x)}
  &  
  \label{eq241} 
  &
\end{flalign}

\begin{flalign}
& \text{\renewcommand{\arraystretch}{0.66}
    \begin{tabular}{@{}c@{}}
    \scriptsize from \\
    \scriptsize (\ref{eq241})
  \end{tabular}} 
  & 
  \max \left( x \right) = \displaystyle \frac{z}{a \cdot b} \cdot \left( \displaystyle \frac{\sqrt{b \cdot (a + b)} - b}{\sqrt{b \cdot (a + b)}} \right)
  &  
  \label{eq242} 
  &
\end{flalign}

\begin{flalign}
& \text{\renewcommand{\arraystretch}{0.66}
    \begin{tabular}{@{}c@{}}
    \scriptsize from \\
    \scriptsize (\ref{eq16})\\\scriptsize (\ref{eq237})
  \end{tabular}} 
  & 
  \displaystyle \frac{\partial y}{\partial x} = - b^{2} = - \displaystyle \frac{\min^{2}(y)}{\displaystyle \frac{z^{2}}{a^{2}} \cdot \left( \displaystyle \frac{\sqrt{b \cdot (a + b)} - b}{\sqrt{b \cdot (a + b)}} \right)^{2}}
  &  
  \label{eq243} 
  &
\end{flalign}

\begin{flalign}
& \text{\renewcommand{\arraystretch}{0.66}
    \begin{tabular}{@{}c@{}}
    \scriptsize from \\
    \scriptsize (\ref{eq243})
  \end{tabular}} 
  & 
  \min \left( y \right) = \displaystyle \frac{z \cdot b}{a} \cdot \left( \displaystyle \frac{\sqrt{b \cdot (a + b)} - b}{\sqrt{b \cdot (a + b)}} \right)
  &  
  \label{eq244} 
  &
\end{flalign}

\begin{flalign}
& \text{\renewcommand{\arraystretch}{0.66}
    \begin{tabular}{@{}c@{}}
    \scriptsize from \\
    \scriptsize (\ref{eq16})\\\scriptsize (\ref{eq237})
  \end{tabular}} 
  & 
  \displaystyle \frac{\partial y}{\partial x} = - (a + b)^{2} = - \displaystyle \frac{{\max^{2}}(y)}{\displaystyle \frac{z^{2}}{a^{2}} \cdot \left( \displaystyle \frac{\sqrt{b \cdot (a + b)} - b}{\sqrt{b \cdot (a + b)}} \right)^{2}}
  &  
  \label{eq245} 
  &
\end{flalign}

\begin{flalign}
& \text{\renewcommand{\arraystretch}{0.66}
    \begin{tabular}{@{}c@{}}
    \scriptsize from \\
    \scriptsize (\ref{eq245})
  \end{tabular}} 
  & 
  \max \left( y \right) = \displaystyle \frac{z \cdot (a + b)}{a} \cdot \left( \displaystyle \frac{\sqrt{b \cdot (a + b)} - b}{\sqrt{b \cdot (a + b)}} \right)
  &  
  \label{eq246} 
  &
\end{flalign}

\begin{figure}[ht]
    \centering
    \includegraphics[width=\textwidth]{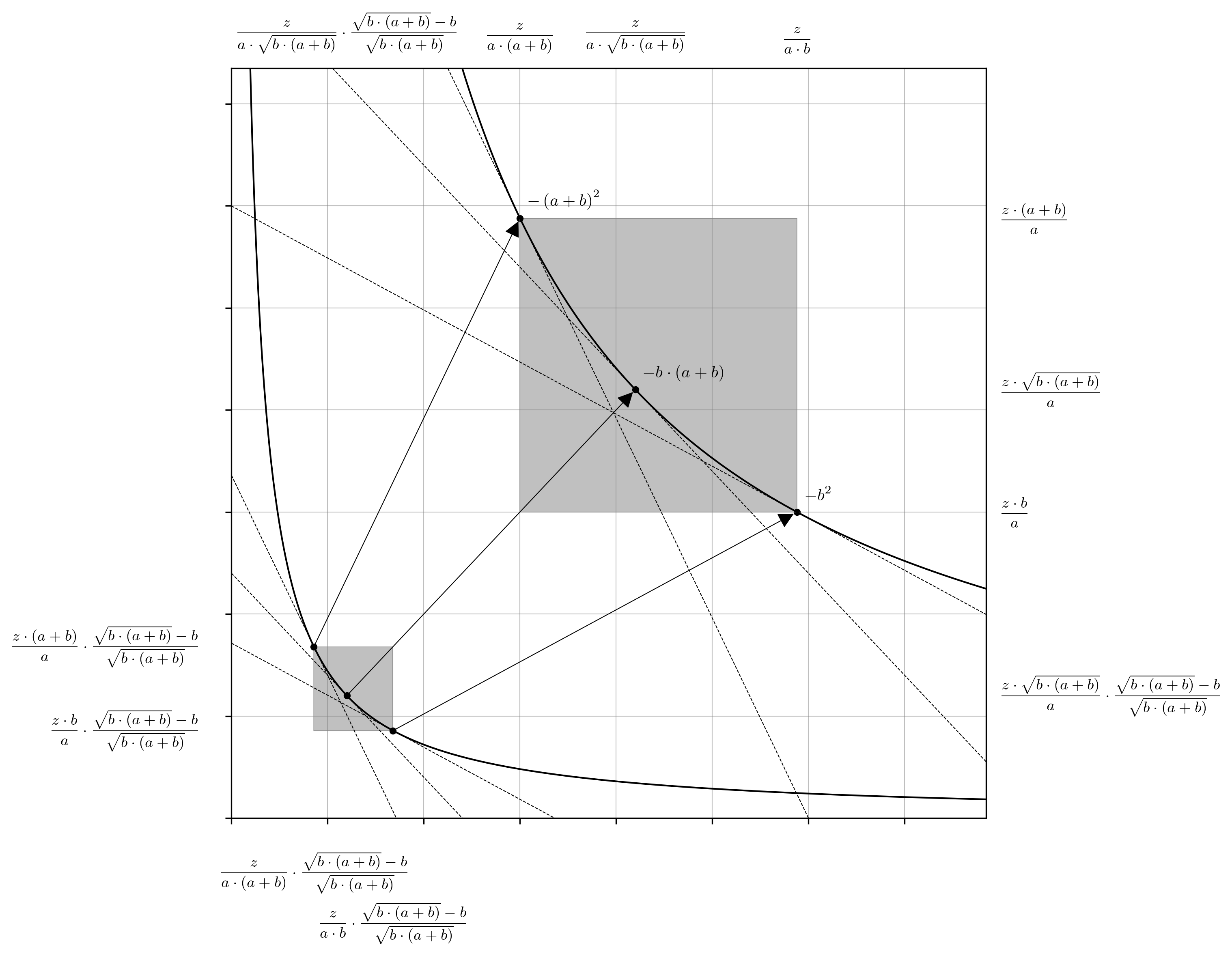}
    \captionsetup{
        justification=raggedright,
        singlelinecheck=false,
        font=small,
        labelfont=bf,
        labelsep=quad,
        format=plain
    }
    \caption{The amplification process is depicted with corresponding shaded areas of the Carbon DeFi reference and virtual curves, respectively. Points where the first derivative of each curve evaluates to the same result are shown as white dots, and the mapping of these points is depicted with white arrows.}
    \label{fig47}
\end{figure}

As with the Uniswap v3 constructions, the Carbon DeFi constructions have been performed in the reverse order compared to the seminal work (Figures \ref{fig1} to \ref{fig21}). However, the liquidity amplification heuristic is still useful, and can be depicted as before with arrows connecting pairs of points on the reference and virtual curves, where the derivatives evaluated at these points are equal (Figure \ref{fig47}).

\begin{figure}[ht]
    \centering
    \includegraphics[width=\textwidth]{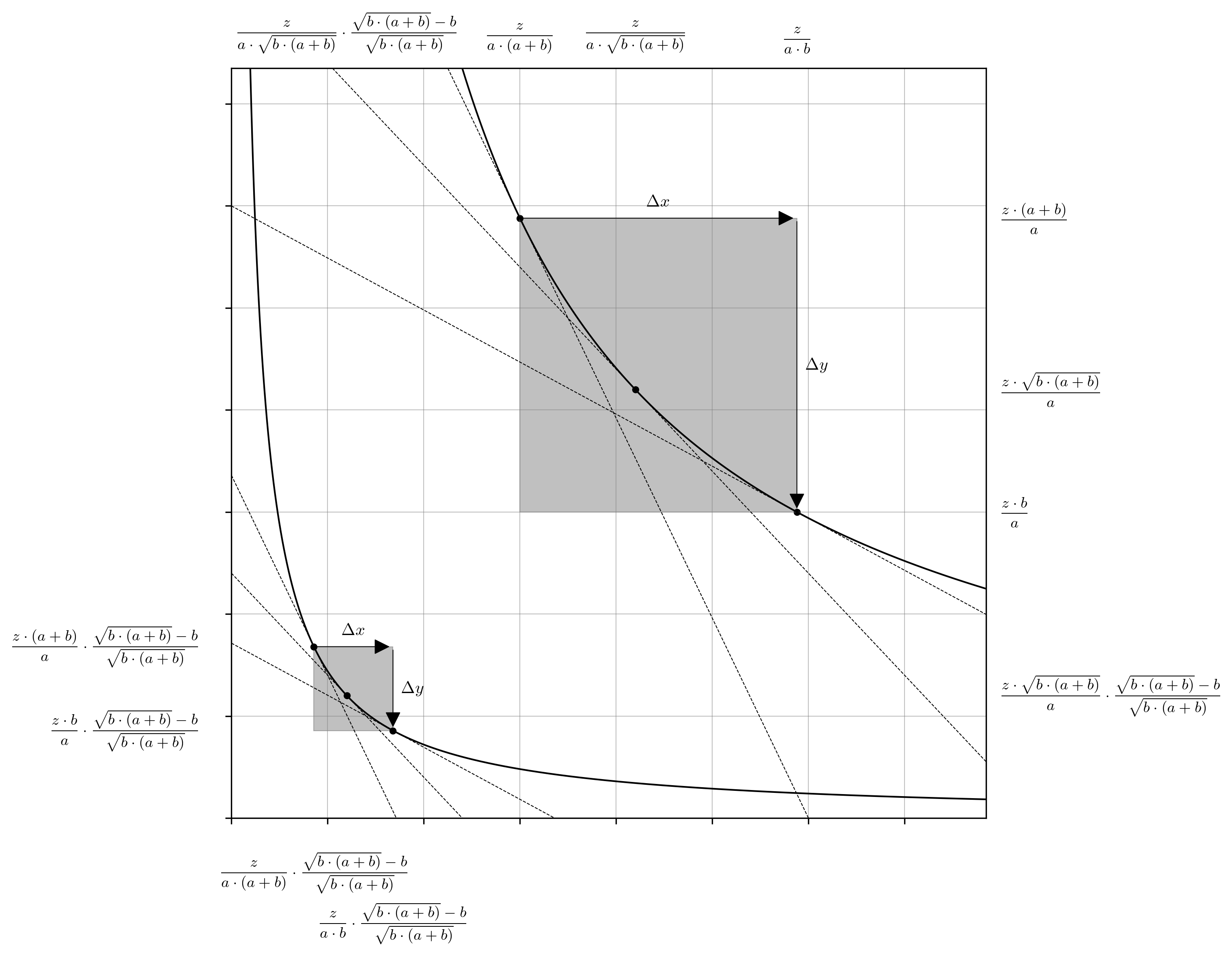}
    \captionsetup{
        justification=raggedright,
        singlelinecheck=false,
        font=small,
        labelfont=bf,
        labelsep=quad,
        format=plain
    }
    \caption{Traversal upon the rectangular hyperbolas $x_{\text{v}} \cdot y_{\text{v}} = z^{2} / a^{2}$ and $x \cdot y = z^{2} \cdot \left( \sqrt{b} \sqrt{a + b} - b \right)^{2} / \left( a^{2} \cdot b \cdot \left( a +b \right) \right)$ (Equations \ref{eq228} and  \ref{eq237}), representing a token swap against the reference and amplified liquidity pool, where $\mathrm{\Delta}x > 0$ and $\mathrm{\Delta}y < 0$. The marginal rates of exchange before and after the swap are identical. The ratio of the $\mathrm{\Delta}x$ and $\mathrm{\Delta}y$ arrow lengths for each curve are also identical. Therefore the overall rate of exchange, $\mathrm{\Delta}y / \mathrm{\Delta}x$ is equal in both cases.}
    \label{fig48}
\end{figure}

\begin{figure}[ht]
    \centering
    \includegraphics[width=\textwidth]{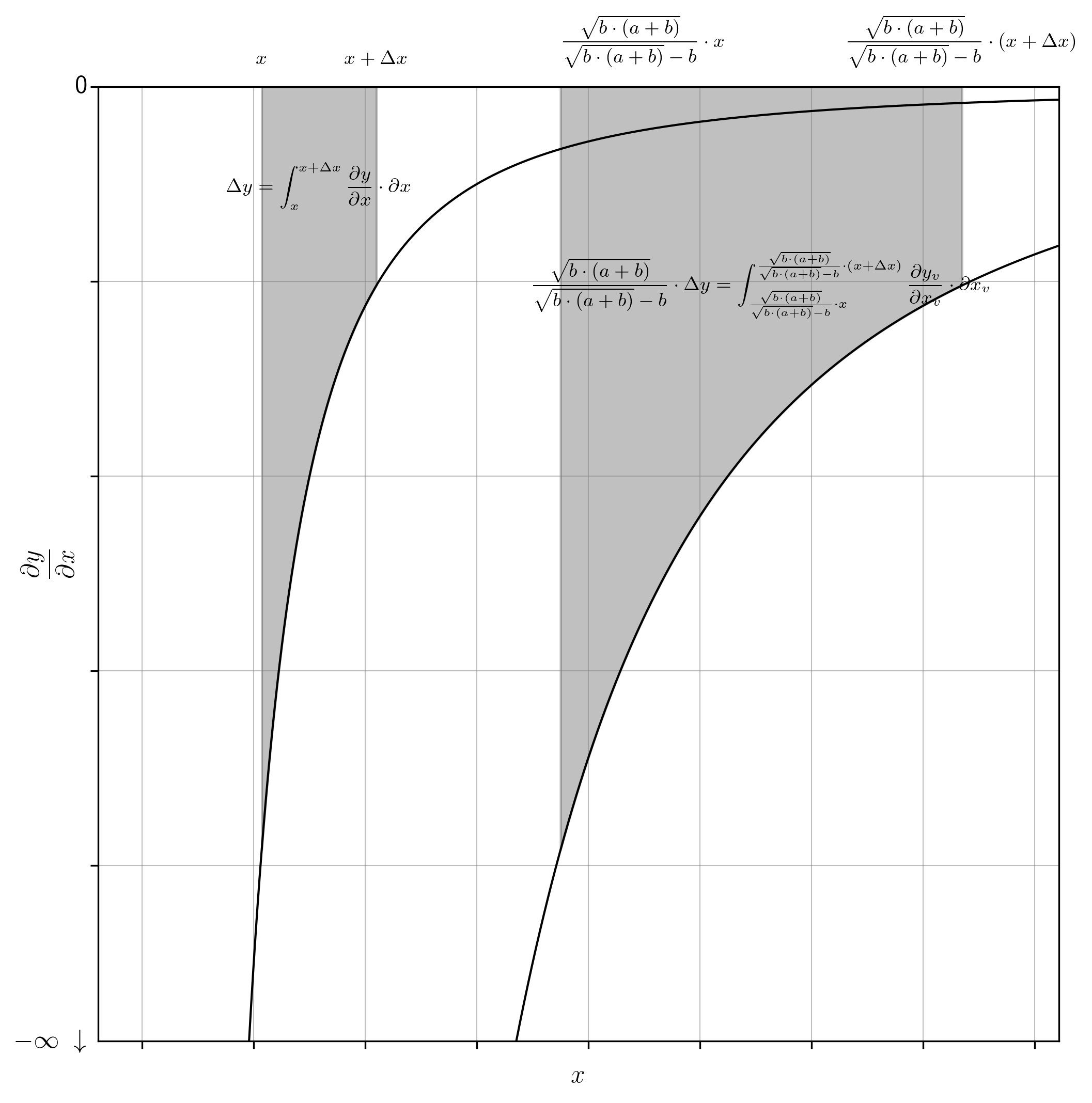}
    \captionsetup{
        justification=raggedright,
        singlelinecheck=false,
        font=small,
        labelfont=bf,
        labelsep=quad,
        format=plain
    }
    \caption{The integration above $\partial y / \partial x = - z^{2} \cdot \left( \sqrt{b} \cdot \sqrt{a + b} - b \right)^{2} / \left( a^{2} \cdot b \cdot \left( a + b \right) \cdot x^{2} \right)$ and $\partial y_{\text{v}} / \partial x_{\text{v}} = - z^{2} / \left( a \cdot x_{\text{v}} \right)^{2}$ over the intervals $x \rightarrow x + \mathrm{\Delta}x$ and $x \cdot \sqrt{b} \cdot \sqrt{a + b} / \left( \sqrt{b} \cdot \sqrt{a + b}  - b\right) \rightarrow \left(x + \mathrm{\Delta}x \right) \cdot \sqrt{b} \cdot \sqrt{a + b} / \left( \sqrt{b} \cdot \sqrt{a + b}  - b\right)$ (i.e. $x_{\text{v}} \rightarrow x_{\text{v}} + \mathrm{\Delta}x$) representing a token swap against both the reference and amplified liquidity pools, where $\mathrm{\Delta}x > 0$ and $\mathrm{\Delta}y < 0$ ($\partial / \partial x$ Equation \ref{eq237} and Equation \ref{eq235}). Note that that the integration areas for the reference and virtual curves are $\mathrm{\Delta}y$ and $\mathrm{\Delta}y \cdot \sqrt{b} \cdot \sqrt{a + b} / \left( \sqrt{b} \cdot \sqrt{a + b} - b \right)$, respectively. By inference, any token swap against the virtual curve can be calculated from the reference curve by applying the amplification term (Equations \ref{eq53} and \ref{eq238}).}
    \label{fig49}
\end{figure}

Trade volumes between reference and virtual curves within the same price range are compared (Figures \ref{fig48} and \ref{fig49}). Despite trading at the same marginal and effective exchange rates, the virtual curve shows significantly higher trade volumes, as seen by longer arrows in Figure \ref{fig48} and a larger integrated area in Figure \ref{fig49}.

This concludes our examination of the Carbon DeFi curves, and the section on “unnatural” parameterizations. The critical algebraic identities for all three protocols examined here (Bancor's v2 and Carbon DeFi, and Uniswap's v3) are tabulated in Table \ref{tab1}, which provides something akin to a concentrated liquidity Rosetta stone, allowing for direct translation of key parameters between the contexts wherein each protocol resides. Remember - from the analytical and geometric perspectives, \textit{all three descriptions are mathematically redundant}. Their apparent differences emerge from the choice of notation and parameterization, which only reflects their smart contract implementation; the underlying pricing algorithm remains unchanged in all three. 

\begin{table}[ht]
    \centering
    \begin{tabular}{c}
    \includegraphics[width=\linewidth]{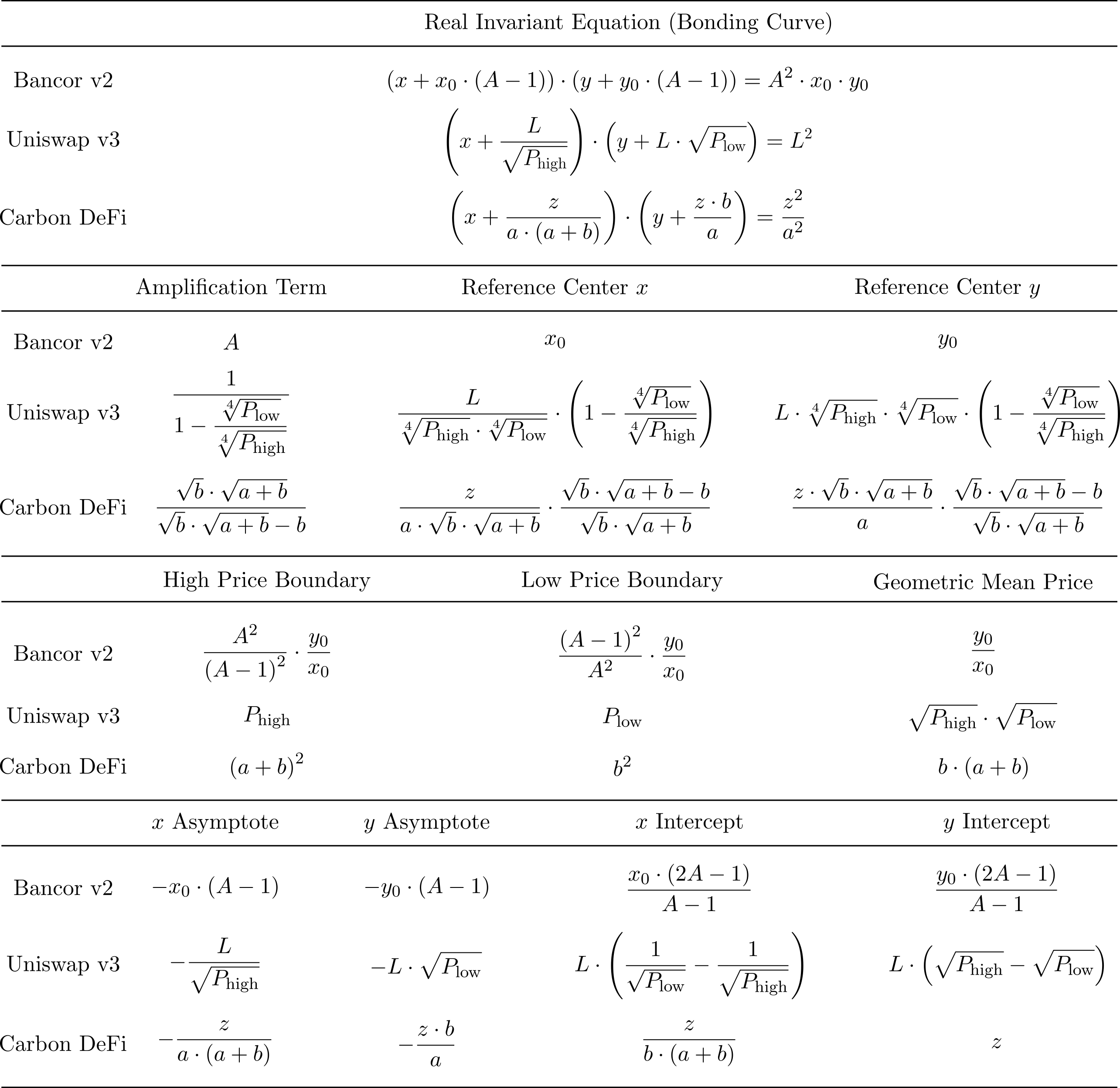}
    \end{tabular}
    \captionsetup{
        justification=raggedright,
        singlelinecheck=false,
        font=small,
        labelfont=bf,
        labelsep=quad,
        format=plain
    }
    \caption{Summary of the real curve capstone identities for Bancor v2, Uniswap v3, and Carbon DeFi, in their native parameterizations.}
    \label{tab1}
\end{table}

\section{The “Natural” Reparameterizations of the Concentrated Liquidity Invariant Function}\label{sec5}

The previous sections are referred to as the “unnatural” parameterizations because the way they are constructed is artificial; they are created by deliberately asserting something, such as the relative amplification of the virtual curve compared to the reference curve, the price bounds, or one of the intercepts of the real curve, whereas a “natural” construction might begin at a more fundamental level, free from the context of the protocol for which it is being defined. 

With that said, it is not immediately clear how to go about obtaining the natural description of these objects. Contributions of an academic nature are often presented as though one already possessed the intuition and depth of understanding to have arrived at what is disclosed, effortlessly. That was not the case here. In fact the opposite is true – a conspicuously compact set of equations for the concentrated real curve was discovered first, and their connection to more fundamental concepts was found later while investigating their significance. 

The following sections are [mostly] presented in chronological order with respect to the investigation of this topic material, which has the benefit of a top-down trajectory towards the underlying hyperbolic trigonometry of these curves. 

\subsection{Three “Pretty” Invariants}\label{subsec5.1}

In Section \ref{subsec3.1}, following the discussion of amplification term, $A$, expressed with respect to the price boundaries, $P_{\text{high}}$ and $P_{\text{low}}$, (Equations \ref{eq51}, \ref{eq52} and \ref{eq53}), it was stated that there is a more conspicuous identity, $C$, which almost seems as though it wants to assert its own existence (Equations \ref{eq54}, \ref{eq55}, \ref{eq56}, \ref{eq57} and \ref{eq58}). This term itches, and its persistence motivates a refactorization which includes it as a native parameter. Beginning from the seminal concentrated liquidity function (Equation \ref{eq74}), and the $x_{\text{int}}$, $y_{\text{int}}$, $x_{\text{asym}}$, and $y_{\text{asym}}$ identities derived therefrom (Equations \ref{eq91}, \ref{eq92}, \ref{eq93} and \ref{eq94}), it is a relatively easy task to identify $A$, $A-1$, and $2A-1$ as key terms which will need to be defined with respect to $C$. Note that this process alone begins to reveal a symmetry that was hidden there all along (Equations \ref{eq247}, \ref{eq248} and \ref{eq249}). 

\begin{flalign}
& \text{\renewcommand{\arraystretch}{0.66}
    \begin{tabular}{@{}c@{}}
    \scriptsize from \\
    \scriptsize (\ref{eq54})\\\scriptsize (\ref{eq55})\\\scriptsize (\ref{eq56})\\\scriptsize (\ref{eq57})\\\scriptsize (\ref{eq58})
  \end{tabular}} 
  & 
  A = \displaystyle \frac{1}{1 - \displaystyle \frac{\sqrt[4]{P_{\text{low}}}}{\sqrt[4]{P_{\text{high}}}}} = \displaystyle \frac{\sqrt[4]{P_{\text{high}}}}{\sqrt[4]{P_{\text{high}}} - \sqrt[4]{P_{\text{low}}}} = \displaystyle \frac{\sqrt{C} + 1}{C - 1} + 1 = \displaystyle \frac{\sqrt{C}}{\sqrt{C} - 1}
  &  
  \label{eq247} 
  &
\end{flalign}

\begin{flalign}
& \text{\renewcommand{\arraystretch}{0.66}
    \begin{tabular}{@{}c@{}}
    \scriptsize from \\
    \scriptsize (\ref{eq54})\\\scriptsize (\ref{eq55})\\\scriptsize (\ref{eq56})\\\scriptsize (\ref{eq57})\\\scriptsize (\ref{eq58})
  \end{tabular}} 
  & 
  \ A - 1 = \displaystyle \frac{1}{\displaystyle \frac{\sqrt[4]{P_{\text{high}}}}{\sqrt[4]{P_{\text{low}}}} - 1} = \displaystyle \frac{\sqrt[4]{P_{\text{low}}}}{\sqrt[4]{P_{\text{high}}} - \sqrt[4]{P_{\text{low}}}} = \displaystyle \frac{\sqrt{C} + 1}{C - 1} = \displaystyle \frac{1}{\sqrt{C} - 1}
  &  
  \label{eq248} 
  &
\end{flalign}

\begin{flalign}
& \text{\renewcommand{\arraystretch}{0.66}
    \begin{tabular}{@{}c@{}}
    \scriptsize from \\
    \scriptsize (\ref{eq247})\\\scriptsize (\ref{eq248})
  \end{tabular}} 
  & 
  2A - 1 = \displaystyle \frac{1}{1 - \displaystyle \frac{\sqrt[4]{P_{\text{low}}}}{\sqrt[4]{P_{\text{high}}}}} + \displaystyle \frac{1}{\displaystyle \frac{\sqrt[4]{P_{\text{high}}}}{\sqrt[4]{P_{\text{low}}}} - 1} = \displaystyle \frac{\sqrt[4]{P_{\text{high}}} + \sqrt[4]{P_{\text{low}}}}{\sqrt[4]{P_{\text{high}}} - \sqrt[4]{P_{\text{low}}}} = \displaystyle \frac{\left( \sqrt{C} + 1 \right)^{2}}{C - 1} = \displaystyle \frac{\sqrt{C} + 1}{\sqrt{C} - 1}
  &  
  \label{eq249} 
  &
\end{flalign}

Incorporating the identities of $A$ and $A-1$ in terms of $C$ (Equations \ref{eq247} and \ref{eq248}) from above into the seminal Bancor v2 invariant (Equation \ref{eq74}) yields Equation \ref{eq250}, and its rearrangement to isolate $C$ yields Equation \ref{eq251}. 

\begin{flalign}
& \text{\renewcommand{\arraystretch}{0.66}
    \begin{tabular}{@{}c@{}}
    \scriptsize from \\
    \scriptsize (\ref{eq74})\\\scriptsize (\ref{eq247})\\\scriptsize (\ref{eq248})
  \end{tabular}} 
  & 
  \left( x + \displaystyle \frac{x_{0}}{\sqrt{C} - 1} \right) \cdot \left( y + \displaystyle \frac{y_{0}}{\sqrt{C} - 1} \right) = \displaystyle \frac{x_{0} \cdot y_{0} \cdot C}{\left( \sqrt{C} - 1 \right)^{2}}
  &  
  \label{eq250} 
  &
\end{flalign}

\begin{flalign}
& \text{\renewcommand{\arraystretch}{0.66}
    \begin{tabular}{@{}c@{}}
    \scriptsize from \\
    \scriptsize (\ref{eq250})
  \end{tabular}} 
  & 
  \displaystyle \frac{\left( x - x_{0} \right)^{2} \cdot \left( y - y_{0} \right)^{2}}{\left( x \cdot y - x_{0} \cdot y_{0} \right)^{2}} = C
  &  
  \label{eq251} 
  &
\end{flalign}

After that, rearrangement of Equations \ref{eq91} and \ref{eq92} to express $x_{0}$ and $y_{0}$ in terms of $A$, $x_{\text{int}}$ and $y_{\text{int}}$ (Equation \ref{eq252}) followed by their substitution into the general invariant (Equation \ref{eq74}) yields Equation \ref{eq253}. Then, substitution of the $A$, $A-1$ and $2A-1$ terms from Equations \ref{eq247}, \ref{eq248} and \ref{eq249} as needed yields Equation \ref{eq254}, which after rearrangement to isolate $C$ yields Equation \ref{eq255}. 

\begin{flalign}
& \text{\renewcommand{\arraystretch}{0.66}
    \begin{tabular}{@{}c@{}}
    \scriptsize from \\
    \scriptsize (\ref{eq91})\\\scriptsize (\ref{eq92})
  \end{tabular}} 
  & 
  x_{0} = x_{\text{int}} \cdot \displaystyle \frac{(A - 1)}{(2A - 1)};\ y_{0} = y_{\text{int}} \cdot \displaystyle \frac{(A - 1)}{(2A - 1)}
  &  
  \label{eq252} 
  &
\end{flalign}

\begin{flalign}
& \text{\renewcommand{\arraystretch}{0.66}
    \begin{tabular}{@{}c@{}}
    \scriptsize from \\
    \scriptsize (\ref{eq74})\\\scriptsize (\ref{eq252})
  \end{tabular}} 
  & 
  \left( x + x_{\text{int}} \cdot \displaystyle \frac{\left( A - 1 \right)^{2}}{(2A - 1)} \right) \cdot \left( y + y_{\text{int}} \cdot \displaystyle \frac{\left( A - 1 \right)^{2}}{(2A - 1)} \right) = x_{\text{int}} \cdot y_{\text{int}} \cdot \displaystyle \frac{A^{2} \cdot \left( A - 1 \right)^{2}}{(2A - 1)^{2}}
  &  
  \label{eq253} 
  &
\end{flalign}

\begin{flalign}
& \text{\renewcommand{\arraystretch}{0.66}
    \begin{tabular}{@{}c@{}}
    \scriptsize from \\
    \scriptsize (\ref{eq247})\\\scriptsize (\ref{eq248})\\\scriptsize (\ref{eq249})\\\scriptsize (\ref{eq253})
  \end{tabular}} 
  & 
  \left( x + \displaystyle \frac{x_{\text{int}}}{C - 1} \right) \cdot \left( y + \displaystyle \frac{y_{\text{int}}}{C - 1} \right) = \displaystyle \frac{x_{\text{int}} \cdot y_{\text{int}} \cdot C}{(C - 1)^{2}}
  &  
  \label{eq254} 
  &
\end{flalign}

\begin{flalign}
& \text{\renewcommand{\arraystretch}{0.66}
    \begin{tabular}{@{}c@{}}
    \scriptsize from \\
    \scriptsize (\ref{eq254})
  \end{tabular}} 
  & 
  \displaystyle \frac{\left( x_{\text{int}} - x \right) \cdot \left( y_{\text{int}} - y \right)}{x \cdot y} = C
  &  
  \label{eq255} 
  &
\end{flalign}

Lastly, rearrangement of Equations \ref{eq93} and \ref{eq94} to express $x_{asym}$ and $y_{asym}$ in terms of $A$, $x_{\text{int}}$ and $y_{\text{int}}$ (Equation \ref{eq256}) followed by their substitution into the general invariant (Equation \ref{eq74}) yields Equation \ref{eq257}. In contrast to Equations \ref{eq74} and \ref{eq253}, Equation \ref{eq257} features $A^{2} / \left( A - 1  \right)^{2}$ as the only term requiring substitution, which has already been proved to be equal to $C$ (Equations \ref{eq54}, \ref{eq55}, \ref{eq56}, \ref{eq57} and \ref{eq58}). Therefore, effective replacement of the $A$ terms to yield Equation \ref{eq258} can be performed without reference to Equations \ref{eq247}, \ref{eq248} and \ref{eq249}. Then, rearrangement to isolate $C$ yields Equation \ref{eq259}.

\begin{flalign}
& \text{\renewcommand{\arraystretch}{0.66}
    \begin{tabular}{@{}c@{}}
    \scriptsize from \\
    \scriptsize (\ref{eq93})\\\scriptsize (\ref{eq94})
  \end{tabular}} 
  & 
  x_{0} = - \displaystyle \frac{x_{\text{asym}}}{(A - 1)};\ y_{0} = - \displaystyle \frac{y_{\text{asym}}}{(A - 1)}
  &  
  \label{eq256} 
  &
\end{flalign}

\begin{flalign}
& \text{\renewcommand{\arraystretch}{0.66}
    \begin{tabular}{@{}c@{}}
    \scriptsize from \\
    \scriptsize (\ref{eq74})\\\scriptsize (\ref{eq256})
  \end{tabular}} 
  & 
  \left( x - x_{\text{asym}} \right) \cdot \left( y - y_{\text{asym}} \right) = x_{\text{asym}} \cdot y_{\text{asym}} \cdot \displaystyle \frac{A^{2}}{\left( A - 1 \right)^{2}}
  &  
  \label{eq257} 
  &
\end{flalign}

\begin{flalign}
& \text{\renewcommand{\arraystretch}{0.66}
    \begin{tabular}{@{}c@{}}
    \scriptsize from \\
    \scriptsize (\ref{eq54})\\\scriptsize (\ref{eq55})\\\scriptsize (\ref{eq56})\\\scriptsize (\ref{eq57})\\\scriptsize (\ref{eq58})\\\scriptsize (\ref{eq257})
  \end{tabular}} 
  & 
  \left( x - x_{\text{asym}} \right) \cdot \left( y - y_{\text{asym}} \right) = x_{\text{asym}} \cdot y_{\text{asym}} \cdot C
  &  
  \label{eq258} 
  &
\end{flalign}

\begin{flalign}
& \text{\renewcommand{\arraystretch}{0.66}
    \begin{tabular}{@{}c@{}}
    \scriptsize from \\
    \scriptsize (\ref{eq258})
  \end{tabular}} 
  & 
  \displaystyle \frac{\left( x - x_{\text{asym}} \right) \cdot \left( y - y_{\text{asym}} \right)}{x_{\text{asym}} \cdot y_{\text{asym}}} = C
  &  
  \label{eq259} 
  &
\end{flalign}

Note that Equations \ref{eq251}, \ref{eq255} and \ref{eq259} are conspicuously simple, even “pretty”, at least compared to the status quo for the “unnatural” archetypes in the previous sections. Moreover, \textit{they are all equal to each other} (Equation \ref{eq260}). Combined with the other identities already proven (and implied) earlier in Equations \ref{eq57}, \ref{eq58}, \ref{eq71}, \ref{eq72}, \ref{eq101}, \ref{eq102}, \ref{eq145}, \ref{eq146}, \ref{eq222} and \ref{eq223} (Equation \ref{eq261}), we are starting to approach something akin to an algebraic universal translator for \textit{any} concentrated real curve, regardless of its parameterization. While this is certainly not the only way to achieve this (e.g. Table \ref{tab1}), there is something peculiarly terse about this one, and its prettiness hints at something more fundamental.

\begin{flalign}
& \text{\renewcommand{\arraystretch}{0.66}
    \begin{tabular}{@{}c@{}}
    \scriptsize from \\
    \scriptsize (\ref{eq258})
  \end{tabular}} 
  & 
  \displaystyle \frac{\left( x - x_{0} \right)^{2} \cdot \left( y - y_{0} \right)^{2}}{\left( x \cdot y - x_{0} \cdot y_{0} \right)^{2}} = \displaystyle \frac{\left( x_{\text{int}} - x \right) \cdot \left( y_{\text{int}} - y \right)}{x \cdot y} = \displaystyle \frac{\left( x - x_{\text{asym}} \right) \cdot \left( y - y_{\text{asym}} \right)}{x_{\text{asym}} \cdot y_{\text{asym}}} = C
  &  
  \label{eq260} 
  &
\end{flalign}

\begin{flalign}
& \text{\renewcommand{\arraystretch}{0.66}
    \begin{tabular}{@{}c@{}}
    \scriptsize from \\
    \scriptsize (\ref{eq258})
  \end{tabular}} 
  & 
  \displaystyle \frac{y_{\text{asym}} - y_{\text{int}}}{y_{\text{asym}}} = \displaystyle \frac{x_{\text{asym}} - x_{\text{int}}}{x_{\text{asym}}} = \displaystyle \frac{\left( y_{\text{asym}} - y_{0} \right)^{2}}{y_{\text{asym}}^{2}} = \displaystyle \frac{\left( x_{\text{asym}} - x_{0} \right)^{2}}{x_{\text{asym}}^{2}} = \displaystyle \frac{\sqrt{P_{\text{high}}}}{\sqrt{P_{\text{low}}}} = C
  &  
  \label{eq261} 
  &
\end{flalign}

\subsection{Transforming and Normalizing the Curves in a New Coordinate System}\label{subsec5.2}

The analysis in the following section employs the hyperbolic trigonometry functions. However, these functions are only defined for a unit hyperbola, which opens to the left and the right, with its vertices at ($-1$, $+1$). Therefore, we have two options. Either we redefine these functions to operate in the native parameter space of the bonding curves, or transform the bonding curves into a compatible representation. There is fundamentally no difference between either option, only a difference of perspective and intuition. Since the hyperbolic trigonometry functions are relatively exotic and given that transforming the curves is a core theme of this exploration, we are going to operate on the curves themselves and then apply the trigonometric analysis on their transformed versions.

This part of the exercise begins by first returning to the standard $x \cdot y = x_{0} \cdot y_{0}$ curve. Prior to the hyperbolic trigonometric analysis, we must first perform a rotation about the origin and then re-scale to create a unit hyperbola. The rotation is $- 45^{{^\circ}}$ (a clockwise rotation is negative), or $- \pi / 4$ radians. Therefore, the transformation matrix, $R_{{- 45}^{{^\circ}}}$, is derived from the general rotation matrix (Equation \ref{eq262}) by substituting $\theta = - \pi / 4$ (Equation \ref{eq263}).

\begin{equation} \label{eq262}
R = \begin{bmatrix} \cos(\theta) & - \sin(\theta) \\ \sin(\theta) & \cos(\theta) \end{bmatrix}
\end{equation}

\begin{flalign}
& \text{\renewcommand{\arraystretch}{0.66}
    \begin{tabular}{@{}c@{}}
    \scriptsize from \\
    \scriptsize (\ref{eq262})
  \end{tabular}} 
  & 
  R_{{- 45}^{{^\circ}}} = \begin{bmatrix} \cos\left( - \displaystyle \frac{\pi}{4} \right) & - \sin\left( - \displaystyle \frac{\pi}{4} \right) \\ & \\ \sin\left( - \displaystyle \frac{\pi}{4} \right) & \cos\left( - \displaystyle \frac{\pi}{4} \right) \end{bmatrix} = \begin{bmatrix} \displaystyle \frac{1}{\sqrt{2}} & - \displaystyle \frac{1}{\sqrt{2}} \\ & \\ \displaystyle \frac{1}{\sqrt{2}} & \displaystyle \frac{1}{\sqrt{2}} \end{bmatrix}
  &  
  \label{eq263} 
  &
\end{flalign}

The above matrix can be used to apply the transformation on all points ($x$, $y$) on the curve $x \cdot y = x_{0} \cdot y_{0}$ as follows (Equation \ref{eq264}):

\begin{flalign}
& \text{\renewcommand{\arraystretch}{0.66}
    \begin{tabular}{@{}c@{}}
    \scriptsize from \\
    \scriptsize (\ref{eq263})
  \end{tabular}} 
  & 
  \begin{bmatrix} t \\ u \end{bmatrix} = R_{{- 45}^{{^\circ}}} \cdot \begin{bmatrix} x \\ y \end{bmatrix} = \begin{bmatrix} \displaystyle \frac{1}{\sqrt{2}} & - \displaystyle \frac{1}{\sqrt{2}} \\ & \\ \displaystyle \frac{1}{\sqrt{2}} & \displaystyle \frac{1}{\sqrt{2}} \end{bmatrix} \cdot \begin{bmatrix} x \\ y \end{bmatrix} = \begin{bmatrix} \displaystyle \frac{x}{\sqrt{2}} + \displaystyle \frac{y}{\sqrt{2}} \\ \\ - \displaystyle \frac{x}{\sqrt{2}} + \displaystyle \frac{y}{\sqrt{2}} \end{bmatrix}
  &  
  \label{eq264} 
  &
\end{flalign}

\begin{figure}[ht]
    \centering
    \includegraphics[width=\textwidth]{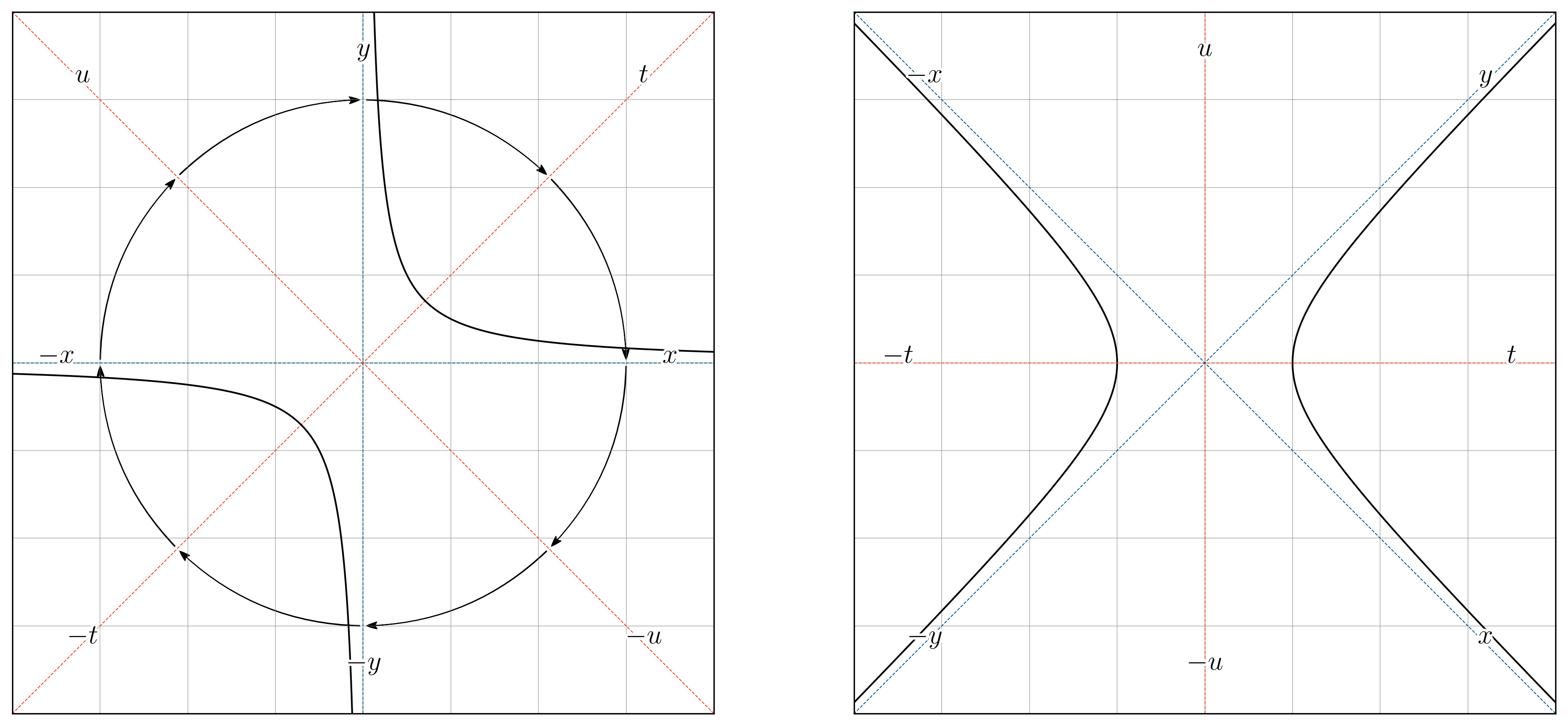}
    \captionsetup{
        justification=raggedright,
        singlelinecheck=false,
        font=small,
        labelfont=bf,
        labelsep=quad,
        format=plain
    }
    \caption{A general depiction of coordinate system transformation and rotation. This figure visualizes the transformation and rotation of a coordinate system from the standard Cartesian x- and y-axes to a rotated system denoted by t- and u- axes. Arrows indicate the direction of rotation from each of the Cartesian axes ($x$, $y$) to the new rotated axes ($t$, $u$), mimicking the motion of a steering wheel as it turns to the right. Note that the hyperbola drawn in the left panel opens towards the bottom-left and top-right quadrants, whereas the hyperbola drawn in the right panel opens towards the left and right. The equation of the hyperbola in the left subplot is of the form $x \cdot y = k$, which after the linear transform becomes $t^{2} - u^{2} = 2 \cdot k$.}
    \label{fig50}
\end{figure}

Therefore, any set of coordinates ($x$, $y$) is transformed to ($t$, $u$), thus rotating the $x \cdot y = x_{0} \cdot y_{0}$ curve $45^{{^\circ}}$ clockwise (Equations \ref{eq265} and \ref{eq266}). The overall process is depicted in Figure \ref{fig50}.

\begin{flalign}
& \text{\renewcommand{\arraystretch}{0.66}
    \begin{tabular}{@{}c@{}}
    \scriptsize from \\
    \scriptsize (\ref{eq264})
  \end{tabular}} 
  & 
  t = f_{x,y \rightarrow t}(x,y) = \displaystyle \frac{x + y}{\sqrt{2}}
  &  
  \label{eq265} 
  &
\end{flalign}

\begin{flalign}
& \text{\renewcommand{\arraystretch}{0.66}
    \begin{tabular}{@{}c@{}}
    \scriptsize from \\
    \scriptsize (\ref{eq264})
  \end{tabular}} 
  & 
  u = f_{x,y \rightarrow u}(x,y) = \displaystyle \frac{y - x}{\sqrt{2}}
  &  
  \label{eq266} 
  &
\end{flalign}

As we have done previously in the context of swap equations and marginal price formulas, we can reduce the dimensions needed for performing the rotation such that the transformation can be performed with reference only to one coordinate in the set by substituting the identities of $x$ and $y$ (Equations \ref{eq6} and \ref{eq8}) as required to eliminate the opposite dimension, denoted by $f_{x \rightarrow t}$ and $f_{y \rightarrow u}$. In other words, $t$ can be expressed entirely as a function of $x$ (Equation \ref{eq267}) and $u$ can be expressed entirely as a function of $y$ (Equation \ref{eq268}).

\begin{flalign}
& \text{\renewcommand{\arraystretch}{0.66}
    \begin{tabular}{@{}c@{}}
    \scriptsize from \\
    \scriptsize (\ref{eq8})\\\scriptsize (\ref{eq265})
  \end{tabular}} 
  & 
  t = f_{x \rightarrow t}(x) = \displaystyle \frac{x + \displaystyle \frac{x_{0} \cdot y_{0}}{x}}{\sqrt{2}} = \displaystyle \frac{x^{2} + x_{0} \cdot y_{0}}{x \cdot \sqrt{2}}
  &  
  \label{eq267} 
  &
\end{flalign}

\begin{flalign}
& \text{\renewcommand{\arraystretch}{0.66}
    \begin{tabular}{@{}c@{}}
    \scriptsize from \\
    \scriptsize (\ref{eq6})\\\scriptsize (\ref{eq266})
  \end{tabular}} 
  & 
  u = f_{y \rightarrow u}(y) = \displaystyle \frac{y - \displaystyle \frac{x_{0} \cdot y_{0}}{y}}{\sqrt{2}} = \displaystyle \frac{y^{2} - x_{0} \cdot y_{0}}{y \cdot \sqrt{2}}
  &  
  \label{eq268} 
  &
\end{flalign}

We can also map $y$ to $t$ ($f_{y \rightarrow t}$), and $x$ to $u$ ($f_{x \rightarrow u}$) (Equations \ref{eq267} and \ref{eq268}); we can use both forms to prove the invariant equation of the transformed hyperbola, which opens to the left and right in the $t$ dimension.

\begin{flalign}
& \text{\renewcommand{\arraystretch}{0.66}
    \begin{tabular}{@{}c@{}}
    \scriptsize from \\
    \scriptsize (\ref{eq6})\\\scriptsize (\ref{eq265})
  \end{tabular}} 
  & 
  t = f_{y \rightarrow t}(y) = \displaystyle \frac{\displaystyle \frac{x_{0} \cdot y_{0}}{y} + y}{\sqrt{2}} = \displaystyle \frac{x_{0} \cdot y_{0} + y^{2}}{y \cdot \sqrt{2}}
  &  
  \label{eq269} 
  &
\end{flalign}

\begin{flalign}
& \text{\renewcommand{\arraystretch}{0.66}
    \begin{tabular}{@{}c@{}}
    \scriptsize from \\
    \scriptsize (\ref{eq8})\\\scriptsize (\ref{eq266})
  \end{tabular}} 
  & 
  u = f_{x \rightarrow u}(x) = \displaystyle \frac{\displaystyle \frac{x_{0} \cdot y_{0}}{x} - x}{\sqrt{2}} = \displaystyle \frac{x_{0} \cdot y_{0} - x^{2}}{x \cdot \sqrt{2}}
  &  
  \label{eq270} 
  &
\end{flalign}

Then, to obtain the identity of the transformed hyperbola, which has the form $t^{2} - u^{2} = 2 \cdot k$, we can substitute the reduced-dimensions expressions for $t$ and $u$ from either Equations \ref{eq267} and \ref{eq270}, or Equations \ref{eq268} and \ref{eq269}; the pair of expressions used must eliminate the remaining $x$ or $y$ dimension, which yields the transformed invariant equation entirely in its native dimensions $t$ and $u$ (Equations \ref{eq271} and \ref{eq272}). Alternatively, the $x_{0} \cdot y_{0}$ terms can be substituted for its identity $x \cdot y$ Equation \ref{eq2}, which delivers the same result. Remember, $x_{0}$ and $y_{0}$ are constants; while the notation references the $x$ and $y$ dimensions, the values themselves are not dependent on them. Therefore Equations \ref{eq271} and \ref{eq272} are truly two-dimensional, \textit{not} four-dimensional expressions.

\begin{flalign}
& \text{\renewcommand{\arraystretch}{0.66}
    \begin{tabular}{@{}c@{}}
    \scriptsize from \\
    \scriptsize (\ref{eq267})\\\scriptsize (\ref{eq270})
  \end{tabular}} 
  & 
  t^{2} - u^{2} = \displaystyle \frac{\left( x^{2} + x_{0} \cdot y_{0} \right)^{2}}{2 \cdot x^{2}} - \displaystyle \frac{\left( x_{0} \cdot y_{0} - x^{2} \right)^{2}}{2 \cdot x^{2}} = 2 \cdot x_{0} \cdot y_{0}
  &  
  \label{eq271} 
  &
\end{flalign}

\begin{flalign}
& \text{\renewcommand{\arraystretch}{0.66}
    \begin{tabular}{@{}c@{}}
    \scriptsize from \\
    \scriptsize (\ref{eq268})\\\scriptsize (\ref{eq269})
  \end{tabular}} 
  & 
  t^{2} - u^{2} = \displaystyle \frac{\left( x_{0} \cdot y_{0} + y^{2} \right)^{2}}{2 \cdot y^{2}} - \displaystyle \frac{\left( y^{2} - x_{0} \cdot y_{0} \right)^{2}}{2 \cdot y^{2}} = 2 \cdot x_{0} \cdot y_{0}
  &  
  \label{eq272} 
  &
\end{flalign}

\begin{figure}[ht]
    \centering
    \includegraphics[width=\textwidth]{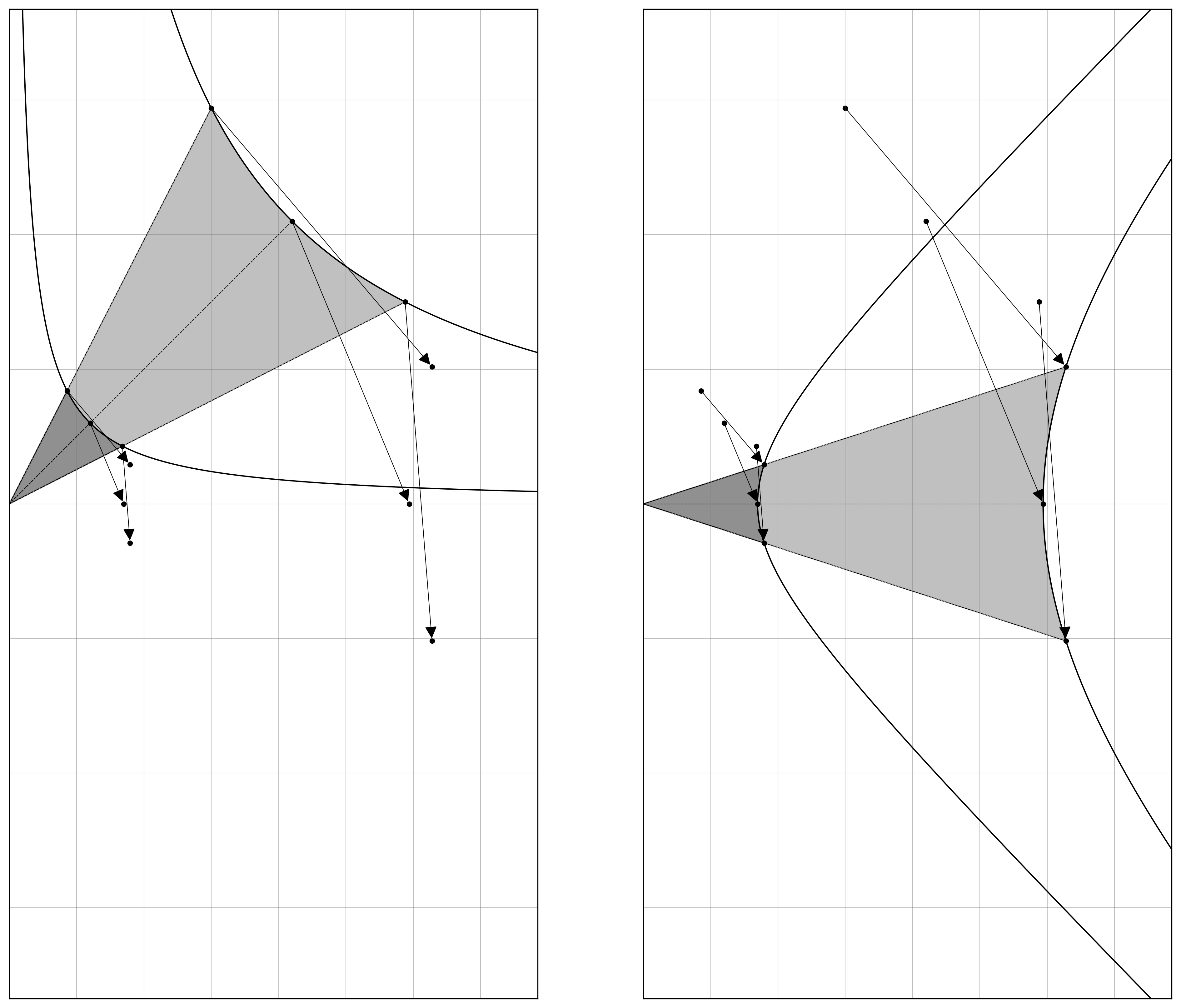}
    \captionsetup{
        justification=raggedright,
        singlelinecheck=false,
        font=small,
        labelfont=bf,
        labelsep=quad,
        format=plain
    }
    \caption{The key identities associated with the reference and virtual curves are drawn with rays emanating from the origin, which enclose a shaded region. The desired linear transformation (Equation \ref{eq264}) is depicted with arrows beginning from these same points in the Cartesian ($x$, $y$) plane, and ending at their implied destinations in the new coordinate system ($t$, $u$). The transformation from the ($x$, $y$) to the ($t$, $u$) coordinates is represented as an apparent ${- 45}^{{^\circ}}$ rotation between the left and right panels of the figure.}
    \label{fig51}
\end{figure}

Our objective is now to apply this rotation to the key points defining the reference and virtual curves (Figure \ref{fig51}). The elements of the real curve that inform this analysis are implicit; to incorporate the real curve into the following derivations, simply scale it down in-place (i.e. without moving the point at ($x_{0}$, $y_{0}$)) to become the reference curve or reverse its horizontal and vertical shifts to become the virtual curve as we have done previously (Equations \ref{eq135}, \ref{eq137}, \ref{eq139}, \ref{eq140}, \ref{eq142}, \ref{eq144}, \ref{eq214}, \ref{eq215}, \ref{eq217}, \ref{eq218}, \ref{eq219} and \ref{eq221}). The analysis being performed here for either of the reference or virtual curves, individually, generalize to each other, and by extension, also to the real curve. Therefore, as with much of the work demonstrated in this document, performing these steps for both the reference and virtual curves is done for its pedagogical benefit and not necessarily because it is required by the analysis. The analysis can be completed with only one of these.

The redundancy produced by the dimension reduction in Equations \ref{eq269}, \ref{eq270}, \ref{eq271} and \ref{eq272} means that the transformation from the Cartesian ($x$, $y$) plane to the new ($t$, $u$) coordinate system can be performed in at least three different ways. Either both the $x$ and $y$ coordinates can be used as arguments (Equations \ref{eq265} and \ref{eq266}), or only one of the $x$ or $y$ coordinates can be used as arguments (Equations \ref{eq267} and \ref{eq270}, and Equations \ref{eq268} and \ref{eq269}, respectively). This is easily demonstrated. Recall that the point ($x_{0}$, $y_{0}$) lies on the reference curve (Figure \ref{fig2}). To obtain its transformed $t$ coordinate, the result of passing both $x_{0}$ and $y_{0}$ as arguments to Equation \ref{eq265} is identical to passing only $x_{0}$ to Equation \ref{eq267} and passing only $y_{0}$ to Equation \ref{eq269} (Equation \ref{eq273}). Similarly, to obtain its transformed $u$ coordinate, the result of passing both $x_{0}$ and $y_{0}$ as arguments to Equation \ref{eq266} is identical to passing only $x_{0}$ to Equation \ref{eq268} and passing only $y_{0}$ to Equation \ref{eq270} (Equation \ref{eq274}).

\begin{flalign}
& \text{\renewcommand{\arraystretch}{0.66}
    \begin{tabular}{@{}c@{}}
    \scriptsize from \\
    \scriptsize (\ref{eq265})\\\scriptsize (\ref{eq267})\\\scriptsize (\ref{eq269})
  \end{tabular}} 
  & 
  f_{x,y \rightarrow t}\left( x_{0},y_{0} \right) = f_{x \rightarrow t}\left( x_{0} \right) = f_{y \rightarrow t}\left( y_{0} \right) = \displaystyle \frac{x_{0} + y_{0}}{\sqrt{2}}
  &  
  \label{eq273} 
  &
\end{flalign}

\begin{flalign}
& \text{\renewcommand{\arraystretch}{0.66}
    \begin{tabular}{@{}c@{}}
    \scriptsize from \\
    \scriptsize (\ref{eq266})\\\scriptsize (\ref{eq268})\\\scriptsize (\ref{eq270})
  \end{tabular}} 
  & 
  f_{x,y \rightarrow u}\left( x_{0},y_{0} \right) = f_{x \rightarrow u}\left( x_{0} \right) = f_{y \rightarrow u}\left( y_{0} \right) = \displaystyle \frac{y_{0} - x_{0}}{\sqrt{2}}
  &  
  \label{eq274} 
  &
\end{flalign}

The remainder of this part of the exercise will use the two-argument functions, Equations \ref{eq265} and \ref{eq266}. Just know this is a completely arbitrary decision and it makes no difference whatsoever if you would prefer to use the single-argument alternatives. Substituting the other key identities previously obtained for the reference curve (Equations \ref{eq59}, \ref{eq60}, \ref{eq61} and \ref{eq62}) into Equations \ref{eq265} and \ref{eq266} yields their transformed ($u$, $t$) coordinates (Equations \ref{eq275}, \ref{eq276}, \ref{eq277} and \ref{eq278}).

\begin{flalign}
& \text{\renewcommand{\arraystretch}{0.66}
    \begin{tabular}{@{}c@{}}
    \scriptsize from \\
    \scriptsize (\ref{eq59})\\\scriptsize (\ref{eq60})\\\scriptsize (\ref{eq265})
  \end{tabular}} 
  & 
  f_{x,y \rightarrow t}\left( \min \left( x \right),\max \left( y \right) \right) = \displaystyle \frac{\displaystyle \frac{x_{0} \cdot \left( A - 1 \right)}{A} + \displaystyle \frac{A \cdot y_{0}}{A - 1}}{\sqrt{2}} = \displaystyle \frac{x_{0} \cdot \left( A - 1 \right)^{2} + {A^{2} \cdot y_{0}}}{\sqrt{2} \cdot A \cdot \left( A - 1 \right)}
  &  
  \label{eq275} 
  &
\end{flalign}

\begin{flalign}
& \text{\renewcommand{\arraystretch}{0.66}
    \begin{tabular}{@{}c@{}}
    \scriptsize from \\
    \scriptsize (\ref{eq59})\\\scriptsize (\ref{eq60})\\\scriptsize (\ref{eq266})
  \end{tabular}} 
  & 
  f_{x,y \rightarrow u}\left( \min \left( x \right),\max \left( y \right) \right) = \displaystyle \frac{\displaystyle \frac{A \cdot y_{0}}{A - 1} - \displaystyle \frac{x_{0} \cdot \left( A - 1 \right)}{A}}{\sqrt{2}} = \displaystyle \frac{A^{2} \cdot y_{0} - x_{0} \cdot \left( A - 1 \right)^{2}}{\sqrt{2} \cdot A \cdot \left( A - 1 \right)}
  &  
  \label{eq276} 
  &
\end{flalign}

\begin{flalign}
& \text{\renewcommand{\arraystretch}{0.66}
    \begin{tabular}{@{}c@{}}
    \scriptsize from \\
    \scriptsize (\ref{eq61})\\\scriptsize (\ref{eq62})\\\scriptsize (\ref{eq265})
  \end{tabular}} 
  & 
  f_{x,y \rightarrow t}\left( \max \left( x \right),\min \left( y \right) \right) = \displaystyle \frac{\displaystyle \frac{{A \cdot x_{0}}}{A - 1} + \displaystyle \frac{y_{0} \cdot \left( A - 1 \right)}{A}}{\sqrt{2}} = \displaystyle \frac{A^{2} \cdot x_{0} + y_{0} \cdot \left( A - 1 \right)^{2}}{\sqrt{2} \cdot A \cdot \left( A - 1 \right)}
  &  
  \label{eq277} 
  &
\end{flalign}

\begin{flalign}
& \text{\renewcommand{\arraystretch}{0.66}
    \begin{tabular}{@{}c@{}}
    \scriptsize from \\
    \scriptsize (\ref{eq61})\\\scriptsize (\ref{eq62})\\\scriptsize (\ref{eq266})
  \end{tabular}} 
  & 
  f_{x,y \rightarrow u}\left( \max \left( x \right),\min \left( y \right) \right) = \displaystyle \frac{\displaystyle \frac{y_{0} \cdot \left( A - 1 \right)}{A} - \displaystyle \frac{{A \cdot x_{0}}}{A - 1}}{\sqrt{2}} = \displaystyle \frac{y_{0} \cdot \left( A - 1 \right)^{2} - A^{2} \cdot x_{0}}{\sqrt{2} \cdot A \cdot \left( A - 1 \right)}
  &  
  \label{eq278} 
  &
\end{flalign}

Then, via the same process, substituting the other key identities previously obtained for the virtual curve (Equations \ref{eq30}, \ref{eq31}, \ref{eq32}, \ref{eq33}, \ref{eq46} and \ref{eq48}) into Equations \ref{eq265} and \ref{eq266} yields their transformed ($u$, $t$) coordinates (Equations \ref{eq279}, \ref{eq280}, \ref{eq281}, \ref{eq282}, \ref{eq283} and \ref{eq284}).

\begin{flalign}
& \text{\renewcommand{\arraystretch}{0.66}
    \begin{tabular}{@{}c@{}}
    \scriptsize from \\
    \scriptsize (\ref{eq46})\\\scriptsize (\ref{eq48})\\\scriptsize (\ref{eq265})
  \end{tabular}} 
  & 
  f_{x_{\text{v}},y_{\text{v}} \rightarrow t_{\text{v}}}\left( {A \cdot x_{0}},\ A \cdot y_{0} \right) = \displaystyle \frac{A \cdot \left( x_{0} + y_{0} \right)}{\sqrt{2}}
  &  
  \label{eq279} 
  &
\end{flalign}

\begin{flalign}
& \text{\renewcommand{\arraystretch}{0.66}
    \begin{tabular}{@{}c@{}}
    \scriptsize from \\
    \scriptsize (\ref{eq46})\\\scriptsize (\ref{eq48})\\\scriptsize (\ref{eq266})
  \end{tabular}} 
  & 
  f_{x_{\text{v}},y_{\text{v}} \rightarrow u_{\text{v}}}\left( {A \cdot x_{0}},\ A \cdot y_{0} \right) = \displaystyle \frac{A \cdot \left( y_{0} - x_{0} \right)}{\sqrt{2}}
  &  
  \label{eq280} 
  &
\end{flalign}

\begin{flalign}
& \text{\renewcommand{\arraystretch}{0.66}
    \begin{tabular}{@{}c@{}}
    \scriptsize from \\
    \scriptsize (\ref{eq30})\\\scriptsize (\ref{eq31})\\\scriptsize (\ref{eq265})
  \end{tabular}} 
  & 
  f_{x_{\text{v}},y_{\text{v}} \rightarrow t_{\text{v}}}\left( \min\left( x_{\text{v}} \right),\max\left( y_{\text{v}} \right) \right) = \displaystyle \frac{x_{0} \cdot \left( A - 1 \right) + \displaystyle \frac{A^{2} \cdot y_{0}}{A - 1}}{\sqrt{2}} = \displaystyle \frac{x_{0} \cdot \left( A - 1 \right)^{2} + {A^{2} \cdot y_{0}}}{\sqrt{2} \cdot \left( A - 1 \right)}
  &  
  \label{eq281} 
  &
\end{flalign}

\begin{flalign}
& \text{\renewcommand{\arraystretch}{0.66}
    \begin{tabular}{@{}c@{}}
    \scriptsize from \\
    \scriptsize (\ref{eq30})\\\scriptsize (\ref{eq31})\\\scriptsize (\ref{eq266})
  \end{tabular}} 
  & 
  f_{x_{\text{v}},y_{\text{v}} \rightarrow u_{\text{v}}}\left( \min\left( x_{\text{v}} \right),\max\left( y_{\text{v}} \right) \right) = \displaystyle \frac{\displaystyle \frac{A^{2} \cdot y_{0}}{A - 1} - x_{0} \cdot \left( A - 1 \right)}{\sqrt{2}} = \displaystyle \frac{A^{2} \cdot y_{0} - x_{0} \cdot \left( A - 1 \right)^{2}}{\sqrt{2} \cdot \left( A - 1 \right)}
  &  
  \label{eq282} 
  &
\end{flalign}

\begin{flalign}
& \text{\renewcommand{\arraystretch}{0.66}
    \begin{tabular}{@{}c@{}}
    \scriptsize from \\
    \scriptsize (\ref{eq32})\\\scriptsize (\ref{eq33})\\\scriptsize (\ref{eq265})
  \end{tabular}} 
  & 
  f_{x_{\text{v}},y_{\text{v}} \rightarrow t_{\text{v}}}\left( \max\left( x_{\text{v}} \right),\min\left( y_{\text{v}} \right) \right) = \displaystyle \frac{\displaystyle \frac{A^{2} \cdot x_{0}}{A - 1} + y_{0} \cdot \left( A - 1 \right)}{\sqrt{2}} = \displaystyle \frac{A^{2} \cdot x_{0} + y_{0} \cdot \left( A - 1 \right)^{2}}{\sqrt{2} \cdot \left( A - 1 \right)}
  &  
  \label{eq283} 
  &
\end{flalign}

\begin{flalign}
& \text{\renewcommand{\arraystretch}{0.66}
    \begin{tabular}{@{}c@{}}
    \scriptsize from \\
    \scriptsize (\ref{eq32})\\\scriptsize (\ref{eq33})\\\scriptsize (\ref{eq266})
  \end{tabular}} 
  & 
  f_{x_{\text{v}},y_{\text{v}} \rightarrow u_{\text{v}}}\left( \max\left( x_{\text{v}} \right),\min\left( y_{\text{v}} \right) \right) = \displaystyle \frac{y_{0} \cdot \left( A - 1 \right) - \displaystyle \frac{A^{2} \cdot x_{0}}{A - 1}}{\sqrt{2}} = \displaystyle \frac{y_{0} \cdot \left( A - 1 \right)^{2} - A^{2} \cdot x_{0}}{\sqrt{2} \cdot \left( A - 1 \right)}
  &  
  \label{eq284} 
  &
\end{flalign}

\begin{figure}[ht]
    \centering
    \includegraphics[width=\textwidth]{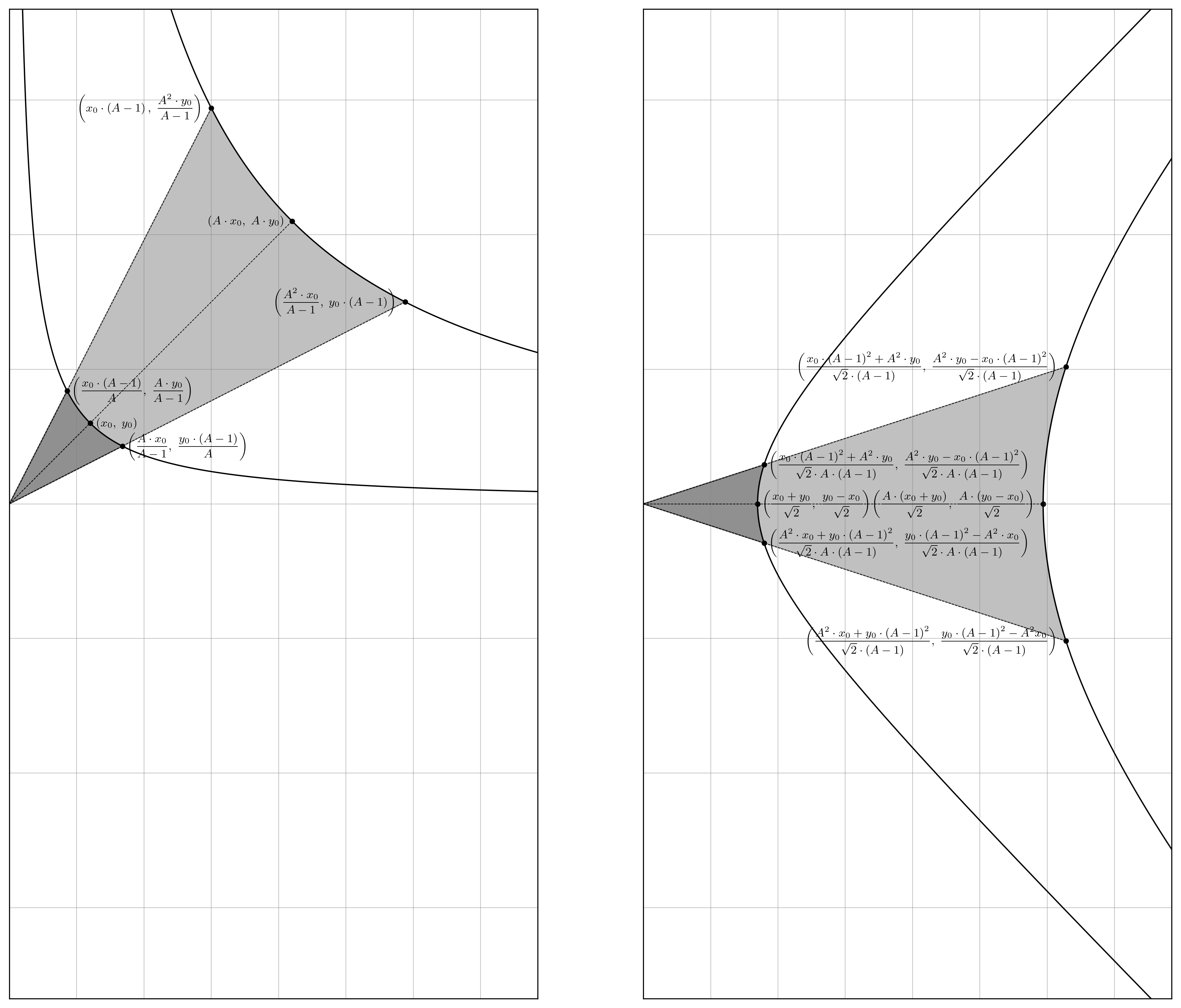}
    \captionsetup{
        justification=raggedright,
        singlelinecheck=false,
        font=small,
        labelfont=bf,
        labelsep=quad,
        format=plain
    }
    \caption{The reference and virtual curves are depicted before and after the linear ${- 45}^{{^\circ}}$ transformation has been applied. The capstone identities from both curves appear as annotations in ($x$, $y$) and ($u$, $t$) format in the left and right panels, respectively (Equations \ref{eq273}, \ref{eq274}, \ref{eq275}, \ref{eq276}, \ref{eq277}, \ref{eq278}, \ref{eq279}, \ref{eq280}, \ref{eq281}, \ref{eq282}, \ref{eq283} and \ref{eq284}).}
    \label{fig52}
\end{figure}

Turning our attention back to the plot, the algebraic identities derived in Equations \ref{eq275}, \ref{eq276}, \ref{eq277}, \ref{eq278}, \ref{eq279}, \ref{eq280}, \ref{eq281}, \ref{eq282}, \ref{eq283} and \ref{eq284} have been appended as point annotations in Figure \ref{fig52}.

Unsurprisingly, even in the transformed coordinate system, the difference between corresponding points on the virtual and reference curves is a factor of $A$ (because of course it is!). For completeness, the invariant for the transformed virtual curve is proved by substitution of the reference curve $x \cdot y = x_{0} \cdot y_{0}$ (Equation \ref{eq2}) for the virtual curve $x \cdot y = A^{2} \cdot x_{0} \cdot y_{0}$ (Equation \ref{eq20}) while repeating the process detailed in Equations \ref{eq267}, \ref{eq268}, \ref{eq269}, \ref{eq270}, \ref{eq271} and \ref{eq272}, which yields Equation \ref{eq285}.

\begin{flalign}
& \text{\renewcommand{\arraystretch}{0.66}
    \begin{tabular}{@{}c@{}}
    \scriptsize from \\
    \scriptsize (\ref{eq267})\\\scriptsize (\ref{eq270})
  \end{tabular}} 
  & 
  t_{\text{v}}^{2} - u_{\text{v}}^{2} = 2 \cdot A^{2} \cdot x_{0} \cdot y_{0}
  &  
  \label{eq285} 
  &
\end{flalign}

Following the rotation, the hyperbola must then be rescaled (i.e. normalized) into unit format such that its vertices lie at $t \pm 1$. To capture this re-scaling in situ, let the re-scaled $t$ and $u$ coordinates be $\hat{t}$ and $\hat{u}$, respectively. Then, the goal is to define $\hat{t}$ and $\hat{u}$ in terms of $x$ and $y$, and $x_{\text{v}}$, and $y_{\text{v}}$, respectively, such that $\hat{t}^{2} - \hat{u}^{2} = 1$ in both cases. This is easily achieved by taking the quotient of the LHS and RHS of Equation \ref{eq271} or Equation \ref{eq272} for $x$ and $y$, and the quotient of the LHS and RHS of Equation \ref{eq271} for $x_{\text{v}}$ and $y_{\text{v}}$ (Equation \ref{eq286}). Distributing the division and leveraging the properties of exponents and radicals yields Equations \ref{eq287} and \ref{eq288}. In both cases, the LHS and RHS of the equality are expressed as a difference of squares and the identities of $\hat{t}$ and $\hat{u}$ can be obtained by asserting a literal 1:1 correspondence between the minuends and subtrahends on either side. In our context this is okay – but note that the abstraction $a - b = c -d$, therefore $a = c$ and $b = d$ is not always true. The equation for the hyperbola in its rescaled form directly leads to the new definitions of $\hat{t}$ and $\hat{u}$, and because we are applying the transformation uniformly to both sides of the equation, the equality and the relationship between the variables is maintained in this case (Equations \ref{eq289} and \ref{eq290}).

\begin{flalign}
& \text{\renewcommand{\arraystretch}{0.66}
    \begin{tabular}{@{}c@{}}
    \scriptsize from \\
    \scriptsize (\ref{eq271})\\\scriptsize (\ref{eq272})\\\scriptsize (\ref{eq285})
  \end{tabular}} 
  & 
  {\hat{t}}^{2} - {\hat{u}}^{2} = 1 = \displaystyle \frac{t^{2} - u^{2}}{2 \cdot x_{0} \cdot y_{0}} = \displaystyle \frac{t_{\text{v}}^{2} - u_{\text{v}}^{2}}{2 \cdot A^{2} \cdot x_{0} \cdot y_{0}}
  &  
  \label{eq286} 
  &
\end{flalign}

\begin{flalign}
& \text{\renewcommand{\arraystretch}{0.66}
    \begin{tabular}{@{}c@{}}
    \scriptsize from \\
    \scriptsize (\ref{eq271})\\\scriptsize (\ref{eq272})\\\scriptsize (\ref{eq286})
  \end{tabular}} 
  & 
  {\hat{t}}^{2} - {\hat{u}}^{2} = \left( \displaystyle \frac{t}{\sqrt{2} \cdot \sqrt{x_{0}} \cdot \sqrt{y_{0}}} \right)^{2} - \left( \displaystyle \frac{u}{\sqrt{2} \cdot \sqrt{x_{0}} \cdot \sqrt{y_{0}}} \right)^{2}
  &  
  \label{eq287} 
  &
\end{flalign}

\begin{flalign}
& \text{\renewcommand{\arraystretch}{0.66}
    \begin{tabular}{@{}c@{}}
    \scriptsize from \\
    \scriptsize (\ref{eq285})\\\scriptsize (\ref{eq286})
  \end{tabular}} 
  & 
  {\hat{t}}^{2} - {\hat{u}}^{2} = \left( \displaystyle \frac{t_{\text{v}}}{\sqrt{2} \cdot A \cdot \sqrt{x_{0}} \cdot \sqrt{y_{0}}} \right)^{2} - \left( \displaystyle \frac{u_{\text{v}}}{\sqrt{2} \cdot A \cdot \sqrt{x_{0}} \cdot \sqrt{y_{0}}} \right)^{2}
  &  
  \label{eq288} 
  &
\end{flalign}

\begin{flalign}
& \text{\renewcommand{\arraystretch}{0.66}
    \begin{tabular}{@{}c@{}}
    \scriptsize from \\
    \scriptsize (\ref{eq287})\\\scriptsize (\ref{eq288})
  \end{tabular}} 
  & 
  \hat{t} = \displaystyle \frac{t}{\sqrt{2} \cdot \sqrt{x_{0}} \cdot \sqrt{y_{0}}} = \displaystyle \frac{t_{\text{v}}}{\sqrt{2} \cdot A \cdot \sqrt{x_{0}} \cdot \sqrt{y_{0}}}
  &  
  \label{eq289} 
  &
\end{flalign}

\begin{flalign}
& \text{\renewcommand{\arraystretch}{0.66}
    \begin{tabular}{@{}c@{}}
    \scriptsize from \\
    \scriptsize (\ref{eq287})\\\scriptsize (\ref{eq288})
  \end{tabular}} 
  & 
  \hat{u} = \displaystyle \frac{u}{\sqrt{2} \cdot \sqrt{x_{0}} \cdot \sqrt{y_{0}}} = \displaystyle \frac{u_{\text{v}}}{\sqrt{2} \cdot A \cdot \sqrt{x_{0}} \cdot \sqrt{y_{0}}}
  &  
  \label{eq290} 
  &
\end{flalign}

By establishing this correspondence, we effectively align the structure of the hyperbola's equation in its original and transformed states. This allows for the direct inference of $\hat{t}$ and $\hat{u}$ from $x$ and $y$, and $x_{\text{v}}$, and $y_{\text{v}}$, respectively, adhering to the normalization criterion $\hat{t}^{2} - \hat{u}^{2} = 1$. From here, the new single-argument mapping functions, $f_{x \rightarrow \hat{t}}$, $f_{y \rightarrow \hat{u}}$, $f_{y \rightarrow \hat{t}}$, $f_{x \rightarrow \hat{u}}$, $f_{x_{\text{v}} \rightarrow \hat{t}}$, $f_{y_{\text{v}} \rightarrow \hat{u}}$, $f_{y_{\text{v}} \rightarrow \hat{t}}$, and $f_{x_{\text{v}} \rightarrow \hat{u}}$ can be derived from the previous mapping functions without any trouble (Equations \ref{eq291}, \ref{eq292}, \ref{eq293}, \ref{eq294}, \ref{eq295}, \ref{eq296}, \ref{eq297} and \ref{eq298}). The two-argument expressions, $f_{x,y \rightarrow \hat{t}}$, $f_{x,y \rightarrow \hat{u}}$, $f_{x_{\text{v}},y_{\text{v}} \rightarrow \hat{t}}$, and $f_{x_{\text{v}},y_{\text{v}} \rightarrow \hat{u}}$ can be recomposed from their single-argument counterparts by substitution of $x_{0} \cdot y_{0}$ for $x \cdot y$ in the reference curve transformation, and substitution of $A^{2} \cdot x_{0} \cdot y_{0}$ for $x \cdot y$ in the virtual curve transformation, via Equations \ref{eq2} and \ref{eq20} (Equations \ref{eq299} and \ref{eq300}). Rather pleasingly, the two-argument reference- and virtual-to-unit hyperbola mapping functions are identical. You won’t have to think about it for too long to convince yourself that this is an obvious result.

\begin{flalign}
& \text{\renewcommand{\arraystretch}{0.66}
    \begin{tabular}{@{}c@{}}
    \scriptsize from \\
    \scriptsize (\ref{eq267})\\\scriptsize (\ref{eq289})
  \end{tabular}} 
  & 
  \hat{t} = f_{x \rightarrow \hat{t}}(x) = \displaystyle \frac{f_{x \rightarrow t}}{\sqrt{2} \cdot \sqrt{x_{0}} \cdot \sqrt{y_{0}}} = \displaystyle \frac{x^{2} + x_{0} \cdot y_{0}}{2 \cdot x \cdot \sqrt{x_{0}} \cdot \sqrt{y_{0}}}
  &  
  \label{eq291} 
  &
\end{flalign}

\begin{flalign}
& \text{\renewcommand{\arraystretch}{0.66}
    \begin{tabular}{@{}c@{}}
    \scriptsize from \\
    \scriptsize (\ref{eq268})\\\scriptsize (\ref{eq290})
  \end{tabular}} 
  & 
  \hat{u} = f_{y \rightarrow \hat{u}}(y) = \displaystyle \frac{f_{y \rightarrow u}}{\sqrt{2} \cdot \sqrt{x_{0}} \cdot \sqrt{y_{0}}} = \displaystyle \frac{y^{2} - x_{0} \cdot y_{0}}{2 \cdot y \cdot \sqrt{x_{0}} \cdot \sqrt{y_{0}}}
  &  
  \label{eq292} 
  &
\end{flalign}

\begin{flalign}
& \text{\renewcommand{\arraystretch}{0.66}
    \begin{tabular}{@{}c@{}}
    \scriptsize from \\
    \scriptsize (\ref{eq269})\\\scriptsize (\ref{eq289})
  \end{tabular}} 
  & 
  \hat{t} = f_{y \rightarrow \hat{t}}(y) = \displaystyle \frac{f_{y \rightarrow t}}{\sqrt{2} \cdot \sqrt{x_{0}} \cdot \sqrt{y_{0}}} = \displaystyle \frac{{x_{0} \cdot y_{0} + y}^{2}}{2 \cdot y \cdot \sqrt{x_{0}} \cdot \sqrt{y_{0}}}
  &  
  \label{eq293} 
  &
\end{flalign}

\begin{flalign}
& \text{\renewcommand{\arraystretch}{0.66}
    \begin{tabular}{@{}c@{}}
    \scriptsize from \\
    \scriptsize (\ref{eq270})\\\scriptsize (\ref{eq290})
  \end{tabular}} 
  & 
  \hat{u} = f_{x \rightarrow \hat{u}}(x) = \displaystyle \frac{f_{x \rightarrow u}}{\sqrt{2} \cdot \sqrt{x_{0}} \cdot \sqrt{y_{0}}} = \displaystyle \frac{x_{0} \cdot y_{0} - x^{2}}{2 \cdot x \cdot \sqrt{x_{0}} \cdot \sqrt{y_{0}}}
  &  
  \label{eq294} 
  &
\end{flalign}

\begin{flalign}
& \text{\renewcommand{\arraystretch}{0.66}
    \begin{tabular}{@{}c@{}}
    \scriptsize from \\
    \scriptsize (\ref{eq269})\\\scriptsize (\ref{eq287})
  \end{tabular}} 
  & 
  \hat{t} = f_{x_{\text{v}} \rightarrow \hat{t}}\left( x_{\text{v}} \right) = \displaystyle \frac{f_{x_{\text{v}} \rightarrow t_{\text{v}}}}{\sqrt{2} \cdot A \cdot \sqrt{x_{0}} \cdot \sqrt{y_{0}}} = \displaystyle \frac{x_{\text{v}}^{2} + A^{2} \cdot x_{0} \cdot y_{0}}{2 \cdot x_{\text{v}} \cdot A \cdot \sqrt{x_{0}} \cdot \sqrt{y_{0}}}
  &  
  \label{eq295} 
  &
\end{flalign}

\begin{flalign}
& \text{\renewcommand{\arraystretch}{0.66}
    \begin{tabular}{@{}c@{}}
    \scriptsize from \\
    \scriptsize (\ref{eq270})\\\scriptsize (\ref{eq287})
  \end{tabular}} 
  & 
  \hat{u} = f_{y_{\text{v}} \rightarrow \hat{u}}\left( y_{\text{v}} \right) = \displaystyle \frac{f_{y_{\text{v}} \rightarrow u_{\text{v}}}}{\sqrt{2} \cdot A \cdot \sqrt{x_{0}} \cdot \sqrt{y_{0}}} = \displaystyle \frac{y_{\text{v}}^{2} - A^{2} \cdot x_{0} \cdot y_{0}}{2 \cdot y_{\text{v}} \cdot A \cdot \sqrt{x_{0}} \cdot \sqrt{y_{0}}}
  &  
  \label{eq296} 
  &
\end{flalign}

\begin{flalign}
& \text{\renewcommand{\arraystretch}{0.66}
    \begin{tabular}{@{}c@{}}
    \scriptsize from \\
    \scriptsize (\ref{eq269})\\\scriptsize (\ref{eq287})
  \end{tabular}} 
  & 
  \hat{t} = f_{y_{\text{v}} \rightarrow \hat{t}}\left( y_{\text{v}} \right) = \displaystyle \frac{f_{y_{\text{v}} \rightarrow t_{\text{v}}}}{\sqrt{2} \cdot A \cdot \sqrt{x_{0}} \cdot \sqrt{y_{0}}} = \displaystyle \frac{A^{2} \cdot x_{0} \cdot y_{0} - y_{\text{v}}^{2}}{2 \cdot y_{\text{v}} \cdot A \cdot \sqrt{x_{0}} \cdot \sqrt{y_{0}}}
  &  
  \label{eq297} 
  &
\end{flalign}

\begin{flalign}
& \text{\renewcommand{\arraystretch}{0.66}
    \begin{tabular}{@{}c@{}}
    \scriptsize from \\
    \scriptsize (\ref{eq270})\\\scriptsize (\ref{eq287})
  \end{tabular}} 
  & 
  \hat{u} = f_{x_{\text{v}} \rightarrow \hat{u}}\left( x_{\text{v}} \right) = \displaystyle \frac{f_{x_{\text{v}} \rightarrow u_{\text{v}}}}{\sqrt{2} \cdot A \cdot \sqrt{x_{0}} \cdot \sqrt{y_{0}}} = \displaystyle \frac{A^{2} \cdot x_{0} \cdot y_{0} - x_{\text{v}}^{2}}{2 \cdot x_{\text{v}} \cdot A \cdot \sqrt{x_{0}} \cdot \sqrt{y_{0}}}
  &  
  \label{eq298} 
  &
\end{flalign}

\begin{flalign}
& \text{\renewcommand{\arraystretch}{0.66}
    \begin{tabular}{@{}c@{}}
    \scriptsize from \\
    \scriptsize (\ref{eq2})\\\scriptsize (\ref{eq20})\\\scriptsize (\ref{eq291})\\\scriptsize (\ref{eq293})\\\scriptsize (\ref{eq295})\\\scriptsize (\ref{eq297})
  \end{tabular}} 
  & 
  \hat{t} = f_{x,y \rightarrow \hat{t}}(x,y) = f_{x_{\text{v}},y_{\text{v}} \rightarrow \hat{t}}\left( x_{\text{v}},y_{\text{v}} \right) = \displaystyle \frac{x + y}{2 \cdot \sqrt{x} \cdot \sqrt{y}} = \displaystyle \frac{x_{\text{v}} + y_{\text{v}}}{2 \cdot \sqrt{x_{\text{v}}} \cdot \sqrt{y_{\text{v}}}}
  &  
  \label{eq299} 
  &
\end{flalign}

\begin{flalign}
& \text{\renewcommand{\arraystretch}{0.66}
    \begin{tabular}{@{}c@{}}
    \scriptsize from \\
    \scriptsize (\ref{eq2})\\\scriptsize (\ref{eq20})\\\scriptsize (\ref{eq292})\\\scriptsize (\ref{eq294})\\\scriptsize (\ref{eq296})\\\scriptsize (\ref{eq298})
  \end{tabular}} 
  & 
  \hat{u} = f_{x,y \rightarrow \hat{u}}(x,y) = f_{x_{\text{v}},y_{\text{v}} \rightarrow \hat{u}}(x,y) = \displaystyle \frac{y - x}{2 \cdot \sqrt{x} \cdot \sqrt{y}} = \displaystyle \frac{y_{\text{v}} - x_{\text{v}}}{2 \cdot \sqrt{x_{\text{v}}} \cdot \sqrt{y_{\text{v}}}}
  &  
  \label{eq300} 
  &
\end{flalign}

To conclude this section, we will derive the algebraic identities for the key points on the unit hyperbola by passing the corresponding points on the reference and virtual curves to the $f_{x,y \rightarrow \hat{t}}$ and $f_{x,y \rightarrow \hat{u}}$ mapping functions, Equations \ref{eq299} and \ref{eq300}. First, the mapping is performed for the reference curve (Equations \ref{eq301}, \ref{eq302}, \ref{eq303}, \ref{eq304}, \ref{eq305} and \ref{eq306}).

\begin{flalign}
& \text{\renewcommand{\arraystretch}{0.66}
    \begin{tabular}{@{}c@{}}
    \scriptsize from \\
    \scriptsize (\ref{eq299})
  \end{tabular}} 
  & 
  \hat{t} = f_{x,y \rightarrow \hat{t}}\left( x_{0},y_{0} \right) = \displaystyle \frac{x_{0} + y_{0}}{2 \cdot \sqrt{x_{0}} \cdot \sqrt{y_{0}}}
  &  
  \label{eq301} 
  &
\end{flalign}

\begin{flalign}
& \text{\renewcommand{\arraystretch}{0.66}
    \begin{tabular}{@{}c@{}}
    \scriptsize from \\
    \scriptsize (\ref{eq300})
  \end{tabular}} 
  & 
  \hat{u} = f_{x,y \rightarrow \hat{u}}\left( x_{0},y_{0} \right) = \displaystyle \frac{y_{0} - x_{0}}{2 \cdot \sqrt{x_{0}} \cdot \sqrt{y_{0}}}
  &  
  \label{eq302} 
  &
\end{flalign}

\begin{flalign}
& \text{\renewcommand{\arraystretch}{0.66}
    \begin{tabular}{@{}c@{}}
    \scriptsize from \\
    \scriptsize (\ref{eq59})\\\scriptsize (\ref{eq60})\\\scriptsize (\ref{eq299})
  \end{tabular}} 
  & 
  {\hat{t} = f}_{x,y \rightarrow \hat{t}}\left( \min \left( x \right),\max \left( y \right) \right) = \displaystyle \frac{\displaystyle \frac{x_{0} \cdot \left( A - 1 \right)}{A} + \displaystyle \frac{A \cdot y_{0}}{A - 1}}{2 \cdot \sqrt{\displaystyle \frac{x_{0} \cdot \left( A - 1 \right)}{A}} \cdot \sqrt{\displaystyle \frac{A \cdot y_{0}}{A - 1}}} = \displaystyle \frac{x_{0} \cdot \left( A - 1 \right)^{2} + A^{2} \cdot y_{0}}{2 \cdot A \cdot \sqrt{x_{0}} \cdot \sqrt{y_{0}} \cdot \left( A - 1 \right)}
  &  
  \label{eq303} 
  &
\end{flalign}

\begin{flalign}
& \text{\renewcommand{\arraystretch}{0.66}
    \begin{tabular}{@{}c@{}}
    \scriptsize from \\
    \scriptsize (\ref{eq59})\\\scriptsize (\ref{eq60})\\\scriptsize (\ref{eq300})
  \end{tabular}} 
  & 
  {\hat{u} = f}_{x,y \rightarrow \hat{u}}\left( \min \left( x \right),\max \left( y \right) \right) = \displaystyle \frac{\displaystyle \frac{A \cdot y_{0}}{A - 1} - \displaystyle \frac{x_{0} \cdot \left( A - 1 \right)}{A}}{2 \cdot \sqrt{\displaystyle \frac{x_{0} \cdot \left( A - 1 \right)}{A}} \cdot \sqrt{\displaystyle \frac{A \cdot y_{0}}{A - 1}}} = \displaystyle \frac{{A^{2} \cdot y_{0} - x_{0}} \cdot \left( A - 1 \right)^{2}}{2 \cdot A \cdot \sqrt{x_{0}} \cdot \sqrt{y_{0}} \cdot \left( A - 1 \right)}
  &  
  \label{eq304} 
  &
\end{flalign}

\begin{flalign}
& \text{\renewcommand{\arraystretch}{0.66}
    \begin{tabular}{@{}c@{}}
    \scriptsize from \\
    \scriptsize (\ref{eq61})\\\scriptsize (\ref{eq62})\\\scriptsize (\ref{eq299})
  \end{tabular}} 
  & 
  \hat{t} = f_{x,y \rightarrow \hat{t}}\left( \max \left( x \right),\min \left( y \right) \right) = \displaystyle \frac{\displaystyle \frac{A \cdot x_{0}}{A - 1} + \displaystyle \frac{y_{0} \cdot \left( A - 1 \right)}{A}}{2 \cdot \sqrt{\displaystyle \frac{A \cdot x_{0}}{A - 1}} \cdot \sqrt{\displaystyle \frac{y_{0} \cdot \left( A - 1 \right)}{A}}} = \displaystyle \frac{{A^{2} \cdot x_{0} + y_{0}} \cdot \left( A - 1 \right)^{2}}{2 \cdot A \cdot \sqrt{x_{0}} \cdot \sqrt{y_{0}} \cdot \left( A - 1 \right)}
  &  
  \label{eq305} 
  &
\end{flalign}

\begin{flalign}
& \text{\renewcommand{\arraystretch}{0.66}
    \begin{tabular}{@{}c@{}}
    \scriptsize from \\
    \scriptsize (\ref{eq61})\\\scriptsize (\ref{eq62})\\\scriptsize (\ref{eq300})
  \end{tabular}} 
  & 
  {\hat{u} = f}_{x,y \rightarrow \hat{u}}\left( \max \left( x \right),\min \left( y \right) \right) = \displaystyle \frac{\displaystyle \frac{y_{0} \cdot \left( A - 1 \right)}{A} - \displaystyle \frac{A \cdot x_{0}}{A - 1}}{2 \cdot \sqrt{\displaystyle \frac{A \cdot x_{0}}{A - 1}} \cdot \sqrt{\displaystyle \frac{y_{0} \cdot \left( A - 1 \right)}{A}}} = \displaystyle \frac{y_{0} \cdot \left( A - 1 \right)^{2} - A^{2} \cdot x_{0}}{2 \cdot A \cdot \sqrt{x_{0}} \cdot \sqrt{y_{0}} \cdot \left( A - 1 \right)}
  &  
  \label{eq306} 
  &
\end{flalign}

The expectation is that both the reference and virtual curves should map to the same coordinates on the unit hyperbola, since the difference in the relative scaling has been appropriately and separately accounted for in the derivations presented in this section. The fact that the final, two-argument mapping functions for both the reference and virtual coordinates are identical is irrelevant; the output for passing key points on the reference and virtual curves should be the same. Indeed, this is now shown to be true (Equations \ref{eq307}, \ref{eq308}, \ref{eq309}, \ref{eq310}, \ref{eq311} and \ref{eq312}). 

\begin{flalign}
& \text{\renewcommand{\arraystretch}{0.66}
    \begin{tabular}{@{}c@{}}
    \scriptsize from \\
    \scriptsize (\ref{eq299})
  \end{tabular}} 
  & 
  \hat{t} = f_{x_{\text{v}},y_{\text{v}} \rightarrow \hat{t}}\left( A \cdot x_{0},A \cdot y_{0} \right) = \displaystyle \frac{A \cdot x_{0} + A \cdot y_{0}}{2 \cdot \sqrt{A \cdot x_{0}} \cdot \sqrt{A \cdot y_{0}}} = \displaystyle \frac{x_{0} + y_{0}}{2 \cdot \sqrt{x_{0}} \cdot \sqrt{y_{0}}}
  &  
  \label{eq307} 
  &
\end{flalign}

\begin{flalign}
& \text{\renewcommand{\arraystretch}{0.66}
    \begin{tabular}{@{}c@{}}
    \scriptsize from \\
    \scriptsize (\ref{eq300})
  \end{tabular}} 
  & 
  \hat{u} = f_{x_{\text{v}},y_{\text{v}} \rightarrow \hat{u}}\left( A \cdot x_{0}, A \cdot y_{0} \right) = \displaystyle \frac{A \cdot y_{0} - A \cdot x_{0}}{2 \cdot \sqrt{A \cdot x_{0}} \cdot \sqrt{A \cdot y_{0}}} = \displaystyle \frac{y_{0} - x_{0}}{2 \cdot \sqrt{x_{0}} \cdot \sqrt{y_{0}}}
  &  
  \label{eq308} 
  &
\end{flalign}

\begin{flalign}
& \text{\renewcommand{\arraystretch}{0.66}
    \begin{tabular}{@{}c@{}}
    \scriptsize from \\
    \scriptsize (\ref{eq30})\\\scriptsize (\ref{eq31})\\\scriptsize (\ref{eq299})
  \end{tabular}} 
  & 
  {\hat{t} = f}_{x_{\text{v}},y_{\text{v}} \rightarrow \hat{t}}\left( \min\left( x_{\text{v}} \right),\max\left( y_{\text{v}} \right) \right) = \displaystyle \frac{x_{0} \cdot \left( A - 1 \right) + \displaystyle \frac{A^{2} \cdot y_{0}}{A - 1}}{2 \cdot \sqrt{x_{0} \cdot \left( A - 1 \right)} \cdot \sqrt{\displaystyle \frac{A^{2} \cdot y_{0}}{A - 1}}} = \displaystyle \frac{x_{0} \cdot \left( A - 1 \right)^{2} + A^{2} \cdot y_{0}}{2 \cdot A \cdot \sqrt{x_{0}} \cdot \sqrt{y_{0}} \cdot \left( A - 1 \right)}
  &  
  \label{eq309} 
  &
\end{flalign}

\begin{flalign}
& \text{\renewcommand{\arraystretch}{0.66}
    \begin{tabular}{@{}c@{}}
    \scriptsize from \\
    \scriptsize (\ref{eq30})\\\scriptsize (\ref{eq31})\\\scriptsize (\ref{eq300})
  \end{tabular}} 
  & 
  \hat{u} = f_{x_{\text{v}},y_{\text{v}} \rightarrow \hat{u}}\left( \min\left( x_{\text{v}} \right),\max\left( y_{\text{v}} \right) \right) = \displaystyle \frac{\displaystyle \frac{A^{2} \cdot y_{0}}{A - 1} - x_{0} \cdot \left( A - 1 \right)}{2 \cdot \sqrt{x_{0} \cdot \left( A - 1 \right)} \cdot \sqrt{\displaystyle \frac{A^{2} \cdot y_{0}}{A - 1}}} = \displaystyle \frac{{A^{2} \cdot y_{0} - x_{0}} \cdot \left( A - 1 \right)^{2}}{2 \cdot A \cdot \sqrt{x_{0}} \cdot \sqrt{y_{0}} \cdot \left( A - 1 \right)}
  &  
  \label{eq310} 
  &
\end{flalign}

\begin{flalign}
& \text{\renewcommand{\arraystretch}{0.66}
    \begin{tabular}{@{}c@{}}
    \scriptsize from \\
    \scriptsize (\ref{eq32})\\\scriptsize (\ref{eq33})\\\scriptsize (\ref{eq299})
  \end{tabular}} 
  & 
  \hat{t} = f_{x_{\text{v}},y_{\text{v}} \rightarrow \hat{t}}\left( \max\left( x_{\text{v}} \right),\min\left( y_{\text{v}} \right) \right) = \displaystyle \frac{\displaystyle \frac{A^{2} \cdot x_{0}}{A - 1} + y_{0} \cdot \left( A - 1 \right)}{2 \cdot \sqrt{\displaystyle \frac{A^{2} \cdot x_{0}}{A - 1}} \cdot \sqrt{y_{0} \cdot \left( A - 1 \right)}} = \displaystyle \frac{{A^{2} \cdot x_{0} + y_{0}} \cdot \left( A - 1 \right)^{2}}{2 \cdot A \cdot \sqrt{x_{0}} \cdot \sqrt{y_{0}} \cdot \left( A - 1 \right)}
  &  
  \label{eq311} 
  &
\end{flalign}

\begin{flalign}
& \text{\renewcommand{\arraystretch}{0.66}
    \begin{tabular}{@{}c@{}}
    \scriptsize from \\
    \scriptsize (\ref{eq32})\\\scriptsize (\ref{eq33})\\\scriptsize (\ref{eq300})
  \end{tabular}} 
  & 
  \hat{u} = f_{x_{\text{v}},y_{\text{v}} \rightarrow \hat{u}}\left( \max\left( x_{\text{v}} \right),\min\left( y_{\text{v}} \right) \right) = \displaystyle \frac{y_{0} \cdot \left( A - 1 \right) - \displaystyle \frac{A^{2} \cdot x_{0}}{A - 1}}{2 \cdot \sqrt{\displaystyle \frac{A^{2} \cdot x_{0}}{A - 1}} \cdot \sqrt{y_{0} \cdot \left( A - 1 \right)}} = \displaystyle \frac{y_{0} \cdot \left( A - 1 \right)^{2} - A^{2} \cdot x_{0}}{2 \cdot A \cdot \sqrt{x_{0}} \cdot \sqrt{y_{0}} \cdot \left( A - 1 \right)}
  &  
  \label{eq312} 
  &
\end{flalign}

\begin{figure}[ht]
    \centering
    \includegraphics[width=\textwidth]{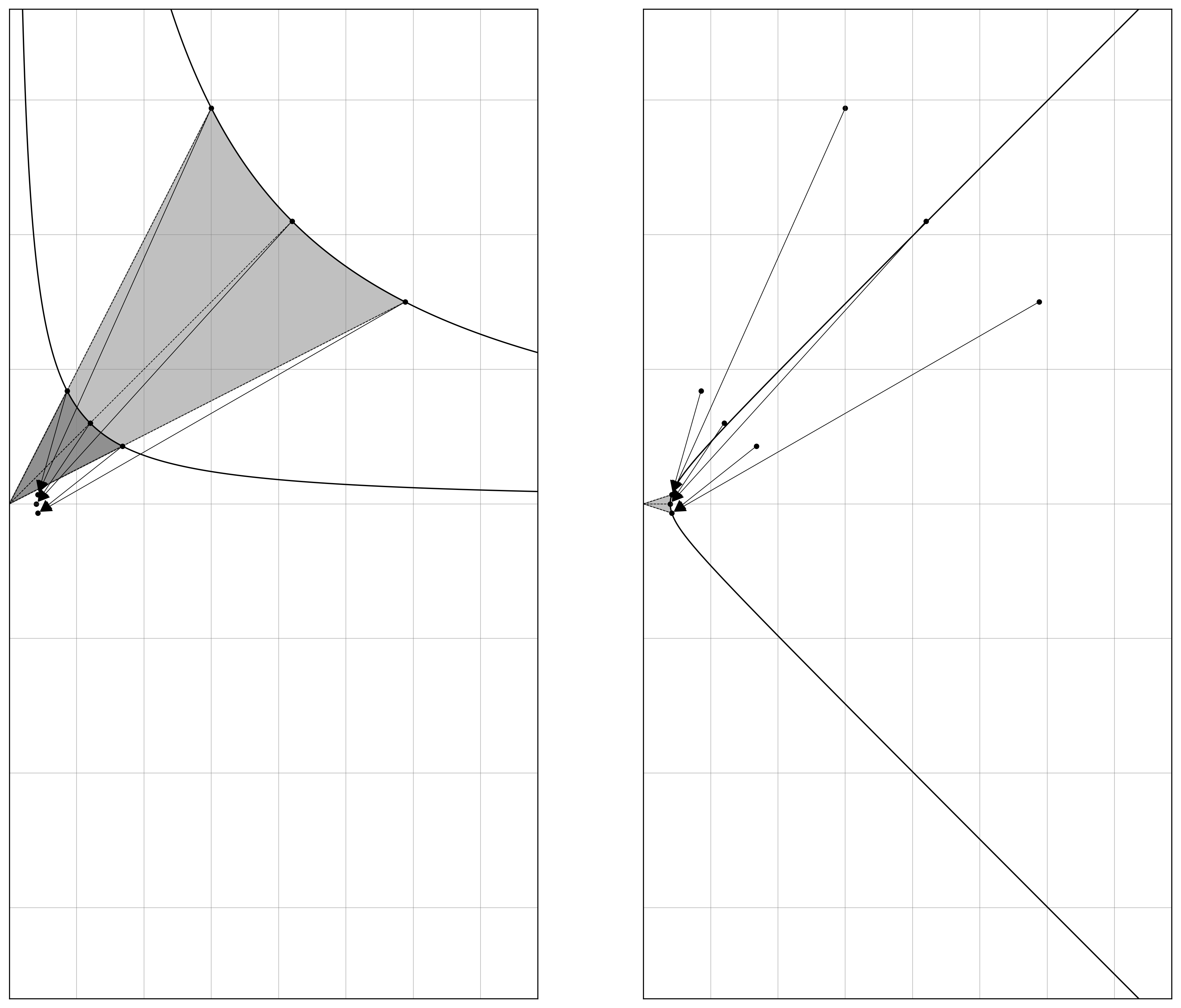}
    \captionsetup{
        justification=raggedright,
        singlelinecheck=false,
        font=small,
        labelfont=bf,
        labelsep=quad,
        format=plain
    }
    \caption{The new linear transformation, which incorporates both a ${- 45}^{{^\circ}}$ rotation and normalization is depicted with arrows beginning from the key points on both the reference and virtual curves in the Cartesian ($x$, $y$) plane, and ending at their implied destinations in the new coordinate system ($\hat{t}$, $\hat{u}$). Both the reference and virtual curves appear in the left panel, which after the ${- 45}^{{^\circ}}$ rotation and normalization converge on the same points, the unit hyperbola, which appears in the right panel.}
    \label{fig53}
\end{figure}

Therefore, we now have the functions which both rotate and normalize either of the reference or virtual curves, and they are the same functions in both cases. After rotating and re-scaling, both the reference and virtual curves are consolidated into each other; the overall process is summarized in Figure \ref{fig53}.  

\begin{figure}[ht]
    \centering
    \includegraphics[width=\textwidth]{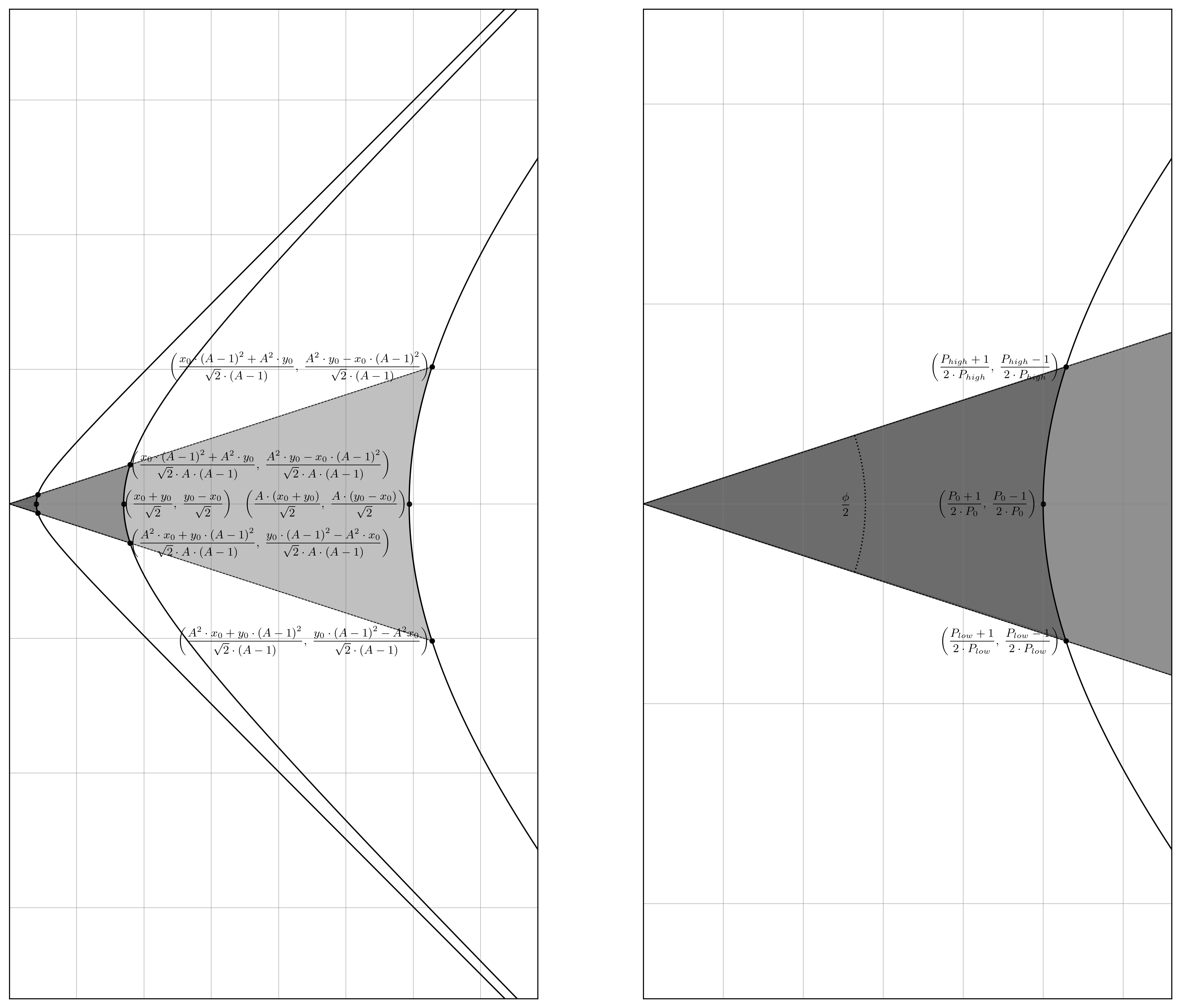}
    \captionsetup{
        justification=raggedright,
        singlelinecheck=false,
        font=small,
        labelfont=bf,
        labelsep=quad,
        format=plain
    }
    \caption{The transformed reference and virtual curves, $t^{2} - u^{2} = 2 \cdot x_{0} \cdot y_{0}$ and $t_{\text{v}}^{2} - u_{\text{v}}^{2} = 2 \cdot A^{2} \cdot x_{0} \cdot y_{0}$ are depicted alongside their normalized counterpart, ${\hat{t}}^{2} - {\hat{u}}^{2} = 1$. The point annotations correspond to the capstone identities for all three curves, and appear in ($u$, $t$) format in both panels. The scale of the first panel is sufficient to include all three curves and their key points, whereas the second panel is zoomed-in such that the detail around the unit curve can be inspected. The area of the shaded region enclosed by the outer-most rays and the unit hyperbola in the second panel is equal to half its hyperbolic angle, $\phi / 2$, annotated accordingly and depicted with a circular arc between its enclosing rays (Equations \ref{eq313}, \ref{eq314}, \ref{eq315}, \ref{eq316}, \ref{eq317}, \ref{eq318}).}
    \label{fig54}
\end{figure}

The expressions obtained that describe the transformed and normalized curve, $\hat{t}^{2} - \hat{u}^{2} - 1$ via Equations \ref{eq301}, \ref{eq302}, \ref{eq303}, \ref{eq304}, \ref{eq305}, \ref{eq306}, \ref{eq307}, \ref{eq308}, \ref{eq309}, \ref{eq310}, \ref{eq311} and \ref{eq312}, are a little cumbersome. The process for refactoring all of these is the same. First, remove a common factor of $1 / 2$ before distributing the division, which produces a sum (for $\hat{t}$) or difference (for $\hat{u}$) of two quotients, where these quotient identities have previously been established as the square roots of the price boundaries, the square root of the geometric mean of the price boundaries, or the reciprocals of the same (Equations \ref{eq37}, \ref{eq39} and \ref{eq41}). Substitution allows for all three parameters, $x_{0}$, $y_{0}$,and $A$ to be replaced for single terms, either $P_{0}$, $P_{high}$ or $P_{low}$, depending on the context (Equations \ref{eq313}, \ref{eq314}, \ref{eq315}, \ref{eq316}, \ref{eq317} and \ref{eq318}) (Figure \ref{fig54}). 

\begin{flalign}
& \text{\renewcommand{\arraystretch}{0.66}
    \begin{tabular}{@{}c@{}}
    \scriptsize from \\
    \scriptsize (\ref{eq39})\\\scriptsize (\ref{eq307})
  \end{tabular}} 
  & 
  {\hat{t}}_{0} = \displaystyle \frac{1}{2} \cdot \left( \displaystyle \frac{\sqrt{y_{0}}}{\sqrt{x_{0}}} + \displaystyle \frac{\sqrt{x_{0}}}{\sqrt{y_{0}}} \right) = \displaystyle \frac{1}{2} \cdot \left( \sqrt{P_{0}} + \displaystyle \frac{1}{\sqrt{P_{0}}} \right) = \displaystyle \frac{P_{0} + 1}{2 \cdot \sqrt{P_{0}}}
  &  
  \label{eq313} 
  &
\end{flalign}

\begin{flalign}
& \text{\renewcommand{\arraystretch}{0.66}
    \begin{tabular}{@{}c@{}}
    \scriptsize from \\
    \scriptsize (\ref{eq39})\\\scriptsize (\ref{eq308})
  \end{tabular}} 
  & 
  {\hat{u}}_{0} = \displaystyle \frac{1}{2} \cdot \left( \displaystyle \frac{\sqrt{y_{0}}}{\sqrt{x_{0}}} - \displaystyle \frac{\sqrt{x_{0}}}{\sqrt{y_{0}}} \right) = \displaystyle \frac{1}{2} \cdot \left( \sqrt{P_{0}} - \displaystyle \frac{1}{\sqrt{P_{0}}} \right) = \displaystyle \frac{P_{0} - 1}{2 \cdot \sqrt{P_{0}}}
  &  
  \label{eq314} 
  &
\end{flalign}

\begin{flalign}
& \text{\renewcommand{\arraystretch}{0.66}
    \begin{tabular}{@{}c@{}}
    \scriptsize from \\
    \scriptsize (\ref{eq37})\\\scriptsize (\ref{eq309})
  \end{tabular}} 
  & 
  \min\left( \hat{t} \right) = \displaystyle \frac{1}{2} \cdot \left( \displaystyle \frac{A}{A - 1} \cdot \displaystyle \frac{\sqrt{y_{0}}}{\sqrt{x_{0}}} + \displaystyle \frac{A - 1}{A} \cdot \displaystyle \frac{\sqrt{x_{0}}}{\sqrt{y_{0}}} \right) = \displaystyle \frac{1}{2} \cdot \left( \sqrt{P_{\text{high}}} + \displaystyle \frac{1}{\sqrt{P_{\text{high}}}} \right) = \displaystyle \frac{P_{\text{high}} + 1}{2 \cdot \sqrt{P_{\text{high}}}}
  &  
  \label{eq315} 
  &
\end{flalign}

\begin{flalign}
& \text{\renewcommand{\arraystretch}{0.66}
    \begin{tabular}{@{}c@{}}
    \scriptsize from \\
    \scriptsize (\ref{eq37})\\\scriptsize (\ref{eq310})
  \end{tabular}} 
  & 
  \max\left( \hat{u} \right) = \displaystyle \frac{1}{2} \cdot \left( \displaystyle \frac{A}{A - 1} \cdot \displaystyle \frac{\sqrt{y_{0}}}{\sqrt{x_{0}}} - \displaystyle \frac{A - 1}{A} \cdot \displaystyle \frac{\sqrt{x_{0}}}{\sqrt{y_{0}}} \right) = \displaystyle \frac{1}{2} \cdot \left( \sqrt{P_{\text{high}}} - \displaystyle \frac{1}{\sqrt{P_{\text{high}}}} \right) = \displaystyle \frac{P_{\text{high}} - 1}{2 \cdot \sqrt{P_{\text{high}}}}
  &  
  \label{eq316} 
  &
\end{flalign}

\begin{flalign}
& \text{\renewcommand{\arraystretch}{0.66}
    \begin{tabular}{@{}c@{}}
    \scriptsize from \\
    \scriptsize (\ref{eq47})\\\scriptsize (\ref{eq311})
  \end{tabular}} 
  & 
  \max\left( \hat{t} \right) = \displaystyle \frac{1}{2} \cdot \left( \displaystyle \frac{A - 1}{A} \cdot \displaystyle \frac{\sqrt{y_{0}}}{\sqrt{x_{0}}} + \displaystyle \frac{A}{A - 1} \cdot \displaystyle \frac{\sqrt{x_{0}}}{\sqrt{y_{0}}} \right) = \displaystyle \frac{1}{2} \cdot \left( \sqrt{P_{\text{low}}} + \displaystyle \frac{1}{\sqrt{P_{\text{low}}}} \right) = \displaystyle \frac{P_{\text{low}} + 1}{2 \cdot \sqrt{P_{\text{low}}}}
  &  
  \label{eq317} 
  &
\end{flalign}

\begin{flalign}
& \text{\renewcommand{\arraystretch}{0.66}
    \begin{tabular}{@{}c@{}}
    \scriptsize from \\
    \scriptsize (\ref{eq41})\\\scriptsize (\ref{eq312})
  \end{tabular}} 
  & 
  \min\left( \hat{u} \right) = \displaystyle \frac{1}{2} \cdot \left( \displaystyle \frac{A - 1}{A} \cdot \displaystyle \frac{\sqrt{y_{0}}}{\sqrt{x_{0}}} - \displaystyle \frac{A}{A - 1} \cdot \displaystyle \frac{\sqrt{x_{0}}}{\sqrt{y_{0}}} \right) = \displaystyle \frac{1}{2} \cdot \left( \sqrt{P_{\text{low}}} - \displaystyle \frac{1}{\sqrt{P_{\text{low}}}} \right) = \displaystyle \frac{P_{\text{low}} - 1}{2 \cdot \sqrt{P_{\text{low}}}}
  &  
  \label{eq318} 
  &
\end{flalign}

The notation above deserves a brief note. The terms ${\hat{t}}_{0}$, ${\hat{u}}_{0}$, $\min(\hat{t})$, $\max(\hat{t})$, $\min(\hat{u})$, and $\max(\hat{u})$ inherit the syntax of their Cartesian counterparts, $x_{0}$, $y_{0}$, $\min(x)$, $\max(x)$, $\min(y)$, and $\max(y)$. The choice of notation here was made to preserve a sense of provenance. For ${\hat{u}}_{0}$, $\min(\hat{u})$, and $\max(\hat{u})$, the implicit meaning of the notation is guaranteed to be preserved after the transformation and scaling; $\min(\hat{u}) \leq {\hat{u}}_{0} \leq \max(\hat{u})$, always and with no exceptions. However, the same cannot be said for ${\hat{t}}_{0}$, $\min(\hat{t})$, and $\max(\hat{t})$. Due to the $-45^\circ$ rotation about the origin, there is no general inequality that applies for the relative sizes of ${\hat{t}}_{0}$, $\min(\hat{t})$, and $\max(\hat{t})$, and all permutations are possible. This is both briefly annoying, but also ultimately irrelevant, as the remainder of the exercise will focus exclusively on $\hat{u}$. The justification for this is coming in due course.

The $\max\left( \hat{u} \right)$ and $\min\left( \hat{u} \right)$ coordinates represent points along the vertical axis of the unit hyperbola, and the area enclosed by the curve and the rays drawn from these points through the origin defines one half the hyperbolic angle, denoted here as $\phi$. This is not the familiar angle from circle trigonometry; it is something else (\textit{vide infra}).

\subsection{The Trigonometric Description of Liquidity Concentration}\label{subsec5.3}

In both Euclidean and hyperbolic geometry, angles serve as fundamental measures of separation between intersecting lines. For the unit circle in Euclidean geometry, the angle $\theta$ is commonly understood to represent a rotation (indeed, we have already used this intuition in the linear transformation of the curves). However, $\theta$ is equally well described as representing an area. Specifically, $\theta$ corresponds to the area of the sector enclosed by the unit circle itself and two rays that originate from the circle's centre and intersect its perimeter.

Consider the unit circle, $x^{2} + y^{2} = 1$. The angle, $\theta$, measured in radians, directly corresponds to the area of the sector formed by the radius of the circle and two rays extending from the origin to two different points on the circle's boundary. This area is equal to one half the angle, $\theta / 2$. This relationship arises because the full circle, having an area of $\pi$ (the area of a circle is $2 \pi \cdot r^{2}$, and $r = 1$ for the unit circle) is spread across $2 \pi$ radians. Thus, the area covered by any angle $\theta$ is proportionally $\theta / \left( 2 \pi \right)$ times the total area of the circle, which simplifies to half the angle, $\theta / 2$, for a unit circle.

\begin{figure}[ht]
    \centering
    \includegraphics[width=\textwidth]{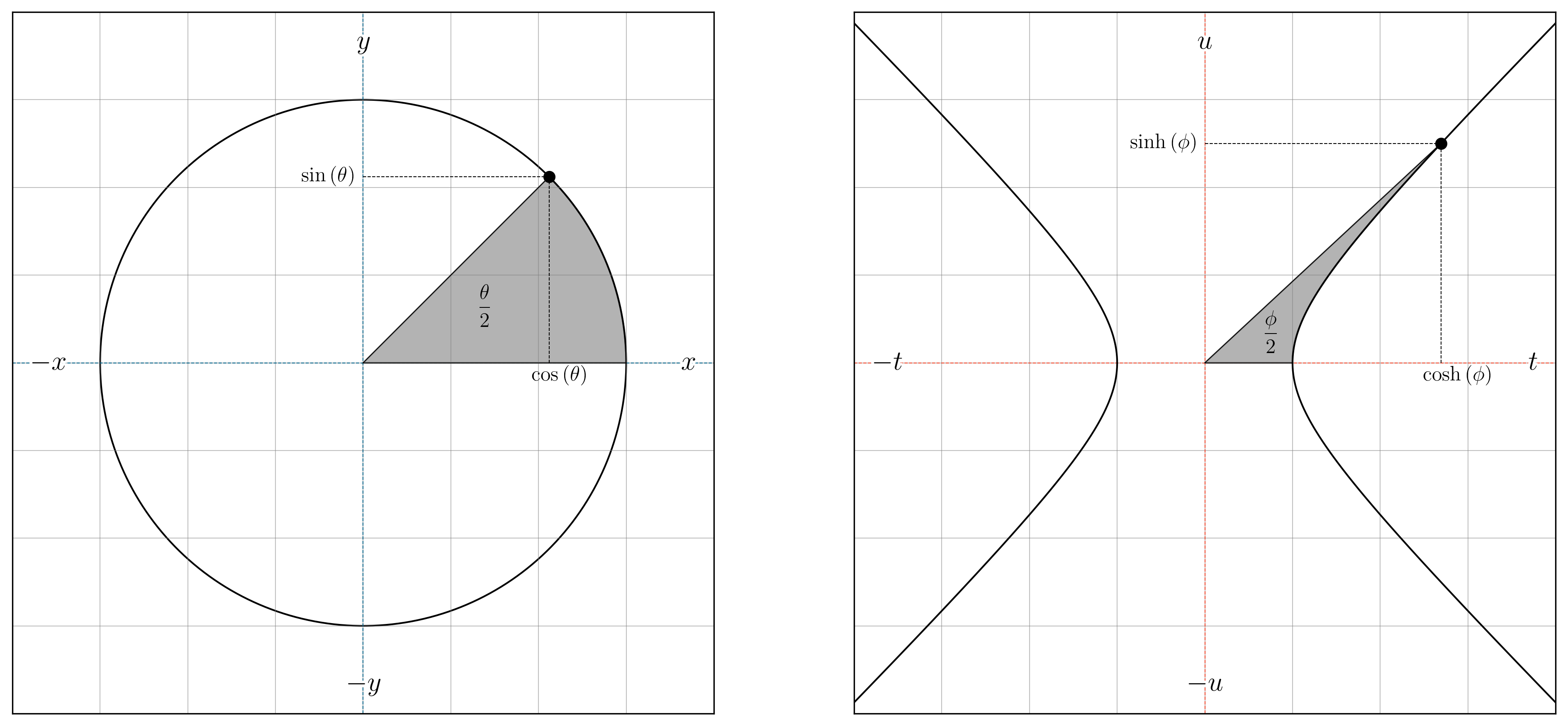}
    \captionsetup{
        justification=raggedright,
        singlelinecheck=false,
        font=small,
        labelfont=bf,
        labelsep=quad,
        format=plain
    }
    \caption{The circle (first panel) and hyperbola (second panel) trigonometric functions, $\sin$ and $\cos$, and $\sinh$ and $\cosh$, respectively, and their relationships to the vertical and horizontal axes of their cognate coordinate systems is presented for clarity’s sake. Note the convention used here is that $\theta$ refers to the angle measured for a circle, and $\phi$ refers to the angle measured for a hyperbola. Elsewhere, you will likely encounter other symbols to represent the hyperbolic angle; $\theta$ and $\psi$ are common, and so are $a$ and $x$, but there is no prevailing standard, it seems.}
    \label{fig55}
\end{figure}

This is the intuition you should have when considering hyperbolic angles – not as a rotation, but as an area. Specifically, half the hyperbolic angle, $\phi / 2$, is precisely the area enclosed by two rays emanating from the origin and intersecting the unit hyperbola, just as half the circular angle, $\theta / 2$, is precisely the area enclosed by two rays emanating from the origin and intersecting the unit circle. Circle and hyperbola trigonometry also share some common ground. The most important for our purpose is that the coordinates for a point on both curves can be expressed with respect to the trigonometric functions, sine and cosine ($\sin$ and $\cos$, respectively) for a circle, and their hyperbola counterparts, the hyperbolic sine and hyperbolic cosine ($\sinh$ and $\cosh$, respectively) (Figure \ref{fig55}). 

The angle that we are interested to determine in the context of concentrated liquidity is the one that measures the separation between $P_{\text{high}}$ and $P_{\text{low}}$; that is, the size of the sector of the hyperbola wherein exchange can occur between tokens. We have the algebraic identities for these points in the ($t$, $u$) plane; to measure $\phi$, we turn to the inverse hyperbolic sine function, $\text{arsinh}$ (or $\sinh^{-1}$), which takes the vertical displacement from the origin, $\hat{u}$ as an argument, and returns the value $\phi$ (Equation \ref{eq319}). 

\begin{equation} \label{eq319}
\text{arsinh}\left( \hat{u} \right) = \sinh^{- 1}\left( \hat{u} \right) = \ln\left( \hat{u} + \sqrt{{\hat{u}}^{2} + 1} \right)
\end{equation}

While Equation \ref{eq319} is the form you are likely to have shown to you as \textit{The Arsinh Function}, it is also an incomplete description. 

The output for any value of $\hat{u}$ passed to Equation \ref{eq319} returns a value for $\phi$ with an implicit reference to $\hat{u} = 0$. It might be helpful to think of the above form of the $\text{arsinh}$ (or $\sinh^{-1}$) function as being the definite integral evaluated from $0$ to some upper bound, $\hat{u}$, rather than evaluated specifically \textit{at} $\hat{u}$. Fortunately, or unfortunately, we need to explore this a little bit before we can progress any further. 

The function $\sinh{\left( \hat{u} \right)}$ is continuously differentiable over the reals; it is smooth everywhere and there exists a derivative at all points in its domain (i.e. it is a $C^{1}$ function). Since $\text{arsinh} \left( \hat{u} \right)$ is the inverse of $\sinh{\left( \hat{u} \right)}$, it inherits similar differentiability properties through the inverse function theorem. Therefore, both $\text{arsinh} \left( \hat{u} \right)$ and its derivative (Equation \ref{eq320}) are well-defined over the real number line as well, $- \infty < \hat{u} < + \infty$. This is the reason we are using $\text{arsinh} \left( \hat{u} \right)$ and not $\text{arcosh} \left( \hat{t} \right)$, which is only well-defined for $\hat{t} > 1$ and is ambiguously multi-valued in this domain. We have already observed this behavior. Recall that there is an annoying lack of a general inequality that applies for the relative sizes of ${\hat{t}_{0}}$, $\min \left( \hat{t} \right)$, and $\max \left( \hat{t} \right)$, mentioned above (Equations \ref{eq313}, \ref{eq315} and \ref{eq317}). The fact that $\text{arsinh} \left( \hat{u} \right)$ provides all the information we require, unambiguously, \textit{is} the justification for why such a lack of a general inequality in the $\hat{t}$ notation can safely be ignored; we have no use for $\hat{t}$ because the functions that accept it as an argument are simply not general enough to be useful in this context.

\begin{equation} \label{eq320}
\displaystyle \frac{\partial\phi}{\partial\hat{u}} = \displaystyle \frac{\partial}{\partial\hat{u}} \cdot \text{arsinh}\left( \hat{u} \right) = \displaystyle \frac{1}{\sqrt{{\hat{u}}^{2} + 1}},\ \forall\ \hat{u}\mathbb{\in R}
\end{equation}

With that out of the way, we can define a more general $\text{arsinh} \left( \hat{u} \right)$ function (which yields the hyperbolic angle, $\phi$) as the definite integral of its derivative (Equation \ref{eq320}) evaluated from $\min \left( \hat{u} \right)$ to $\max \left( \hat{u} \right)$ (Equation \ref{eq321}). Note that forcing $\min \left( \hat{u} \right)$ in Equation \ref{eq321} to zero yields Equation \ref{eq319}; however, we can now pass any value for $\min \left( \hat{u} \right)$ and $\max \left( \hat{u} \right)$ and measure the hyperbolic angle, $\phi$, relative to any positions we want. Much better!

\begin{equation} \label{eq321}
\phi = \int_{\min\left( \hat{u} \right)}^{\max\left( \hat{u} \right)}{\displaystyle \frac{1}{\sqrt{{\hat{u}}^{2} + 1}} \cdot}{\partial\hat{u} = \ }\left\lbrack \ln\left( \hat{u} + \sqrt{{\hat{u}}^{2} + 1} \right) \right\rbrack_{\min\left( \hat{u} \right)}^{\max\left( \hat{u} \right)}
\end{equation}

Then, evaluation of the definite integral begins with substitution of the identities for $\max \left( \hat{u} \right)$ and $\min \left( \hat{u} \right)$ from Equations \ref{eq316} and \ref{eq318}, respectively, into the integration bounds in Equation \ref{eq321} yields a rather grotesque expression for $\phi$ (Equation \ref{eq322}). However, this can be refactored and simplified into at least three very pleasing redundant forms (Equation \ref{eq324}). 

\begin{flalign}
& \text{\renewcommand{\arraystretch}{0.66}
    \begin{tabular}{@{}c@{}}
    \scriptsize from \\
    \scriptsize (\ref{eq316})\\\scriptsize (\ref{eq318})\\\scriptsize (\ref{eq321})
  \end{tabular}} 
  & 
  \phi = \ \ln\left( \displaystyle \frac{P_{\text{high}} - 1}{2 \cdot \sqrt{P_{\text{high}}}} + \sqrt{\left( \displaystyle \frac{P_{\text{high}} - 1}{2 \cdot \sqrt{P_{\text{high}}}} \right)^{2} + 1} \right) - \ln\left( \displaystyle \frac{P_{\text{low}} - 1}{2 \cdot \sqrt{P_{\text{low}}}} + \sqrt{\left( \displaystyle \frac{P_{\text{low}} - 1}{2 \cdot \sqrt{P_{\text{low}}}} \right)^{2} + 1} \right)
  &  
  \label{eq322} 
  &
\end{flalign}

\begin{flalign}
& \text{\renewcommand{\arraystretch}{0.66}
    \begin{tabular}{@{}c@{}}
    \scriptsize from \\
    \scriptsize (\ref{eq322})
  \end{tabular}} 
  & 
  \phi = \ \ln\left( \displaystyle \frac{\displaystyle \frac{P_{\text{high}} - 1}{2 \cdot \sqrt{P_{\text{high}}}} + \sqrt{\left( \displaystyle \frac{P_{\text{high}} - 1}{2 \cdot \sqrt{P_{\text{high}}}} \right)^{2} + 1}}{\displaystyle \frac{P_{\text{low}} - 1}{2 \cdot \sqrt{P_{\text{low}}}} + \sqrt{\left( \displaystyle \frac{P_{\text{low}} - 1}{2 \cdot \sqrt{P_{\text{low}}}} \right)^{2} + 1}} \right) = \ln\left( \displaystyle \frac{\sqrt{P_{\text{low}}} \cdot \left( P_{\text{high}} + \sqrt{\left( P_{\text{high}} + 1 \right)^{2}} - 1 \right)}{\sqrt{P_{\text{high}}} \cdot \left( P_{\text{low}} + \sqrt{\left( P_{\text{low}} + 1 \right)^{2}} - 1 \right)} \right)
  &  
  \label{eq323} 
  &
\end{flalign}

\begin{flalign}
& \text{\renewcommand{\arraystretch}{0.66}
    \begin{tabular}{@{}c@{}}
    \scriptsize from \\
    \scriptsize (\ref{eq323})
  \end{tabular}} 
  & 
  \phi = \ \ln\left( \displaystyle \frac{\sqrt{P_{\text{high}}}}{\sqrt{P_{\text{low}}}} \right) = \ln{\left( \sqrt{P_{\text{high}}} \right) -}\ln{\left( \sqrt{P_{\text{low}}} \right) = \displaystyle \frac{\ln\left( P_{\text{high}} \right) - \ln\left( P_{\text{low}} \right)}{2}}
  &  
  \label{eq324} 
  &
\end{flalign}

At long last, the “natural” parametrization for concentrated liquidity becomes evident. Passing $\phi$ as an argument to the exponential function, (i.e. $e^{\phi}$) yields the square root of the quotient of $P_{\text{high}}$ and $P_{\text{low}}$, which is equal to the [no longer a mystery] constant $C$, from which the three “pretty” invariant equations emerged (see Section \ref{subsec5.1}), which brings this part of the exercise full-circle (Equations \ref{eq325}, \ref{eq326} and \ref{eq327}).

\begin{flalign}
& \text{\renewcommand{\arraystretch}{0.66}
    \begin{tabular}{@{}c@{}}
    \scriptsize from \\
    \scriptsize (\ref{eq324})
  \end{tabular}} 
  & 
  e^{\phi} = C
  &  
  \label{eq325} 
  &
\end{flalign}

\begin{flalign}
& \text{\renewcommand{\arraystretch}{0.66}
    \begin{tabular}{@{}c@{}}
    \scriptsize from \\
    \scriptsize (\ref{eq260})\\\scriptsize (\ref{eq325})
  \end{tabular}} 
  & 
  e^{\phi} = \displaystyle \frac{\left( x - x_{0} \right)^{2} \cdot \left( y - y_{0} \right)^{2}}{\left( x \cdot y - x_{0} \cdot y_{0} \right)^{2}} = \displaystyle \frac{\left( x_{\text{int}} - x \right) \cdot \left( y_{\text{int}} - y \right)}{x \cdot y} = \displaystyle \frac{\left( x - x_{\text{asym}} \right) \cdot \left( y - y_{\text{asym}} \right)}{x_{\text{asym}} \cdot y_{\text{asym}}}
  &  
  \label{eq326} 
  &
\end{flalign}

\begin{flalign}
& \text{\renewcommand{\arraystretch}{0.66}
    \begin{tabular}{@{}c@{}}
    \scriptsize from \\
    \scriptsize (\ref{eq261})\\\scriptsize (\ref{eq325})
  \end{tabular}} 
  & 
  e^{\phi} = \displaystyle \frac{y_{\text{asym}} - y_{\text{int}}}{y_{\text{asym}}} = \displaystyle \frac{x_{\text{asym}} - x_{\text{int}}}{x_{\text{asym}}} = \displaystyle \frac{\left( y_{\text{asym}} - y_{0} \right)^{2}}{y_{\text{asym}}^{2}} = \displaystyle \frac{\left( x_{\text{asym}} - x_{0} \right)^{2}}{x_{\text{asym}}^{2}} = \displaystyle \frac{\sqrt{P_{\text{high}}}}{\sqrt{P_{\text{low}}}}
  &  
  \label{eq327} 
  &
\end{flalign}

The remaining trigonometric identities are easily elucidated from the defining relationships of the hyperbolic sine, cosine, and tangent functions ($\sinh$, $\cosh$, and $\tanh$, respectively) (Equations \ref{eq328}, \ref{eq329}, \ref{eq330} and \ref{eq331}). Curiously, $\cosh \left( \phi \right)$ is equal to the quotient of the arithmetic mean and the geometric mean of the price boundaries, which despite being a meaningless coincidence, is a surprising and potentially useful result in and of itself.

\begin{flalign}
& \text{\renewcommand{\arraystretch}{0.66}
    \begin{tabular}{@{}c@{}}
    \scriptsize from \\
    \scriptsize (\ref{eq324})
  \end{tabular}} 
  & 
  e^{\phi} = \sinh{(\phi) + \cosh(\phi)} = \displaystyle \frac{\sqrt{P_{\text{high}}}}{\sqrt{P_{\text{low}}}}
  &  
  \label{eq328} 
  &
\end{flalign}

\begin{flalign}
& \text{\renewcommand{\arraystretch}{0.66}
    \begin{tabular}{@{}c@{}}
    \scriptsize from \\
    \scriptsize (\ref{eq324})
  \end{tabular}} 
  & 
  \sinh(\phi) = \displaystyle \frac{e^{\phi} - e^{- \phi}}{2} = \displaystyle \frac{1}{2} \cdot \left( \displaystyle \frac{\sqrt{P_{\text{high}}}}{\sqrt{P_{\text{low}}}} - \displaystyle \frac{\sqrt{P_{\text{low}}}}{\sqrt{P_{\text{high}}}} \right) = \displaystyle \frac{P_{\text{high}} - P_{\text{low}}}{2 \cdot \sqrt{P_{\text{high}}} \cdot \sqrt{P_{\text{low}}}}
  &  
  \label{eq329} 
  &
\end{flalign}

\begin{flalign}
& \text{\renewcommand{\arraystretch}{0.66}
    \begin{tabular}{@{}c@{}}
    \scriptsize from \\
    \scriptsize (\ref{eq324})
  \end{tabular}} 
  & 
  \cosh(\phi) = \displaystyle \frac{e^{\phi} + e^{- \phi}}{2} = \displaystyle \frac{1}{2} \cdot \left( \displaystyle \frac{\sqrt{P_{\text{high}}}}{\sqrt{P_{\text{low}}}} + \displaystyle \frac{\sqrt{P_{\text{low}}}}{\sqrt{P_{\text{high}}}} \right) = \displaystyle \frac{P_{\text{high}} + P_{\text{low}}}{2 \cdot \sqrt{P_{\text{high}}} \cdot \sqrt{P_{\text{low}}}}
  &  
  \label{eq330} 
  &
\end{flalign}

\begin{flalign}
& \text{\renewcommand{\arraystretch}{0.66}
    \begin{tabular}{@{}c@{}}
    \scriptsize from \\
    \scriptsize (\ref{eq324})
  \end{tabular}} 
  & 
  \tanh(\phi) = \displaystyle \frac{\sinh(\phi)}{\cosh(\phi)} = \displaystyle \frac{e^{\phi} - e^{- \phi}}{e^{\phi} + e^{- \phi}} = \displaystyle \frac{P_{\text{high}} - P_{\text{low}}}{P_{\text{high}} + P_{\text{low}}}
  &  
  \label{eq331} 
  &
\end{flalign}

This concludes both our examination of the “natural” concentrated liquidity curve reparameterizations, and our examination of concentrated liquidity in general. 

\section{Conclusion}\label{sec6}

This document provides a detailed elaboration of the three preeminent descriptions of concentrated liquidity from Bancor and Uniswap, as well as three additional descriptions informed by their trigonometric analysis in a transformed and normalized coordinate system. Importantly, the equivalence of all six parameterizations is rigorously demonstrated with a view to establish necessary reference standards, and ameliorate an apparent confusion exhibited by fledgling developers as to how to wield the underlying theory.        

\end{document}